\RequirePackage{lineno}
\documentclass[aps,prd,onecolumn,showpacs,superscriptaddress,groupedaddress, nofootinbib,11pt]{revtex4}  
\usepackage{graphicx}
\usepackage{epstopdf}
\usepackage{amsmath}
\usepackage{amsfonts}
\usepackage{amssymb}
\usepackage{appendix}
\usepackage{mathtools}
\usepackage{comment}
\usepackage{bbold}
\usepackage{color}
\usepackage{slashed}
\usepackage{subfigure}
\usepackage{setspace}
\usepackage{footnote}
\usepackage[T1]{fontenc}
\usepackage{multirow}
\usepackage{makecell}
\usepackage{siunitx}
\usepackage{pbox}
\usepackage{verbatim}

\begin{document}

\singlespacing


\title{Simulation of $e^+e^-\rightarrow\mbox{Hadrons}$ and Comparison to ALEPH Data at Full Detector Simulation with an Emphasis on Four-Jet States}

\author{Jennifer Kile}
\address{Centro de F\'isica Te\'orica de Part\'iculas--CFTP\\Instituto Superior T\'ecnico--IST, Universidade de Lisboa, Av. Rovisco Pais,\\P-1049-001 Lisboa, Portugal\\{\normalfont email:}  jennifer.kile@cftp.tecnico.ulisboa.pt}


\author{Julian von Wimmersperg-Toeller}
\address{{\normalfont email:}  jvonwimm@gmail.com}

\date{\today}

\begin{abstract}
We use the SHERPA Monte Carlo generator to simulate the process $e^+e^-\rightarrow\mbox{hadrons}$ using matrix elements with up to six partons in the final state.  Two samples of SHERPA events are generated.  In the ``LO'' sample, all final states are generated with leading order matrix elements; in the ``NLO'' sample, matrix elements for final states with up to four partons are generated at next-to-leading order, while matrix elements for final states with five or six partons are generated at leading order.  The resulting samples are then passed through the ALEPH detector simulation.  We compare the Monte Carlo samples to each other, to samples generated using the KK2f generator interfaced with PYTHIA, and to the archived ALEPH data at both LEP1 and LEP2 energies.  We focus on four-jet observables with particular attention given to dijet masses.  The LO and NLO SHERPA samples show significant improvement over the KK2f generation for observables directly related to clustering events into four jets, while maintaining similar performance to KK2f for event-shape variables.  We additionally reweight the dijet masses using LEP1 data and find that this greatly improves the agreement between the three Monte Carlo samples at LEP2 energies for these observables.  
\end{abstract}

\pacs{12.38.-t, 12.38.Qk, 13.66.Bc, 13.66.Hk}
\maketitle


\section{Introduction}
\label{intro}

The LEP experiments were immensely important in empirically establishing the properties of quantum chromodynamics (QCD) \cite{Heister:2003aj,Abreu:1996na,Achard:2004sv,Pfeifenschneider:1999rz}.  During the LEP era, the generation of QCD Monte Carlo (MC) was typically accomplished using a parton shower generated with PYTHIA \cite{Sjostrand:2000wi}, HERWIG \cite{Corcella:2000bw,Corcella:2002jc}, or ARIADNE \cite{Lonnblad:1992tz} interfaced to the $q\bar{q}$ matrix element (ME); matching to the $q\bar{q}g$ ME improved the description of three-jet states.  PYTHIA and HERWIG were also used to produce samples using the four-parton ME, but without accompanying contributions from two- and three-parton states needed for a general-purpose MC sample.

In recent years, however, new techniques have been developed which match the multi-parton ME to the parton shower.  Among these is the SHERPA \cite{Hoeche:2011fd,Hoeche:2009rj,Gleisberg:2008ta,Schonherr:2008av,Gleisberg:2008fv,Berger:2008sj,Gleisberg:2007md,Schumann:2007mg,Krauss:2001iv} package, which can generate $e^+e^-\rightarrow\mbox{hadrons}$ with full ME merging and matching to a parton shower.  Additionally, the option exists to produce some of these final states using next-to-leading order (NLO) MEs.  This opens up the possibility to compare the LEP dataset to QCD MC with a description of multi-parton final states that is much improved compared to those available when the LEP experiments were in operation.  Of particular importance are four-jet states, as they constituted an important background to analyses such as Higgs searches and $W^+W^-$ production; such final states would also be of interest at future lepton colliders.

In an accompanying paper \cite{paper3}, we describe an excess observed in hadronic events in the archived LEP2 data from ALEPH\footnote{We refer the reader to the ALEPH Archived Data Statement \cite{alephdata}.}.  In that paper, hadronic events are forced into four jets and then the jets are paired to minimize the difference in the two dijet masses.  An excess of events is seen near $M_1+M_2\sim 110\mbox{ GeV}$, where $M_1$ is the mass of the dijet containing the most energetic jet, and $M_2$ is the mass of the other dijet in the event.  About half of the events are clustered around $M_1\sim 80\mbox{ GeV}$, $M_2\sim 25\mbox{ GeV}$; the excess in this region has a local significance between $4.7\sigma$ and $5.5\sigma$, depending on hadronization uncertainty assumptions.  The rest of the events are located in a broad excess around $M_1\sim M_2\sim 55\mbox{ GeV}$, with a local significance between $4.1\sigma$ and $4.5\sigma$.  For details, we direct readers to Ref. \cite{paper3}.  Some mass distributions can also be seen here when we discuss four-jet observables and the effects of different jet-clustering algorithms in Sec. \ref{jetclus}.

The Standard Model (SM) background in the region of the excess is dominated by QCD.  It is thus important that the MC gives a good description of the QCD background; in particular, it is especially desirable that the MC includes the four-parton ME.  In Ref. \cite{paper1}, we describe  the process by which we generated SM QCD Monte Carlo using the SHERPA generator.  In one simulation, we generated events using the LO MEs for final states with up to six partons.  In another, we generated events with MEs for up to six final-state partons, but MEs for states with up to four partons were generated at NLO.  For each of these MC generations, we tuned the generator parameters to LEP1 data unfolded from detector effects, selection cuts, and ISR using the publicly-available Rivet v. 2.0.0 \cite{Buckley:2010ar,Cacciari:2011ma} and Professor v. 1.3.3 \cite{Buckley:2009bj} packages.  Using the resulting tune parameters, we generated QCD MC samples at both LEP1 and LEP2 energies and compared these samples to unfolded data using Rivet.  Those readers requiring details of our tune and MC generation are encouraged to see Ref. \cite{paper1}.

In this paper, we apply this MC generation to an experimental context.  We pass the aforementioned MC samples through the full ALEPH detector simulation and compare them to the archived ALEPH LEP1 and LEP2 datasets.  In addition to being necessary for physics analyses, running MC through full simulation also allows us to look at arbitrary observables, including distributions of variables particularly relevant for studying the excess in Ref. \cite{paper3}.  We compare various distributions from these LO and NLO MC samples to data, to each other, and to those obtained from KK2f v. 4.19 \cite{Jadach:2000ir,Jadach:1999vf,Jadach:1998jb} interfaced with PYTHIA 6.156\footnote{For simplicity, we will just refer to this generation as being done with KK2f below.} \cite{Sjostrand:2000wi}.  We place an emphasis on variables directly relevant to four-jet states and dijet masses where we expect to see a large improvement in the SHERPA performance relative to that of KK2f, as the latter does not utilize the four-parton ME.   We also examine the effects on all three MCs at LEP2 obtained by reweighting the MC such that agreement between data and simulation is achieved at LEP1.  Given our interest in four-jet states, we also compare the behavior of the samples using the DURHAM \cite{Stirling:1991ds} and LUCLUS \cite{Sjostrand:1986hx} jet-clustering algorithms.

Our results here reflect those seen in Ref. \cite{paper1}.  We find significant improvement in the LO SHERPA samples with respect to the KK2f MC in distributions directly related to four-jet final states.  Improvement is less consistent in other distributions, and in some cases KK2f more closely matches the data.  Overall, the SHERPA LO MC more accurately describes the data at both LEP1 and LEP2 energies.  The NLO SHERPA MC, on the other hand, does not reproduce the success of the LO generation, but is still superior to KK2f for four-jet observables and comparable to KK2f for other variables.  Differences between SHERPA v. 2.2.0 and SHERPA v.2.0.beta, the inclusion of the $b$ quark mass in the ME and PS for the LO tune but not for the NLO tune, and differences in the choice of merging scale may complicate comparisons between our LO and NLO tunes, however.  Additionally, we find that by reweighting the three MC samples to agree with data at LEP1, the dijet mass distributions from the three samples come into better agreement with each other at LEP2.  We provide plots of many such distributions in the text and in the Appendices.

Our main purpose in this work is to establish improved QCD background MC samples for use in Ref. \cite{paper3}, but the utility of our results is not limited to this analysis.  It also demonstrates the potential usefulness of using SHERPA for other analyses at ALEPH and the other LEP experiments, particularly for those analyses where four-jet events are important.  Additionally, the improved MC may be useful for future LEP QCD studies\footnote{While we do provide some data-MC comparisons for standard LEP QCD variables, we note that we are not attempting to reproduce or update results of the ALEPH QCD group \cite{Heister:2003aj,Barate:1996fi}.  In particular, we do not perform an unfolding of the data.}, and its utility would additionally extend to future linear collider four-jet and QCD studies.  Lastly, we also hope that the numerous distributions we give here will provide useful feedback for the authors of QCD MC generators.  

The remaining sections of this paper are organized as follows.  In Section \ref{samples}, we describe our Monte Carlo and data samples and preselection.  We study some event-shape observables in Section \ref{evtshp} and concentrate on variables related to clustering to four jets in Section \ref{jetclus}.  The reweighting procedure and its effects on some distributions are described in Section \ref{reweight}.    We discuss our MC generation choices relevant for Ref. \cite{paper3} in Section \ref{disc} and consider hadronization uncertainties in Section \ref{hadunc}.  We conclude in Section \ref{conc}.  A few technical details and distributions of additional variables can be found in the Appendices.

\section{Monte Carlo and Data Samples}
\label{samples}

\subsection{The ALEPH Detector}

A detailed description of the detector and its performance can be found elsewhere \cite{Decamp:1990jra,Buskulic:1994wz}.  Charged particles are tracked through a vertex detector, then a cylindrical drift chamber, and lastly a time projection chamber, all in the presence of a $1.5$T axial magnetic field provided by a superconducting solenoidal coil. A charged track transverse momentum resolution of $\sigma(p_t)/p_t = 0.0006 p_t/(\mbox{GeV}/c) \oplus 0.005$ is achieved.  Of importance for our preselection are ``good tracks''; these are tracks from charged particles which emanate from a cylindrical volume of radius $2$ cm and length $20$ cm centered on the nominal collision point and which have at least four associated hits in the time projection chamber.

Electrons and photons are identified by their characteristic shower development in a 45-layer lead/wire-chamber sampling electromagnetic calorimeter situated between the time projection chamber and the coil. For isolated photons, a relative energy resolution of $\sigma(E)/E=0.18/\sqrt{E/\mbox{GeV}}$ + 0.009 is achieved. The hadron calorimeter is formed by the iron return yoke, instrumented with 23 layers of streamer tubes; a relative energy resolution of $\sigma(E)/E=0.85/\sqrt{E/\mbox{GeV}}$ for charged and neutral hadrons is attained. Muons are identified by their penetration pattern in the hadron calorimeter together with two surrounding muon chambers, each composed of a double-layer of streamer tubes.

Information from the tracking detectors and calorimeters is used by an energy-flow \cite{Buskulic:1994wz} algorithm to construct a list of charged and neutral objects, called energy-flow objects, which go into our jet reconstruction.  Angular resolutions of jets are typically ${\mathcal O}(20\mbox{ mrad})$ in $\theta$ and $\phi$; the inter-jet angles are important for jet rescaling, discussed as part of our preselection below.

\subsection{Data Samples}

The data described here are the archived LEP1 and LEP2 data from the ALEPH detector.  For convenience, we do not use the entire LEP1 data set, but only that from 1994, which amounts to approximately $58 \mbox{ pb}^{-1}$ at $\sqrt{s}=91.2$ GeV.  We use the entire LEP2 data set, $\sim 735$ $\mbox{pb}^{-1}$ collected at center-of-mass energies in the range $130-208$ GeV, approximated here as $18$ discrete values.  The luminosities of the data for each value of $\sqrt{s}$, along with those of our MC generation, are shown in Table \ref{tab:luminosities}.

\begin{table}
\begin{tabular}{| c| r| c| c| c| c|}
\hline
$\sqrt{s}$ & \makecell{ALEPH Archived\\Data$/\mbox{pb}^{-1}$}& \makecell{LO\\SHERPA/data} & \makecell{NLO\\SHERPA/data} & KK2f/data & KORALW/data \\
\hline
$130.0$ GeV & 3.30\hphantom{111}& 92& & 92& 382\\
\cline{1-3}\cline{5-6}
$130.3$ GeV & 2.88\hphantom{111} & 107 & & 107 & 439\\
\cline{1-3}\cline{5-6}
$136.0$ GeV & 3.50\hphantom{111} & 104 & & 104 & 351\\
\cline{1-3}\cline{5-6}
$136.3$ GeV & 2.86\hphantom{111} & 129 & &129 & 429 \\
\cline{1-3}\cline{5-6}
$140.0$ GeV & 0.05\hphantom{111} & 0 & & 3930  &5860\\
\cline{1-3}\cline{5-6}
$161.3$ GeV & 11.08\hphantom{111} & 60 & & 60& 226\\
\cline{1-3}\cline{5-6}
$164.5$ GeV & 0.04\hphantom{111} & 0 & & 8580& 5610\\
\cline{1-3}\cline{5-6}
$170.3$ GeV & 1.11\hphantom{111} & 0 & & 346& 367\\
\cline{1-3}\cline{5-6}
$172.3$ GeV & 9.54\hphantom{111} & 84 & $2\times$ LO&84 & 208\\
\cline{1-3}\cline{5-6}
$182.6$ GeV & 59.37\hphantom{111} & 79 & &159 &122\\
\cline{1-3}\cline{5-6}
$188.6$ GeV & 177.08\hphantom{111} & 58 & & 116& 157\\
\cline{1-3}\cline{5-6}
$191.6$ GeV & 29.01\hphantom{111} & 74 & & 147&147\\
\cline{1-3}\cline{5-6}
$195.5$ GeV & 82.62\hphantom{111} & 68 & & 136&162\\
\cline{1-3}\cline{5-6}
$199.5$ GeV & 87.85\hphantom{111} & 67 & & 135& 151\\
\cline{1-3}\cline{5-6}
$201.6$ GeV & 42.14\hphantom{111} & 144 & &202 & 117\\
\cline{1-3}\cline{5-6}
$204.9$ GeV & 84.03\hphantom{111} & 75 & & 151& 116\\
\cline{1-3}\cline{5-6}
$206.5$ GeV & 130.59\hphantom{111} & 99 & & 198&149\\
\cline{1-3}\cline{5-6}
$208.0$ GeV & 7.73\hphantom{111} & 170 & & 679&209\\
\hline
\end{tabular}
\caption{Luminosities of data and MC generated at each LEP2 center-of-mass energy.  See text for details.}
\label{tab:luminosities}
\end{table}

\subsection{SHERPA MC generation}
\label{mcrun}

We generate QCD samples at LEP1 and LEP2 energies using the SHERPA tunes described in Ref. \cite{paper1}.  The first of these, the ``LO'' tune, produces $e^+e^-\rightarrow\mbox{hadrons}$ using the MEs for final states with up to six partons, all at LO.  The second, ``NLO'' tune also generates events using the MEs for final states containing up to six partons, but MEs for final states with up to four partons are generated at NLO with BlackHat v. 0.9.9 \cite{Berger:2008sj}.  In both cases, hadronization is performed with PYTHIA 6.4.18 \cite{Sjostrand:2006za}.  To generate events for full simulation, a few particles must be set stable in the MC generation, as they are decayed by the ALEPH detector simulation.  These are listed in Appendix \ref{sec:rundat}.  ISR is turned off and is discussed below.

At LEP1, we generate $2\times 10^6$ unweighted events using the LO tune, and $3\times 10^6$ partially unweighted events using the NLO tune at $\sqrt{s}=91.25$ GeV.  The effective luminosity of the SHERPA MC generated at each LEP2 center-of-mass energy $\sqrt{s}$ is shown in Table \ref{tab:luminosities}.  There are a few LEP2 center-of-mass energies ($\sqrt{s}=140.0, 164.5, 170.3$ GeV) with very small luminosities whose generation was omitted; we compensate by slightly reweighting MC events at nearby values of $\sqrt{s}$.  In all other cases, the SHERPA LO generation has an effective luminosity at least $50$ times that of the data.  The SHERPA NLO generation was done with an effective luminosity twice that of the LO generation.  As was the case for the LEP1 samples, the LO generation is done with unweighted events, and the NLO events are partially unweighted; a plot of the NLO event weights can be found in Appendix \ref{sec:nloevtwts}.  Thus, the sizes of the statistical samples cannot be directly compared, but should nonetheless be illustrative.

LEP1 events are then passed through the full ALEPH detector simulation; we only use the 1994 geometry.  For LEP2 events, we add in the effects of initial state radiation (ISR) before detector simulation.  For the total hadronic cross-section, we use that obtained from the KK2f generation described below.

\subsubsection{ISR}

For LEP2, initial-state radiation cannot be neglected.  However, SHERPA only generates ISR parallel to the incoming $e^+$ or $e^-$.  In order to have a more accurate description of ISR which includes photons at nonzero angles, we take the ISR spectrum from KK2f for our SHERPA samples.  This proceeds as follows.  For each value of $\sqrt{s}$, we generate QCD events using KK2f with ISR effects included, and with final-state emission switched off.  We then discard the hadronic system in each of these events, retaining the four-momenta of any photons.  These sets of four-momenta define an effective center-of-mass energy, $\sqrt{s'}$, for the discarded hadronic system.  

We then generate SHERPA event samples, with ISR turned off, at a series of center-of-mass energies ranging from $20$ GeV\footnote{For hadronic events where the hadronic system has a $\sqrt{s'}$ less than $20$ GeV, we forego SHERPA generation and simply generate the events with KK2f.} up to $\sqrt{s}$.  For each KK2f event we choose a SHERPA event whose center-of-mass energy is closest to $\sqrt{s'}$ of the discarded KK2f hadronic system.  Then, the SHERPA event and the KK2f ISR photon(s) are merged into a single, new event; the photon four-momenta are slightly rescaled to account for the difference between $\sqrt{s'}$ and the center-of-mass energy of the SHERPA event, and the SHERPA hadronic system is boosted to bring the total energy to $\sqrt{s}$ and the total three-momentum to zero.  This procedure is followed for both the LO and NLO samples.  All events are then passed through the full ALEPH detector simulation.

\subsection{Other MC Samples}

In addition to the SHERPA generation described above, we also produce a few more MC samples.  First, we use KK2f to produce $e^+e^-\rightarrow\mbox{hadrons}$ samples for comparison to SHERPA.  We also obtain the total hadronic cross-sections for each center-of-mass energy and for the above-mentioned ISR production using KK2f, which utilizes the DIZET 6.21 library \cite{Bardin:1999yd} for higher-order corrections.  Additionally, we also need to simulate $e^+e^-\rightarrow\tau^+\tau^-$ as a small number of these events can pass the preselection cuts.  For LEP2 energies, we additionally need MC for the four-fermion and two-photon SM processes.  

At LEP1, we generate $2\times 10^6$ QCD MC events with KK2f using PYTHIA 6.156, using the standard ALEPH tune, including the effects of ISR and final-state radiation; this can be used for comparisons with the SHERPA MC.  The effective luminosity of this generation is comparable to that of the 1994 data set.  At LEP2, we generate KK2f MC at each of the $18$ values of $\sqrt{s}$.  The effective luminosities of these samples are given in Table \ref{tab:luminosities}; most have an effective luminosity at least $100$ times that of the corresponding data sample.

We produce four-fermion MC using KORALW 1.53.3 \cite{Skrzypek:1995wd,Skrzypek:1995ur,Jadach:1998gi,Jadach:2001mp,Jadach:2001uu} with showering and hadronization done by JETSET 7.4 \cite{Sjostrand:1993yb}\footnote{Here, we generate all four-fermion MC using KORALW in one step and with loose generator-level cuts, changing the minimum visible $p_T^2$ from default $600\mbox{ GeV}^2$ to $100\mbox{ GeV}^2$.  At preselection level we see no discernable difference between our generation and the official ALEPH production which generated $W^+W^-$-like and $ZZ$-like states separately.}.  The luminosities of the 4-fermion generation are shown for each value of $\sqrt{s}$ in Table \ref{tab:luminosities}; in all cases, we generate MC with an effective luminosity at least $100$ times that of the data.  Additionally, we use the official ALEPH production of two-photon MC, which we augment by generating additional events using PYTHIA 6.156 in a fashion identical to the official generation. The large 2-photon cross-section somewhat restricts the effective luminosity of this generation; for all LEP2 center-of-mass energies, an effective luminosity at least equal to and, in most cases, $\gtrsim 2\times$ that of the data is obtained.  Lastly, we also use KK2f to generate $e^+e^-\rightarrow\tau^+\tau^-$.  At LEP1, we generate $e^+e^-\rightarrow\tau^+\tau^-$ events with an effective luminosity $\sim 5\times$ that of the data; at LEP2 energies, the effective luminosity is $\geq 100\times$ that of the data.

We then pass all MC events through the full ALEPH detector simulation.  MC distributions in all plots in this paper are at the level of full detector simulation.

\subsection{Preselection}

The purpose of our preselection is to select hadronic events while cutting away the two-photon background and events with hard ISR; it is very similar to that used for other ALEPH four-jet analyses\footnote{Many analyses apply additional cuts to suppress SM hadronic events, unlike what is done here.}.  We begin by requiring at least $7$ good charged tracks in the event; this reduces non-hadronic backgrounds.  Next, we force the event into four jets and require that each jet contains at least one good charged track; this reduces events with ISR photons in the detector and events with large energy deposits at low angle where tracking is less effective.  We then require that the sum of the jet transverse momenta $p_{tsum}$ be greater than $25\%\sqrt{s}$; this reduces the two-photon background.  We then rescale the energy and momenta of the four jets, keeping their directions and masses fixed, so that the sum of their four-momenta is $(\sqrt{s},0,0,0)$; we select only events where all of the jet rescaling factors are positive.  

At LEP2 energies, there are two additional cuts to reduce ISR returns to the $Z$.  First, the electromagnetic energy for each jet is calculated using energy flow objects corresponding to identified photons, including photon conversions, neutral particles which traverse an electromagnetic calorimeter crack and are detected in the hadron calorimeter, and particles detected by the luminosity calorimeters.  We require that no jet have more than $80\%$ of its electromagnetic energy contained in a one-degree cone around any electromagnetic energy-flow object.  Secondly, we require $|p_{zmis}|<1.5(m_{vis}-90)$, where $p_{zmis}$ is the missing momentum in the direction of the beampipe, and $m_{vis}$ is the visible mass in the event.  These cuts have also been used for other ALEPH four-jet analyses such as the SM Higgs search \cite{Barate:1997mb,Barate:2000na}.

All distributions shown in this work are with the above preselection cuts applied.  Unless otherwise noted, events are clustered into four jets using the LUCLUS algorithm.  The number of data events and the SM expectation obtained from MC at preselection are shown in Table \ref{tab:preselbkg} for both LEP1 and LEP2.  The total in the last column is that obtained if the SHERPA LO sample is used.  The column marked ``Other'' includes the two-photon (at LEP2) and $e^+e^-\rightarrow\tau^+\tau^-$ (at both LEP1 and LEP2) expectation.  

\begin{table}
\begin{tabular}{| c| c| c| c| c| c| c| c|}
\hline
Dataset & \makecell{ALEPH\\Archived Data} & \makecell{LO\\SHERPA} & \makecell{NLO\\SHERPA} & KK2f & KRLW03-4F & Other & Total MC (LO)\\
\hline
LEP1 &1307068 &1312789 &1309110 &1297974 &N/A &265 & 1313054\\
\hline
LEP2 &17602  &11226 & 11090& 10974&6564 &17 & 17807\\
\hline
\end{tabular}
\caption{Number of data events and expected backgrounds after preselection for both LEP1 and LEP2 energies.  In the case of LEP2, all center-of-mass energies have been added together.  In the case of LEP1, the number in the column marked ``Other'' consists of $e^+e^-\rightarrow\tau^+\tau^-$ events; for LEP2, it includes $e^+e^-\rightarrow\tau^+\tau^-$ and two-photon events.  The total for the SM prediction in the last column assumes the LO SHERPA MC has been used for the QCD prediction.}
\label{tab:preselbkg}
\end{table}

\section{Event-shape Observables}
\label{evtshp}

\begin{figure}[h]
\begin{center}
\subfigure[]{\includegraphics[width=2.5in,bb=80 150 520 720]{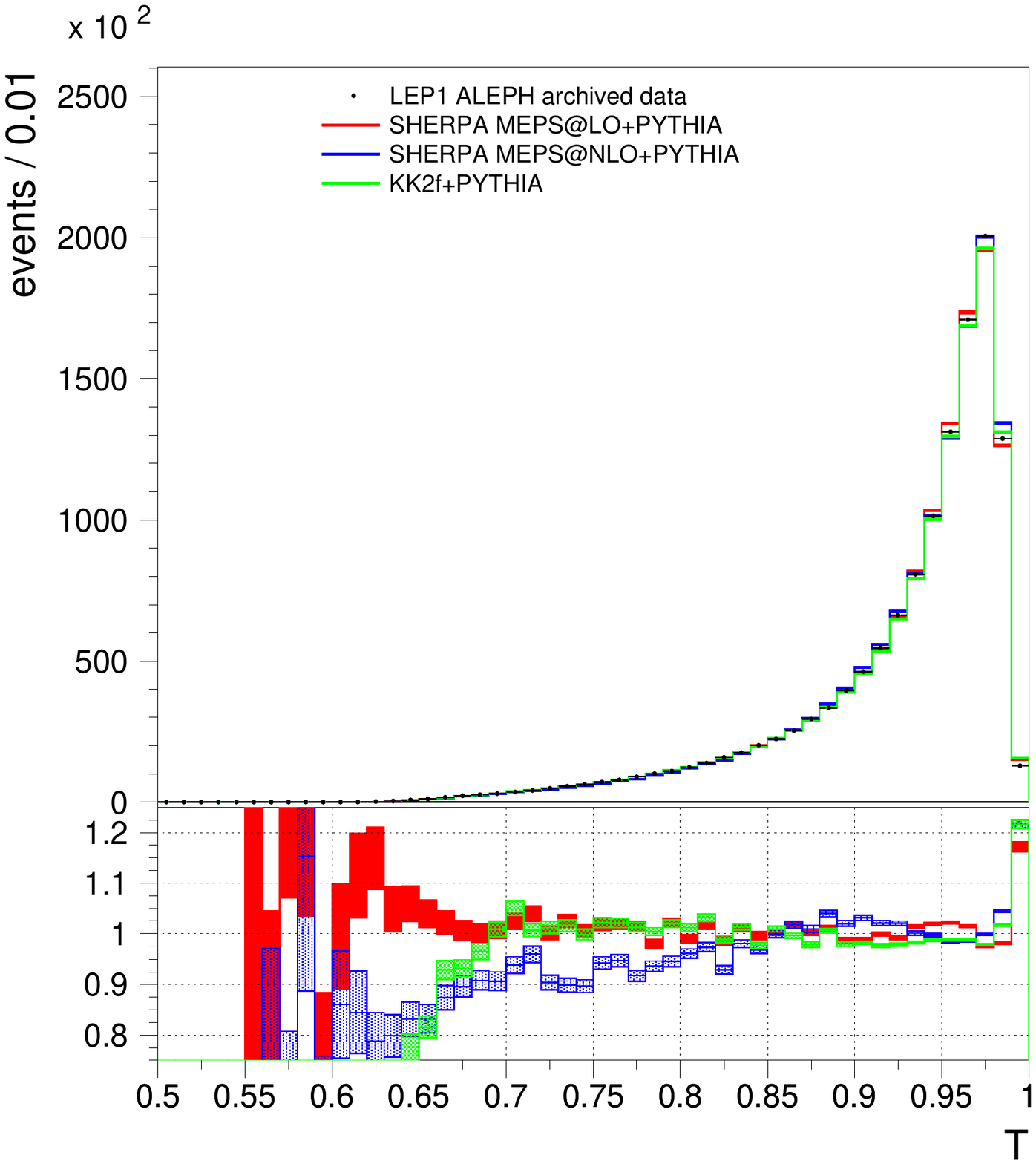}}\hspace{.4in}
\subfigure[]{\includegraphics[width=2.5in,bb=80 150 520 720]{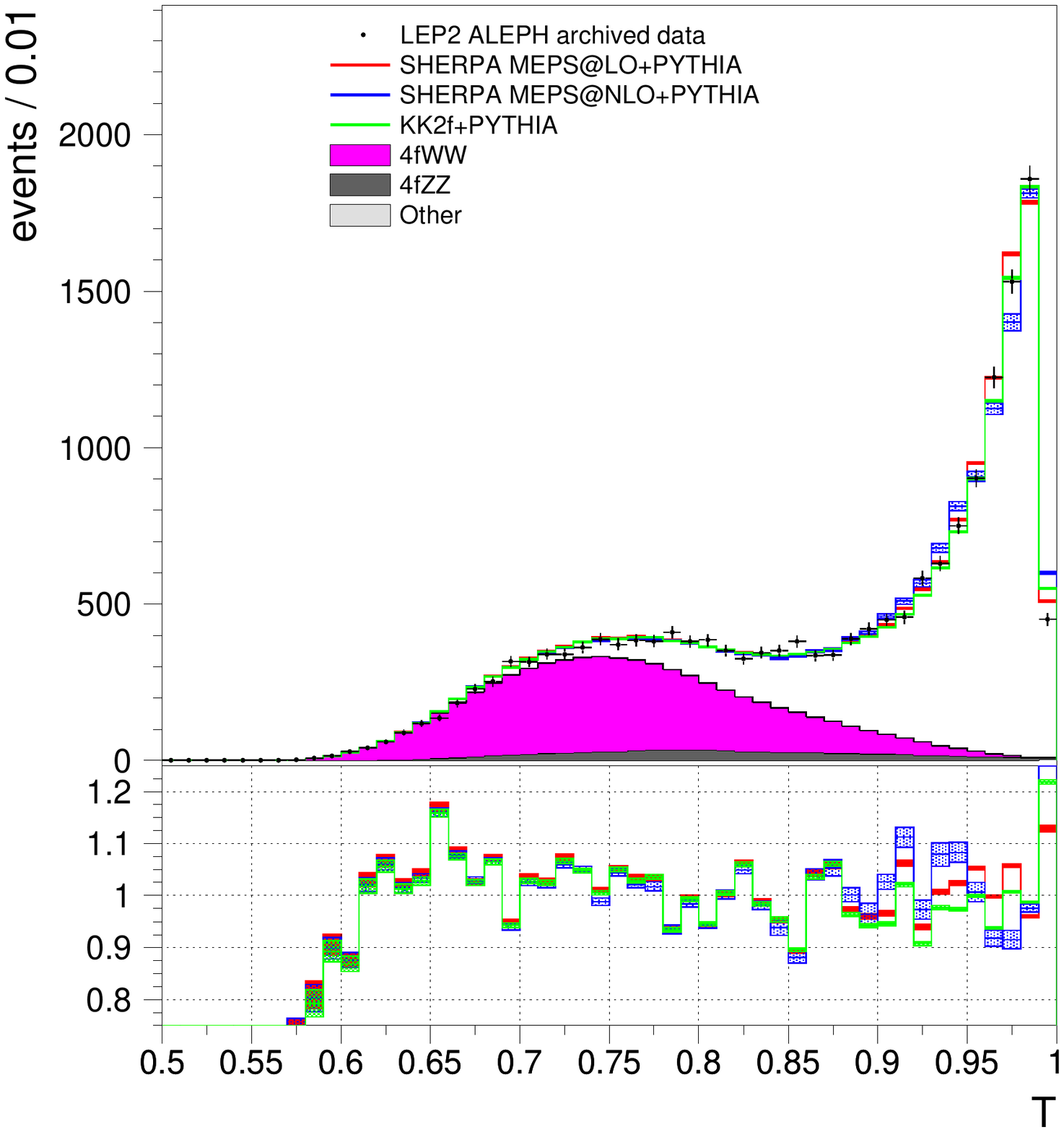}}
\end{center}
\caption{The thrust $T$ plotted (a) at LEP1 and (b) at LEP2.  Data from ALEPH is compared to SHERPA LO, SHERPA NLO, and KK2f MC.  In (b), all LEP2 center-of-mass energies are shown together, and four-fermion and two-photon MC samples are included.  The bottom of each panel shows the ratio of MC to data for each sample.   The widths of the colored bands reflect the uncertainty on the simulation from MC statistics only.}
\label{fig:thr}
\end{figure}

\begin{figure}[h]
\begin{center}
\subfigure[]{\includegraphics[width=2.5in,bb=80 150 520 720]{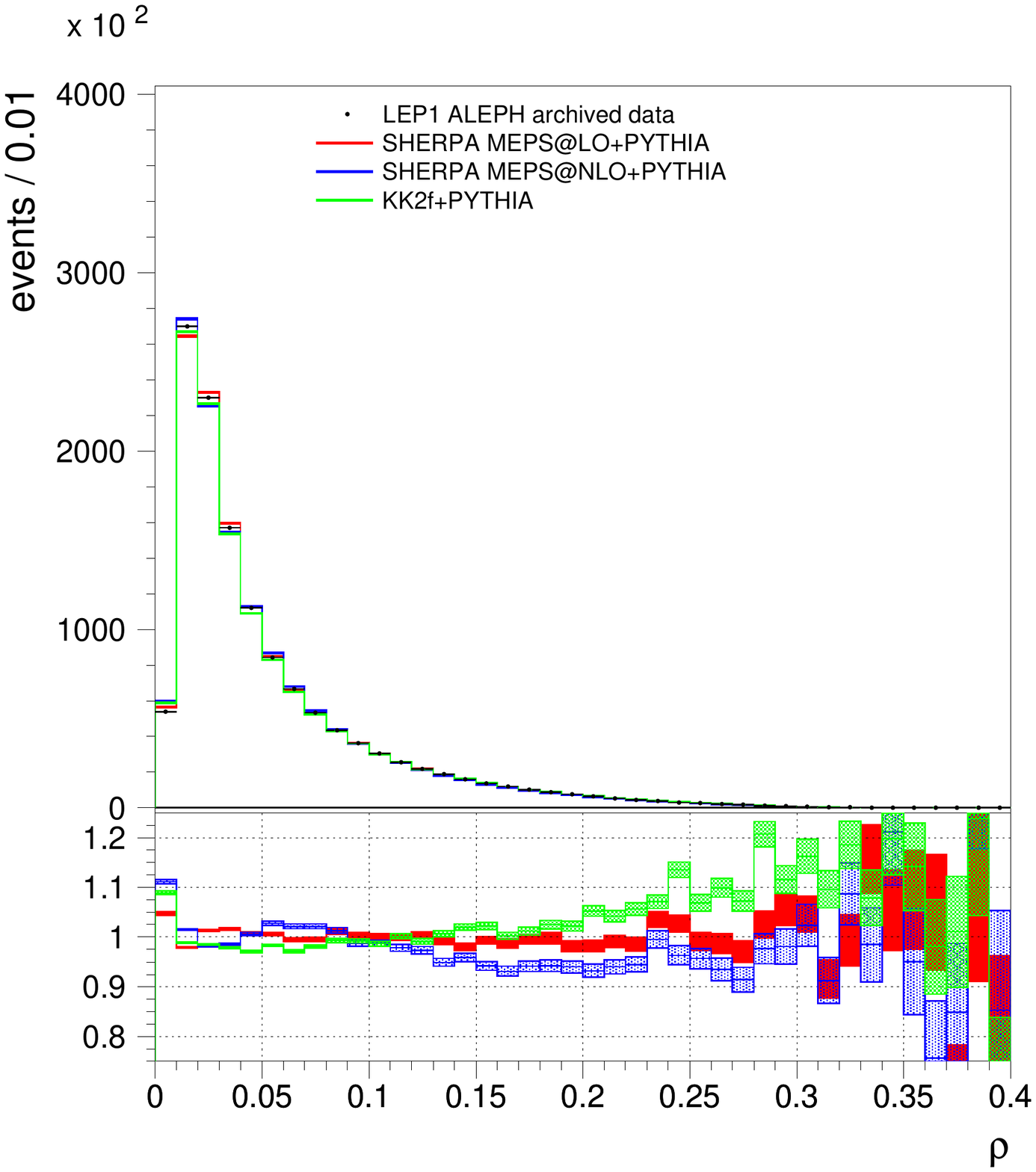}}\hspace{.4in}
\subfigure[]{\includegraphics[width=2.5in,bb=80 150 520 720]{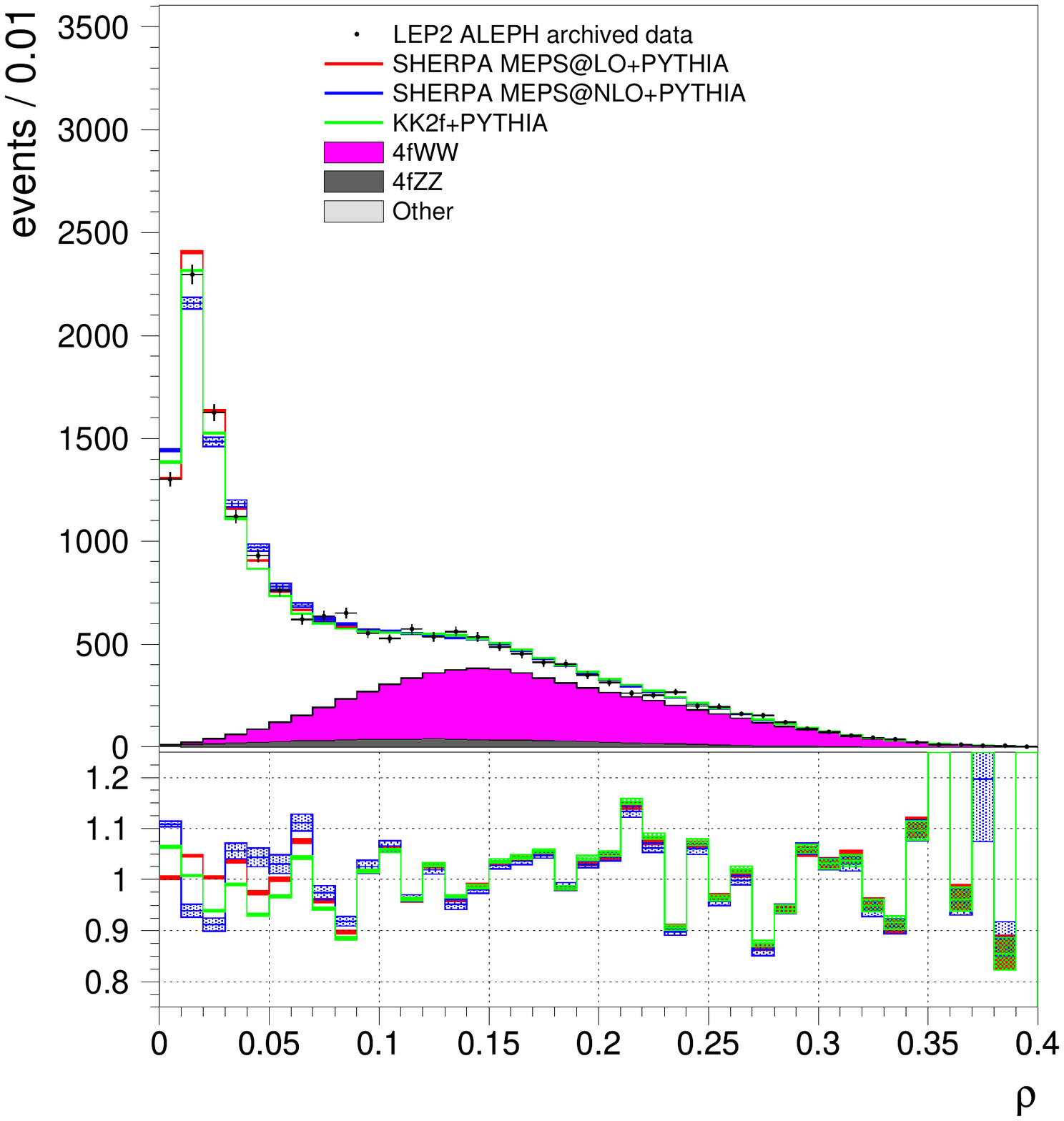}}\\
\subfigure[]{\includegraphics[width=2.5in,bb=80 150 520 720]{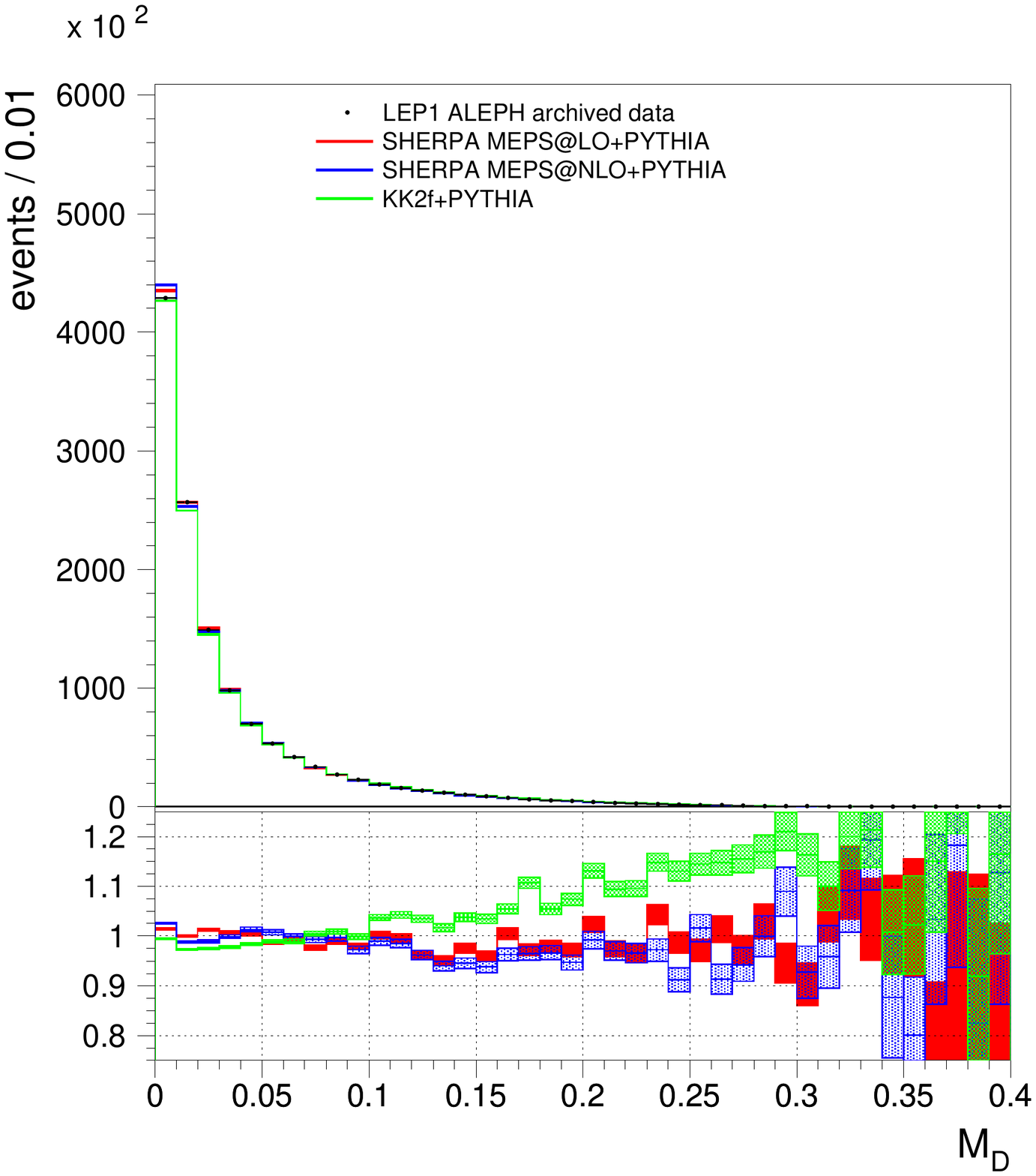}}\hspace{.4in}
\subfigure[]{\includegraphics[width=2.5in,bb=80 150 520 720]{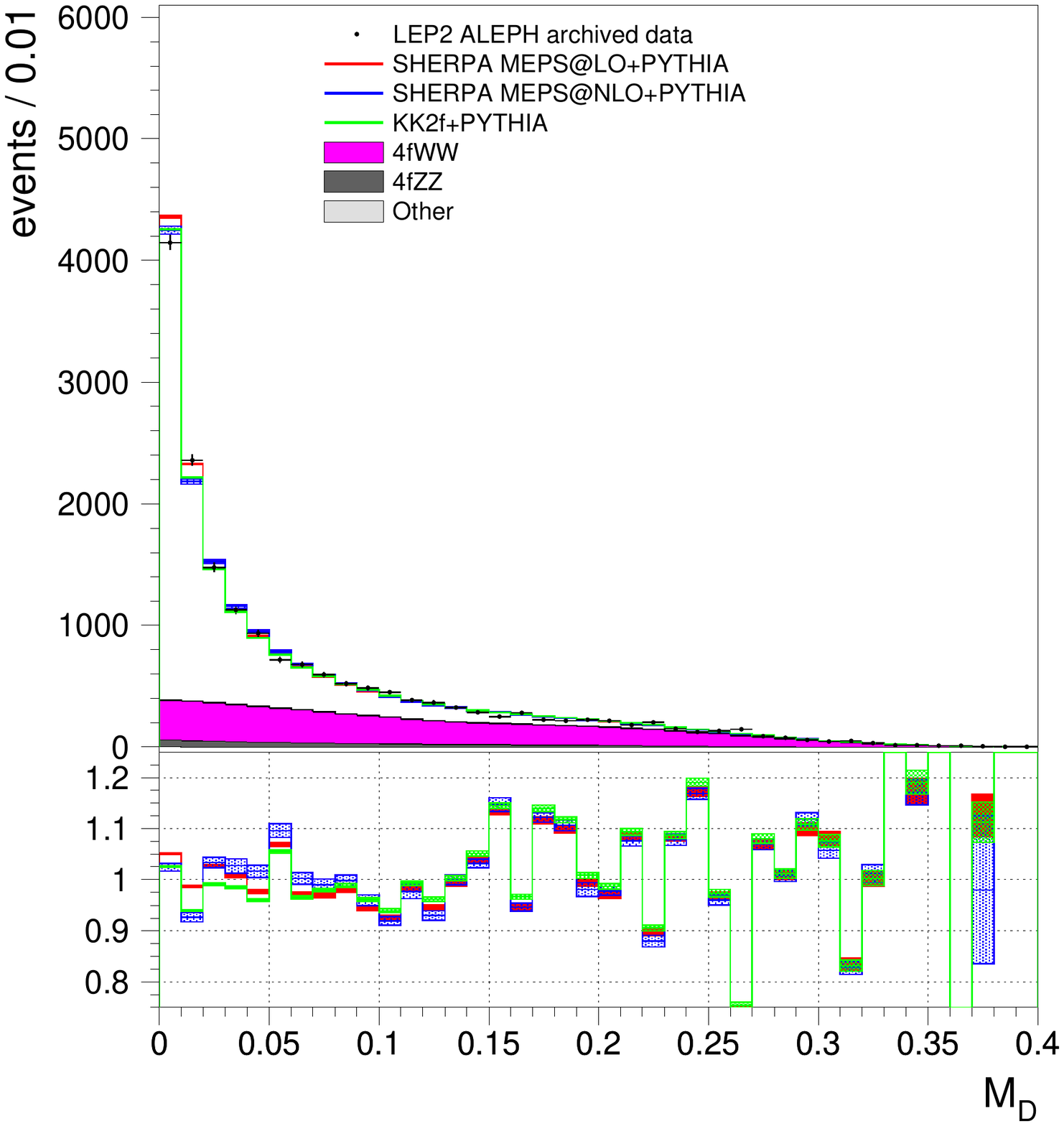}}
\end{center}
\caption{Heavy jet mass $\rho$ and jet mass difference $M_D$.  Conventions are the same as those in Fig. \ref{fig:thr}. }
\label{fig:jetmass}
\end{figure}

\begin{figure}[h]
\begin{center}
\subfigure[]{\includegraphics[width=2.5in,bb=80 150 520 720]{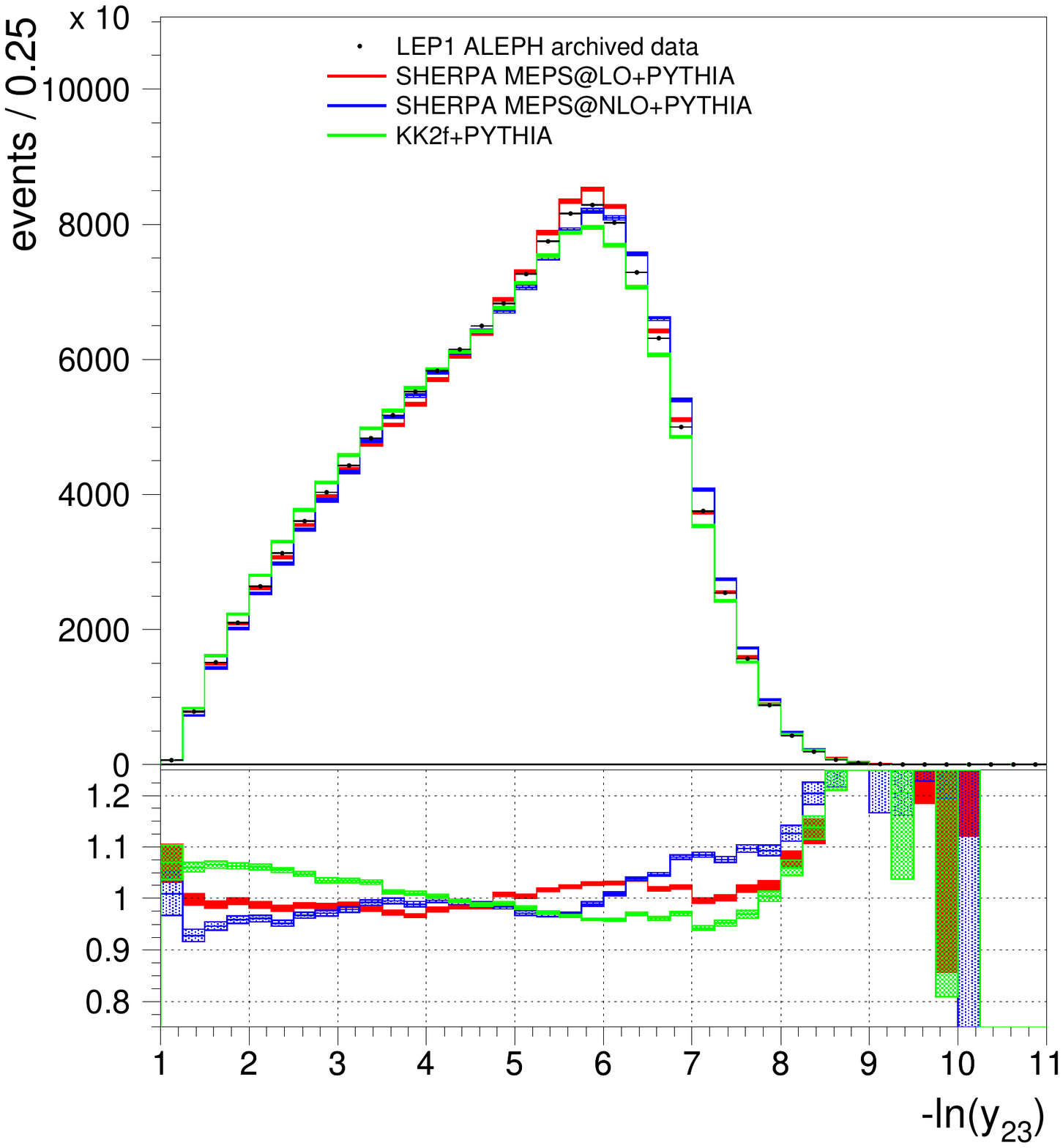}}\hspace{.4in}
\subfigure[]{\includegraphics[width=2.5in,bb=80 150 520 720]{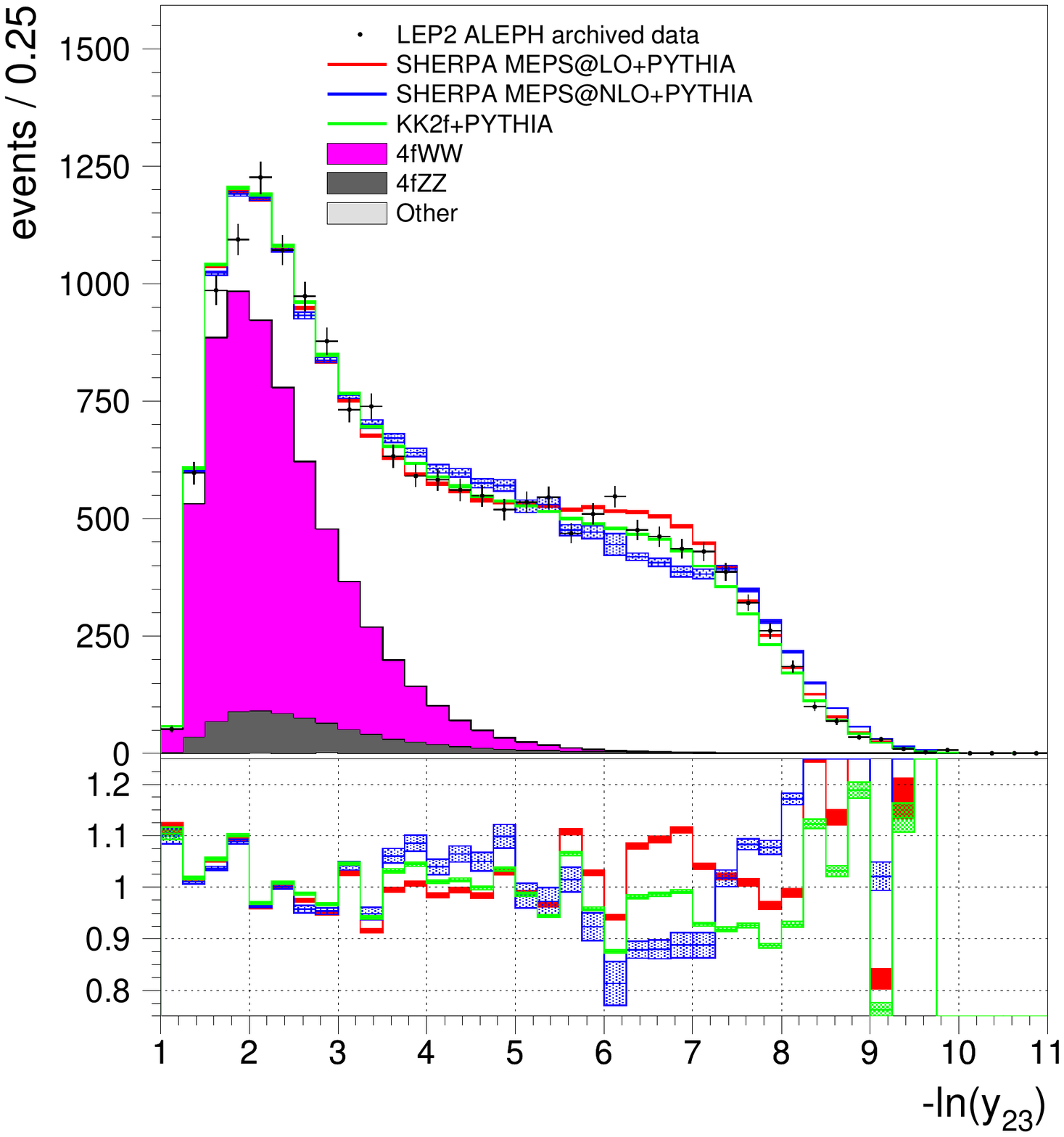}}\\
\subfigure[]{\includegraphics[width=2.5in,bb=80 150 520 720]{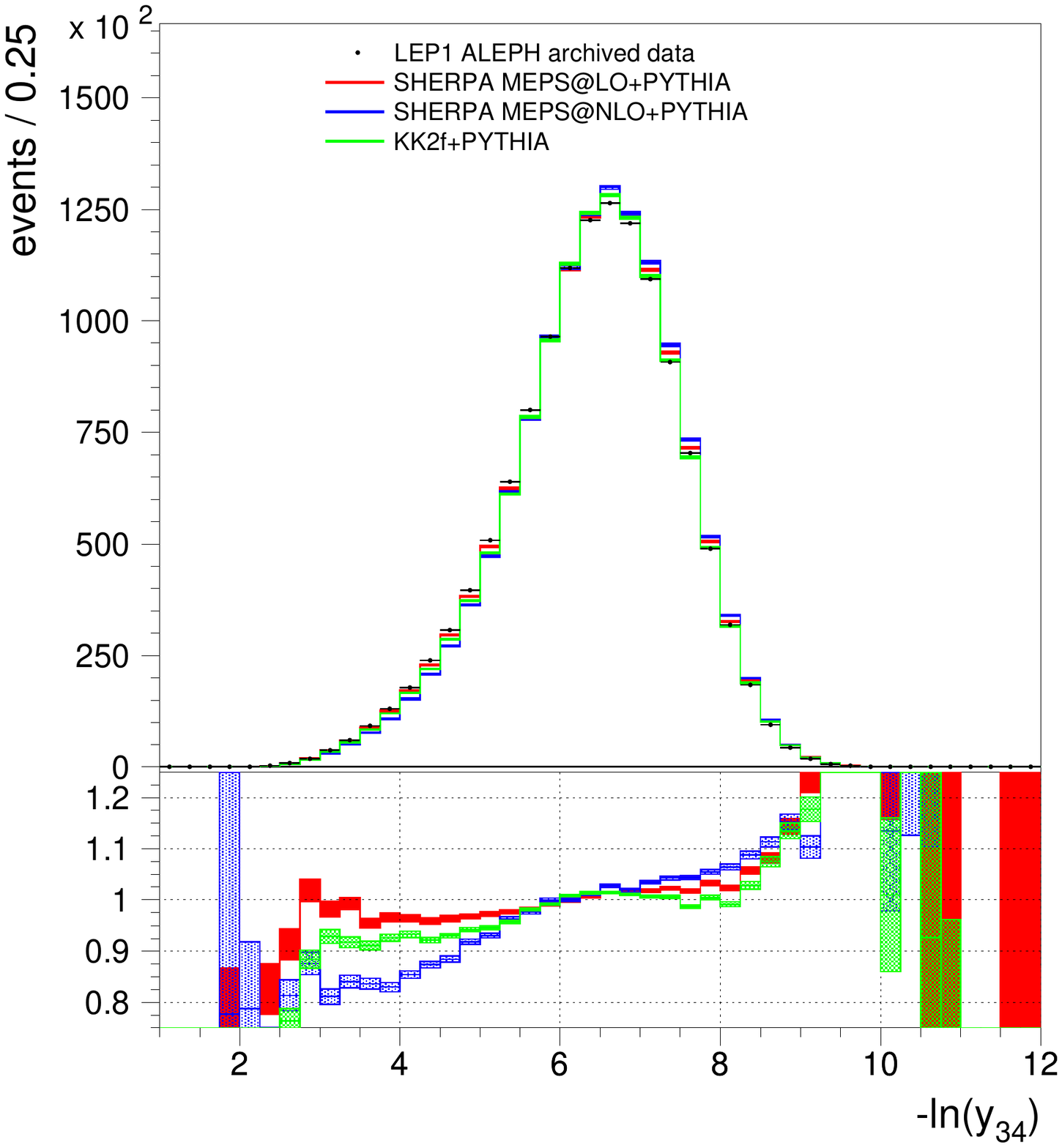}}\hspace{.4in}
\subfigure[]{\includegraphics[width=2.5in,bb=80 150 520 720]{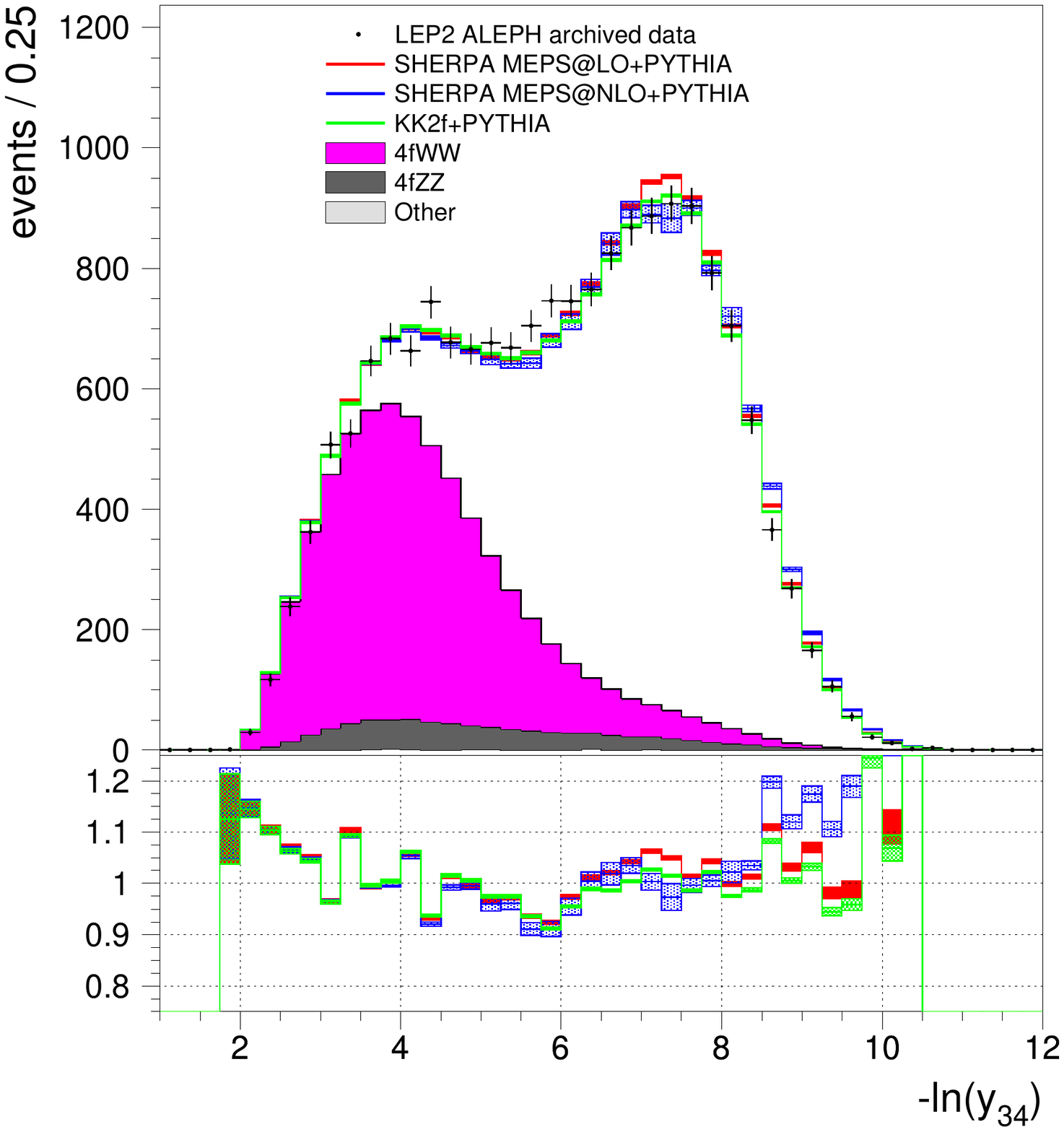}}
\end{center}
\caption{Plots of $-\ln{y_{23}}$ and $-\ln{y_{34}}$ at LEP1 and LEP2.  Conventions are the same as those in Fig. \ref{fig:thr}.}
\label{fig:y23y34}
\end{figure}

\begin{figure}[h]
\begin{center}
\subfigure[]{\includegraphics[width=2.5in,bb=80 150 520 720]{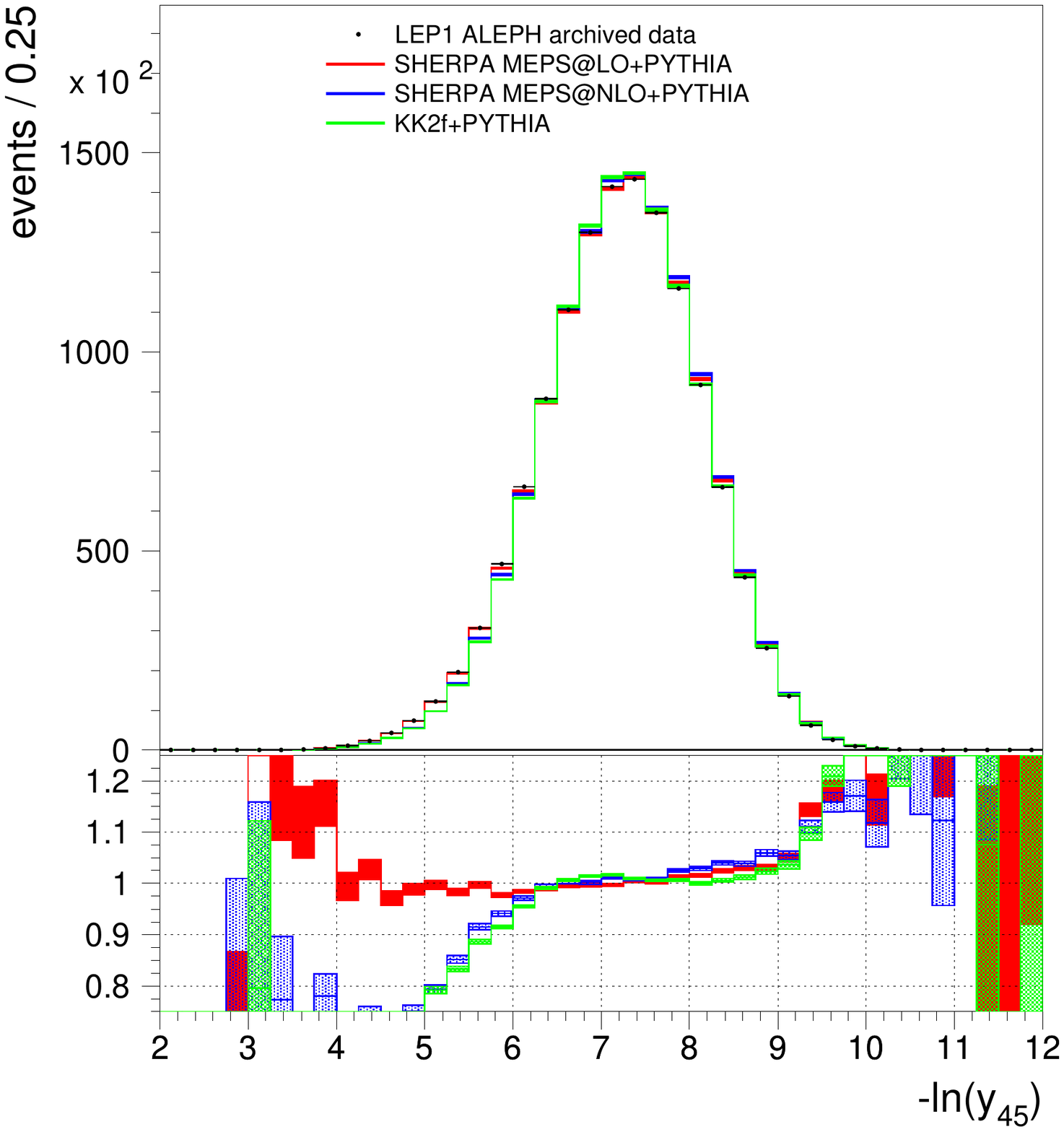}}\hspace{.4in}
\subfigure[]{\includegraphics[width=2.5in,bb=80 150 520 720]{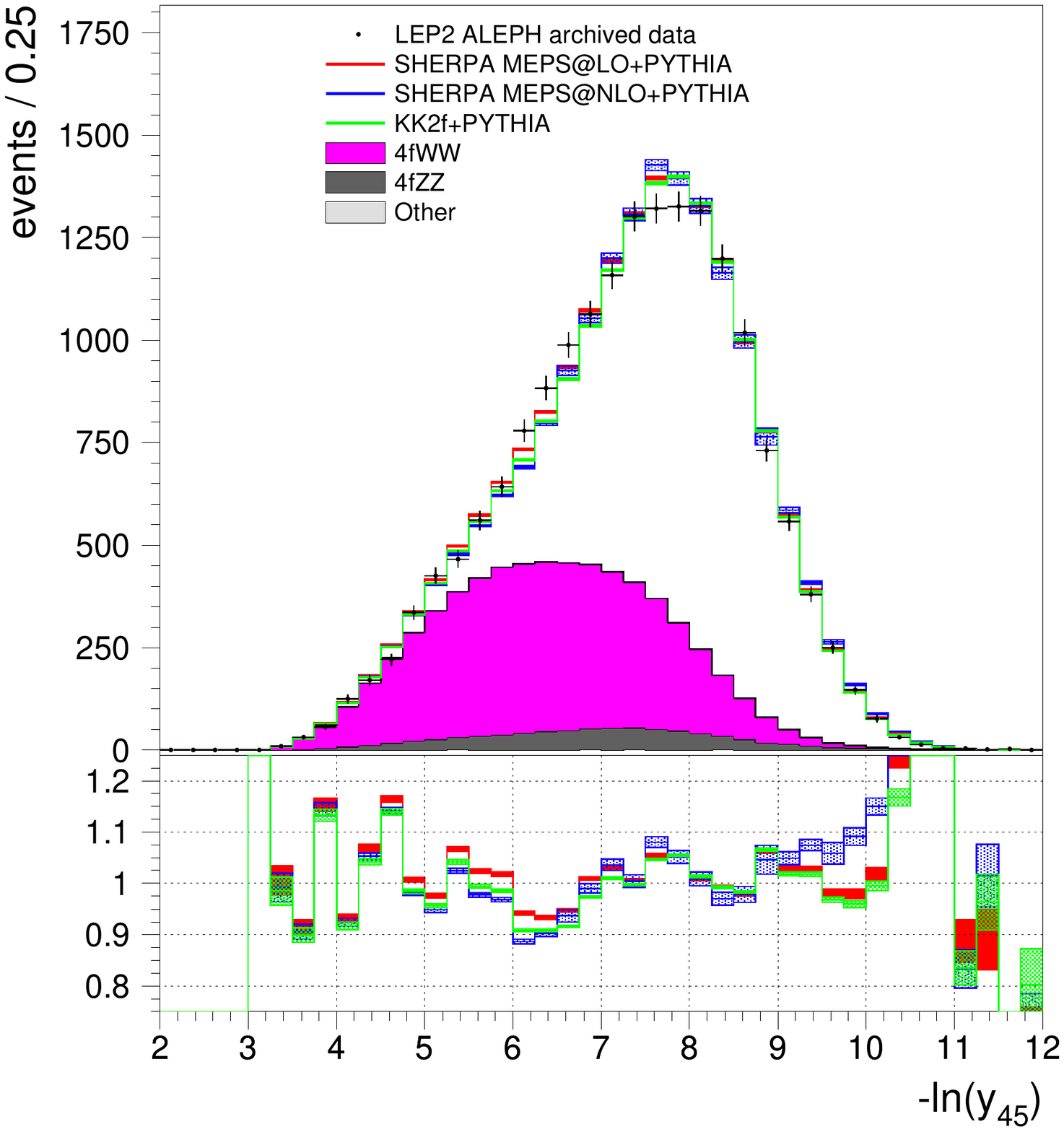}}\\
\subfigure[]{\includegraphics[width=2.5in,bb=80 150 520 720]{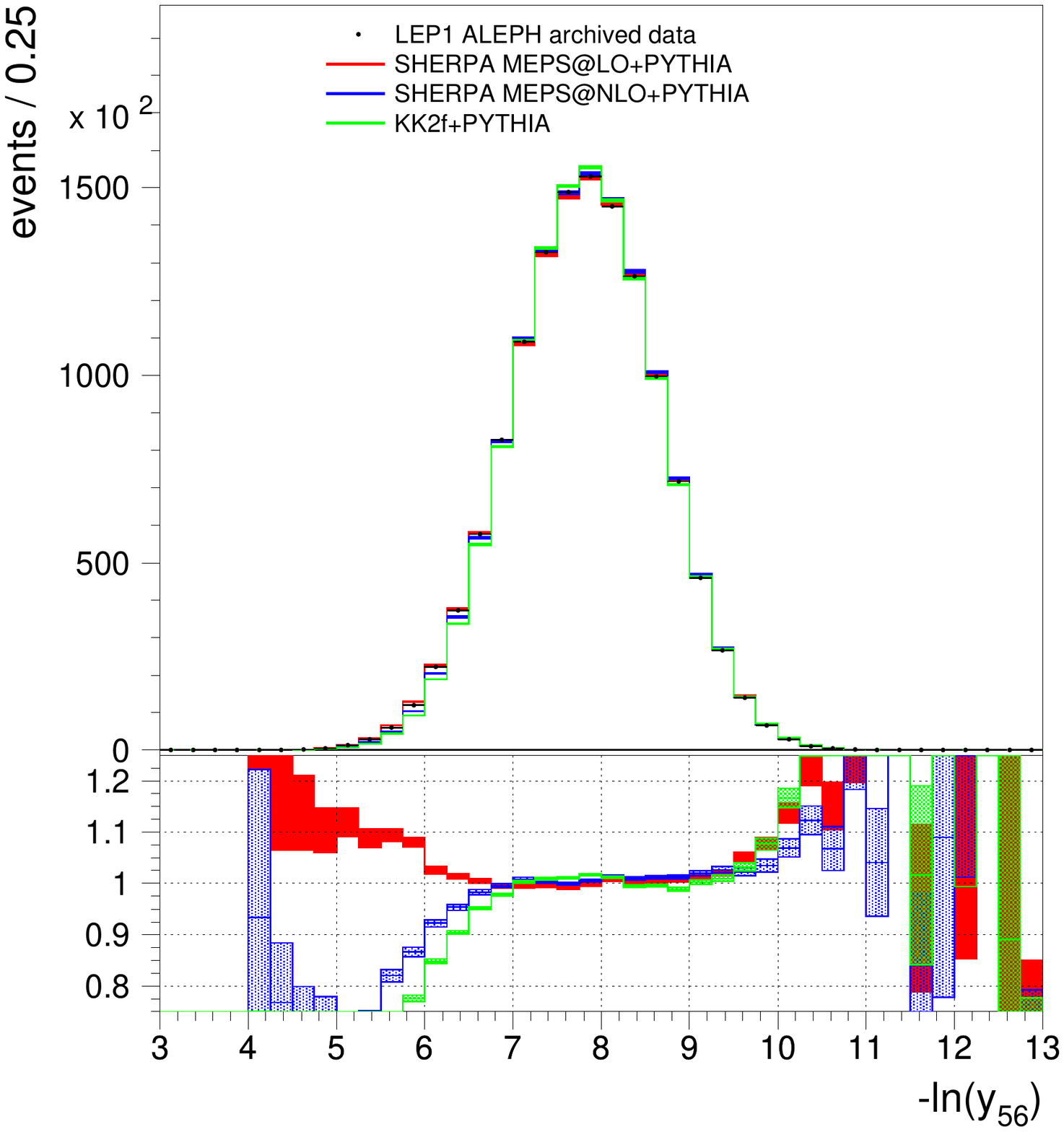}}\hspace{.4in}
\subfigure[]{\includegraphics[width=2.5in,bb=80 150 520 720]{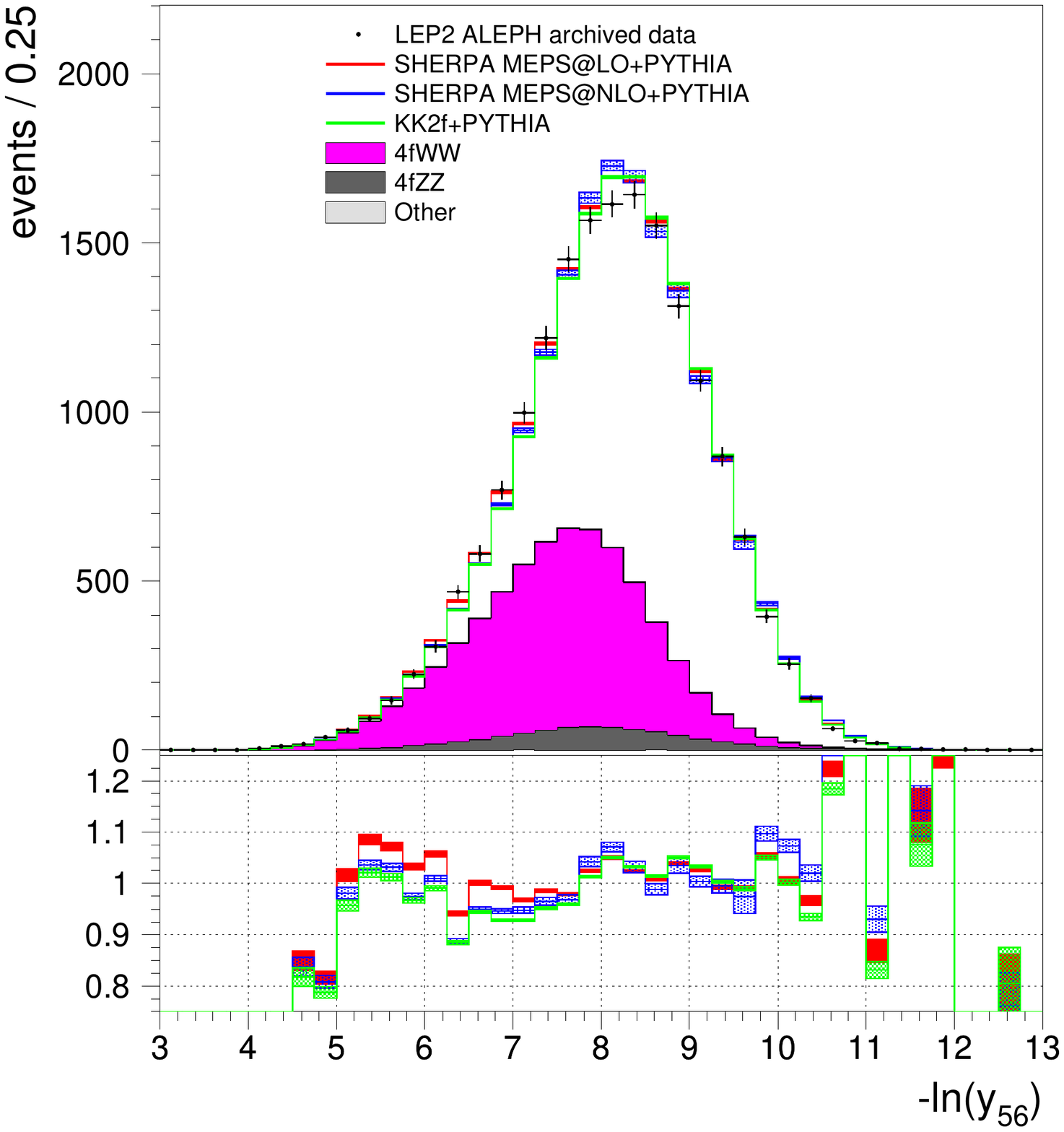}}
\end{center}
\caption{Plots of $-\ln{y_{45}}$ and $-\ln{y_{56}}$ at LEP1 and LEP2.  Conventions are the same as those in Fig. \ref{fig:thr}.}
\label{fig:y45y56}
\end{figure}

Here, we compare data and MC for some event-shape variables commonly used in QCD analyses \cite{Heister:2003aj,Abreu:1996na,Achard:2004sv,Pfeifenschneider:1999rz,Barate:1996fi} and also studied at Rivet level in Ref. \cite{paper1}.  We note that these distributions are all plotted at detector level with the above preselection.  These quantities are shown in Figs. \ref{fig:thr}--\ref{fig:y45y56}; definitions can be found in, e.g., \cite{Heister:2003aj}.  We have limited the discussion here to the event-shape variables which are most relevant for the analysis in Ref. \cite{paper3}, but plots of additional quantities can be found in Appendix \ref{moredists}.  We note that all of our data-MC comparisons are done without normalizing the MC curves to data, unlike what was done in our previous Rivet analyses \cite{paper1}.

The thrust $T$ at LEP1 is shown in Fig. \ref{fig:thr} (a).  The analogous plot for the entire LEP2 dataset, $\sqrt{s}=130-208$ GeV is given in Fig. \ref{fig:thr} (b).   Each plot shows separate lines for the KK2f, SHERPA LO, and SHERPA NLO MC samples.  Additionally, the four-fermion and two-photon MC are included in Fig. \ref{fig:thr} (b).  The bottom of each panel gives the ratio of MC to data for each sample.  In both the top and bottom panels, the widths of the colored bands reflect the uncertainty on the simulation from MC statistics only.  At both LEP1 and LEP2, the three MC samples describe the data similarly well, except that the NLO SHERPA sample displays some discrepancy in the tail of the distribution at LEP1.  

Fig. \ref{fig:jetmass} shows the heavy jet mass $\rho$ and the jet mass difference $M_D$.     At LEP1, we can see that the LO SHERPA sample best describes $\rho$, although a similar determination at LEP2 is constrained by statistics.  Both SHERPA samples outperform KK2f in the LEP1 $M_D$ distribution.  The three samples perform similarly within the statistics of the LEP2 sample.

We plot the DURHAM jet resolution parameters $-\ln{y_{ij}}$ in Figs. \ref{fig:y23y34}-\ref{fig:y45y56}.   In Fig. \ref{fig:y23y34} (a) and (b), we show $-\ln{y_{23}}$.  At LEP1, we see significant improvement in going from KK2f to the SHERPA LO MC sample; the NLO sample shows disagreement with the data for $\ln{y_{23}}\lesssim -6$.  The SHERPA LO and KK2f MC samples describe the LEP2 $y_{23}$ distribution comparably, while the NLO shows significant discrepancies for $\ln{y_{23}}\lesssim -6$. 

In Fig. \ref{fig:y23y34} (c) and (d), we show $-\ln{y_{34}}$.  At LEP1, the SHERPA LO and KK2f describe the distribution well, with each being the best description of the data in different regions of the plot; the NLO sample does not perform as well as the other two throughout the plot.  At LEP2, the KK2f sample describes the data most accurately; although all of the samples perform reasonably within the limits of the available statistics, some deviation is seen in the NLO sample for $\ln{y_{34}}\lesssim -8.5$ .

The remaining two DURHAM jet parameters are shown in Fig. \ref{fig:y45y56}.  The differences in the three MC samples at LEP1 are small in the case of $-\ln{y_{45}}$, except for the region $\ln{y_{45}}\gtrsim -6$, where the LO SHERPA sample shows significant improvement over the other two; the three samples perfom comparably within the LEP2 statistics. The central region of the $-\ln{y_{56}}$ distribution at LEP1 is described well by all three MC samples, with the tails best simulated by the two SHERPA generations; at LEP2, the three samples are comparable.  

We note that, for many of these event-shape variables, the NLO and LO SHERPA MC do not perform substantially better than the KK2f MC, particularly within the statistical limits of the LEP2 dataset.  We will now move on to a discussion of four-jet variables.  A few additional event-shape variable plots can be found in  Appendix \ref{moredists}.

\section{Behavior of Jet-Clustering Algorithms and 4-Jet Observables}
\label{jetclus}

We now specifically discuss some observables directly related to clustering the events into $4$ jets.  The concentration on these observables is motivated by several considerations.  First, our analysis in Ref. \cite{paper3}  relies heavily on clustering the events into $4$ jets.  Second, the accurate simulation  of four-jet states additionally has relevance to many other LEP analyses.  Third, it is expected that the SHERPA LO and NLO MC samples, which are produced using the four-parton ME, should model these observables more accurately than KK2f, which uses only the parton shower, so comparing the performance of the three MC samples for these variables is of interest.  Lastly, most of these distributions do not exist in the publicly-available Rivet analyses, so study of them requires input directly from the LEP data\footnote{For example, the variable $\theta_{14}$ has been suggested to compare phenomenological QCD models \cite{Fischer:2014bja}, although with different cuts than used here.}.

\begin{figure}[h]
\begin{center}
\subfigure[]{\includegraphics[width=2.5in,bb=80 150 520 720]{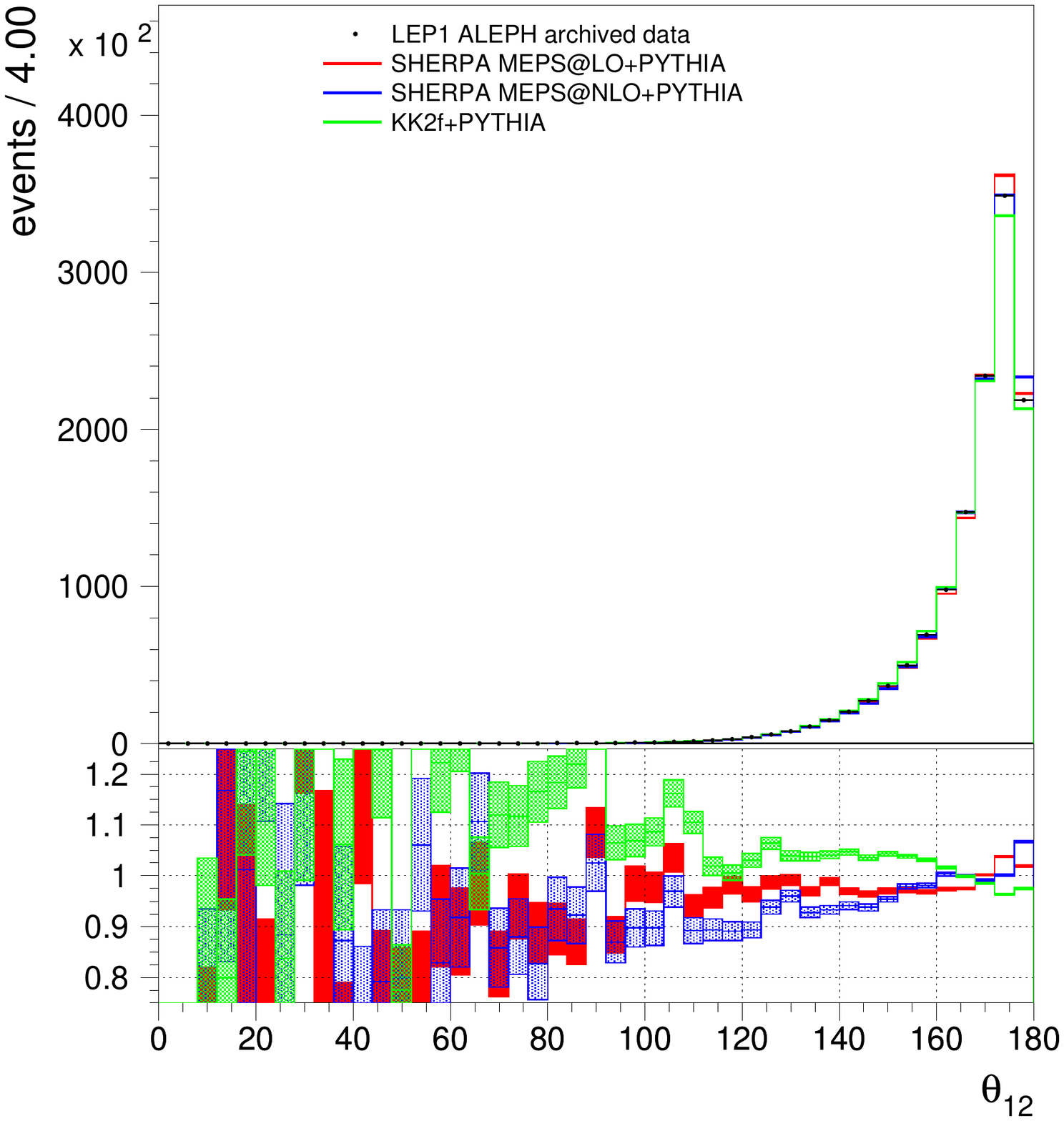}}\hspace{.35in}
\subfigure[]{\includegraphics[width=2.5in,bb=80 150 520 720]{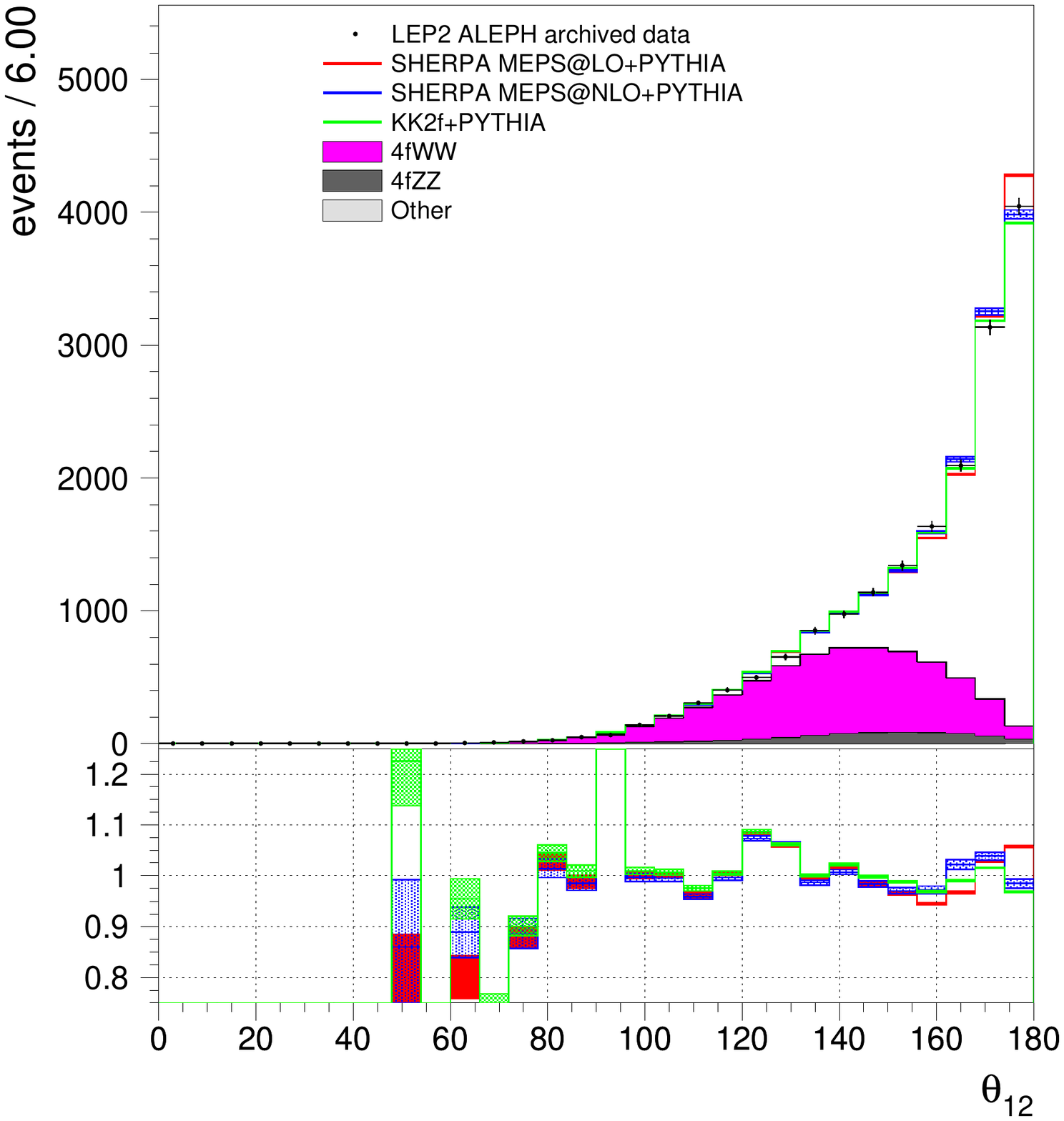}}\\
\subfigure[]{\includegraphics[width=2.5in,bb=80 150 520 720]{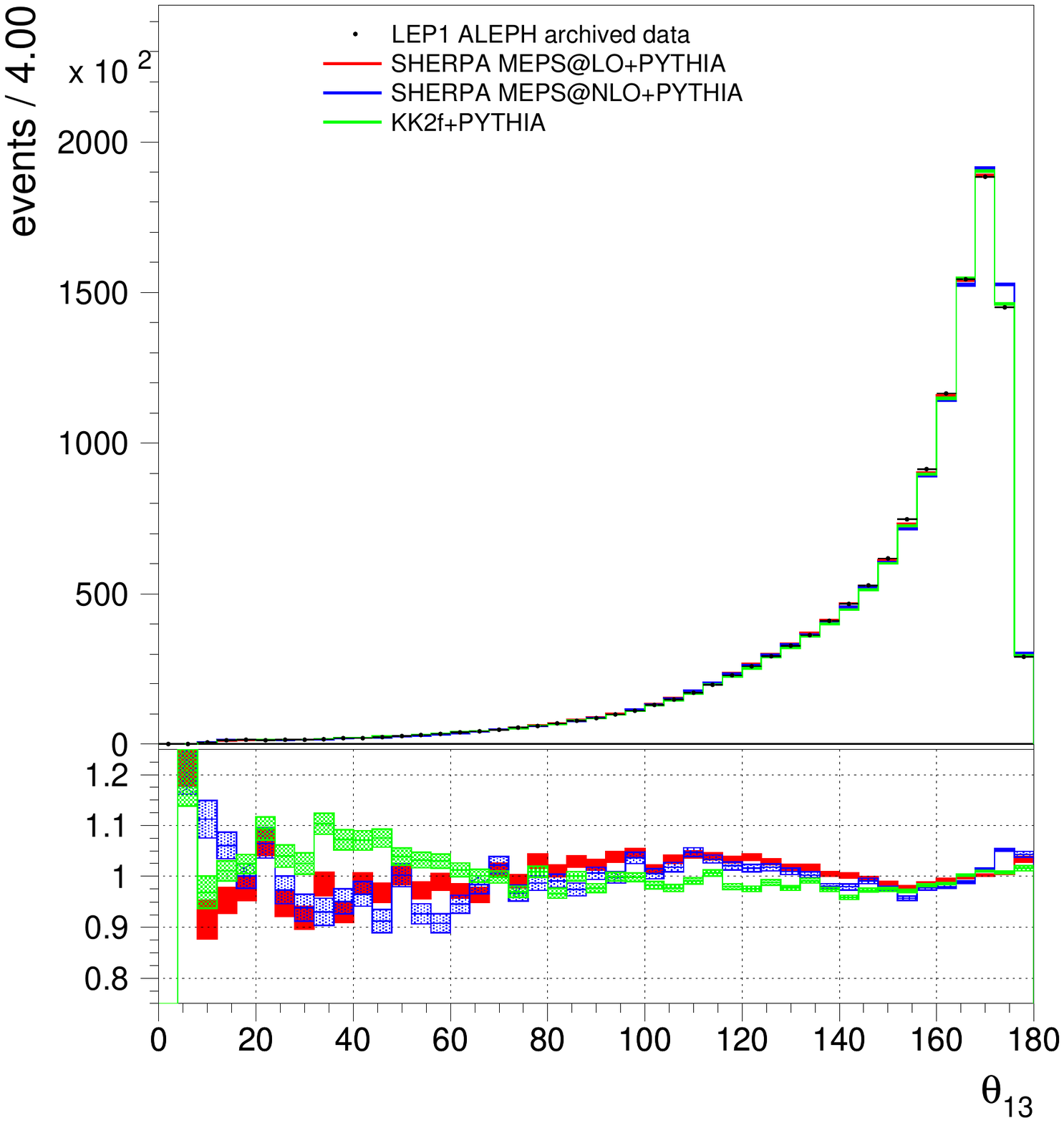}}\hspace{.35in}
\subfigure[]{\includegraphics[width=2.5in,bb=80 150 520 720]{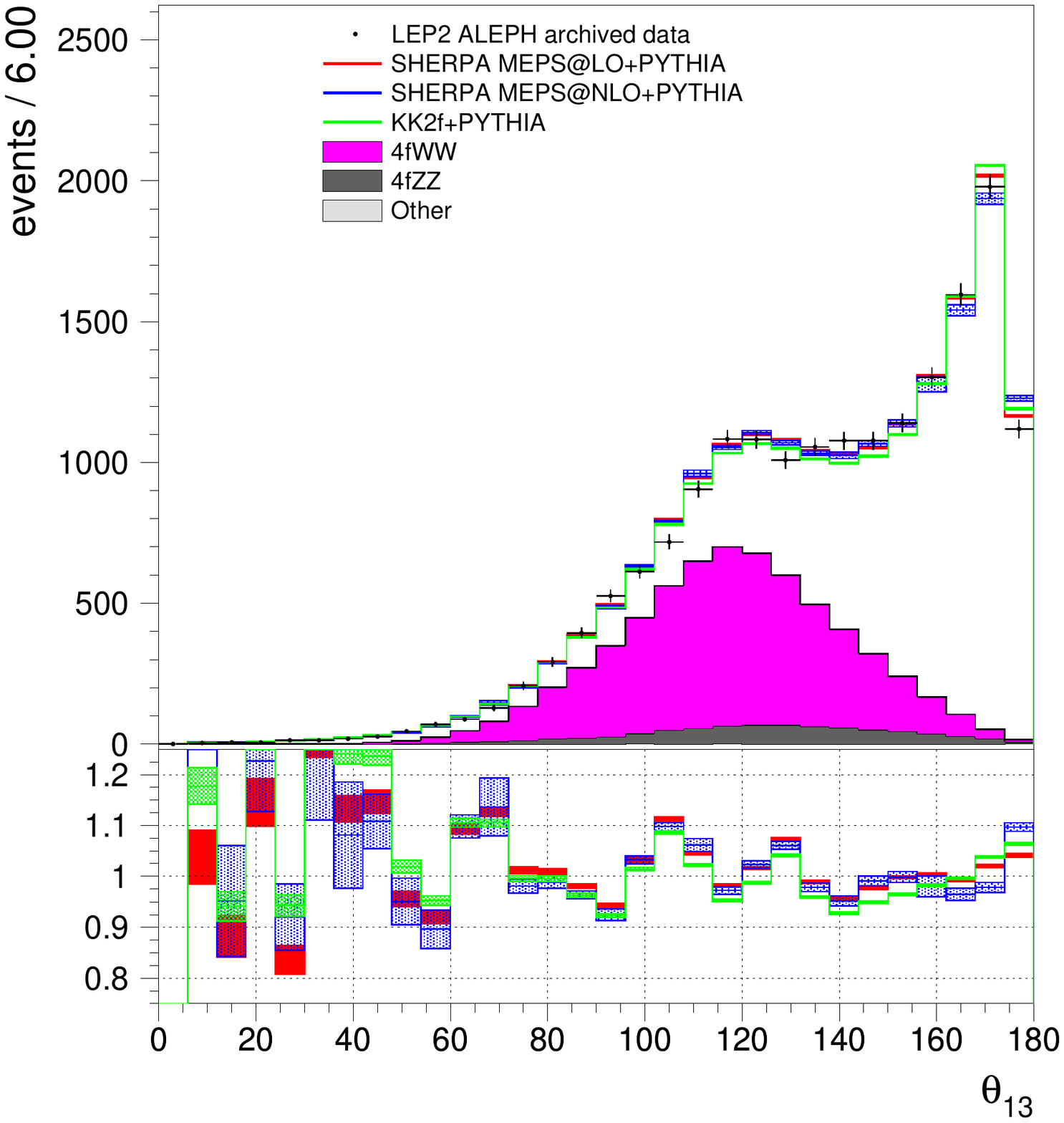}}
\end{center}
\caption{Inter-jet angles $\theta_{12}$ and $\theta_{13}$ at LEP1 and LEP2 using the LUCLUS algorithm.  Data from ALEPH is compared to KK2f, SHERPA LO, and SHERPA NLO MC.}
\label{fig:t12t13}
\end{figure}

Unless otherwise noted, our plots here are made with the LUCLUS jet-clustering algorithm.  While a few plots made with DURHAM are given below, DURHAM and other jet-clustering algorithms, such as JADE \cite{Bartel:1986ua} and DICLUS \cite{Lonnblad:1992qd} will be studied in more detail in \cite{paper3}.  Additionally, distributions of some Durham angular variables can be found in Appendix \ref{moredists}.  The four-jet variables below are divided into two main groups:  angular variables and dijet masses.

\begin{figure}[h]
\begin{center}
\subfigure[]{\includegraphics[width=2.5in,bb=80 150 520 720]{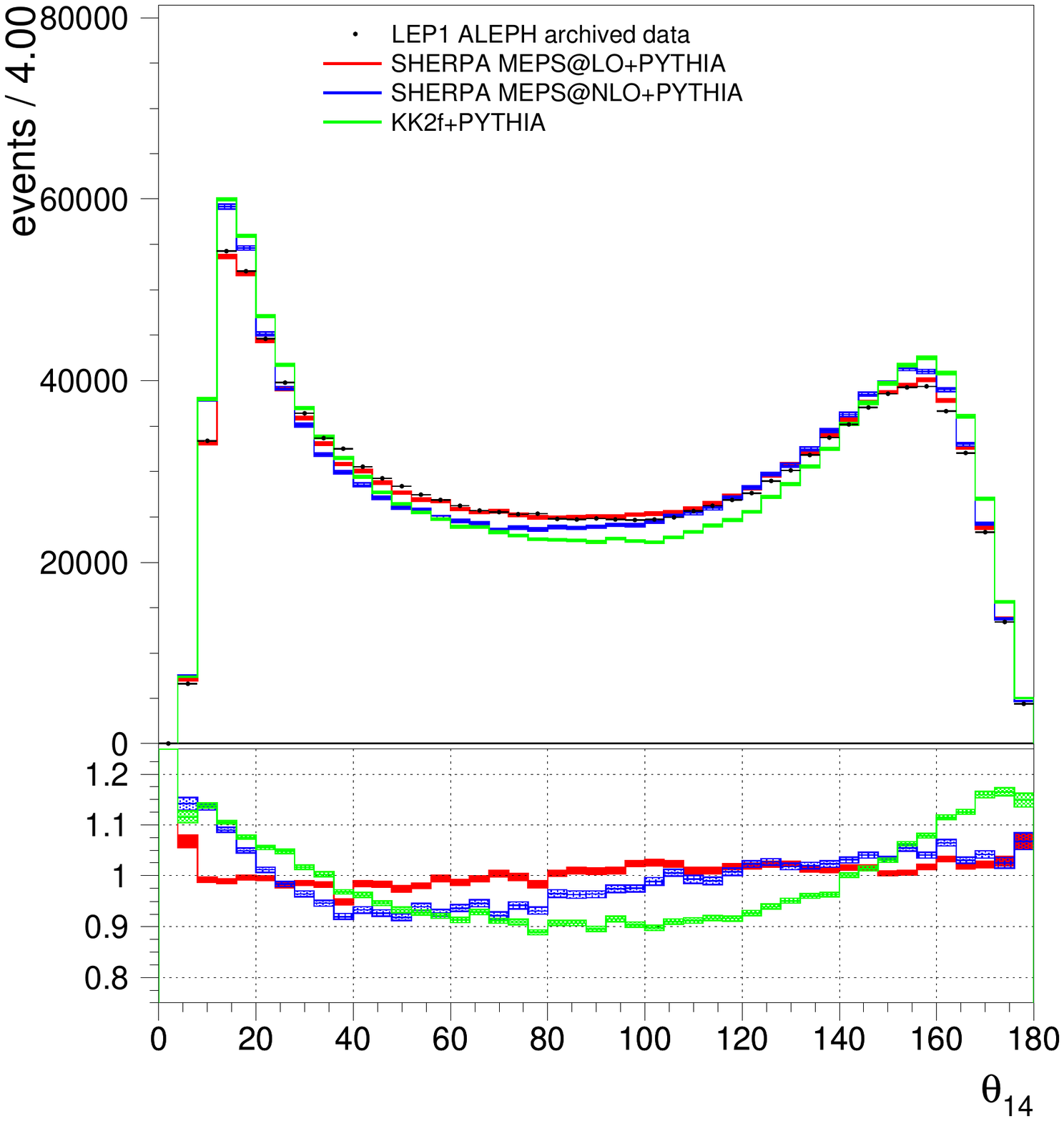}}\hspace{.35in}
\subfigure[]{\includegraphics[width=2.5in,bb=80 150 520 720]{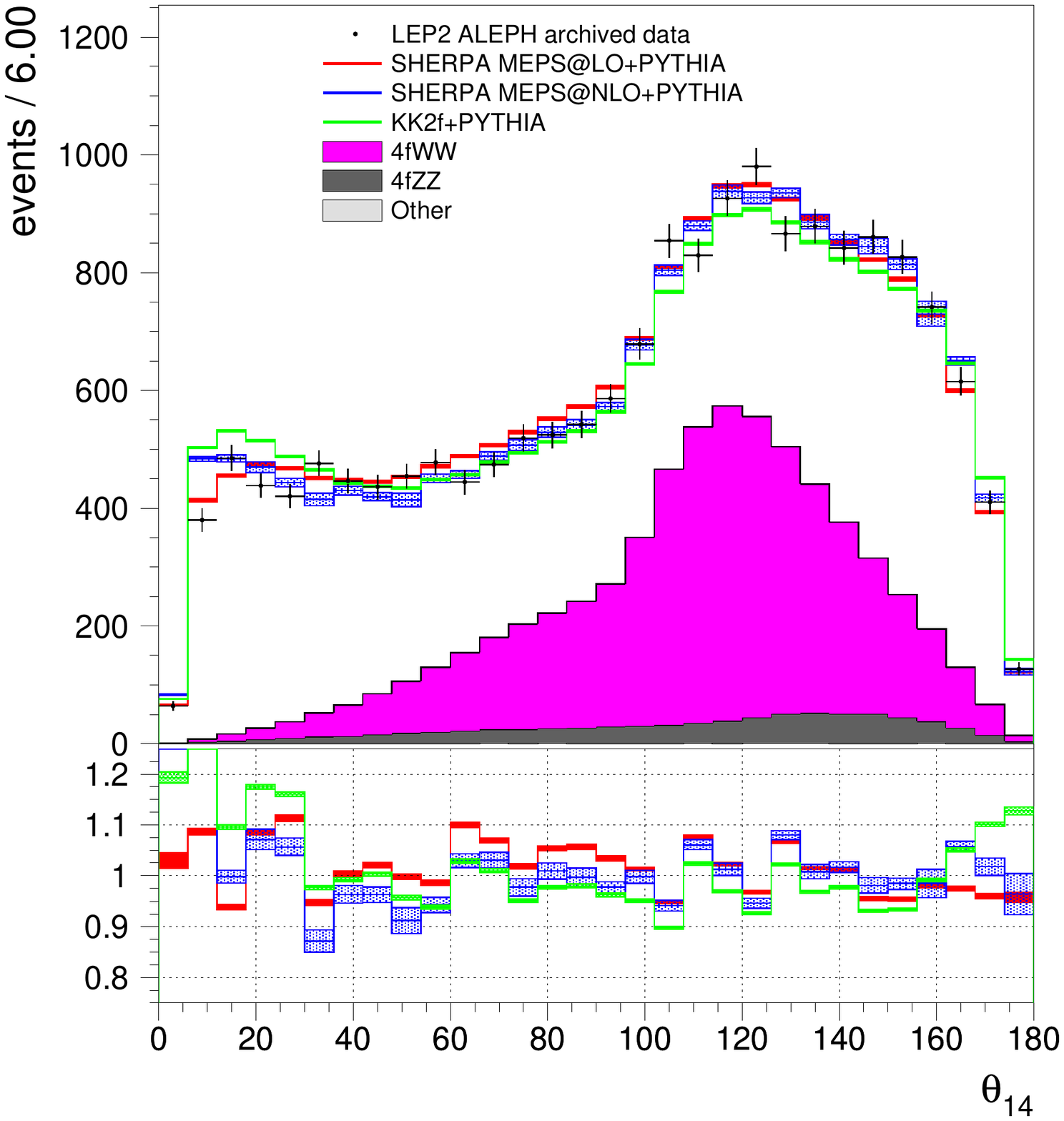}}\\
\subfigure[]{\includegraphics[width=2.5in,bb=80 150 520 720]{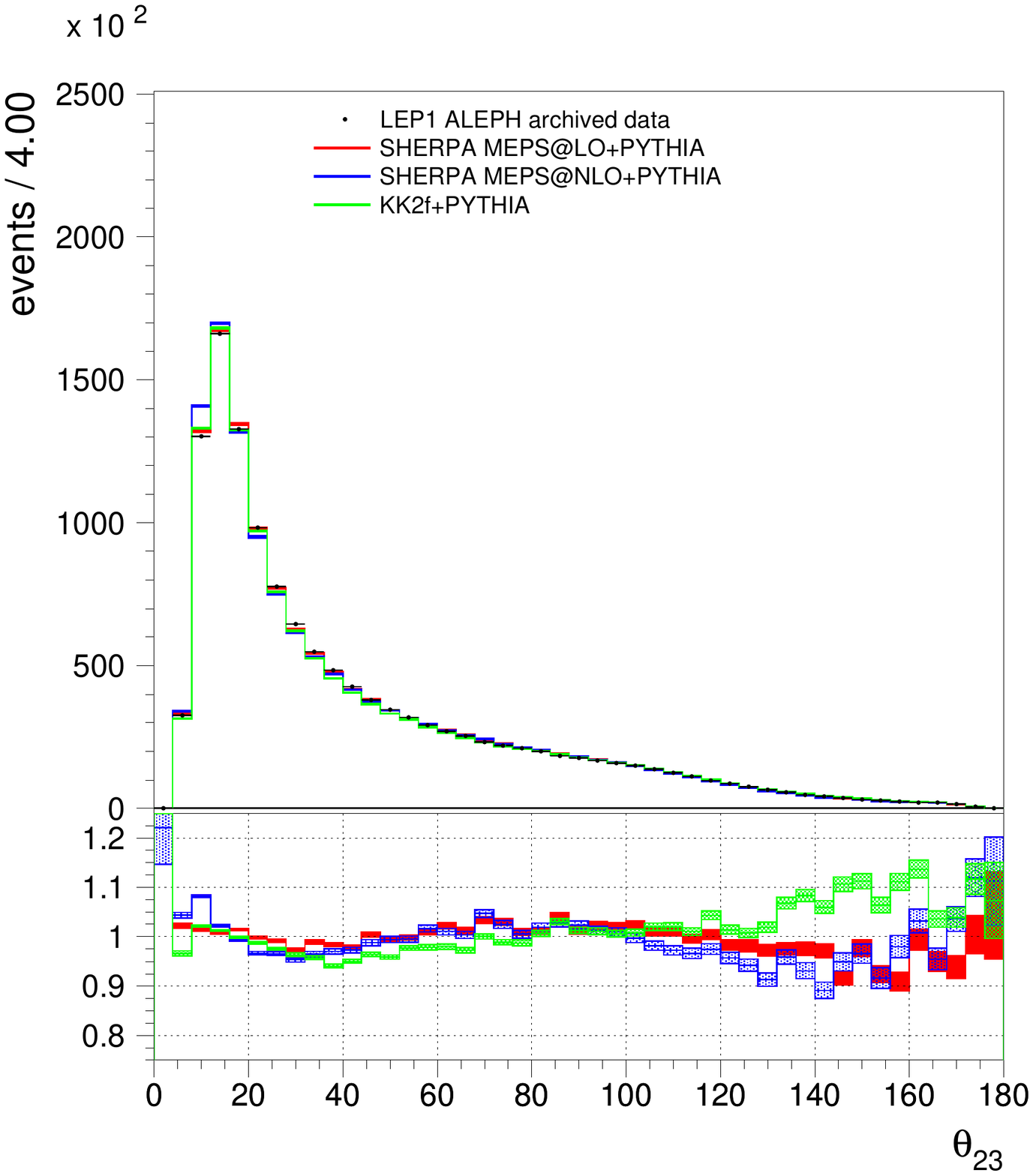}}\hspace{.35in}
\subfigure[]{\includegraphics[width=2.5in,bb=80 150 520 720]{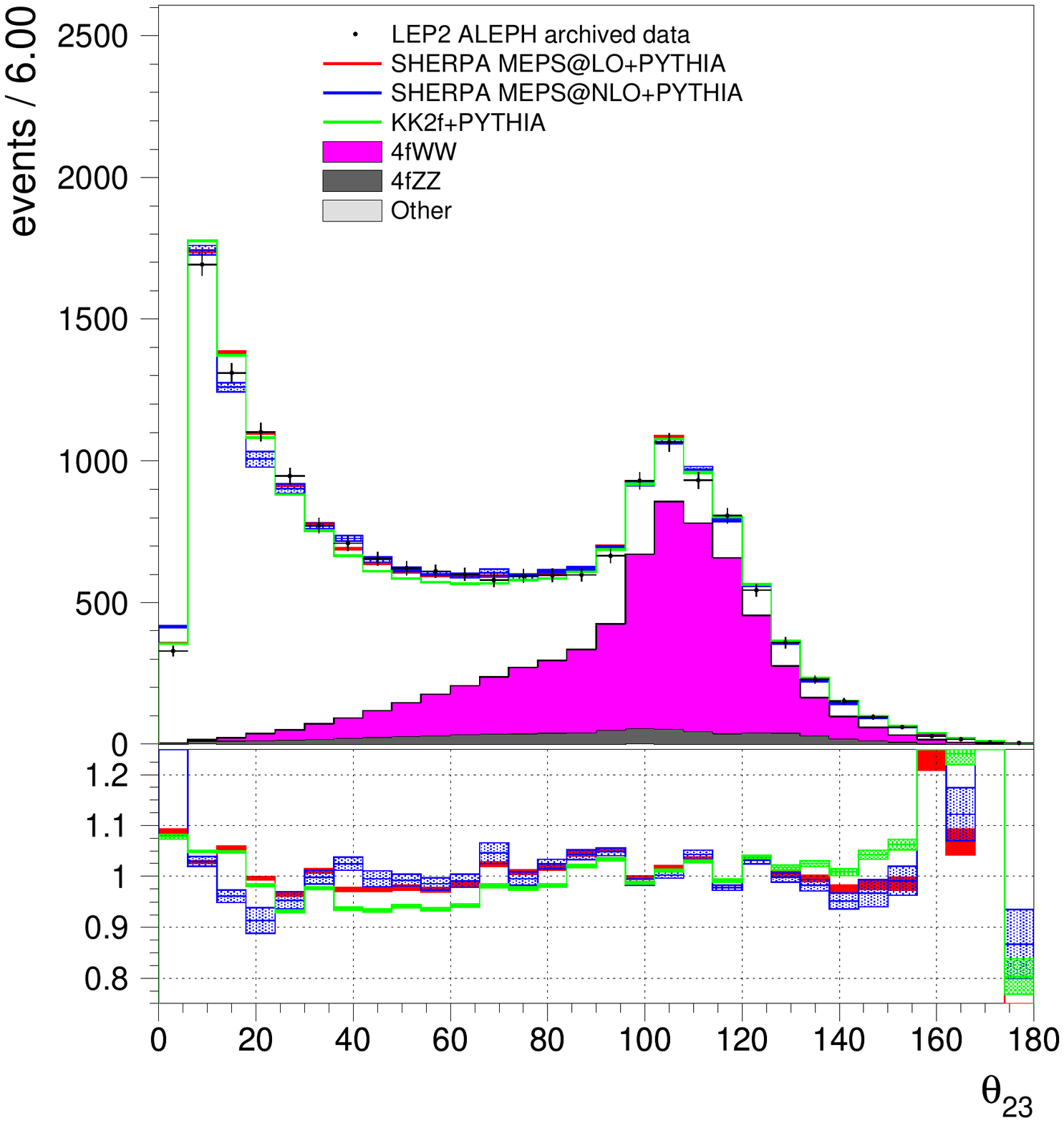}}
\end{center}
\caption{Inter-jet angles $\theta_{14}$ and $\theta_{23}$ at LEP1 and LEP2 using the LUCLUS algorithm.  Data from ALEPH is compared to KK2f, SHERPA LO, and SHERPA NLO MC.}
\label{fig:t14t23}
\end{figure}

\subsection{Angular Variables}

Of particular interest are the inter-jet angles $\theta_{ij}$; here, the jets are numbered in energy, with jet $1$ being the most energetic.  These are important as the jet rescaling algorithm determines the jet energies using the inter-jet angles and the jet masses and enforcing energy and momentum conservation.  As jet masses are typically small compared to their energies, the rescaled jet energies are mostly determined by the inter-jet angles.  Thus if there is good agreement between data and MC among the inter-jet angles, good agreement is also expected for the rescaled jet energies and quantities derived from them.

We plot the first two of these variables, $\theta_{12}$ and $\theta_{13}$ for LEP1 and LEP2 in Fig. \ref{fig:t12t13}.  For QCD events, $\theta_{12}$ is peaked at $180^\circ$, and all three MC samples reproduce the data reasonably well at LEP1, with the LO SHERPA sample showing a mild improvement over the other two.  At LEP2, we see reasonable agreement between data and all three samples.  All three samples reproduce the $\theta_{13}$ distributions at LEP1 and LEP2 rather well.

The situation is rather different for $\theta_{14}$, shown in Fig.  \ref{fig:t14t23} (a) and (b).  At LEP1, we can clearly see that the SHERPA LO sample makes a remarkable improvement over KK2f for this variable.  While the NLO SHERPA also shows a clear improvement over KK2f, it does not share the success of the LO sample.  Within the LEP2 statistics, the SHERPA samples are comparable and an improvement over that of KK2f.  We can see at both LEP1 and LEP2 KK2f predicts too many events at the endpoints of the distributions and too few in the central region.  $\theta_{23}$, shown in Fig.  \ref{fig:t14t23} (c) and (d), displays a smaller but still visible improvement in moving from KK2f to the SHERPA samples at both LEP1 and LEP2.

\begin{figure}[h]
\begin{center}
\subfigure[]{\includegraphics[width=2.5in,bb=80 150 520 720]{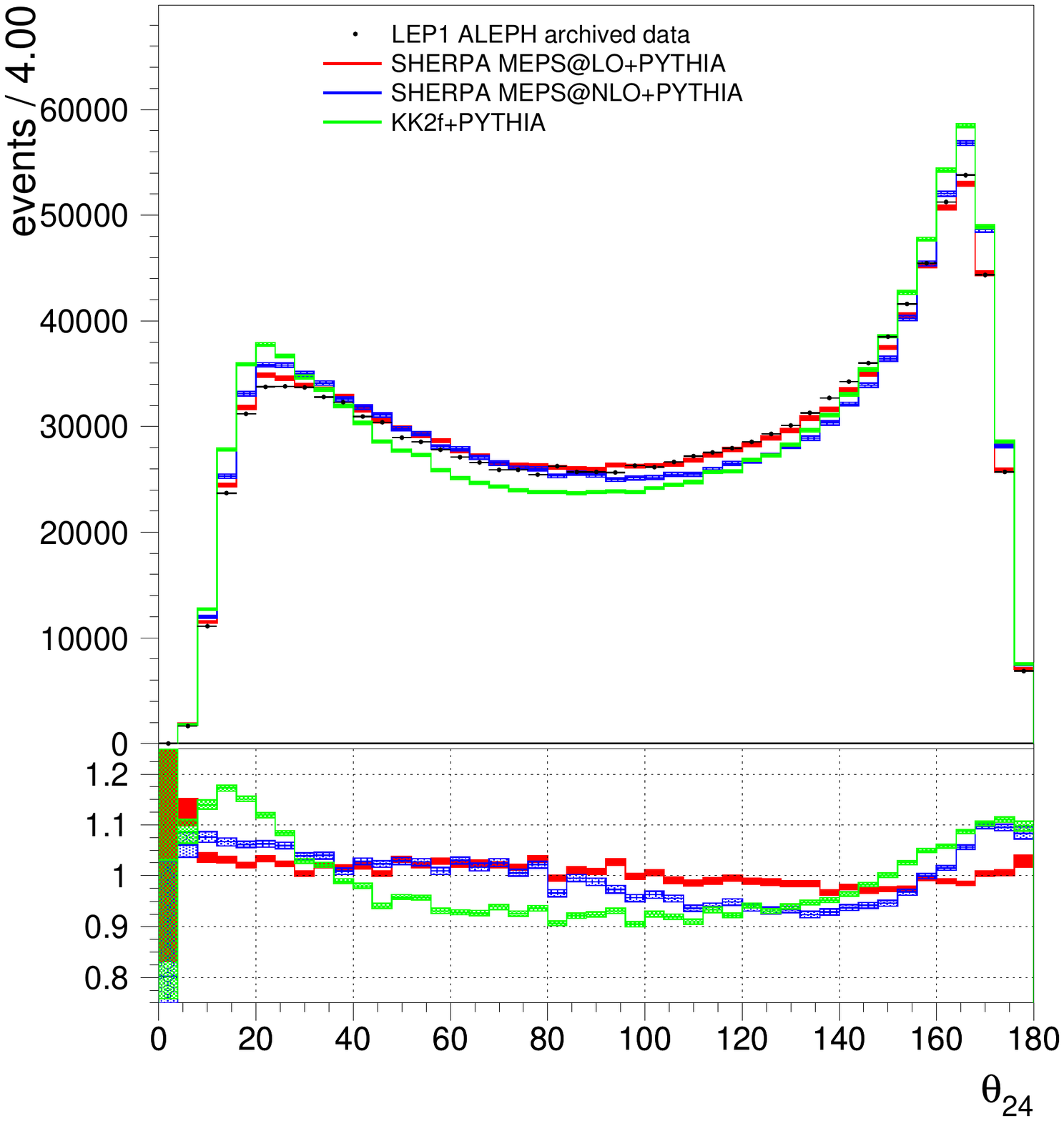}}\hspace{.35in}
\subfigure[]{\includegraphics[width=2.5in,bb=80 150 520 720]{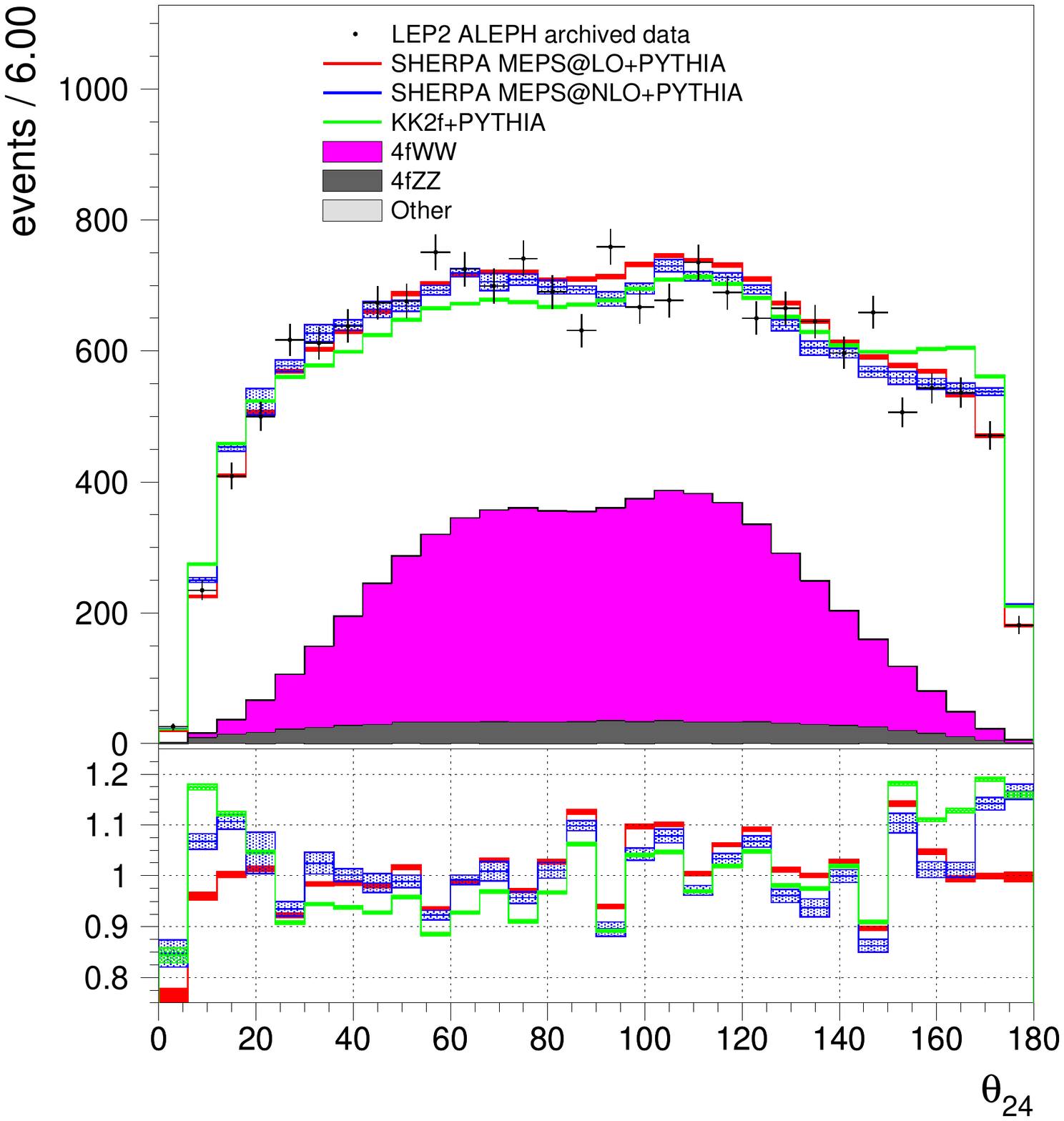}}\\
\subfigure[]{\includegraphics[width=2.5in,bb=80 150 520 720]{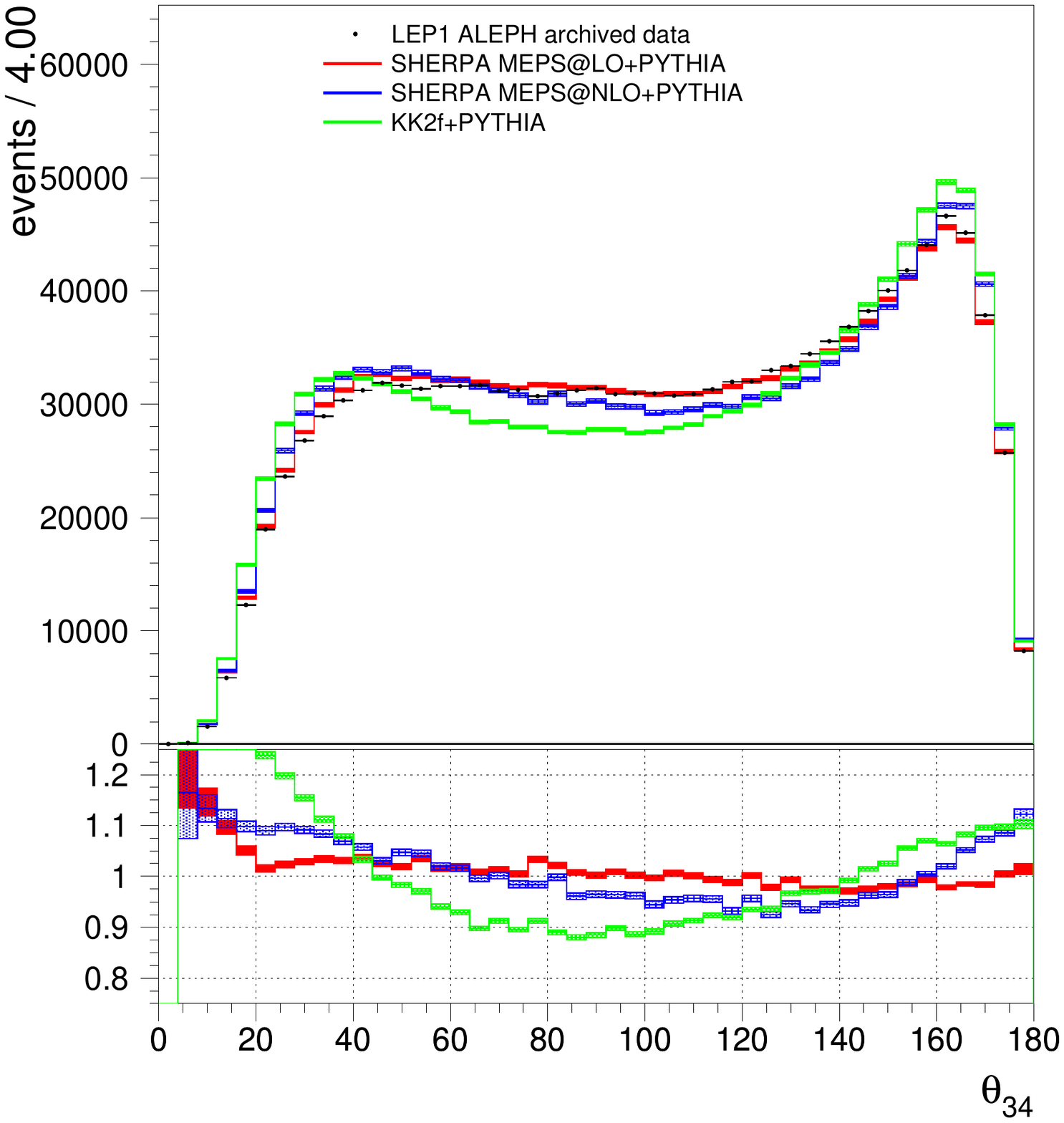}}\hspace{.35in}
\subfigure[]{\includegraphics[width=2.5in,bb=80 150 520 720]{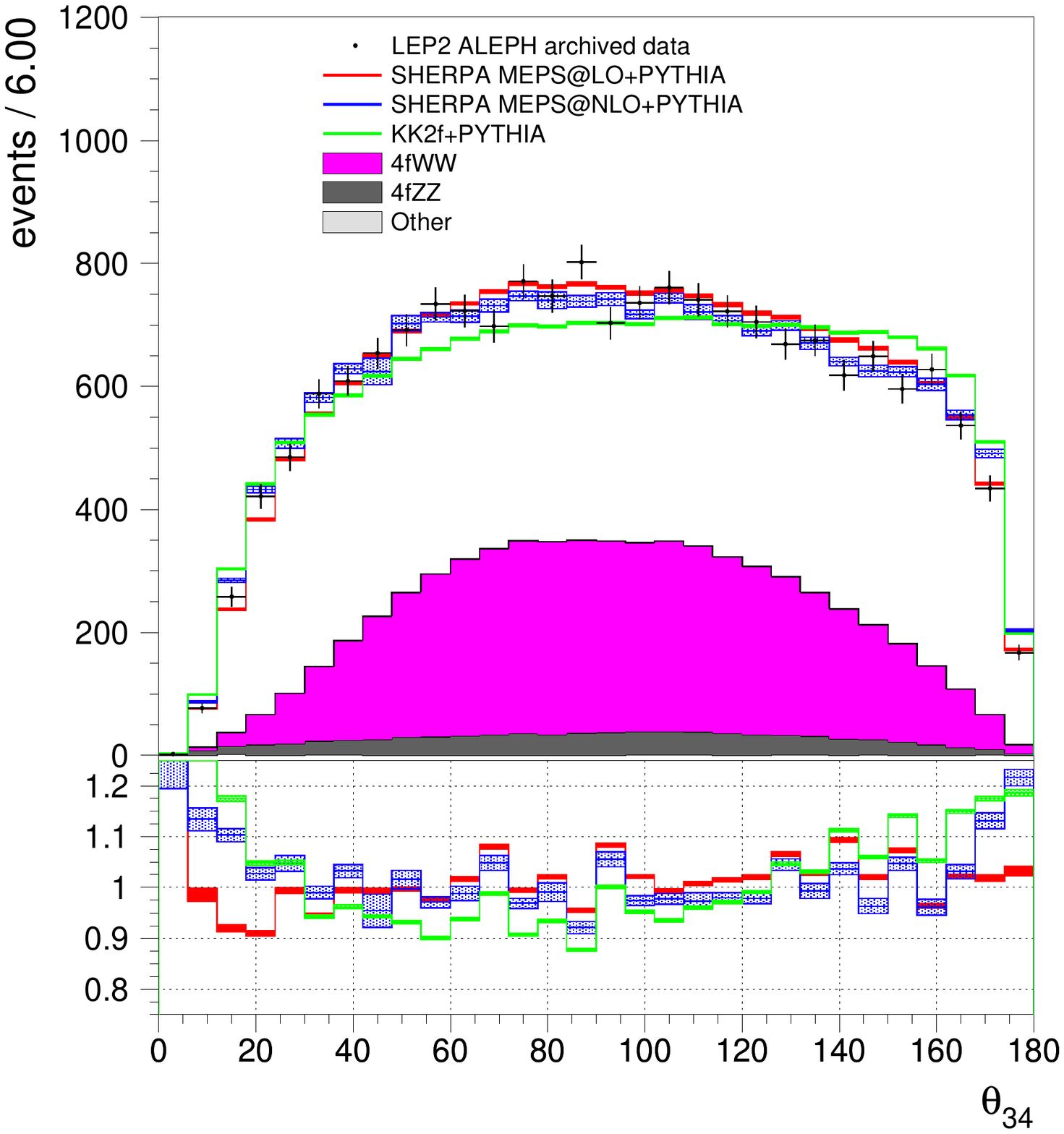}}
\end{center}
\caption{Inter-jet angles $\theta_{24}$ and $\theta_{34}$ at LEP1 and LEP2 using the LUCLUS algorithm.  Data from ALEPH is compared to KK2f, SHERPA LO, and SHERPA NLO MC.}
\label{fig:t24t34}
\end{figure}

The last two inter-jet angles are plotted in Fig. \ref{fig:t24t34}.  The behavior of $\theta_{24}$ in Fig. \ref{fig:t24t34} (a) and (b) for the different MC samples is much like that of $\theta_{14}$; at LEP1, we see significant improvement in moving from KK2f to the SHERPA LO sample; we also see significant improvement in the NLO sample with respect to KK2f, but not at the level of agreement with data that we see for the LO sample.  Both the LO and NLO SHERPA samples show significant improvement over KK2f at LEP2.  As with $\theta_{14}$, KK2f overestimates the number of events near the endpoints of the distributions and predicts too few events in the middle regions.  This pattern is displayed also in $\theta_{34}$ in Fig. \ref{fig:t24t34} (c) and (d).  At LEP1, the LO SHERPA sample shows excellent agreement with data; the NLO sample shows worse agreement than the LO sample, but is still better than that of KK2f.  At LEP2, both the LO and NLO samples show significant improvement in comparison with KK2f which overestimates the data for very large and very small angles while underestimating the data in the middle region of the plot.

As we have seen significant improvement in the modelling of $\theta_{ij}$ in moving from KK2f to the SHERPA samples, it is worth explicitly pointing out that the angles $\theta_{ij}$ were not used in the SHERPA tunes; the difference in performance of the MC generators with respect to these variables represents true improvement in the modelling of four-jet states.

\begin{figure}[h]
\begin{center}
\subfigure[]{\includegraphics[width=2.5in,bb=80 150 520 720]{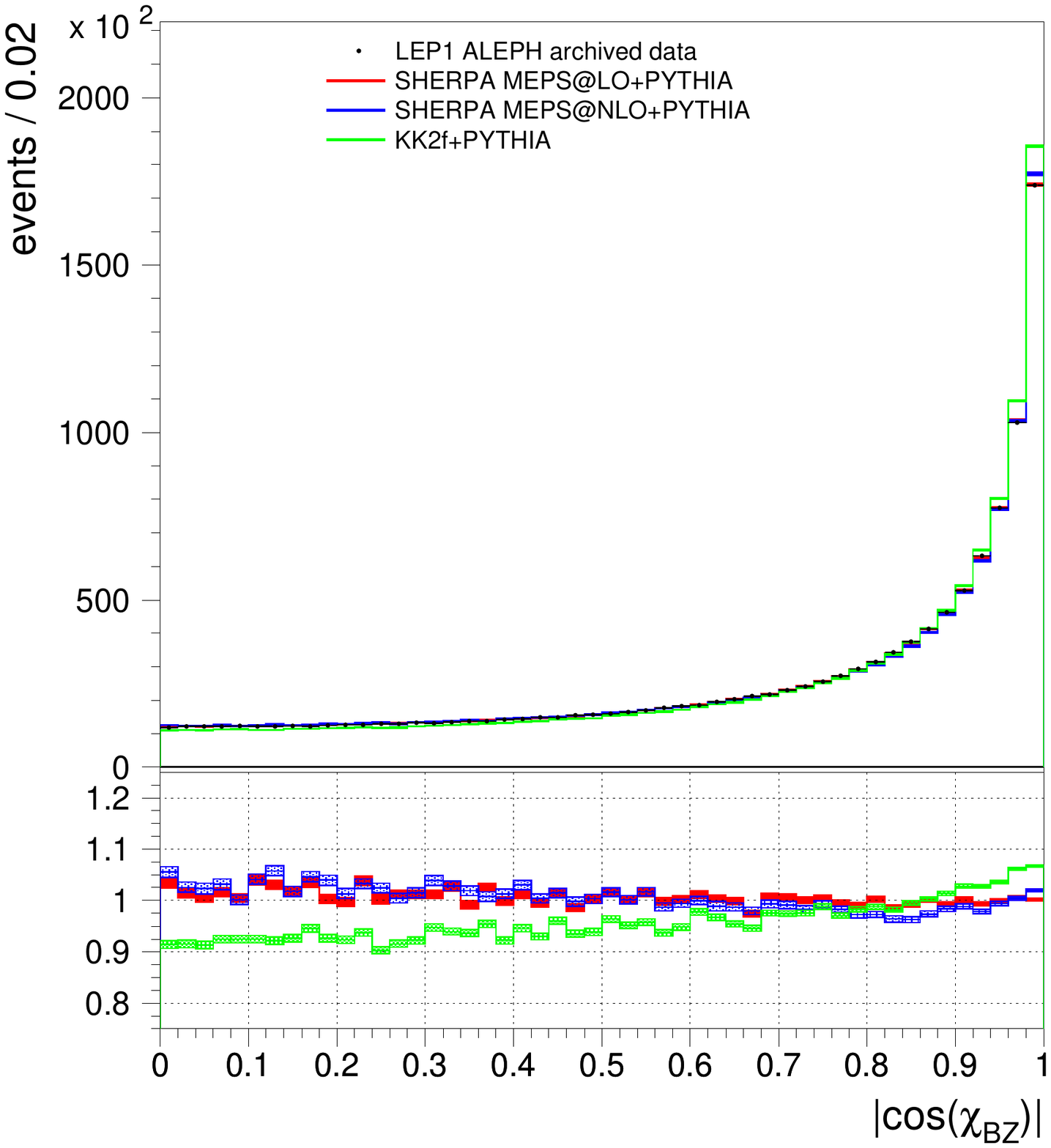}}\hspace{.35in}
\subfigure[]{\includegraphics[width=2.5in,bb=80 150 520 720]{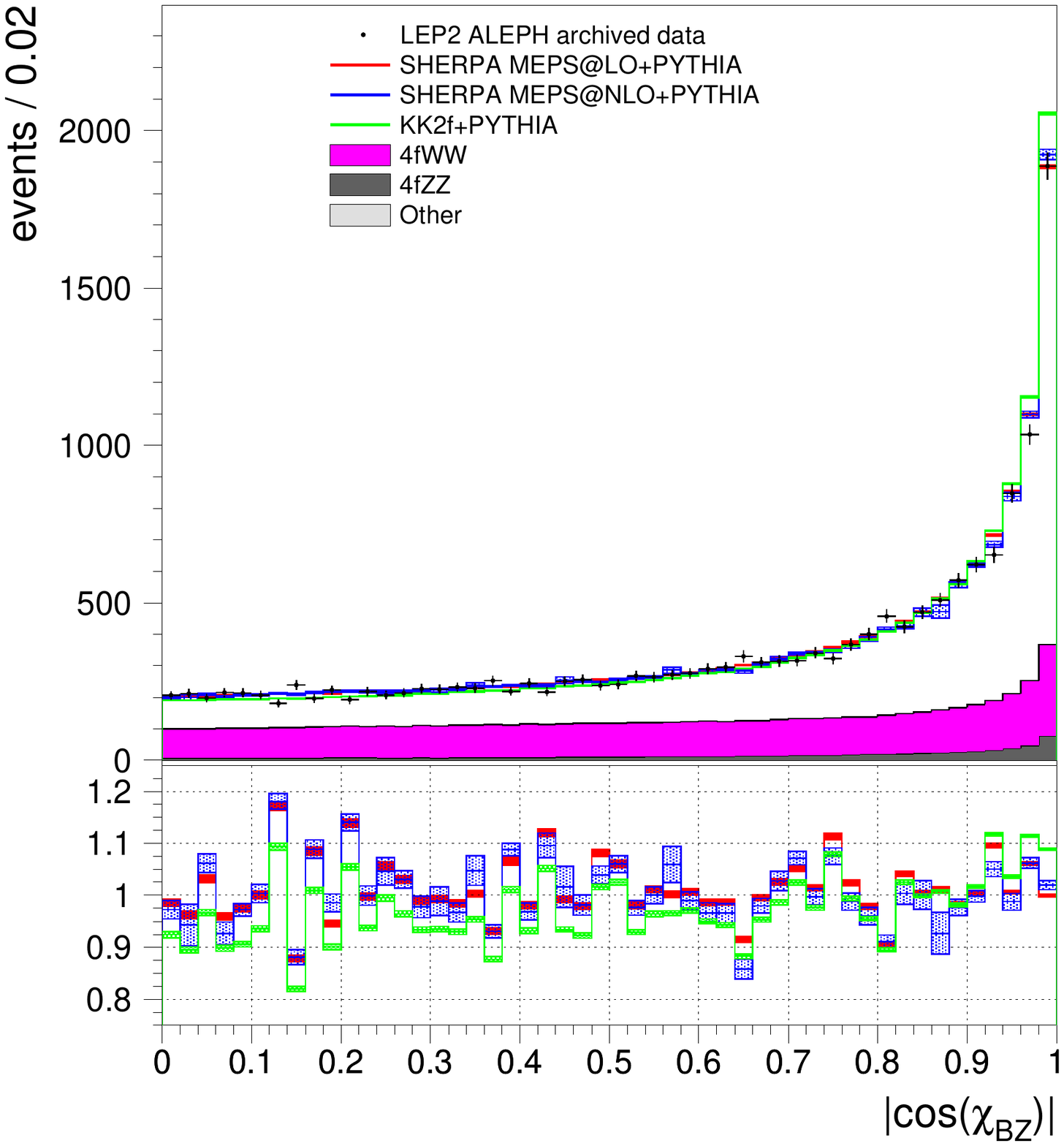}}\\
\subfigure[]{\includegraphics[width=2.5in,bb=80 150 520 720]{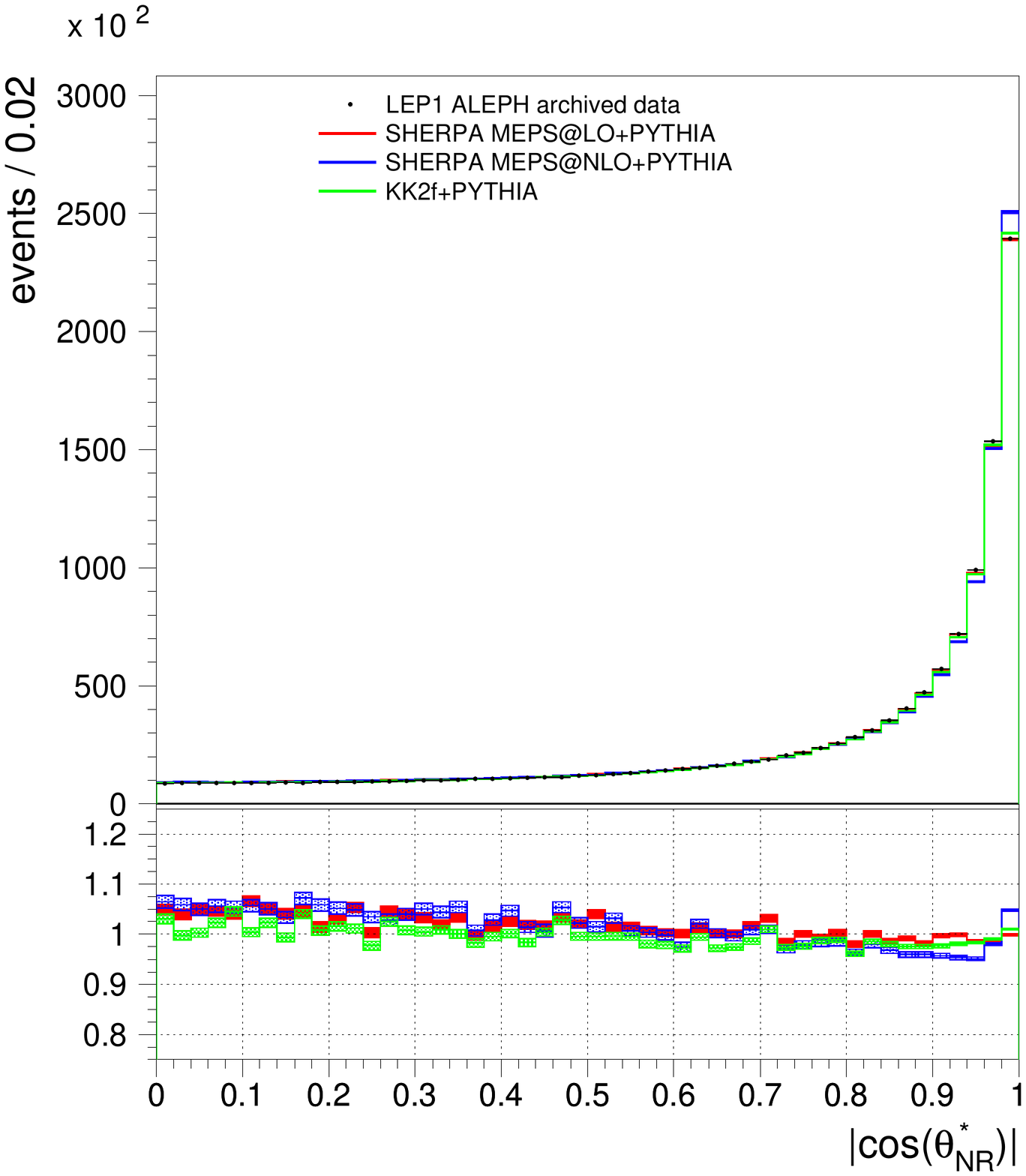}}\hspace{.35in}
\subfigure[]{\includegraphics[width=2.5in,bb=80 150 520 720]{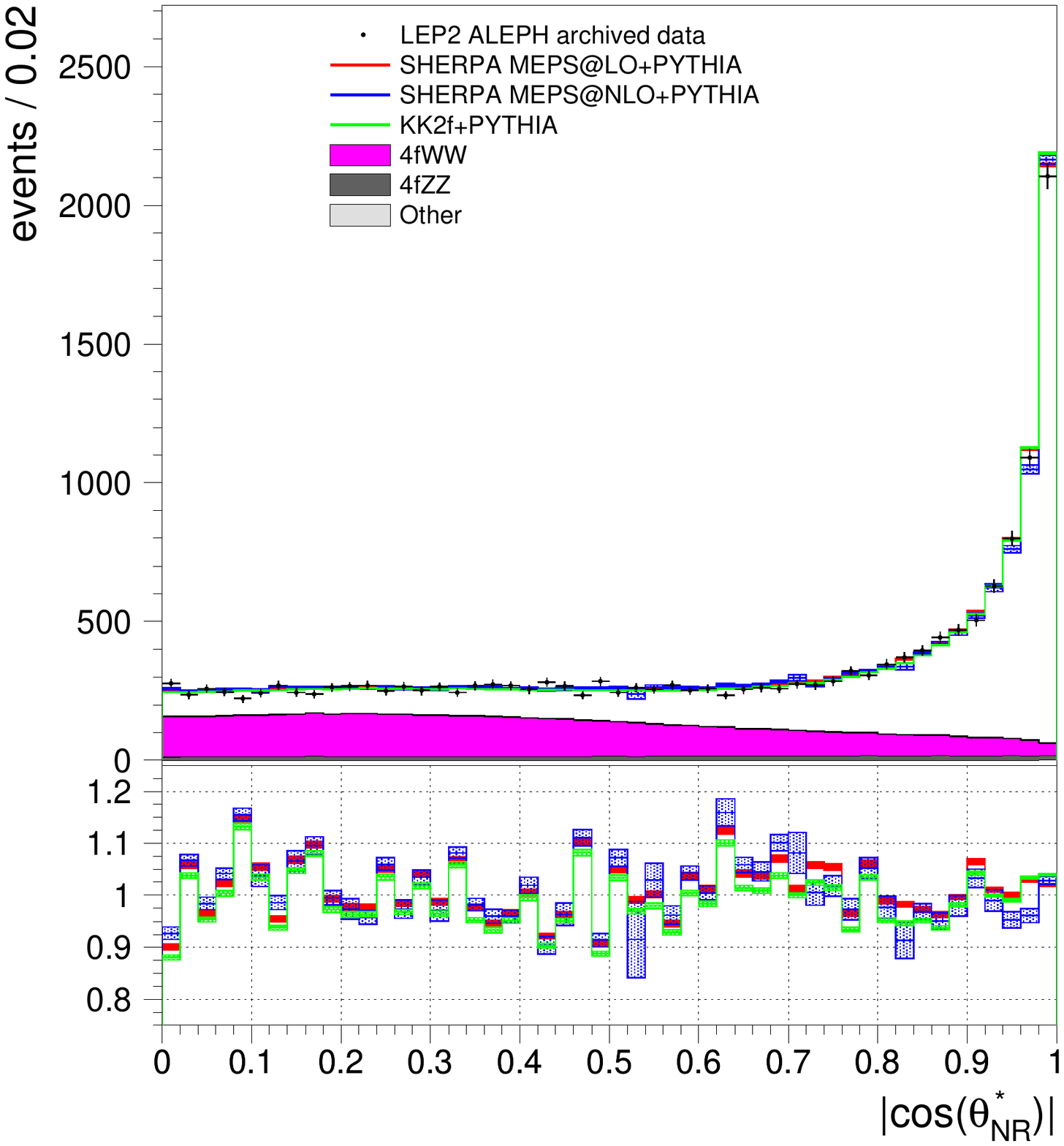}}
\end{center}
\caption{Bengtsson-Zerwas (top) and modified Nachtman-Reiter (bottom) angles.  ALEPH data is compared to KK2f, SHERPA LO, and SHERPA NLO MC.  Plots are from LEP1 (left) and LEP2 (right).}
\label{fig:bznrang}
\end{figure}

Next, we consider three angular variables important for QCD studies\footnote{A fourth angular variable often shown with these is $\theta_{34}$, discussed above.}.  The first of these, the Bengtsson-Zerwas angle \cite{Bengtsson:1988qg}, is shown in Fig. \ref{fig:bznrang} (a) and (b).  At LEP1, both the LO and the NLO SHERPA samples show significant improvement over KK2f over the entire range.  Despite the reduced statistics, the discrepancy in the KK2f distribution at LEP1 can also be seen in the LEP2 plot.  In Fig. \ref{fig:bznrang} (c) and (d), we give the modified Nachtman-Reiter angle \cite{Nachtmann:1982xr}.  At both LEP1 and LEP2, we see that all three samples describe the data very well for this variable.    

\begin{figure}[h]
\begin{center}
\subfigure[]{\includegraphics[width=2.5in,bb=80 150 520 720]{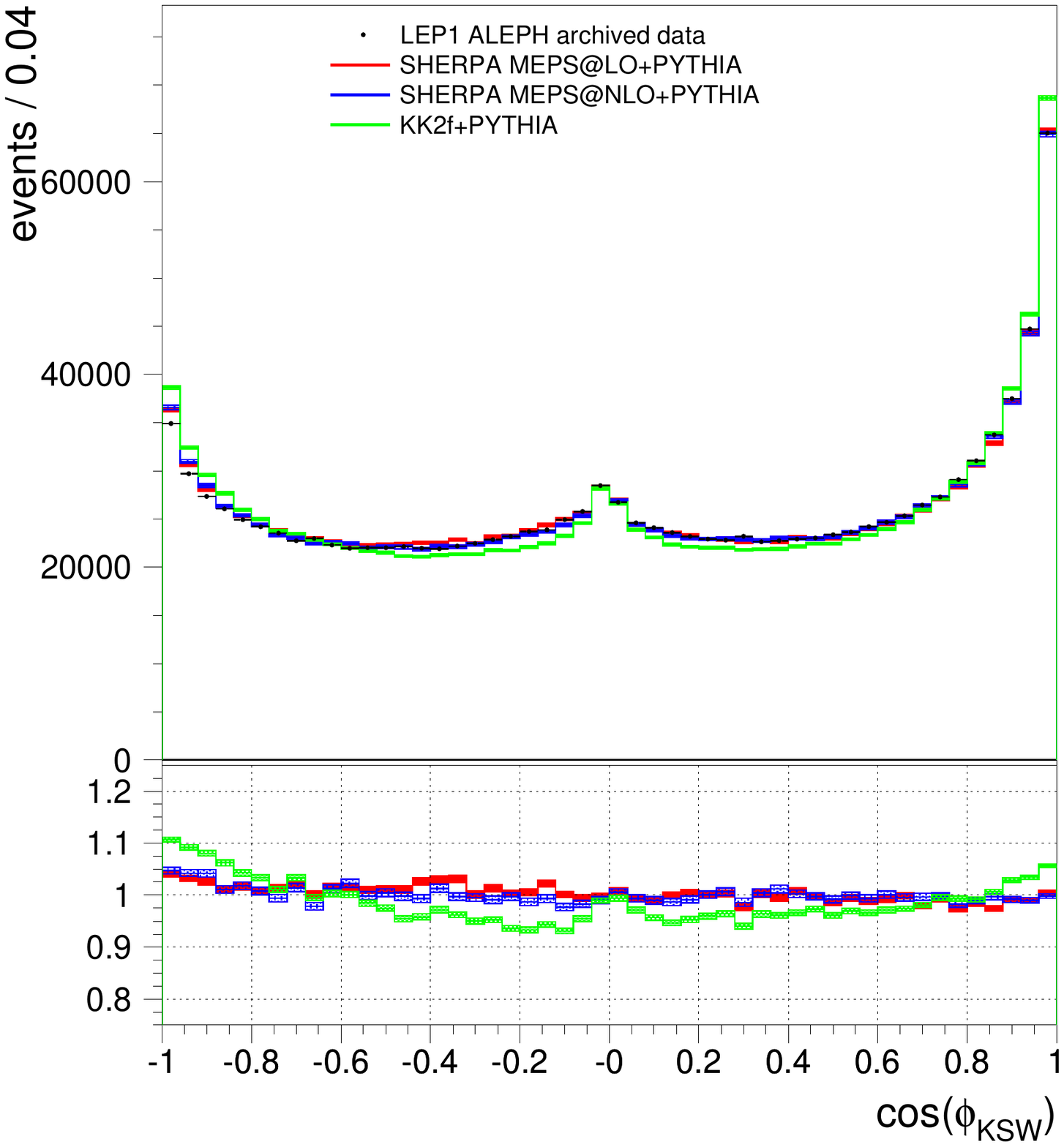}}\hspace{.35in}
\subfigure[]{\includegraphics[width=2.5in,bb=80 150 520 720]{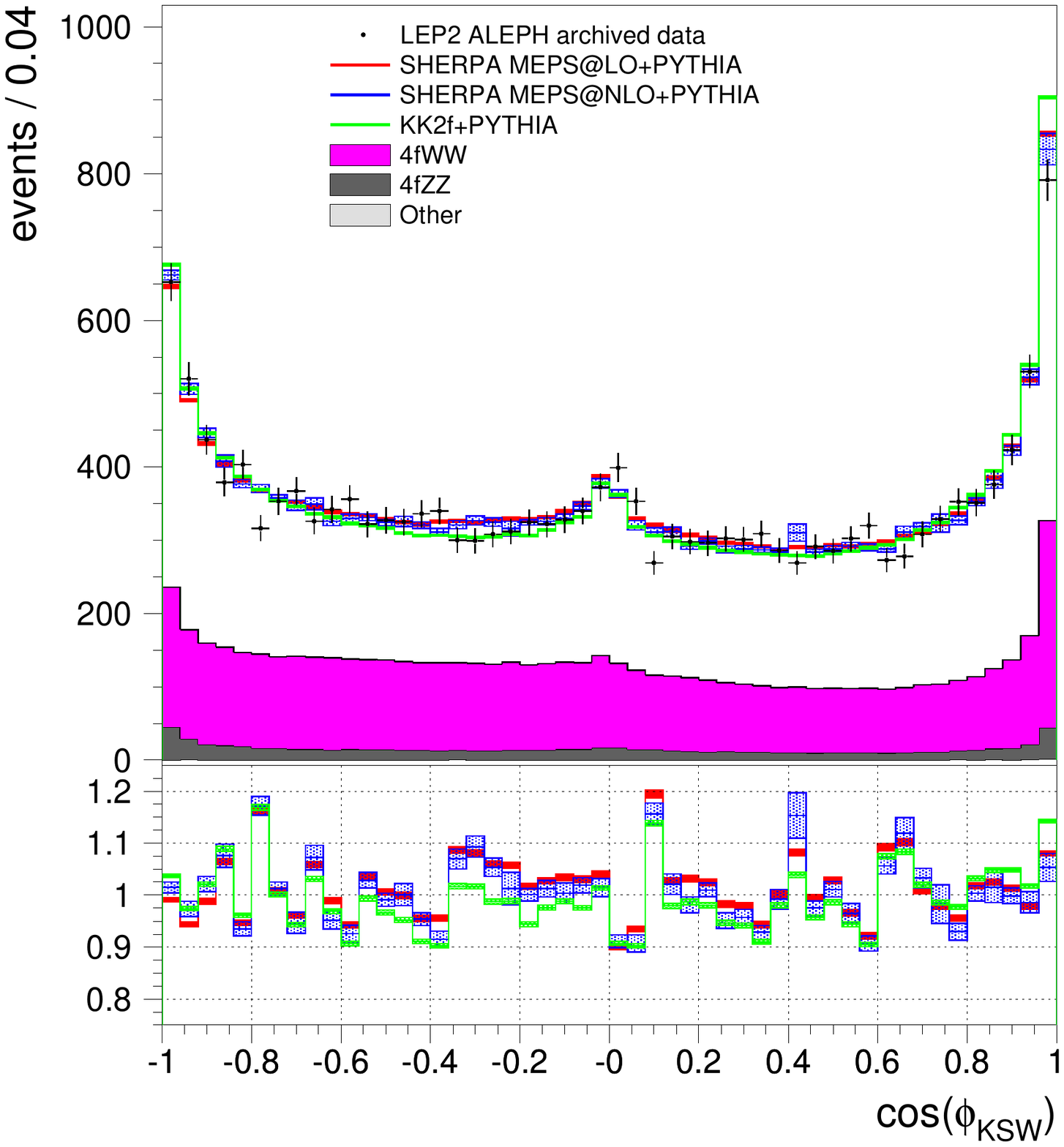}}
\end{center}
\caption{Plots of the K\"{o}rner-Schierholz-Willrodt angle.  Data from ALEPH is compared to KK2f, SHERPA LO, and SHERPA NLO MC.  Plot (a) is from LEP1; plot (b) is from LEP2.}
\label{fig:kswang}
\end{figure}

In Fig. \ref{fig:kswang} we show the K\"{o}rner-Schierholz-Willrodt angle \cite{Korner:1980pv}.  At LEP1, we see significant improvement in both the LO and NLO SHERPA samples relative to the KK2f across the entire range; the KK2f sample overestimates the data near $\pm 1$, and underestimates it elsewhere.  Hints of this behavior can also be seen at LEP2, but with reduced statistics.
 
\subsection{Dijet Masses}

We now move on to observables which are particularly important for the analysis in Ref. \cite{paper3}, those related to dijet masses.  As we wish to compare our MC samples to data, to avoid confusion with the excess explored in those works, we will confine our discussion in this section to distributions from LEP1.  We will return to LEP2 mass distributions after we discuss our reweighting procedure in Section \ref{reweight}.

First, we consider an observable which we refer to as $\Sigma$.  To construct this variable, we pair the rescaled jets such that the difference in mass between the two dijet masses is minimized; $\Sigma$ is then the sum of the two resulting dijet masses, divided by two. We additionally denote the difference between the two dijet masses as $\Delta=M_1-M_2$, where $M_1$ is defined to contain the most energetic jet in the event.  At LEP2 energies, the minimum-mass pairing is $(14)(23)$ approximately $92\%$ of the time, with the remainder being almost entirely the $(13)(24)$ combination.  

\begin{figure}[!h]
\begin{center}
\subfigure[]{\includegraphics[width=2.5in,bb=80 150 520 720]{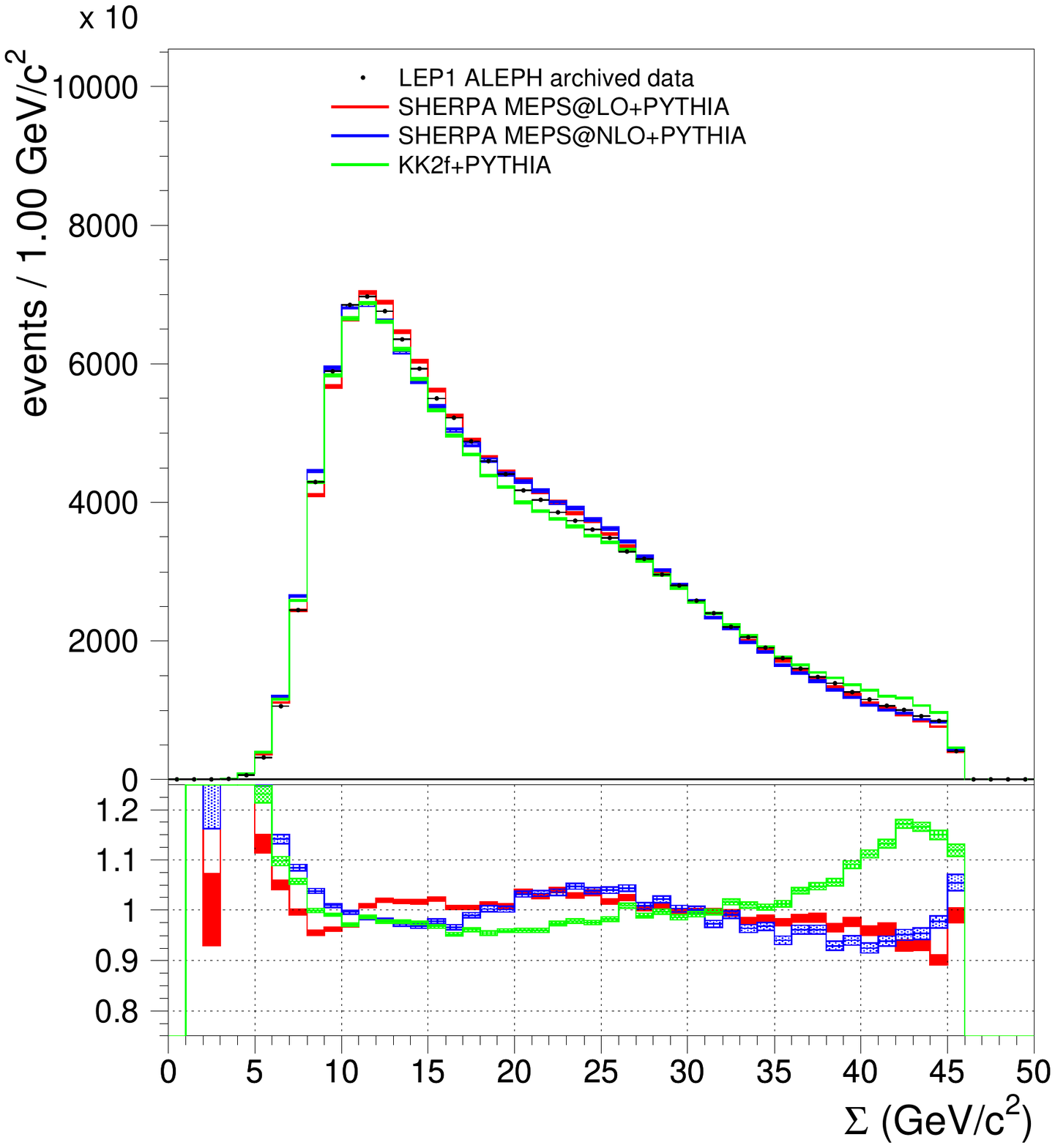}}\hspace{.35in}
\subfigure[]{\includegraphics[width=2.5in,bb=80 150 520 720]{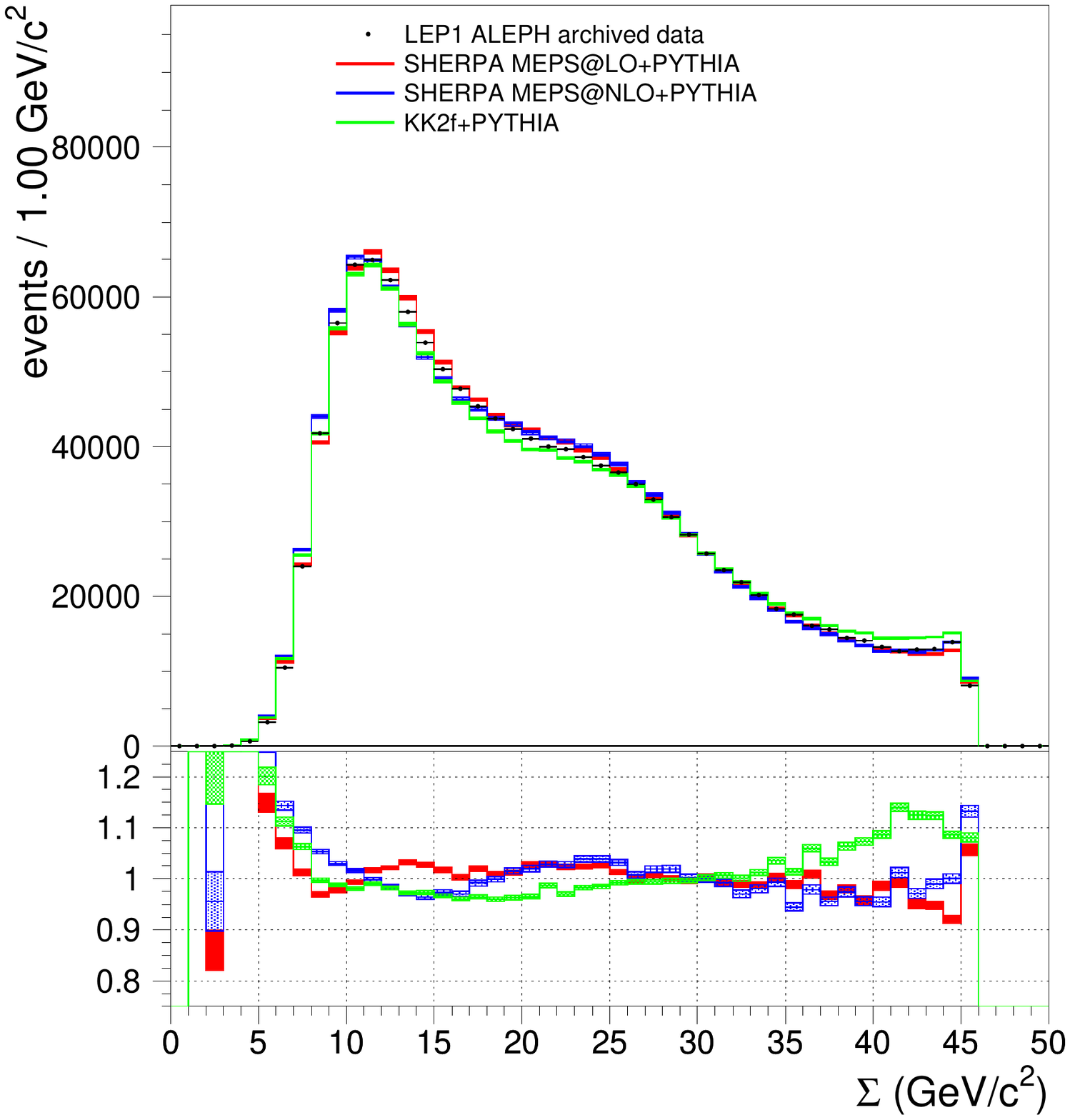}}\\
\subfigure[]{\includegraphics[width=2.5in,bb=80 150 520 720]{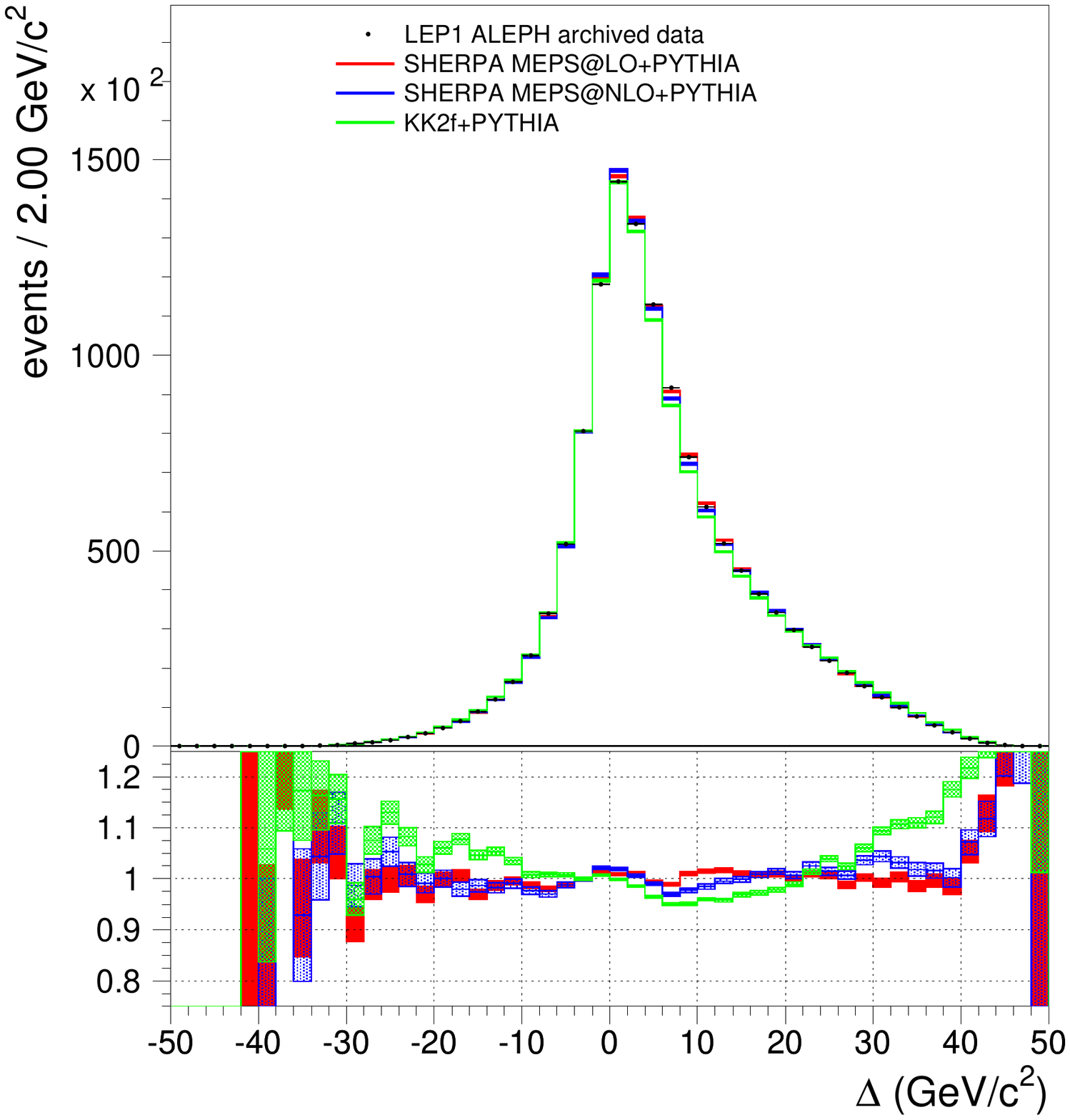}}\hspace{.35in}
\subfigure[]{\includegraphics[width=2.5in,bb=80 150 520 720]{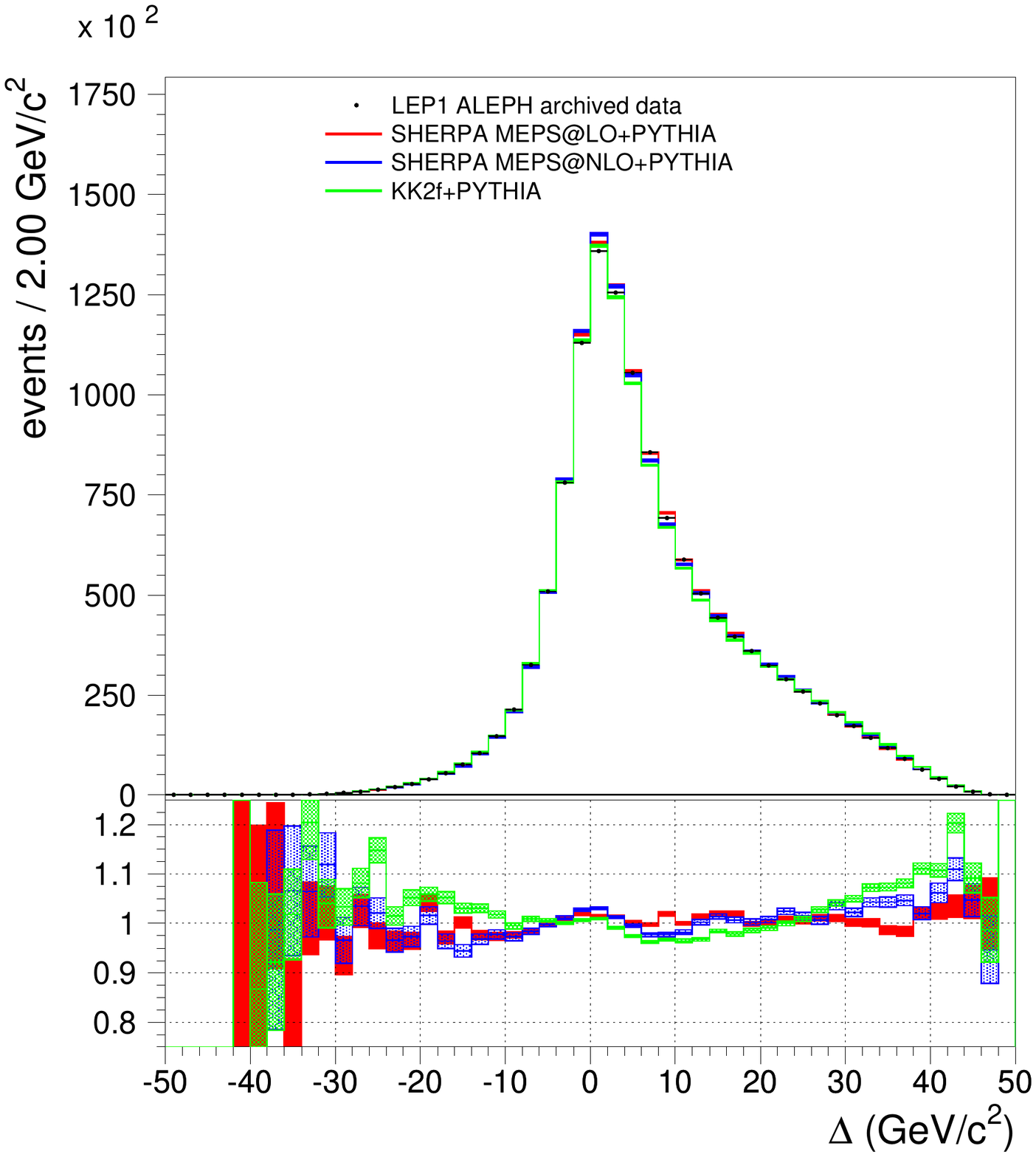}}
\end{center}
\caption{$\Sigma$ (top) and $\Delta$ (bottom) at LEP1.  Plots on the left-hand side are for jets made with LUCLUS jets; those on the right are made with DURHAM jets.}
\label{fig:masssumdiff}
\end{figure}

We plot $\Sigma$ and $\Delta$ at LEP1 in Fig. \ref{fig:masssumdiff}.  In (a), we plot  $\Sigma$ made with jets clustered with the LUCLUS algorithm, while in (b) we give the corresponding plots for the DURHAM algorithm\footnote{In the case of plots made with DURHAM-clustered jets, these jets were also those used at preselection.  For this reason, the events in the LUCLUS and DURHAM plots are not exactly the same.}.  For both jet-clustering algorithms, we see that both SHERPA samples show significant improvement over KK2f for high values of $\Sigma$.  All three MC samples struggle to reproduce the data on the far low end of the spectrum, where there are larger effects from hadronization.  $\Delta$ is plotted in Fig. \ref{fig:masssumdiff} (c) and (d) at LEP1 for the two jet-clustering algorithms.   We see that both SHERPA samples have better data-MC agreement than KK2f does, for both LUCLUS and DURHAM.  The two SHERPA samples perform comparably.

\begin{figure}[h]
\begin{center}
\subfigure[]{\includegraphics[width=2.5in,bb=80 150 520 720]{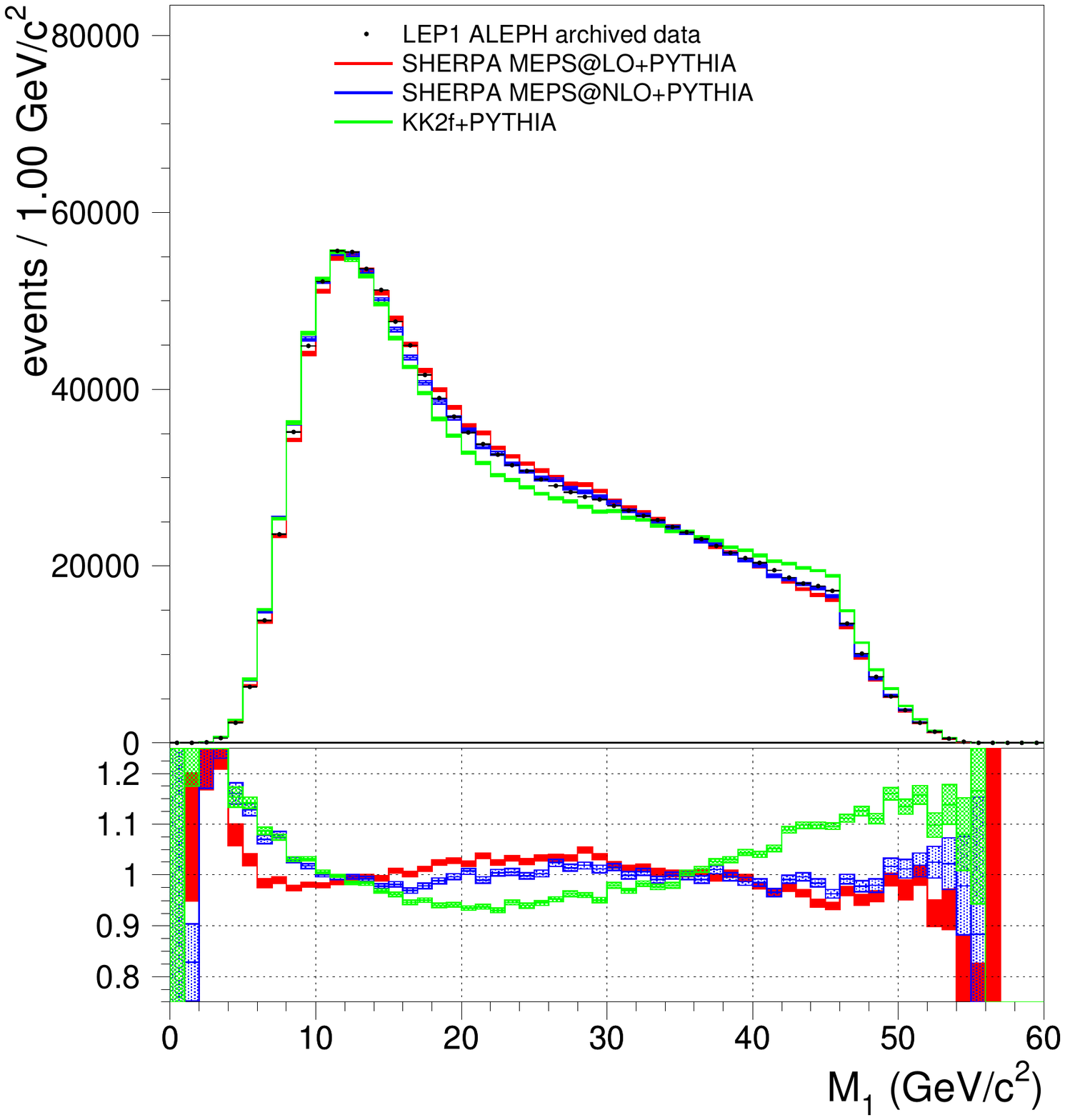}}\hspace{.35in}
\subfigure[]{\includegraphics[width=2.5in,bb=80 150 520 720]{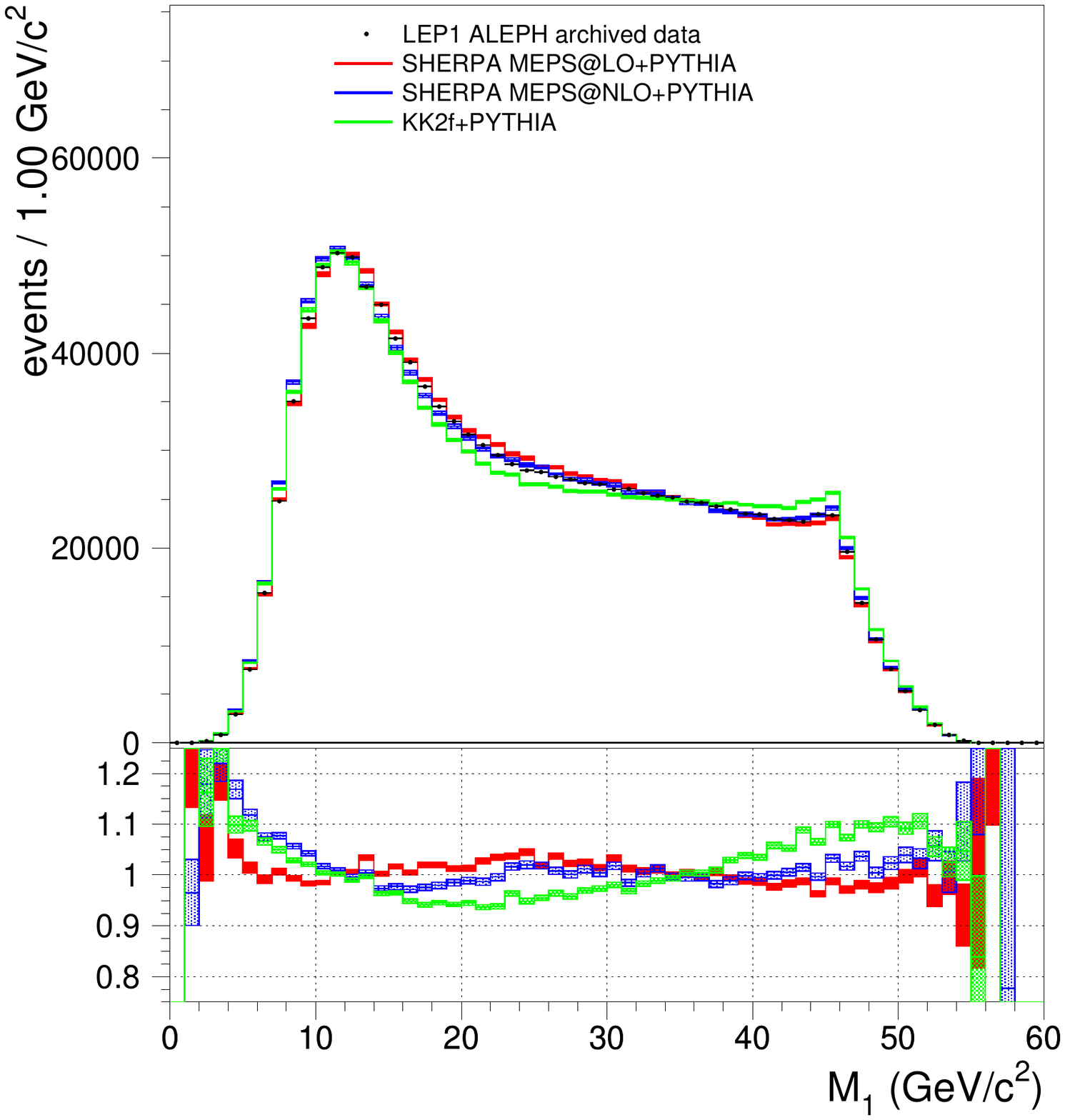}}\\
\subfigure[]{\includegraphics[width=2.5in,bb=80 150 520 720]{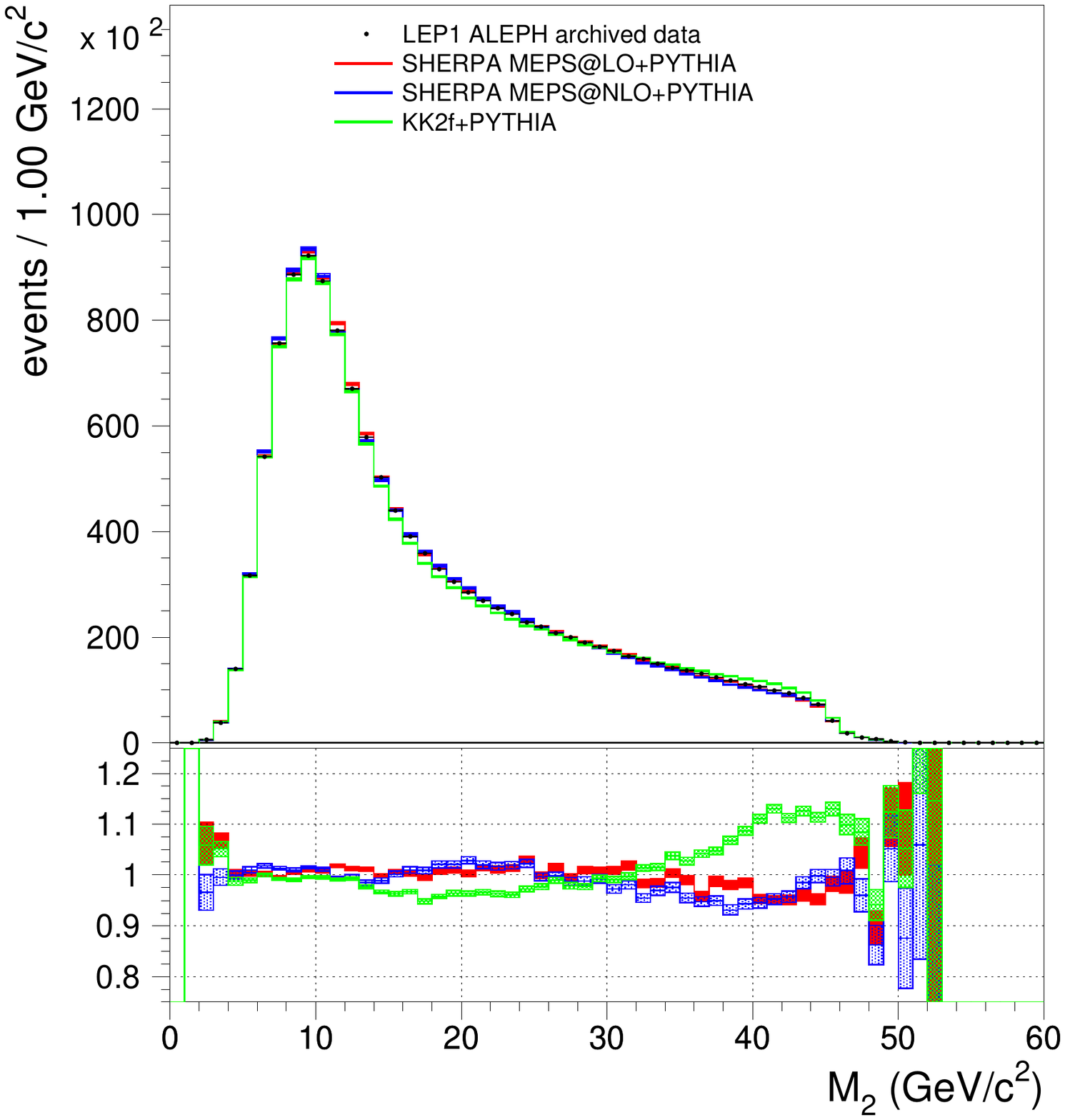}}\hspace{.35in}
\subfigure[]{\includegraphics[width=2.5in,bb=80 150 520 720]{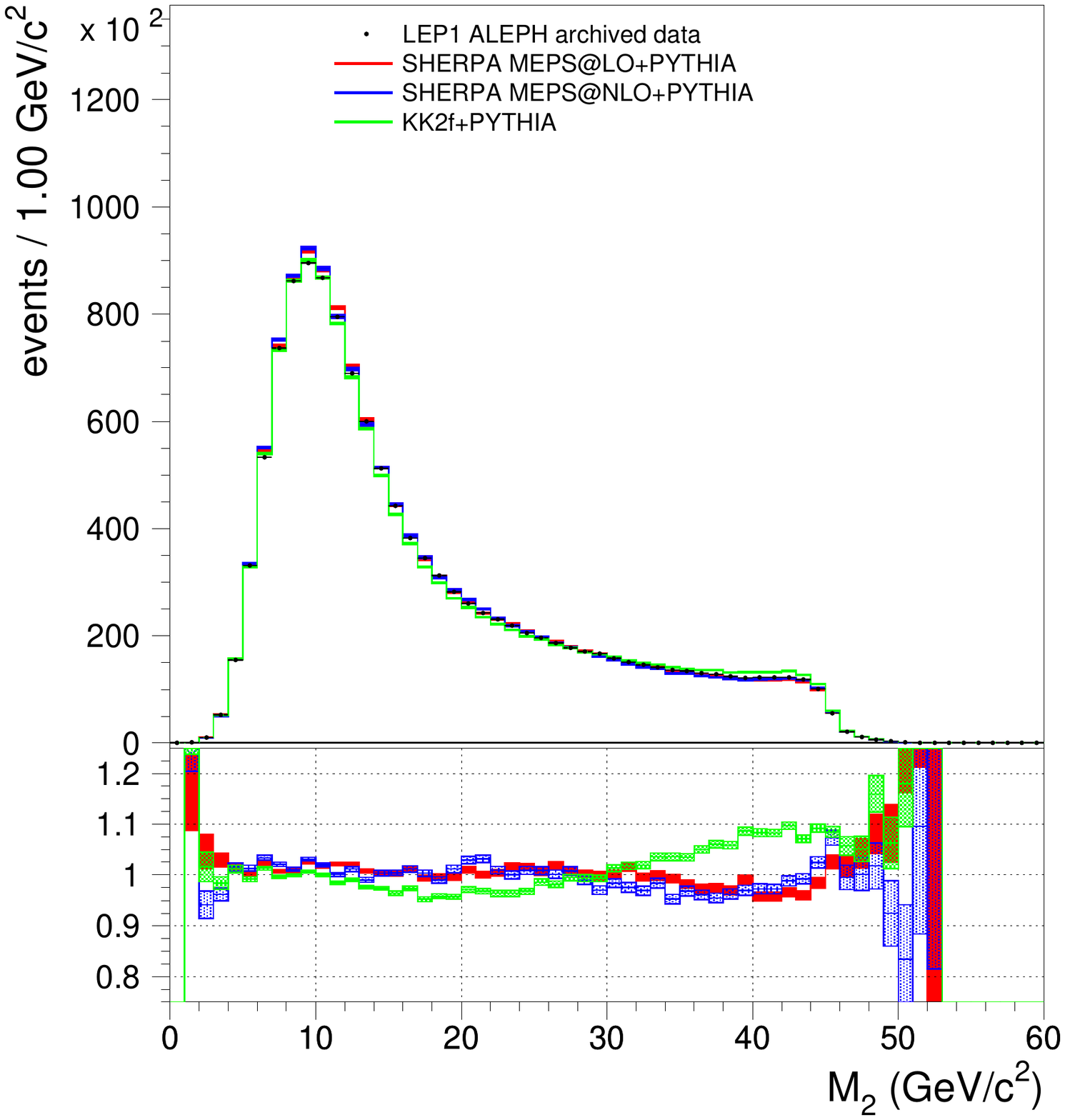}}
\end{center}
\caption{$M_1$ (top) and $M_2$ (bottom) at preselection level at LEP1.  Plots on the left are from LUCLUS jets, while those on the right are from DURHAM jets.}
\label{fig:m1m2l1}
\end{figure}

Lastly, we plot the masses $M_1$ and $M_2$ at preselection level.  These are shown in Fig. \ref{fig:m1m2l1} for both LUCLUS and DURHAM jet clustering at LEP1.  We see significant improvement in the $M_1$ distributions in Fig. \ref{fig:m1m2l1} (a) and (b) in both SHERPA samples compared to KK2f.  Similar comments also apply to the $M_2$ distribution in Fig. \ref{fig:m1m2l1} (c) and (d).  Deviations of the SHERPA MC samples from the data are at the level of a few percent over most of the range.

In summary, we see almost universal improvement for the four-jet variables in going from KK2f to the SHERPA samples, particularly the LO SHERPA sample.  In particular, we see improvement in the dijet masses important for many previous LEP analyses and the studies we perform in Ref. \cite{paper3}.  Additional data-MC plots can be found in Appendix \ref{moredists}.

\section{Reweighting Procedure}
\label{reweight}

Above, we have seen significant improvement in the modelling for four-jet variables in moving from KK2f to the SHERPA samples, particularly the SHERPA LO sample.  Of prime importance for our purposes is to have proper modelling of $M_1$ and $M_2$ and, if possible, their correlations.  For this reason, it is clear that, for our purposes, the LO SHERPA sample is the most reliable of the available choices.

While the SHERPA MC produces improved distributions for $M_1$ and $M_2$, discrepancies do remain at both LEP1 and LEP2.  To try to optimally model the QCD contribution to these distributions at LEP2, we take advantage of the fact that we can directly compare our LEP1 MC samples with the LEP1 dataset and with each other.  We thus explore the possibility to remove some of the data-MC disagreement in these variables at LEP2 by reweighting MC events with the same correction factors as would be needed to achieve data-MC agreement at LEP1.  In this section, we construct our reweighting procedure and show results before and after the procedure is applied.  A discussion of the interpretation and the limits of the validity of our reweighting procedure are postponed until Section \ref{disc}.  Unless otherwise specified, all dijet masses described in this section are constructed from jets clustered using the LUCLUS algorithm.

For a given variable, e.g. $M_1$, we can correct any of our LEP1 MC samples to agree with LEP1 data through multiplicative factors.  We then correct the LEP2 MC by applying these same correction factors.  However, to do this, we need a mapping from a variable at LEP1 to that same variable at a given LEP2 value of $\sqrt{s}$.  We construct such a map as detailed below.  Our map is motivated by the observation that the distributions for $\Sigma$, $\Delta$, $M_1$, and $M_2$ fairly closely scale with $\sqrt{s}$.  To test our mapping and reweighting procedure, we correct our three LEP2 MC samples and compare the MC samples before and after reweighting to see if the procedure improves the agreement among them.  

\begin{figure}[h]
\begin{center}
\subfigure[]{\includegraphics[width=1.7in,bb=80 150 520 720]{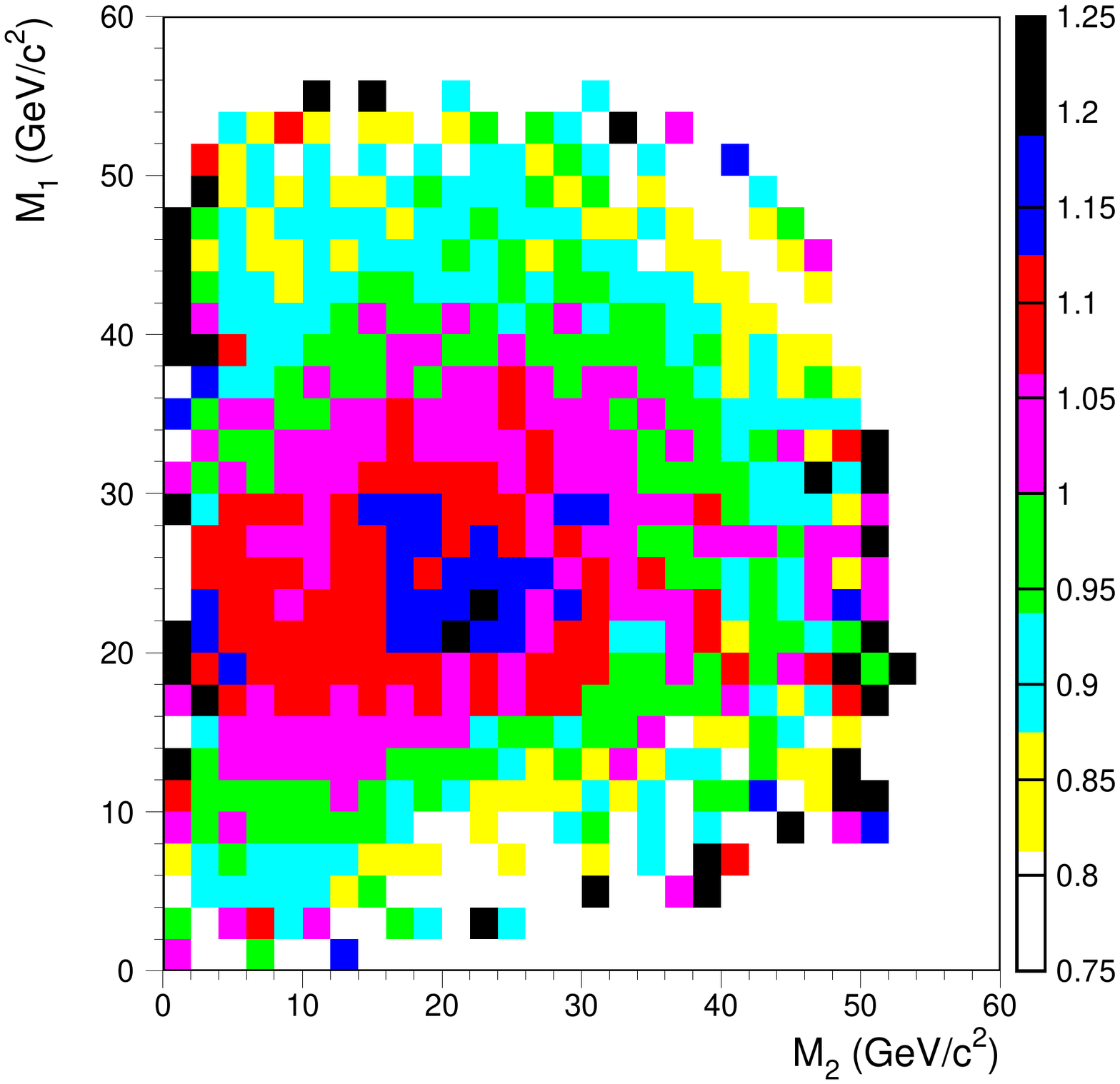}}\hspace{.45in}
\subfigure[]{\includegraphics[width=1.7in,bb=80 150 520 720]{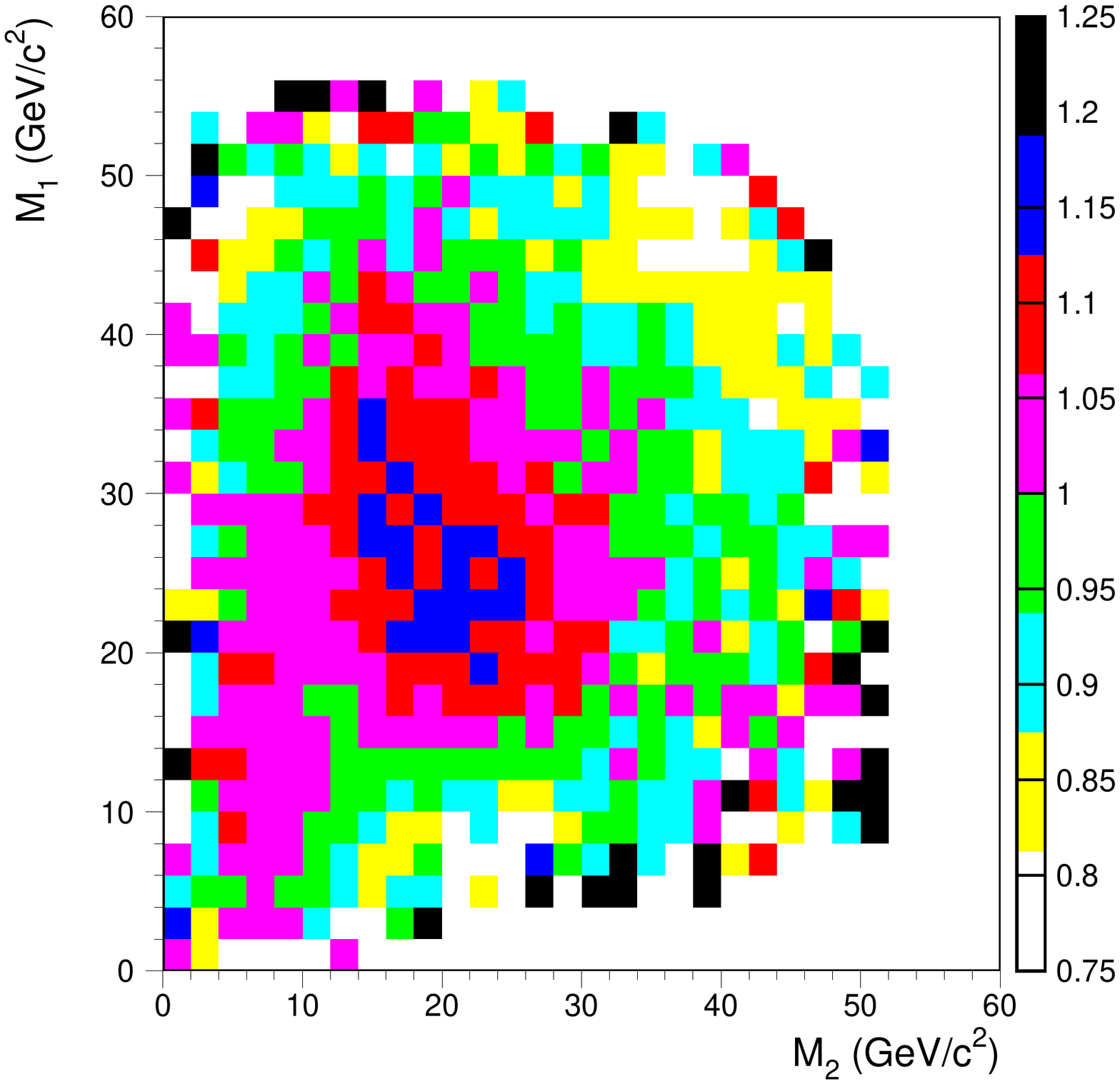}}\hspace{.45in}
\subfigure[]{\includegraphics[width=1.7in,bb=80 150 520 720]{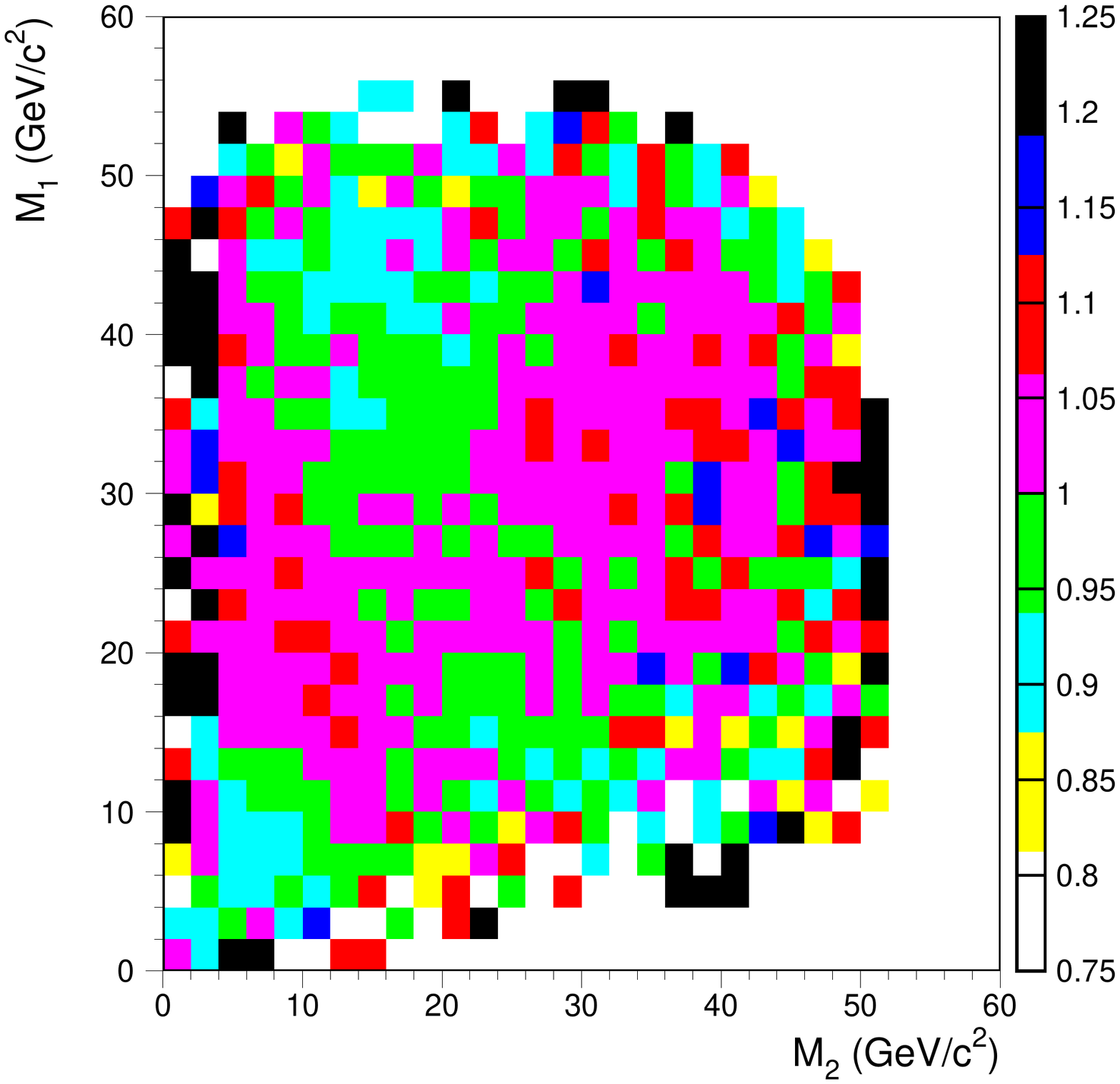}}\\
\subfigure[]{\includegraphics[width=1.7in,bb=80 150 520 720]{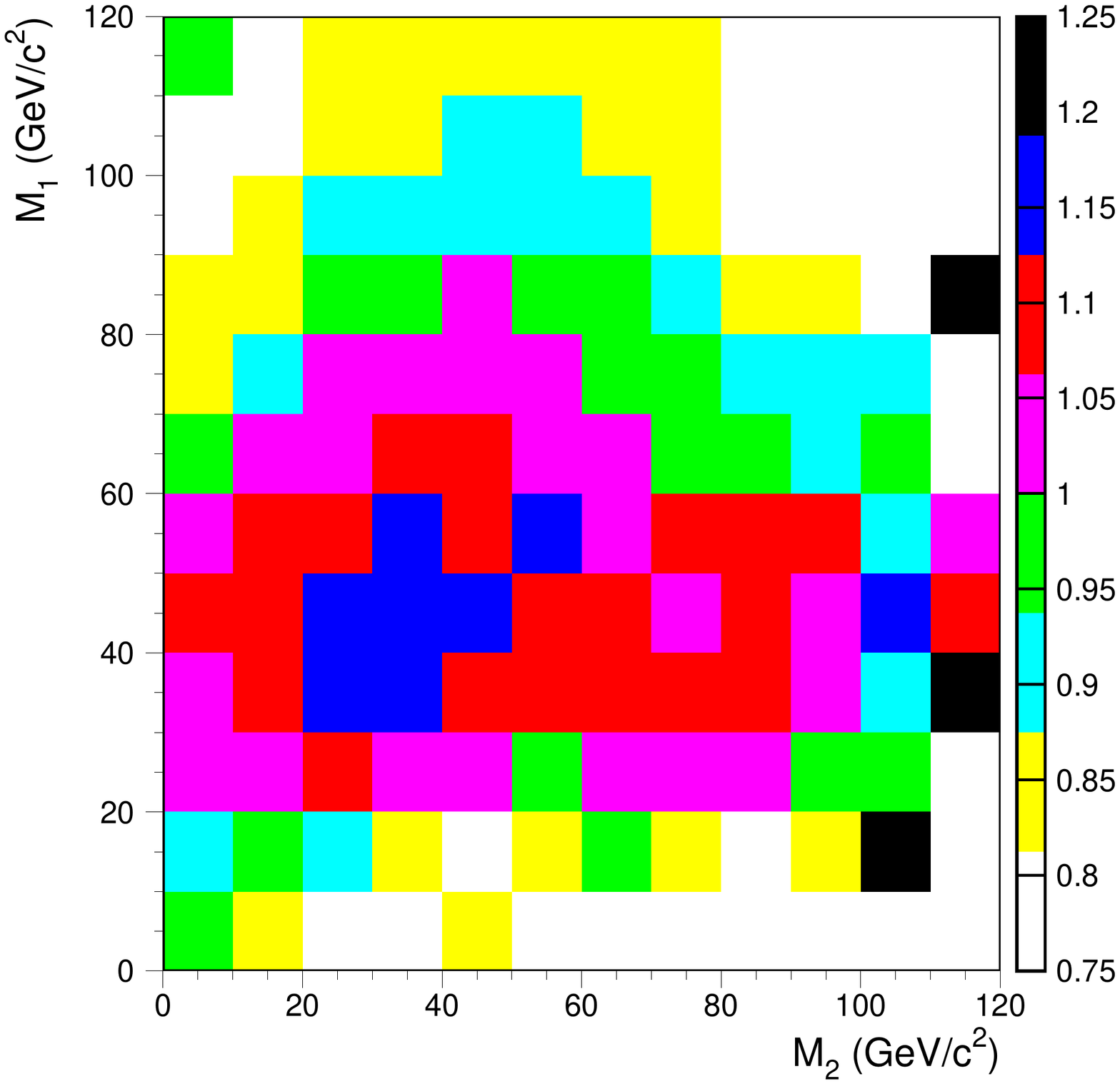}}\hspace{.45in}
\subfigure[]{\includegraphics[width=1.7in,bb=80 150 520 720]{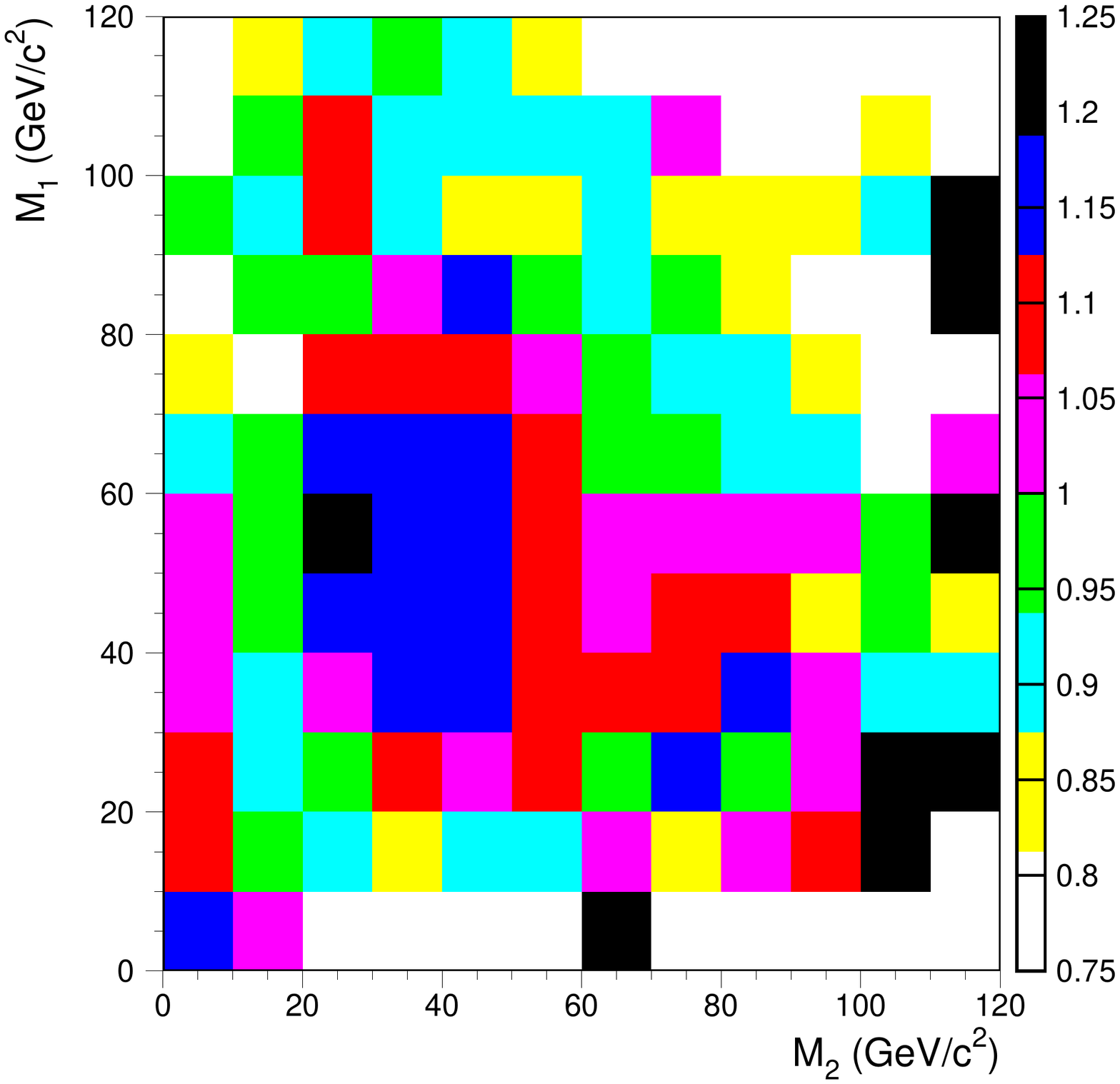}}\hspace{.45in}
\subfigure[]{\includegraphics[width=1.7in,bb=80 150 520 720]{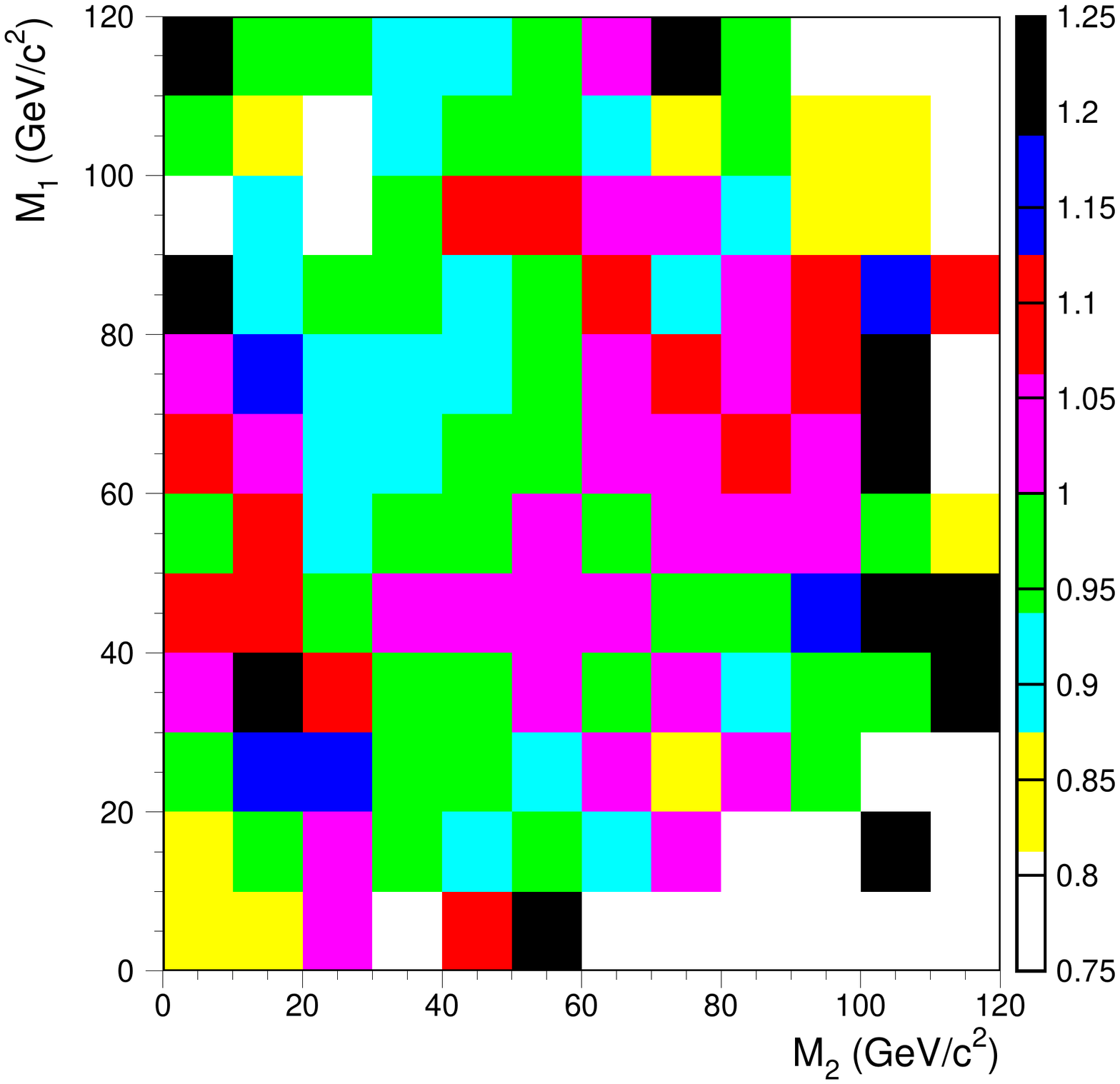}}
\end{center}
\caption{The ratio of the QCD expectations from different MC samples in the $M_1$-$M_2$ plane at preselection level at LEP1 (top) and LEP2 (bottom) energies.  The first column ((a) and (d)) gives the ratio of LO SHERPA MC to KK2f, the second column ((b) and (e)) gives the ratio of NLO SHERPA to KK2f, and the last column ((c) and (f)) gives the ratio of LO SHERPA to NLO SHERPA.  All LEP2 center-of-mass energies are included. }
\label{fig:mcratios}
\end{figure}

We begin by studying the agreement of the $M_1$ and $M_2$ distributions among the three MC samples at LEP1 and LEP2.  This is displayed in the two-dimensional plots shown in Fig. \ref{fig:mcratios}.  In each of these plots, we plot the ratio of two MC expectations; all center-of-mass energies are included at LEP2.  In the first column of Fig. \ref{fig:mcratios}, we plot the ratio of the LO SHERPA to KK2f, in the second column, we plot the ratio of the NLO SHERPA to KK2f, and in the third column we plot the ratio of the two SHERPA samples.  In each case, we show these two-dimensional planes for LEP1 and LEP2 separately.  It should be noted that these plots do not include non-QCD samples, such as the four-fermion or two-photon MC.

We first notice that the differences between KK2f and the two SHERPA samples are larger than the differences between the SHERPA samples themselves.  In particular, we can see ``islands'' of disagreement in the first two columns of Fig. \ref{fig:mcratios}.  Additionally, for each of the pairs of MC samples, the behavior in the plane is qualitatively similar for LEP1 and LEP2, up to an overall scale in $M_1$ and $M_2$ that is very similar to $\sqrt{s}/(91.2\mbox{ GeV})$.  This scaling is not exact, however; while some quantities closely related to $M_1$ and $M_2$ (e.g., the energy of the most energetic jet) scale with $\sqrt{s}$, other quantities (such as jet masses) do not.

We thus perform a mapping of LEP1 MC to LEP2 MC as follows.  Let us consider the mapping of the LO SHERPA sample expectation for $M_1$ at LEP1, $M_{1,LO,LEP1}$ to its analogue at the LEP2 center-of-mass energy $\sqrt{s}=188.6$ GeV, $M_{1,LO,188.6}$ We take the (${\cal O}(10^5)$) LO SHERPA events which pass preselection in the $188.6$ GeV sample and order them in increasing value of $M_{1,LO,188.6}$.  We take an equal number of events from our $91.2$ GeV LO sample and order them similarly.  We then map events from the LEP1 sample to the $188.6$ GeV sample according to their order in the samples.  This defines a map from $M_{1,LO,LEP1}$ to $M_{1,LO,188.6}$.

\begin{figure}[h]
\begin{center}
\subfigure[]{\includegraphics[width=2.5in,bb=80 150 520 720]{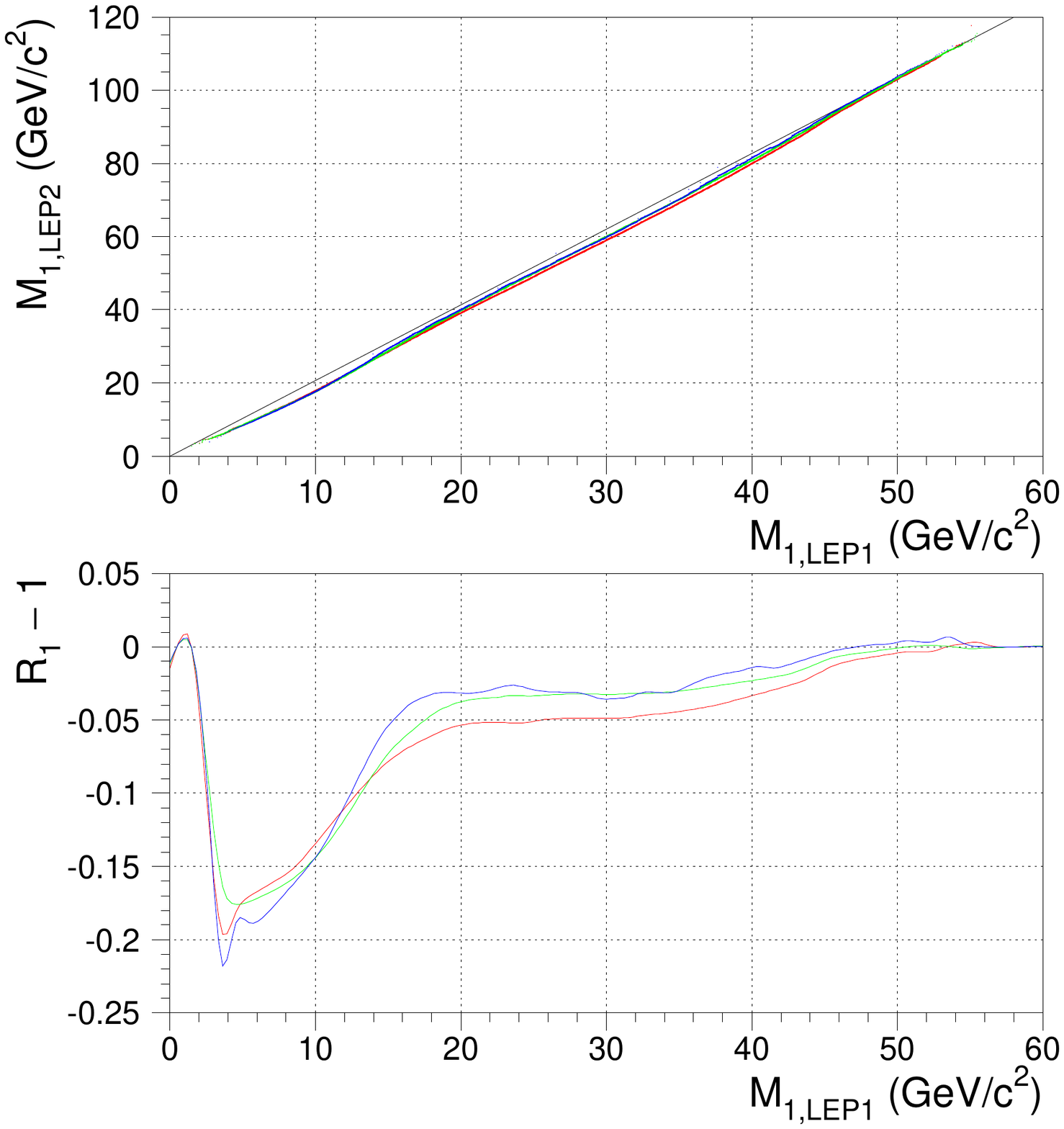}}\hspace{.35in}
\subfigure[]{\includegraphics[width=2.5in,bb=80 150 520 720]{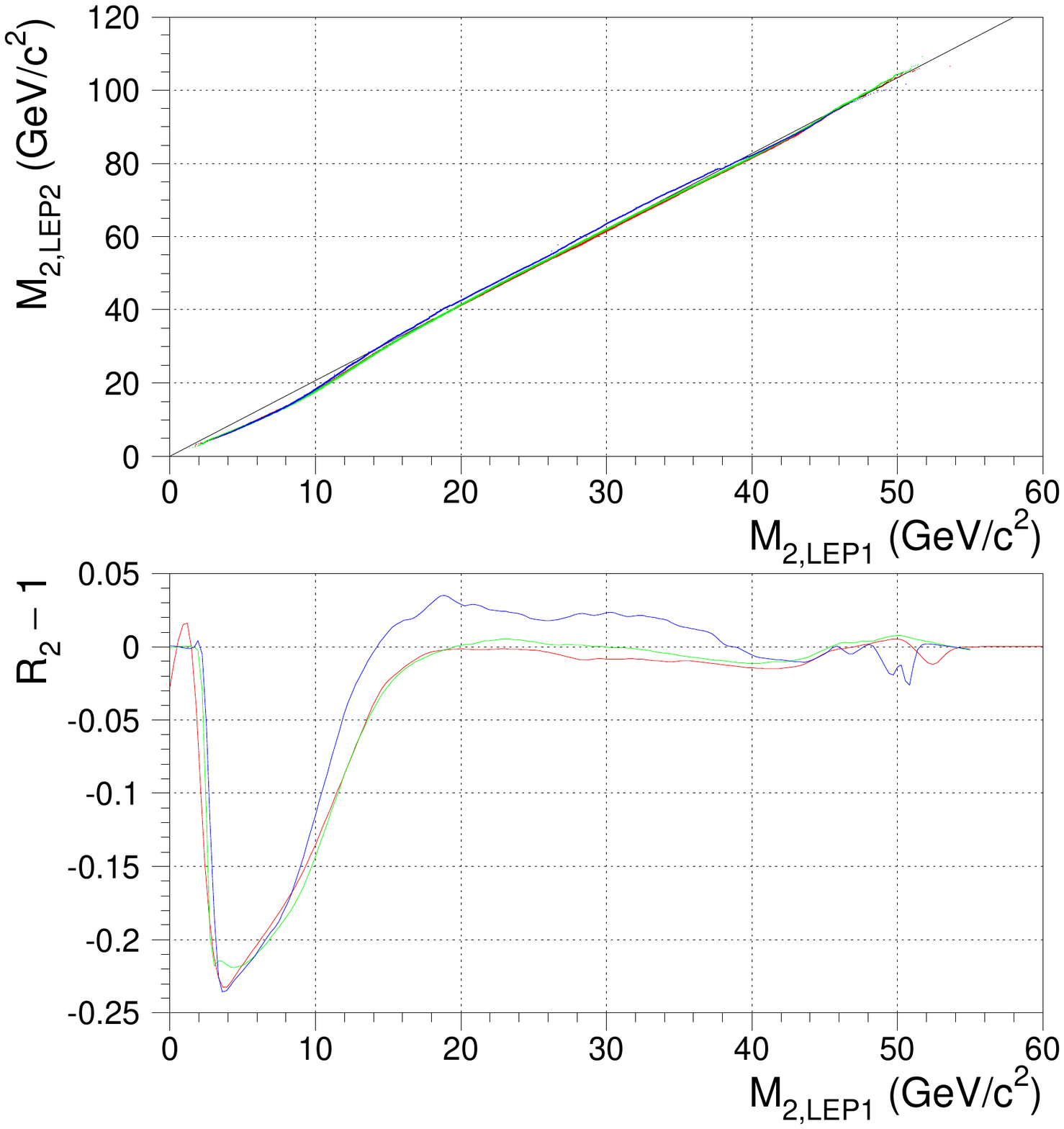}}
\end{center}
\caption{The (a) $M_1$ and (b) $M_2$ mapping functions for the LO (red) and NLO (blue) SHERPA and the KK2f (green) samples, for $\sqrt{s}=188.6$ GeV.  We also plot, for comparison, the naive scaling with center-of-mass energy. The top of each plot shows the mapping functions, while the bottom shows the fractional difference between the mapping function and the naive scaling.}
\label{fig:mapfuncs}
\end{figure}

%
We define analogous maps for the NLO and KK2f generations and also for $M_2$.  We plot the mapping functions for $\sqrt{s}=188.6$ GeV in Fig. \ref{fig:mapfuncs}.  These are all compared with a naive linear map, where $M_{i,LEP1}\rightarrow M_{i,188.6}=M_{i,LEP1}\times (188.6\mbox{ GeV})/(91.2\mbox{ GeV})$.  We also plot $R_i-1$, where $R_i=M_{i,188.6}/M_{i,LEP1}\times (91.2\mbox{ GeV})/(188.6\mbox{ GeV})$ to better illustrate the relation to the linear map.

We smooth the two-dimensional $M_1$ and $M_2$ distributions for both the LO MC, $\rho_{LO,LEP1}(M_{1},M_{2})$ and the data, $\rho_{data,LEP1}(M_{1},M_{2})$ at LEP1.  We then correct the LO SHERPA sample at LEP2 as follows.  For a LO SHERPA MC event generated at $\sqrt{s}=188.6$ GeV, which has values for $M_1$ and $M_2$ which we denote by $M_{1,LO,188.6}$ and $M_{2,LO,188.6}$, we multiply this event by the reweighting factor 
\begin{equation}
\frac{\rho_{data,LEP1}(M_{1,LO,LEP1},M_{2,LO,LEP1})}{\rho_{LO,LEP1}(M_{1,LO,LEP1},M_{2,LO,LEP1})}
\end{equation}
where $M_{i,LO,LEP1}$ and $M_{i,LO,188.6}$ ($i=1,2$) are related by our mapping functions.  We correct the NLO and KK2f samples similarly.

\begin{figure}[h]
\begin{center}
\subfigure[]{\includegraphics[width=1.5in,bb=80 150 520 720]{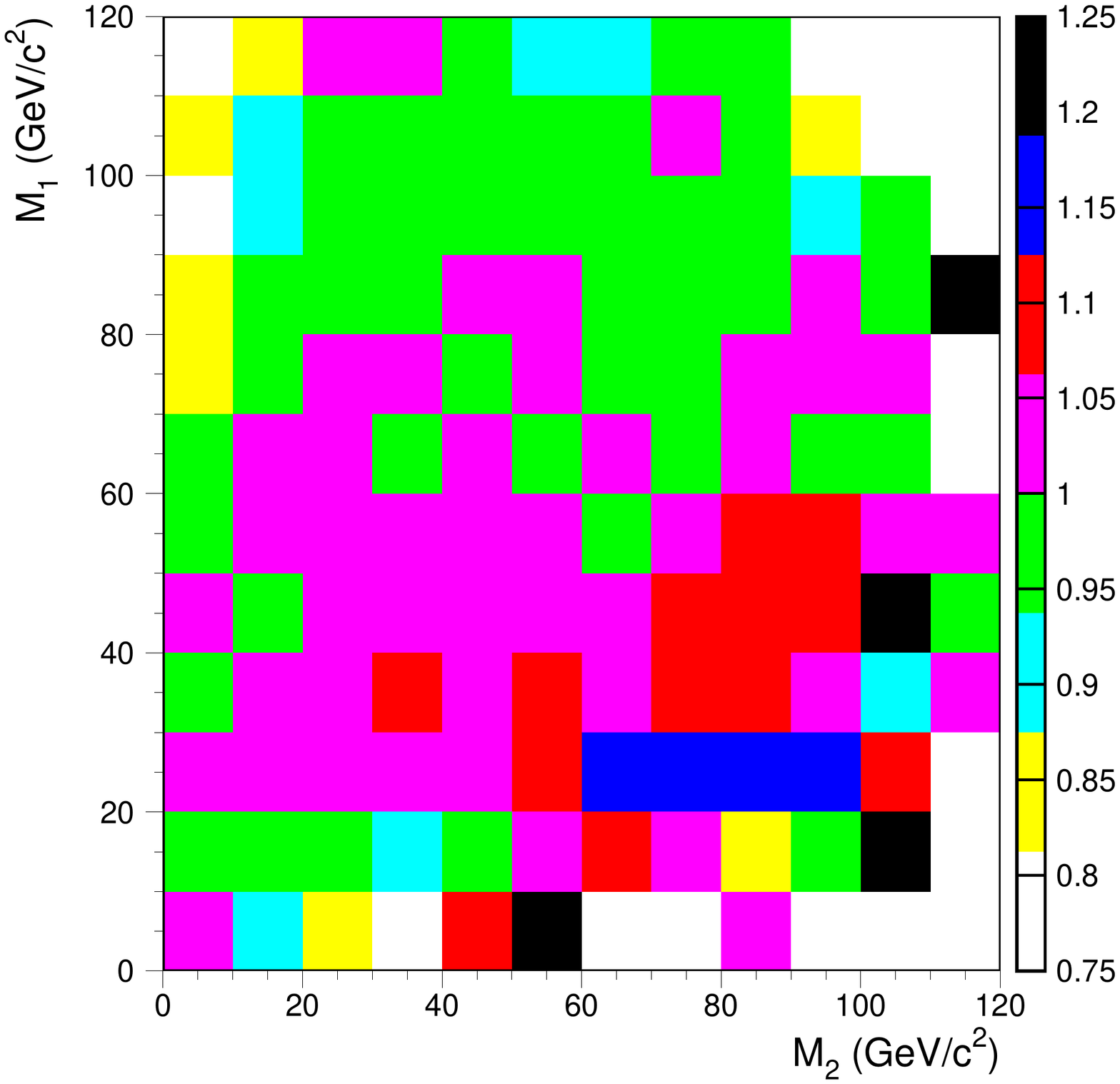}}\hspace{.35in}
\subfigure[]{\includegraphics[width=1.5in,bb=80 150 520 720]{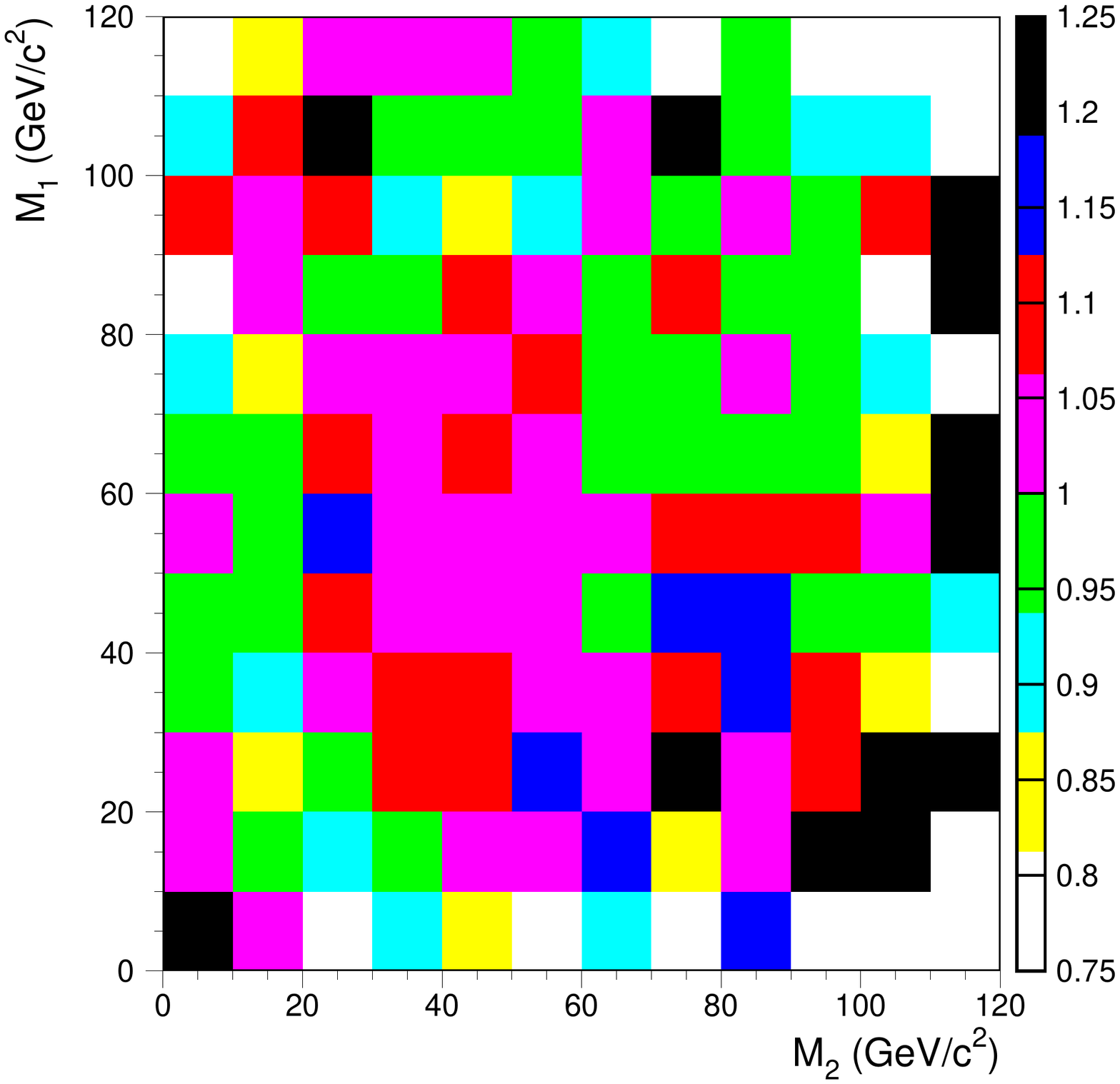}}\hspace{.35in}
\subfigure[]{\includegraphics[width=1.5in,bb=80 150 520 720]{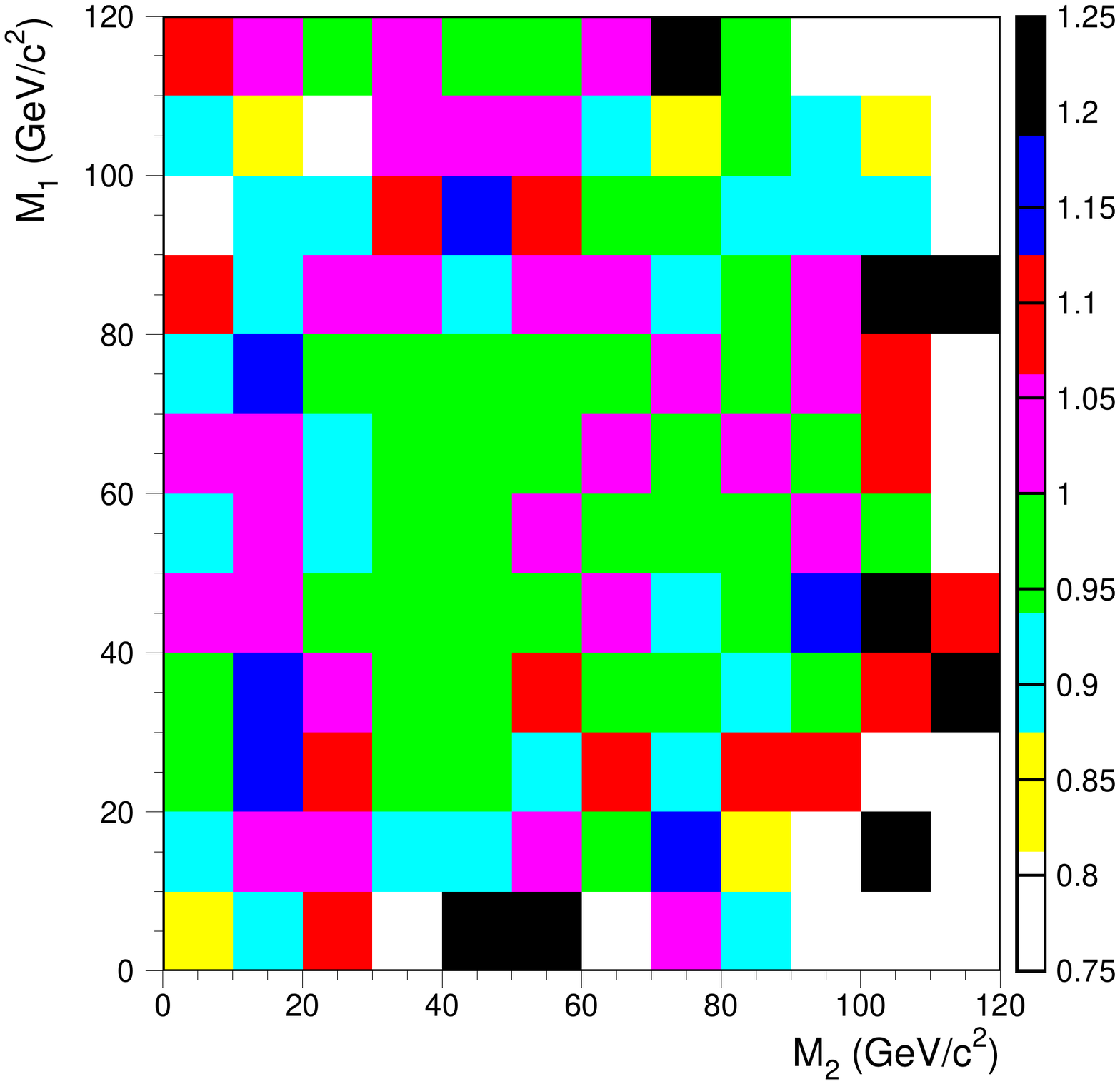}}\hspace{.35in}\\
\subfigure[]{\includegraphics[width=1.5in,bb=80 150 520 720]{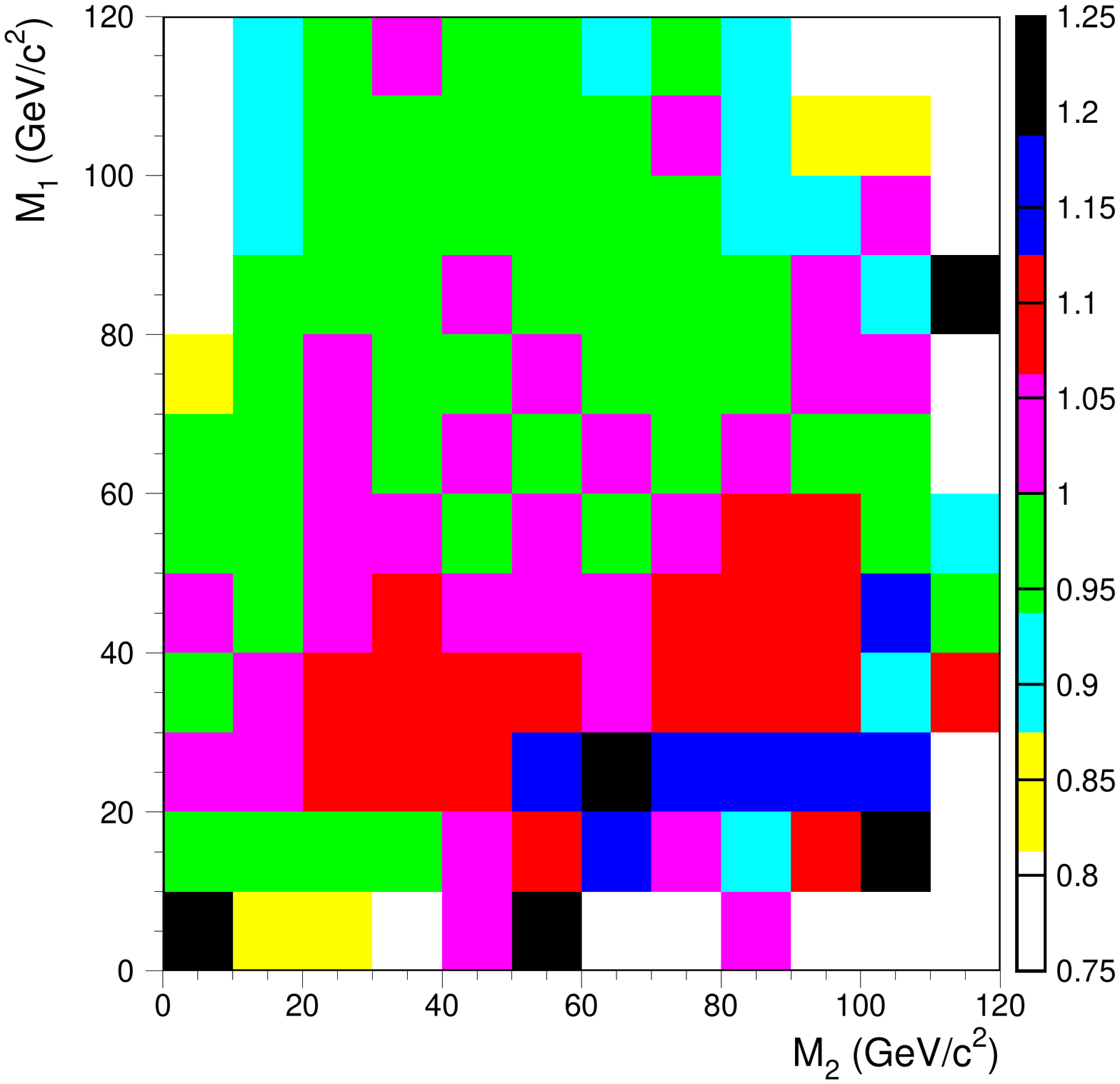}}\hspace{.35in}
\subfigure[]{\includegraphics[width=1.5in,bb=80 150 520 720]{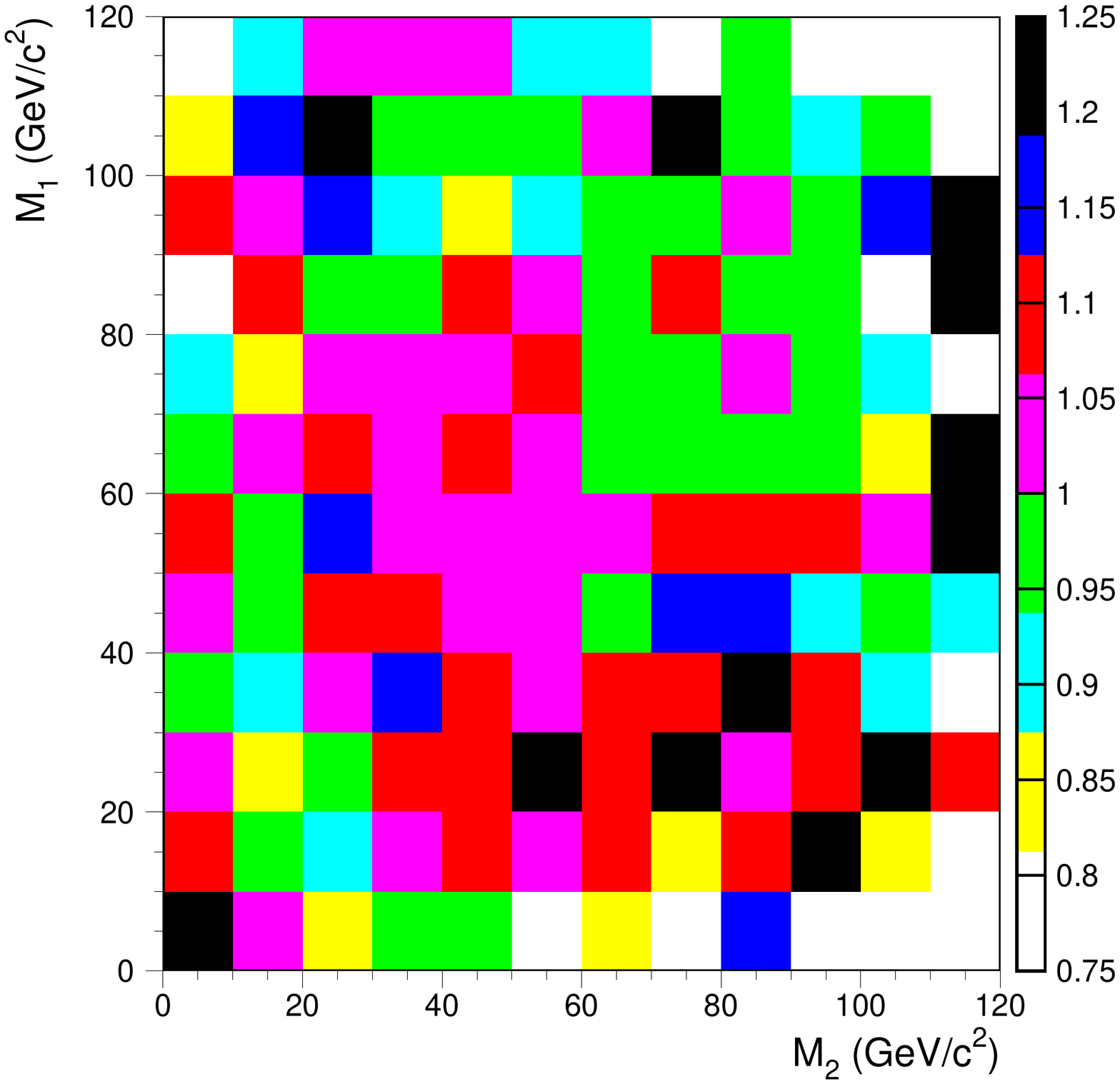}}\hspace{.35in}
\subfigure[]{\includegraphics[width=1.5in,bb=80 150 520 720]{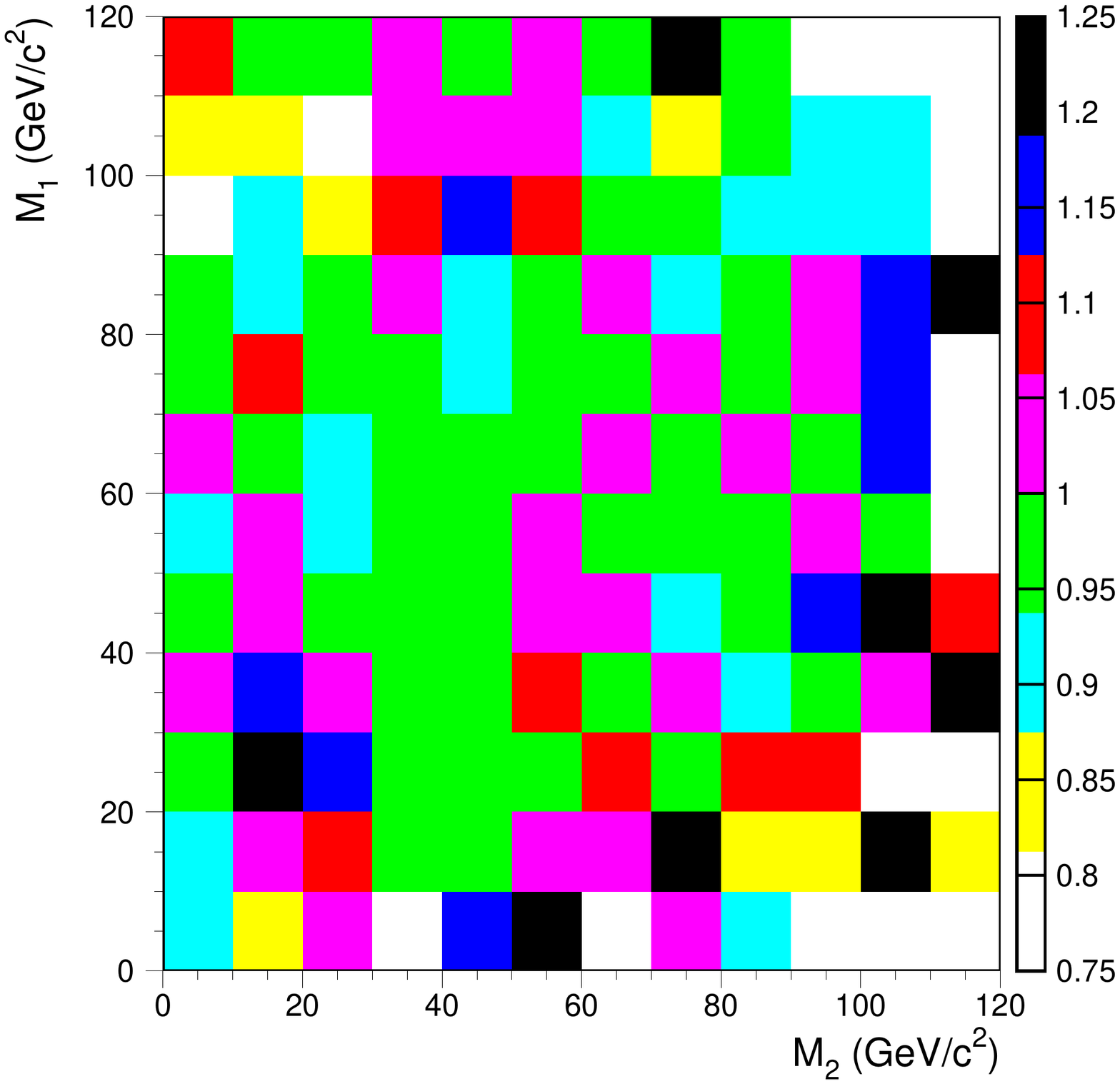}}\hspace{.35in}\\
\subfigure[]{\includegraphics[width=1.5in,bb=80 150 520 720]{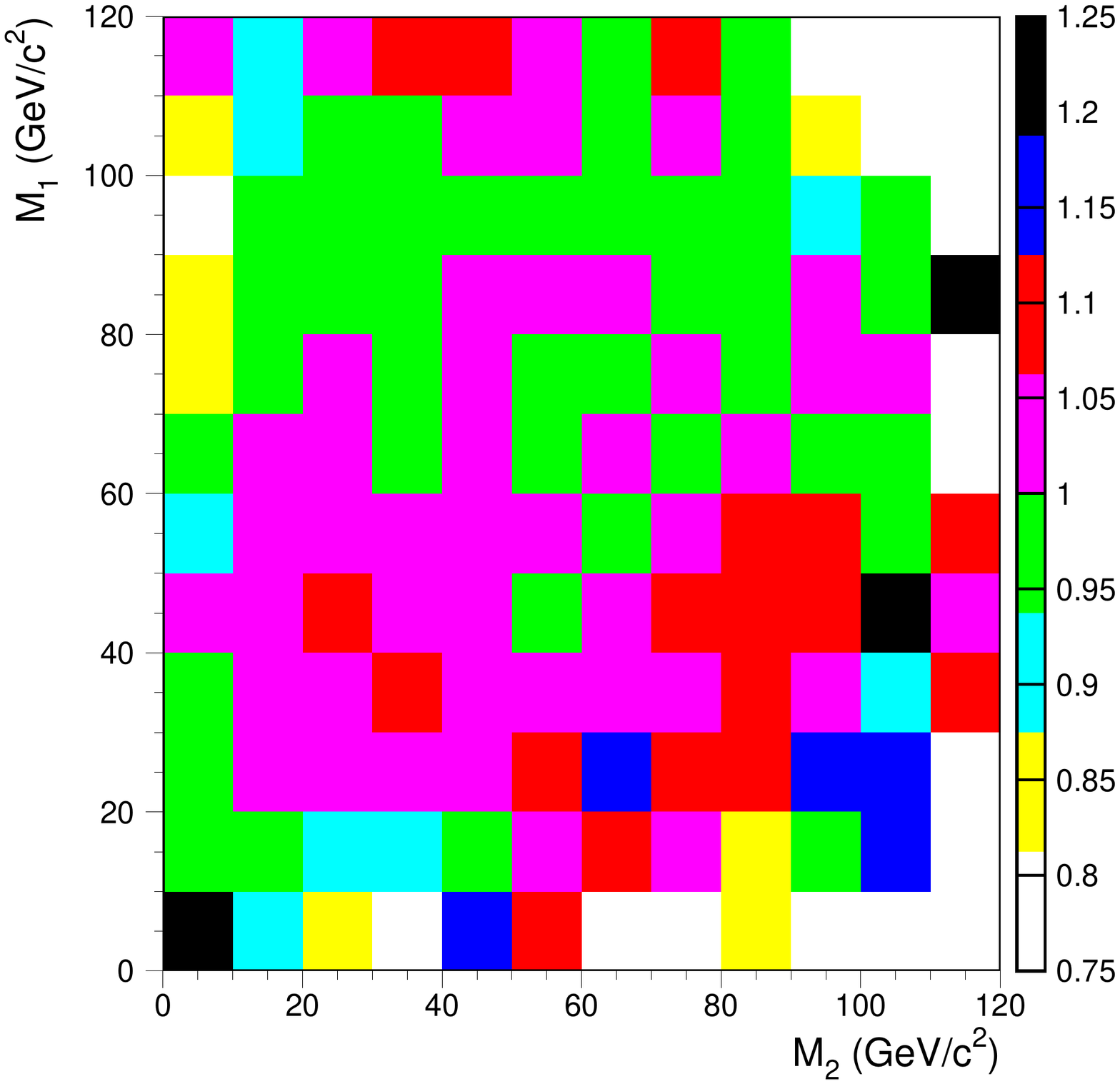}}\hspace{.35in}
\subfigure[]{\includegraphics[width=1.5in,bb=80 150 520 720]{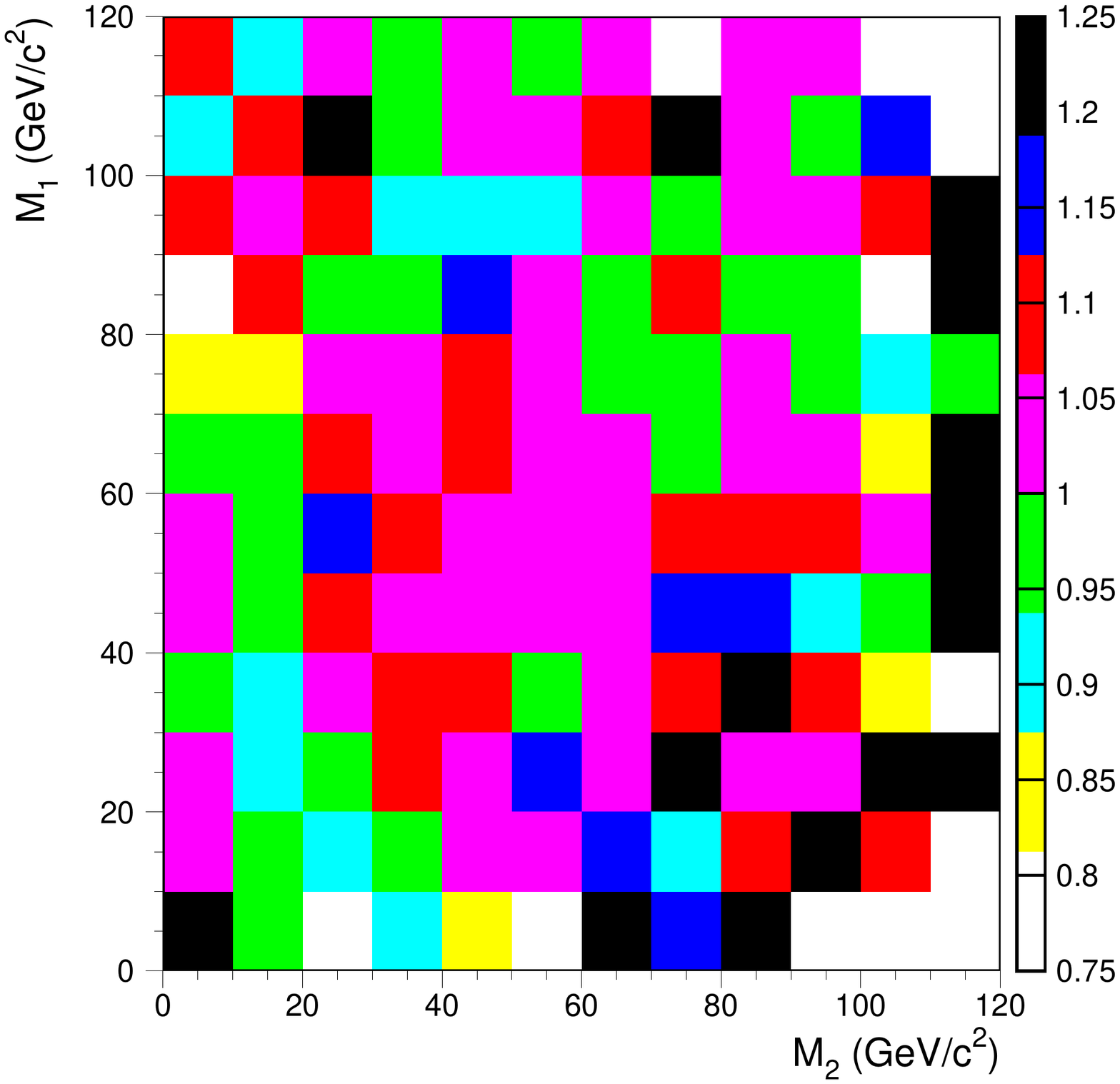}}\hspace{.35in}
\subfigure[]{\includegraphics[width=1.5in,bb=80 150 520 720]{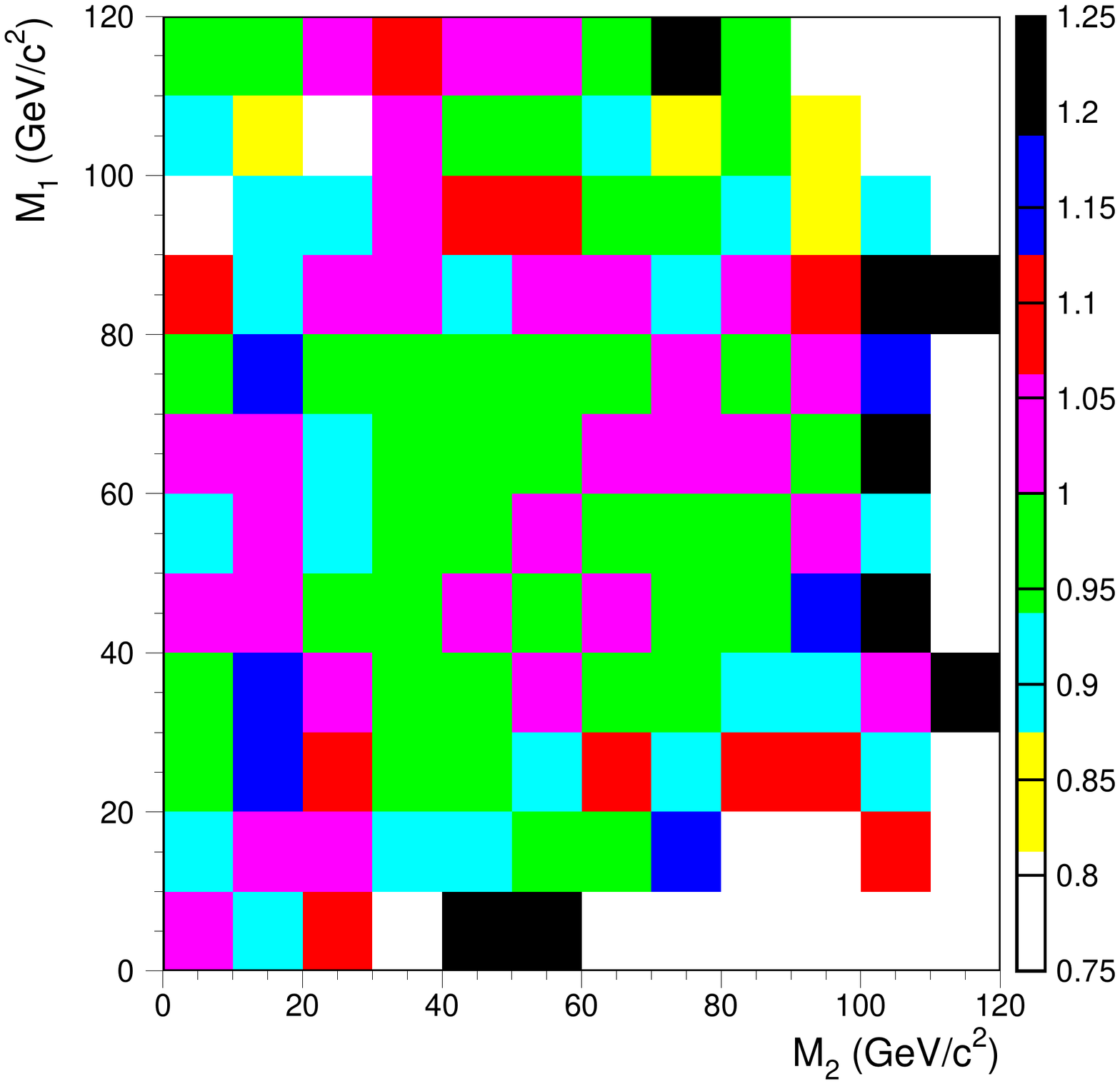}}\hspace{.35in}\\
\end{center}
\caption{The relative disagreement between LO SHERPA and KK2f (left column), NLO SHERPA and KK2f (middle column) and LO and NLO SHERPA (right column) samples in the $M_1$-$M_2$ plane after reweighting with the mapping defined by ordering in $M_1$ and $M_2$ (top line),  using a naive linear map (middle), and repeating the map with $\Sigma$ and $\Delta$ instead of $M_1$ and $M_2$ (bottom).}
\label{fig:rwmcratios}
\end{figure}

We plot the results of our reweighting procedure applied among the MC samples in Fig. \ref{fig:rwmcratios}.  In each plot, we show the agreement of two MC samples in the $M_1$-$M_2$ plane at LEP2 after reweighting.  The first column gives the ratio of the LO SHERPA sample to KK2f, the second column gives the ratio of the NLO SHERPA sample to KK2f, and the last column gives the ratio of the LO and NLO SHERPA samples.  In the top line, we show the results after reweighting using the map described above using the ordering of $M_1$ and $M_2$.  The middle line shows the results of a naive linear map.  Lastly, we note that our mapping procedure maps $M_1$ and $M_2$ separately and does not include any correlations between $M_1$ and $M_2$.  As a check we repeat our reweighting procedure, but using $\Sigma$ and $\Delta$ instead of $M_1$ and $M_2$.  The ratios for the reweighted MC samples for this rotated map are given in the last line of Fig. \ref{fig:rwmcratios}.

In each case, we see that the mapping defined by ordering in $M_1$ and $M_2$ performs better than the linear map, and both significantly reduce the discrepancy between MC samples compared to the unreweighted comparisons in Fig. \ref{fig:mcratios}.  The discrepancy between the MC samples is now $\lesssim 5\%$ over most of the plane.  Additionally, the structures in the LO/KK2f and NLO/KK2f comparisons in Fig. \ref{fig:mcratios} are now greatly reduced or absent in Fig. \ref{fig:rwmcratios}. Additionally, we see that the map using the rotated masses $\Sigma$ and $\Delta$ gives very similar results to that using $M_1$ and $M_2$, which implies that our reweighting procedure is not terribly harmed by neglecting correlations between $M_1$ and $M_2$.

\begin{figure}[h]
\begin{center}
\subfigure[]{\includegraphics[width=2.4in,bb=80 150 520 720]{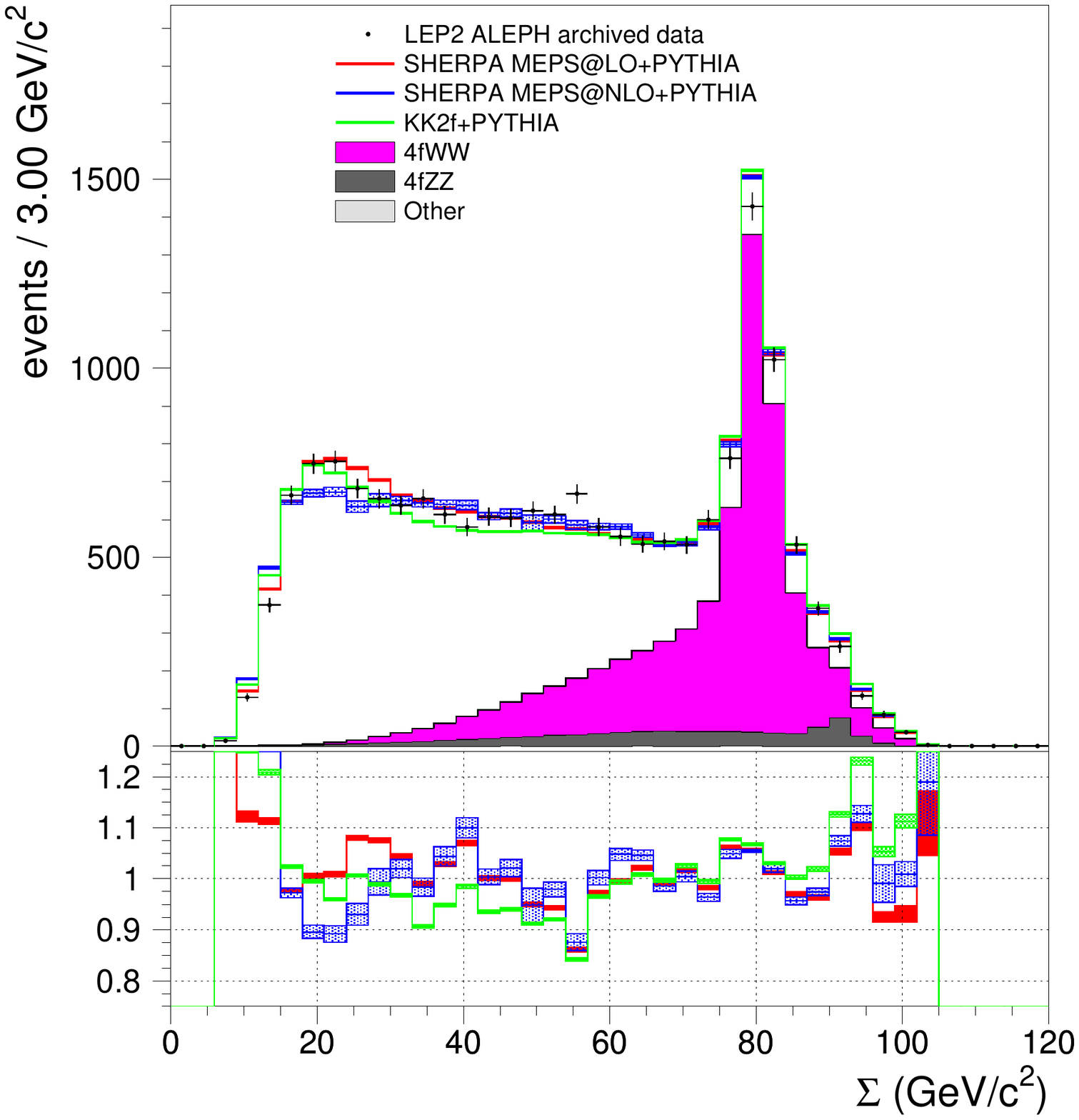}}\hspace{.35in}
\subfigure[]{\includegraphics[width=2.4in,bb=80 150 520 720]{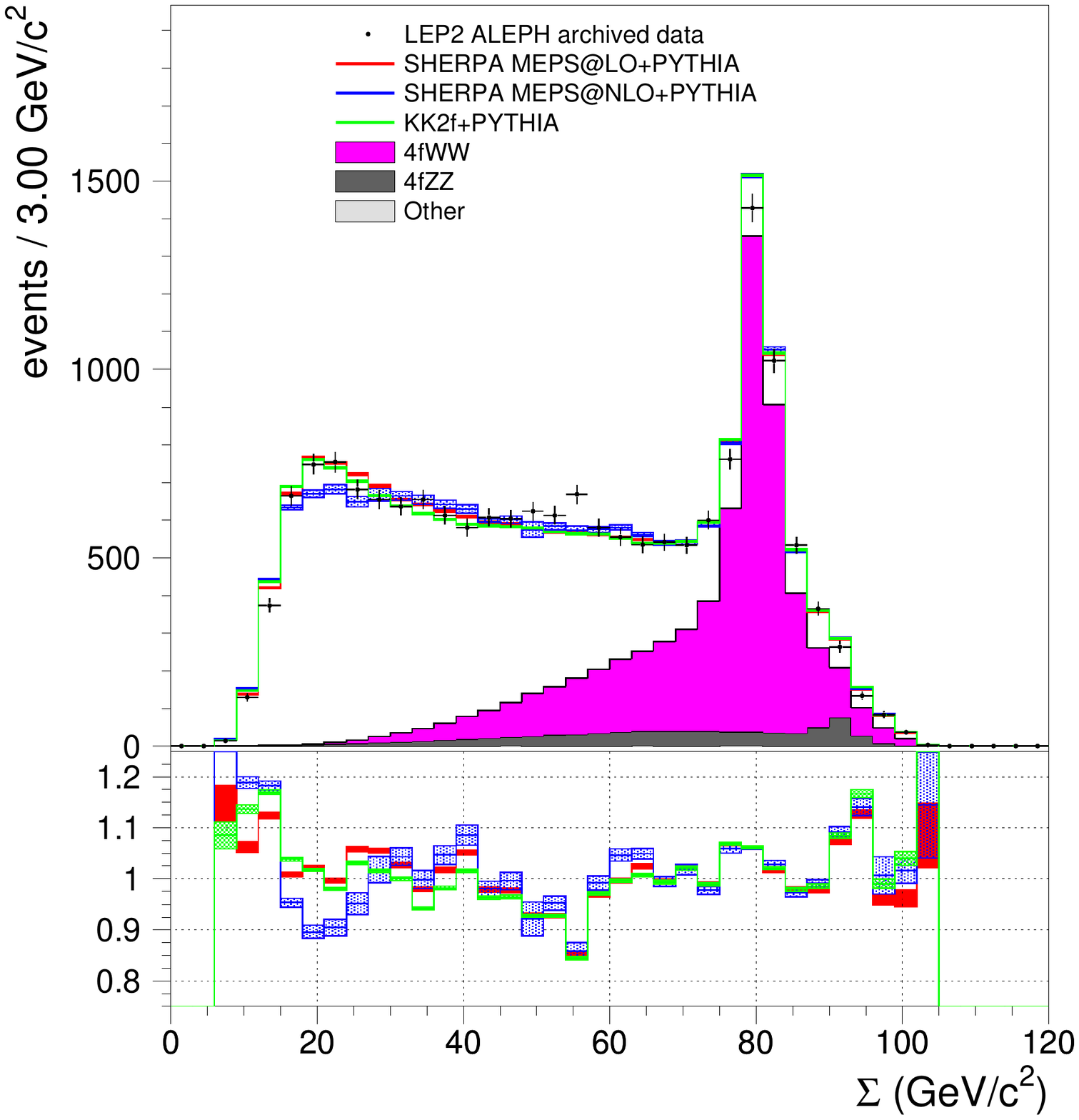}}\\
\subfigure[]{\includegraphics[width=2.4in,bb=80 150 520 720]{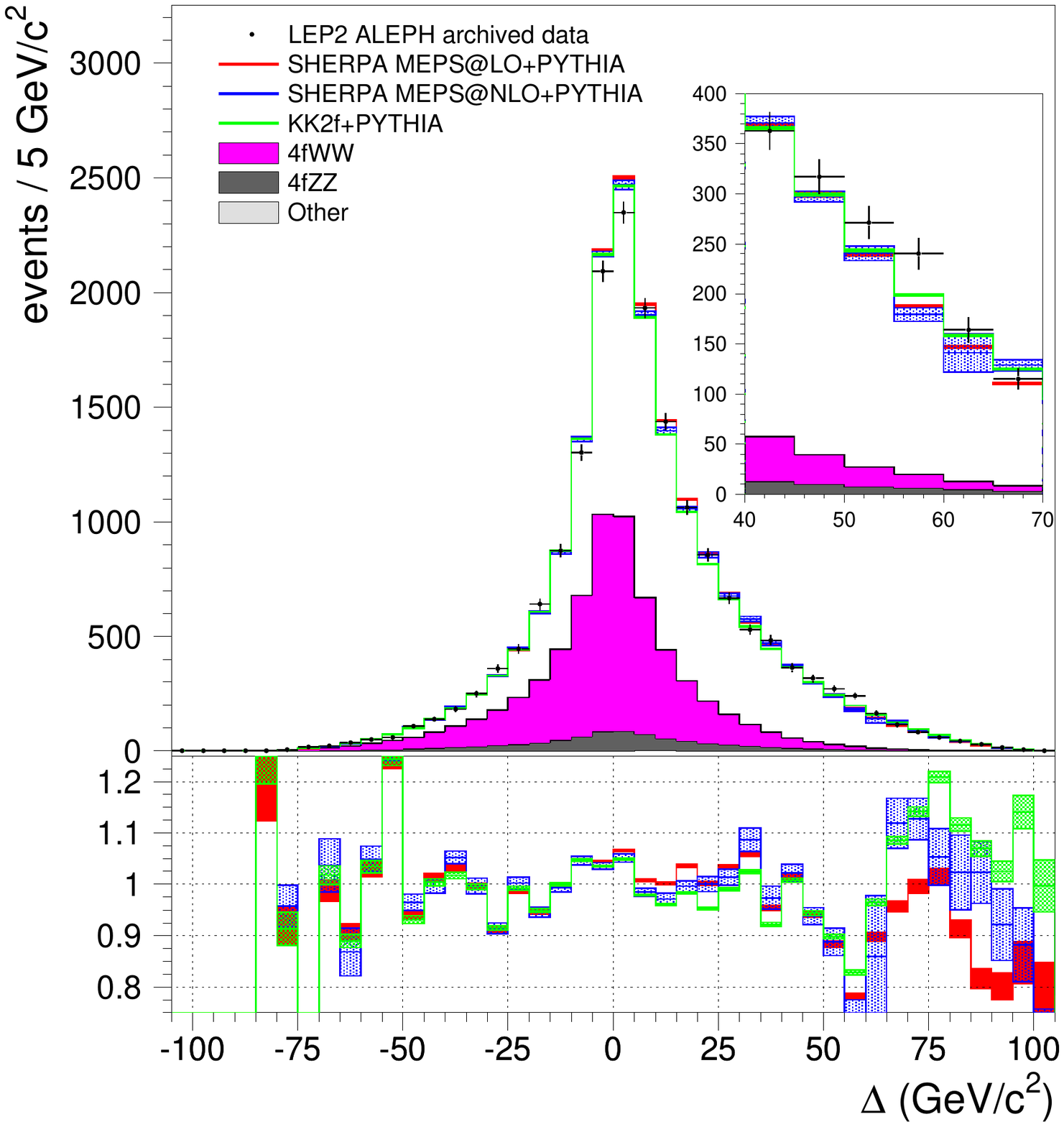}}\hspace{.35in}
\subfigure[]{\includegraphics[width=2.4in,bb=80 150 520 720]{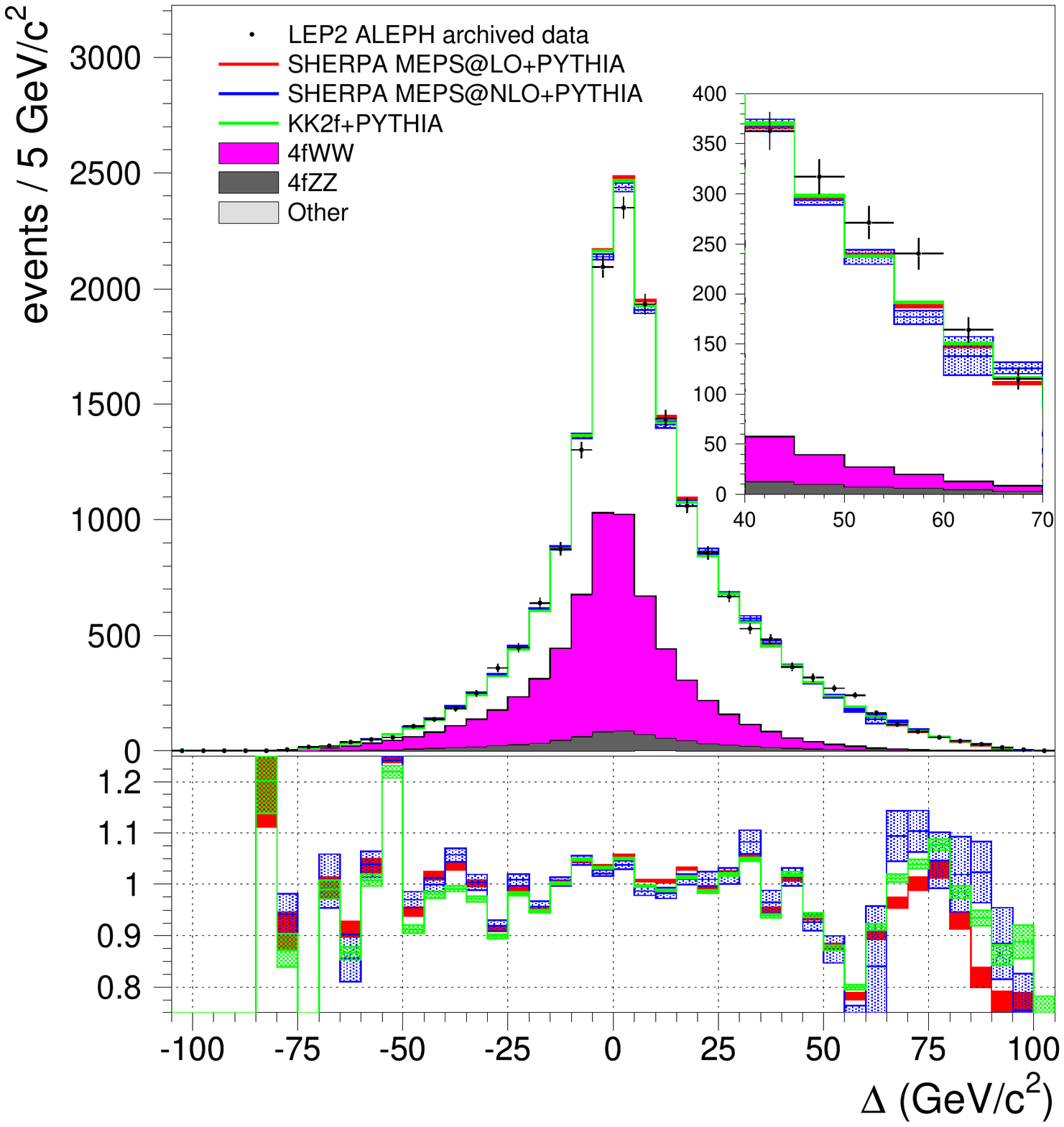}}
\end{center}
\caption{Comparison between LEP2 data and MC samples for (top) $\Sigma$ and (bottom) $\Delta$ for LUCLUS-clustered jets.  Plots include all LEP2 energies and show the LO SHERPA, NLO SHERPA, and KK2f predictions (left) before and (right) after reweighting. Insets in (c) and (d) show the excess region.}
\label{fig:datavsrewmasssumdiff}
\end{figure}

We have seen from our results above that of the three MC samples, the LO SHERPA best describes the data before reweighting.  While the NLO SHERPA sample often improves on the KK2f for four-jet variables, it typically underperforms relative to the LO tune.  At the same time, the fact that the three LEP2 MC samples come into better agreement when reweighted according to LEP1 data indicates that reweighting is capable of removing at least some systematic errors from the MC samples.  For these reasons, we use the reweighted LO SHERPA sample as our primary MC in continued studies.  The unreweighted LO MC as well as the NLO SHERPA and KK2f samples will be retained for systematic studies.

\begin{figure}[h]
\begin{center}
\subfigure[]{\includegraphics[width=2.3in,bb=80 150 520 720]{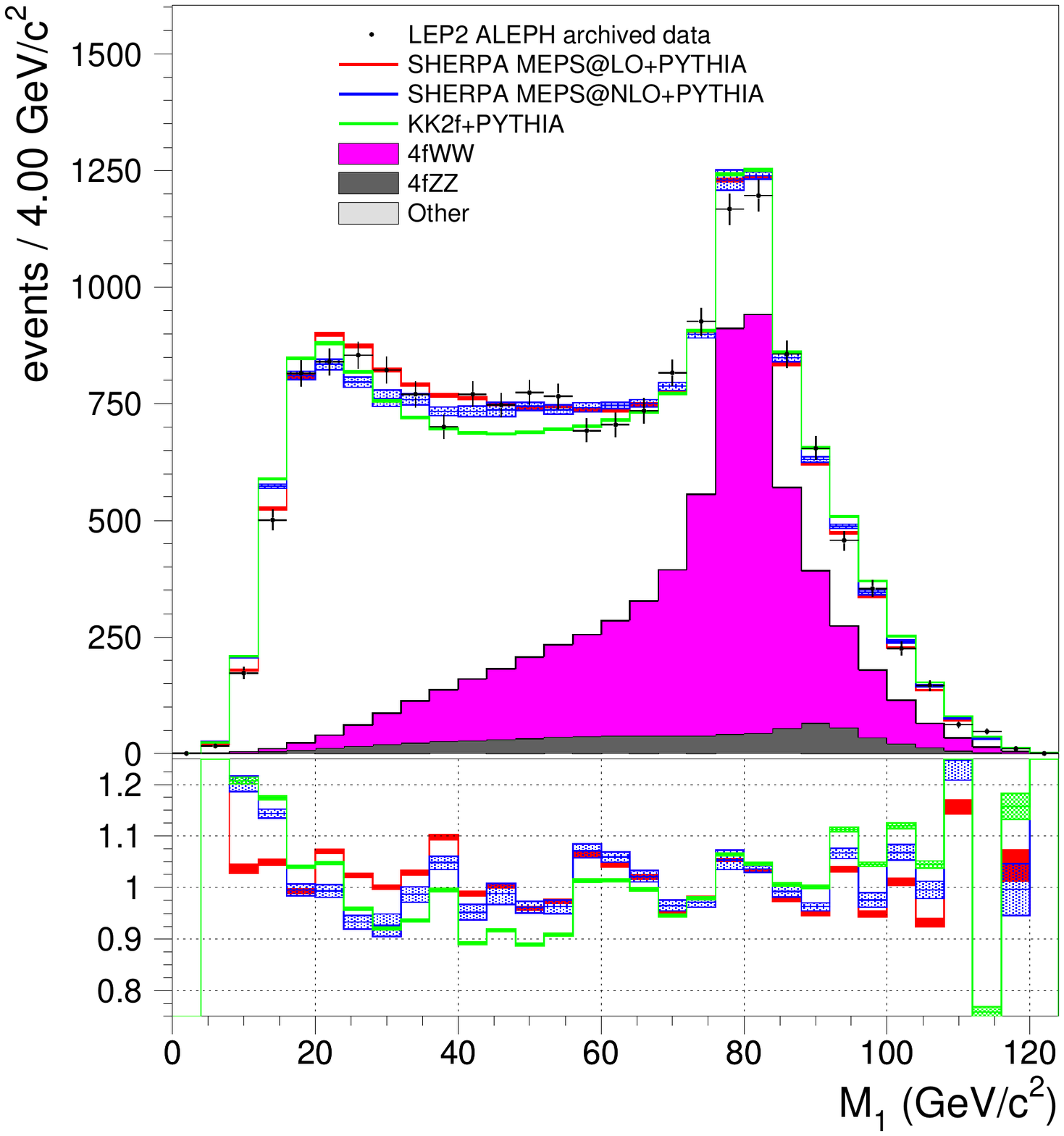}}\hspace{.35in}
\subfigure[]{\includegraphics[width=2.3in,bb=80 150 520 720]{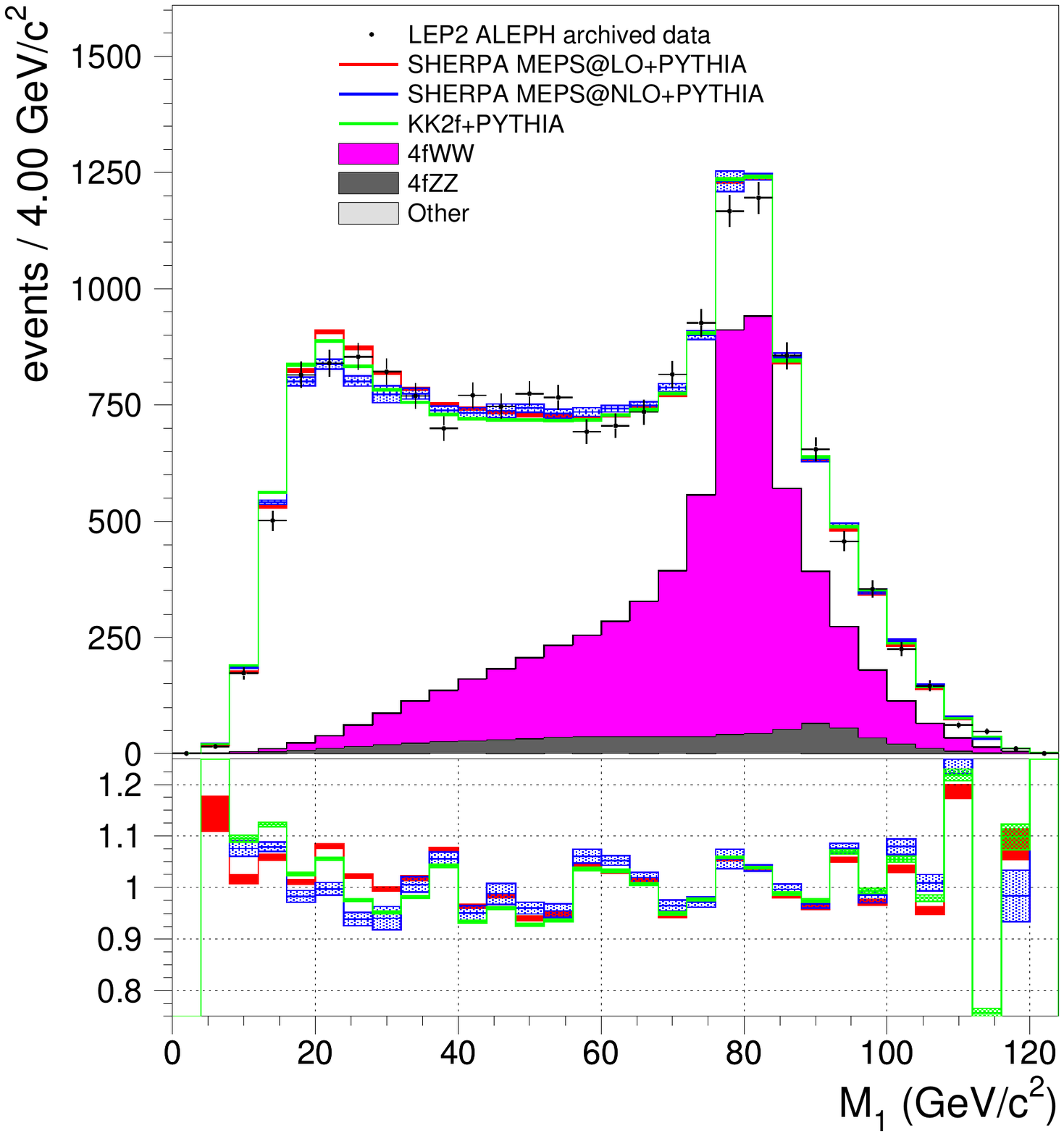}}\\
\subfigure[]{\includegraphics[width=2.3in,bb=80 150 520 720]{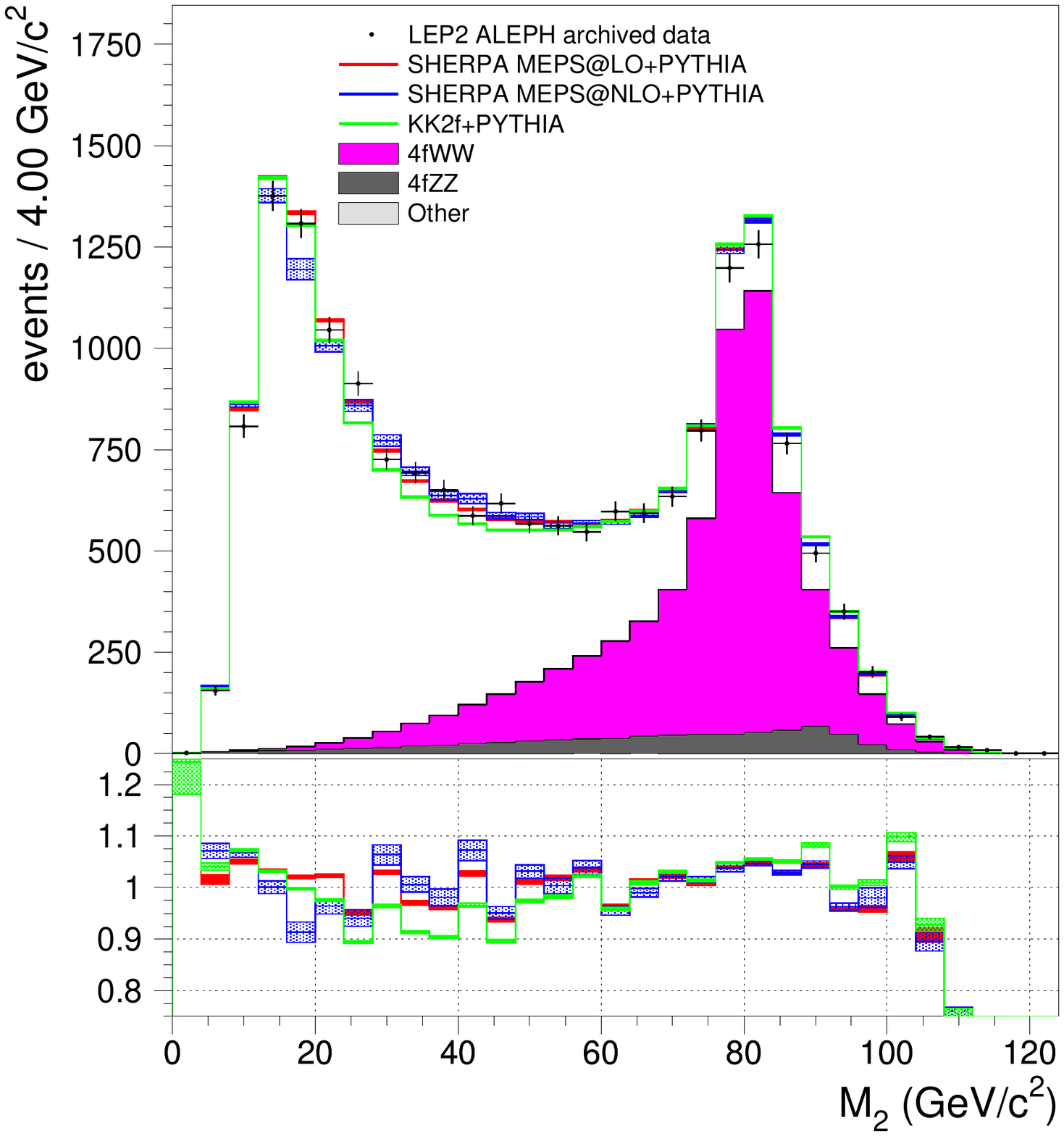}}\hspace{.35in}
\subfigure[]{\includegraphics[width=2.3in,bb=80 150 520 720]{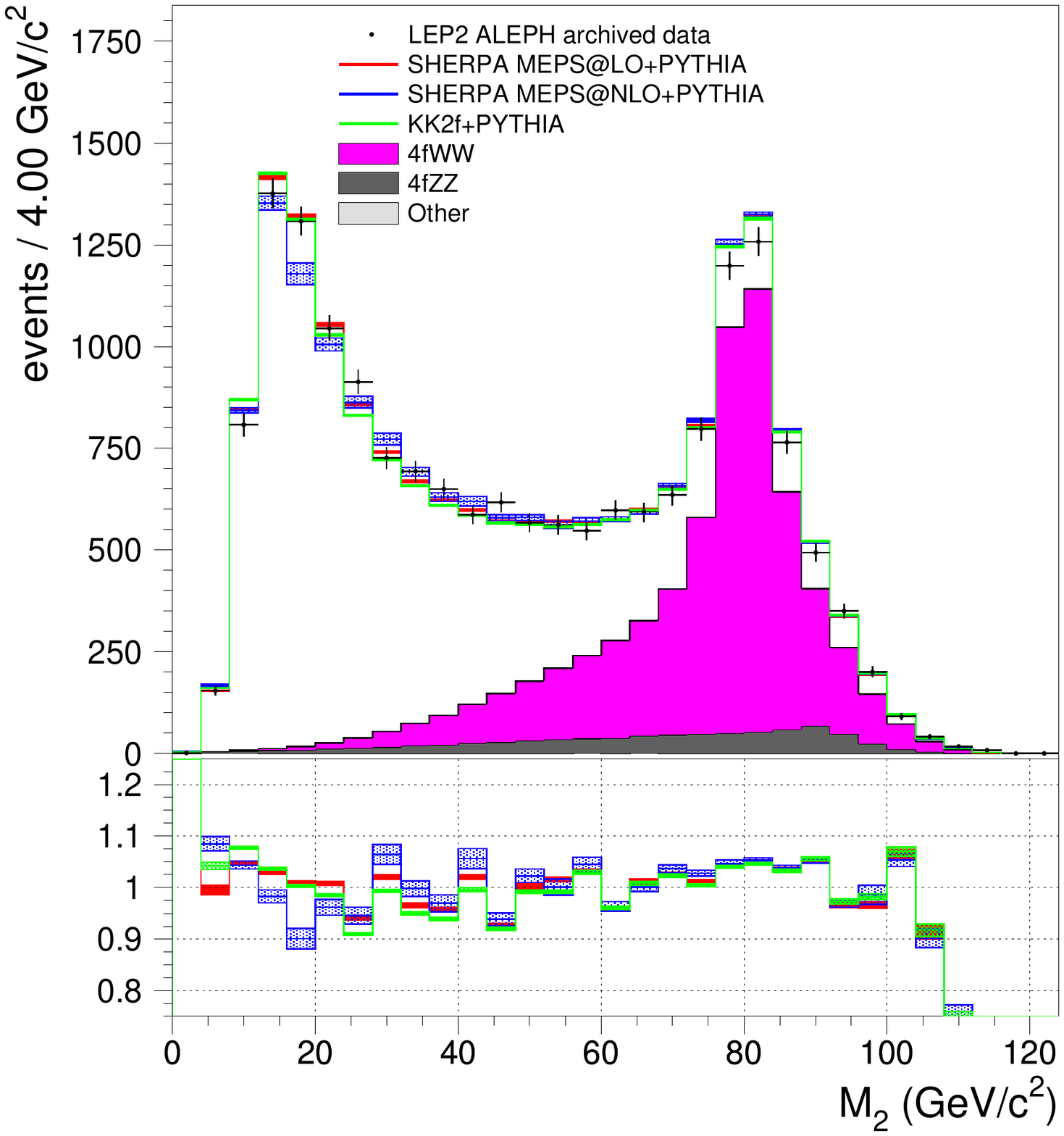}}
\end{center}
\caption{Comparison between LEP2 data and MC samples for (top) $M_1$ and (bottom) $M_2$ for LUCLUS-clustered jets.  Plots include all LEP2 energies and show the LO SHERPA, NLO SHERPA, and KK2f predictions (left) before reweighting and (right) after reweighting.}
\label{fig:datavsrewm1m2}
\end{figure}

We now compare the data to the reweighted LO MC at LEP2.   In Fig. \ref{fig:datavsrewmasssumdiff}, we show $\Sigma$ and $\Delta$, constructed with LUCLUS jets, before and after MC reweighting.  All LEP2 energies are included.  We see in Fig. \ref{fig:datavsrewmasssumdiff} (a) and (b) that the MC predictions for $\Sigma$ come into much better agreement after reweighting.  We see an excess in the data in the region of $45\mbox{ GeV}\lesssim\Sigma\lesssim 60\mbox{ GeV}$ relative to all of the reweighted MC samples.  We note that relative to the unreweighted KK2f sample (generated with the standard ALEPH tune), this excess extended down to $\Sigma\sim 30 \mbox{ GeV}$, as can be seen by comparing the data points to the KK2f curve in Fig. \ref{fig:datavsrewmasssumdiff} (a).  This region $30\mbox{ GeV}\lesssim\Sigma\lesssim 45\mbox{ GeV}$ comes into much better agreement with the LEP2 data by reweighting the KK2f MC, switching to either of the SHERPA samples, or both.

We see similar improvement in the plots of $\Delta$ in Fig. \ref{fig:datavsrewmasssumdiff} (c) and (d).  We note that all MC samples, both reweighted and unreweighted, yield an excess in the data in the region $\Delta\sim 55\mbox{ GeV}$.  We note that both the excess near $45\mbox{ GeV}\lesssim\Sigma\lesssim 60\mbox{ GeV}$ and the excess near $\Delta\sim 55\mbox{ GeV}$ do not have analogues in the LEP1 plots in Fig. \ref{fig:masssumdiff}.

We similarly plot $M_1$ and $M_2$ in Fig. \ref{fig:datavsrewm1m2}.  We again see that the three MC samples come into closer agreement with each other after reweighting.  In particular, the data shows a large excess compared to KK2f MC in the region $40\mbox{ GeV}\lesssim M_1\lesssim 55\mbox{ GeV}$ in Fig. \ref{fig:datavsrewm1m2} (a) and (b); this excess is greatly reduced by reweighting or switching to either of the SHERPA MC samples.  A smaller, but similar effect can also be seen in the distributions of $M_2$ in Fig. \ref{fig:datavsrewm1m2} (c) and (d).

\begin{figure}[h]
\begin{center}
\subfigure[]{\includegraphics[width=2.5in,bb=80 150 520 720]{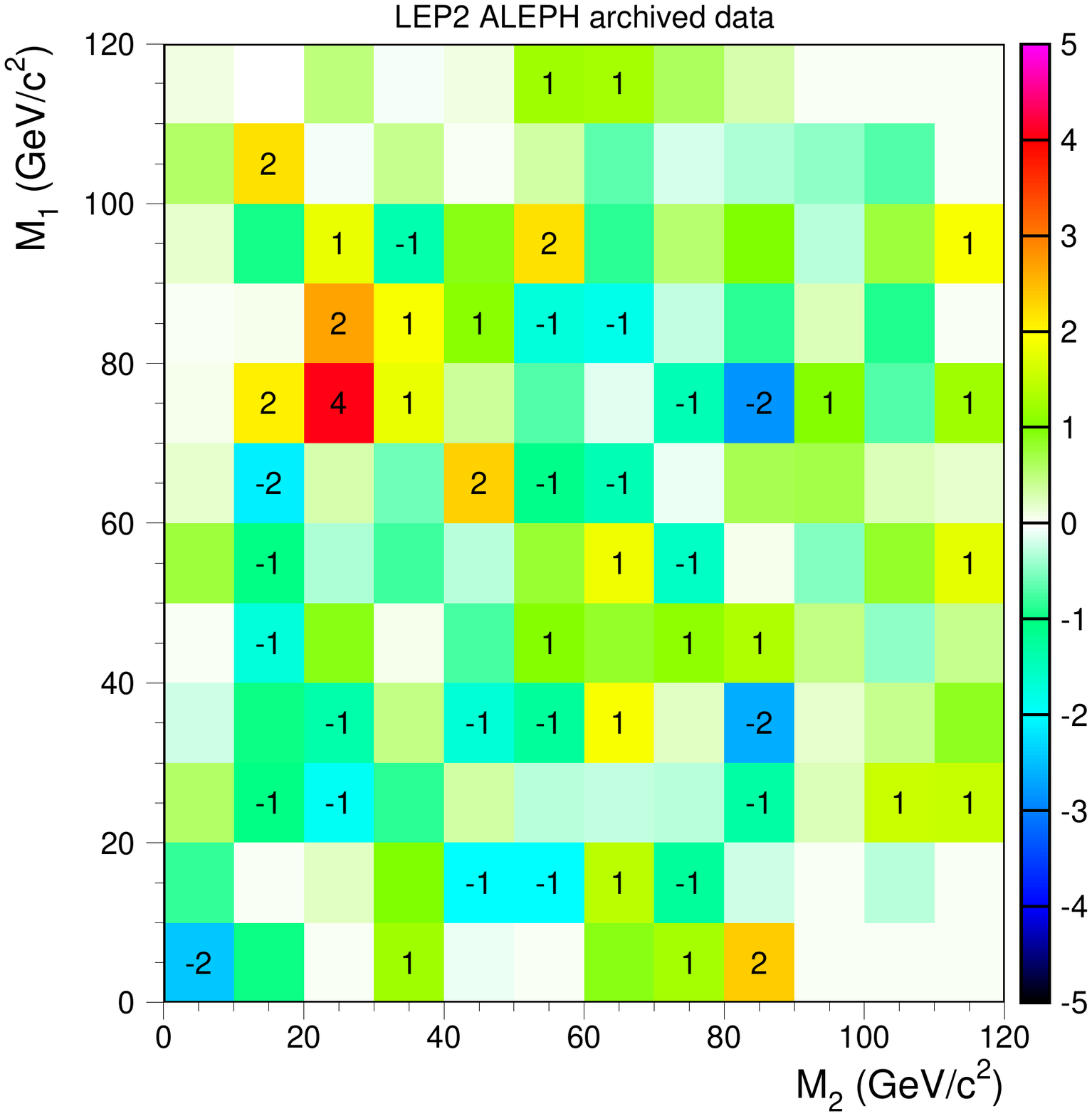}}\hspace{.5in}
\subfigure[]{\includegraphics[width=2.5in,bb=80 150 520 720]{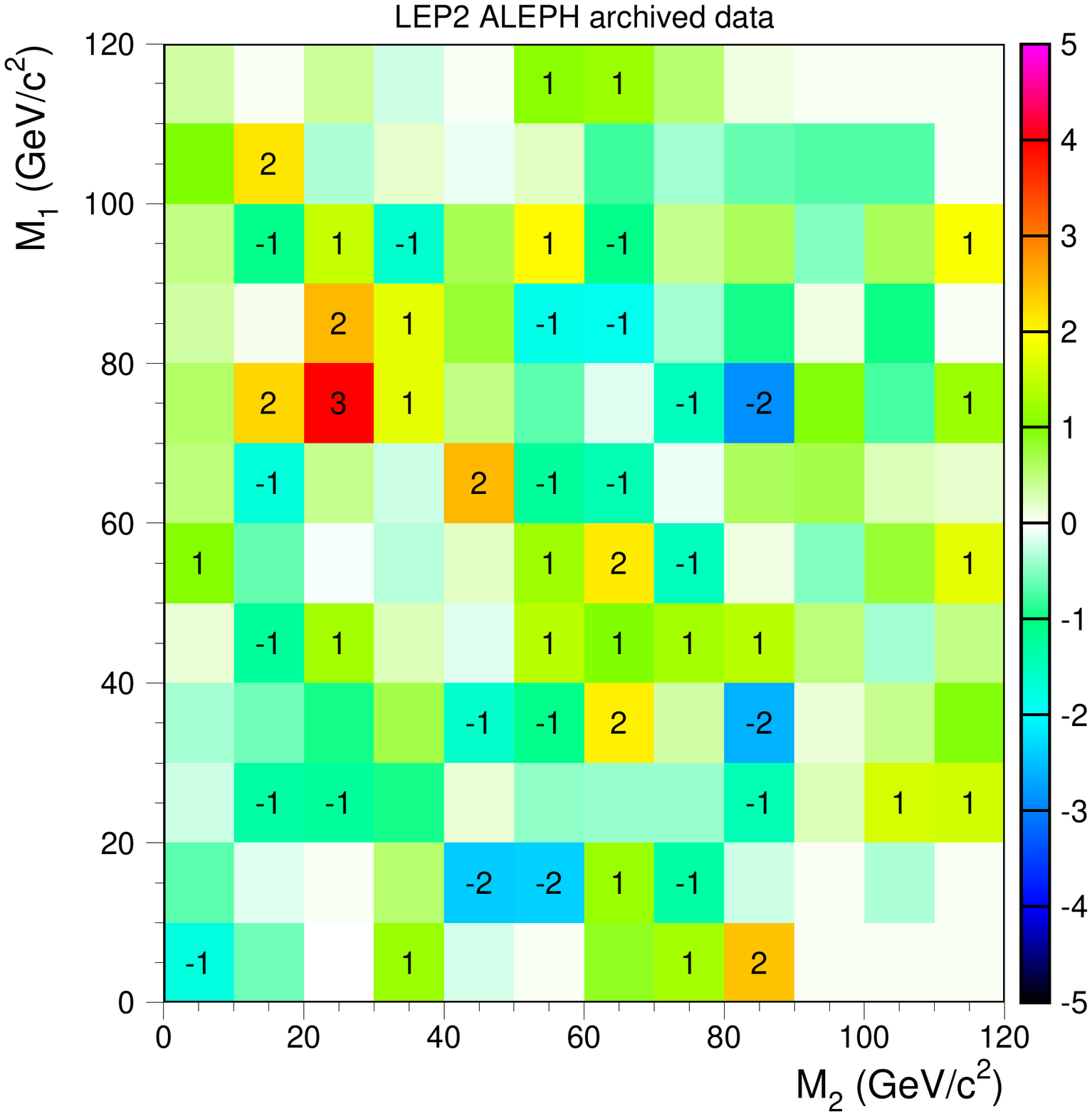}}
\end{center}
\caption{Comparison between LEP2 data and LO SHERPA for LUCLUS-clustered jets in the $M_1$-$M_2$ plane (a) before reweighting (b) after reweighting.  Numbers in the individual squares give the integer part of the significance for bins where the difference between data and MC is more than one standard deviation.}
\label{fig:datavsrewlo2d}
\end{figure}

In Fig. \ref{fig:datavsrewlo2d}, we show the significance of the difference between LEP2 data and the LO SHERPA MC\footnote{In many bins, the SM expectation is small, and gaussian statistics do not apply.  In each bin, we calculate the Poisson probability for the SM expectation to fluctuate as high or higher (in the case of an excess) or as low or lower (in the case of a deficit) than the number of events observed in the data.  For readability, we then convert this probability to the corresponding significance for a gaussian distribution.} in the $M_1$-$M_2$ plane, before and after reweighting.  All LEP2 energies are included; systematic uncertainties are not included.  Each square on the plot where the data differs from the MC expectation by more than one standard deviation is marked with the integer part of its significance.  (A square marked with a $2$ notes that the data shows an excess of two to three $\sigma$ in that bin, while a square marked with $-1$ has a deficit of between one and two $\sigma$.)  We see a substantial excess near $M_1\sim 80\mbox{ GeV}, M_2\sim 25\mbox{ GeV}$ in both the unreweighted (a) and reweighted (b) plots.  We note that this corresponds to both the $45\mbox{ GeV}\lesssim\Sigma\lesssim 60\mbox{ GeV}$ excess and the $\Delta\sim 55\mbox{ GeV}$ excess seen in previous plots.  These excesses are not visible in the plots in Fig. \ref{fig:datavsrewm1m2} due to large background due to the $W^\pm$ peak (in the case of $M_1\sim 80\mbox{ GeV}$) and the SM QCD (in the case of $M_2\sim 25\mbox{ GeV}$).  Additionally, we also see an excess in the region of $M_1\sim M_2\sim 55\mbox{ GeV}$ in both plots.

\begin{figure}[h]
\begin{center}
\subfigure[]{\includegraphics[width=2.5in,bb=80 150 520 720]{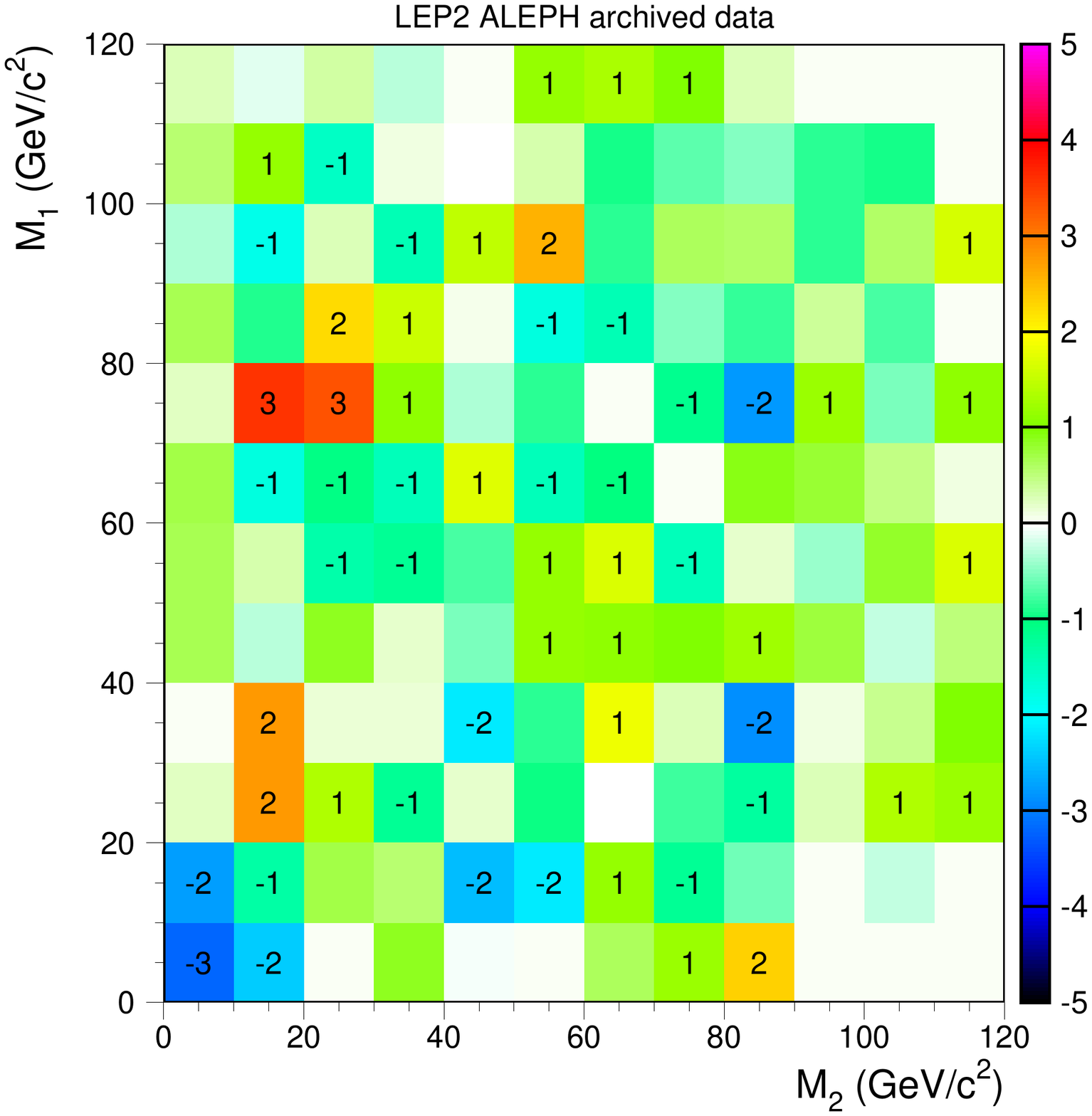}}
\end{center}
\caption{Comparison between LEP2 data and NLO SHERPA for LUCLUS-clustered jets in the $M_1$-$M_2$ plane.  No MC reweighting has been done.  Significances are marked as in Fig. \ref{fig:datavsrewlo2d}.}
\label{fig:datavsunrewnlo2d}
\end{figure}

While we discuss in detail the change in significance of the excess seen in Figs. \ref{fig:datavsrewmasssumdiff}-\ref{fig:datavsrewlo2d} under changes in the MC sample or reweighting in Ref. \cite{paper3}, we make a few brief comments here.  To do this, we return to the plots of $\Sigma$ in Fig. \ref{fig:datavsrewmasssumdiff}.  While reweighting decreased the size of the excess (in number of events) in this plot with respect to the KK2f MC sample, it increased the size of the excess with respect to the SHERPA samples, and particularly the NLO sample; this might cause concern that our reweighting procedure is artificially increasing the size of this excess.  We plot the significance of data-MC at LEP2 using the unreweighted NLO MC as our SM QCD expectation in Fig. \ref{fig:datavsunrewnlo2d} and see that the excess near $\Sigma\sim 45-60$ GeV, and particularly near $M_1\sim 80$ GeV, $M_2\sim 25$ GeV, remains.  We emphasize that our data-MC comparisons at LEP1 and the encouraging agreement of the MC samples after reweighting make using the unreweighted NLO MC hard to justify.  Nonetheless, we show Fig. \ref{fig:datavsunrewnlo2d} to demonstrate that this excess is robust against our choice of MC samples and reweighting.  A detailed discussion of the significance of the excess, the effects of our MC sample choices and reweighting, as well as other systematics, will be given in Ref. \cite{paper3}.

\begin{figure}[h]
\begin{center}
\subfigure[]{\includegraphics[width=2.3in,bb=80 150 520 720]{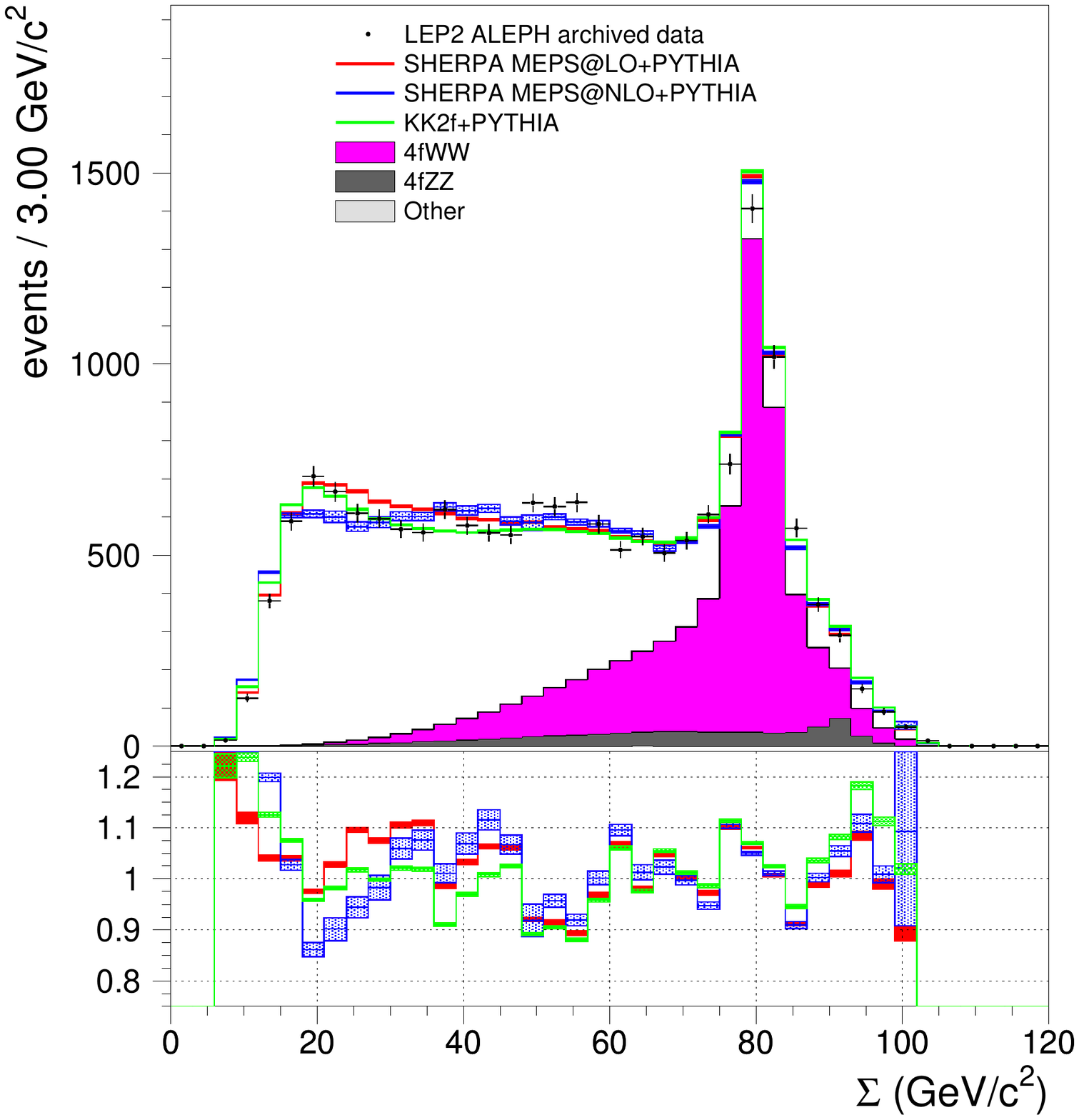}}\hspace{.35in}
\subfigure[]{\includegraphics[width=2.3in,bb=80 150 520 720]{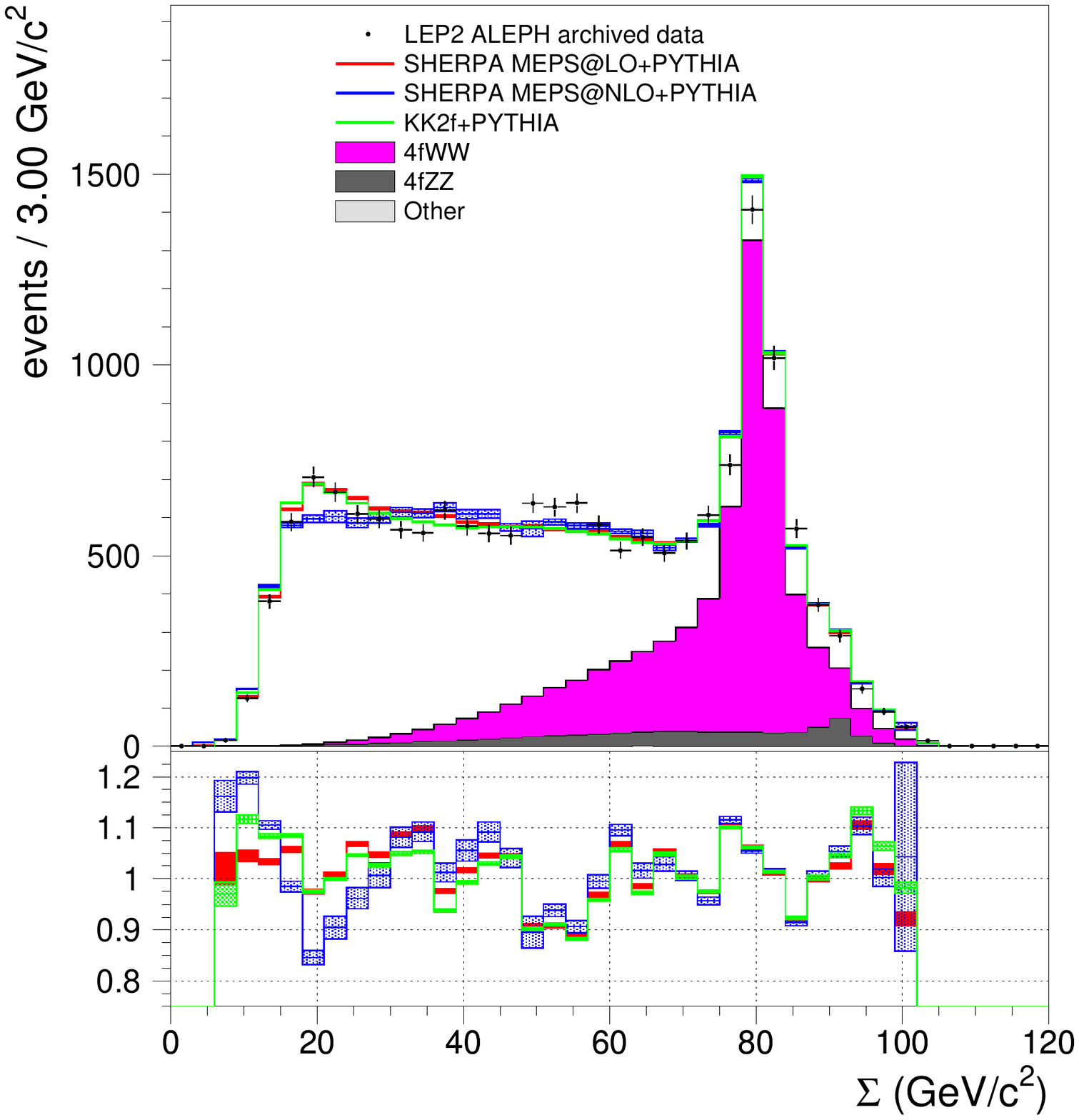}}\\
\subfigure[]{\includegraphics[width=2.3in,bb=80 150 520 720]{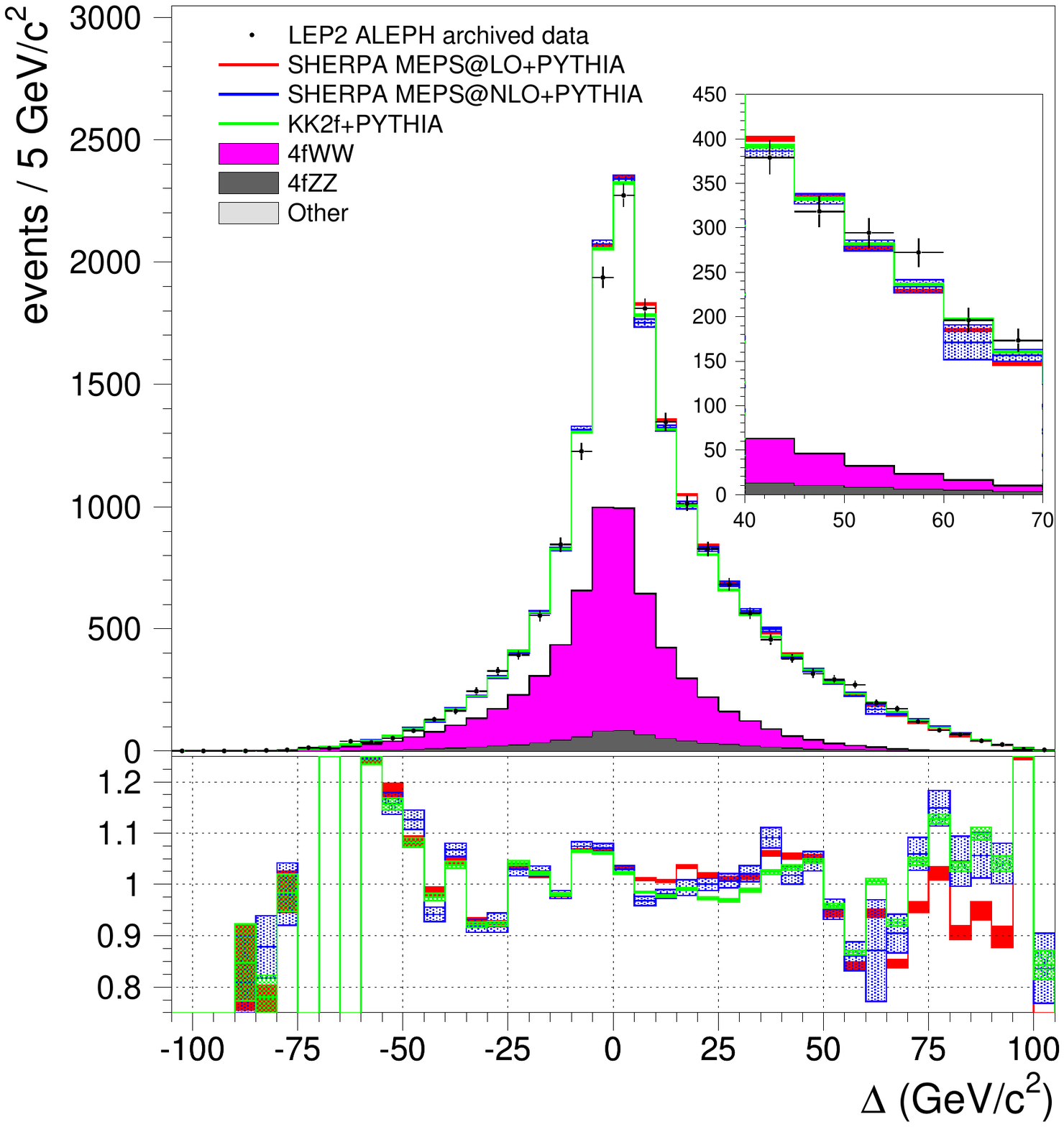}}\hspace{.35in}
\subfigure[]{\includegraphics[width=2.3in,bb=80 150 520 720]{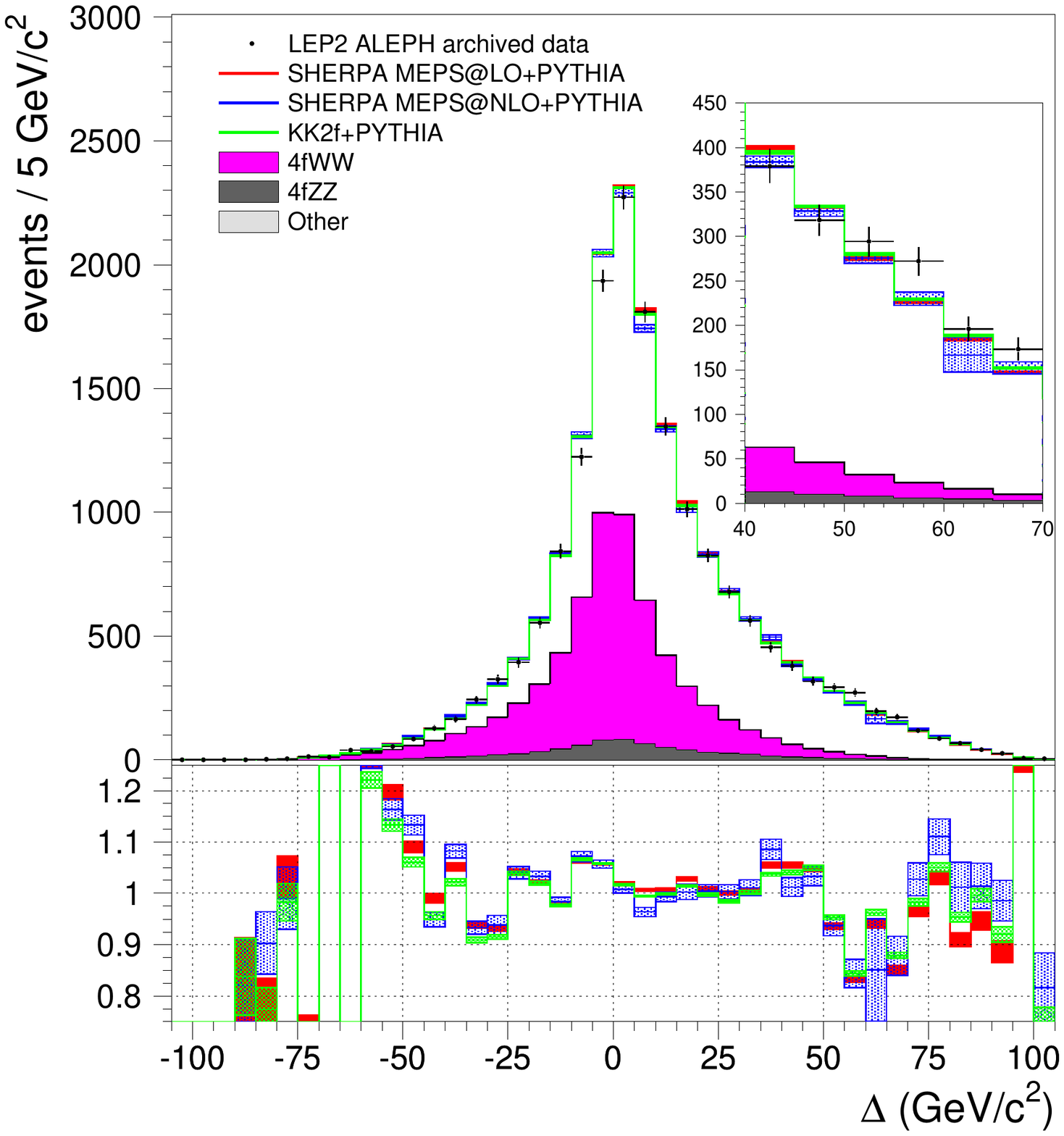}}
\end{center}
\caption{Comparison between LEP2 data and MC samples for (top) $\Sigma$ and (bottom) $\Delta$ for DURHAM-clustered jets.  Plots include all LEP2 energies and show the LO SHERPA, NLO SHERPA, and KK2f predictions (left) before reweighting and (right) after reweighting.  Insets in (c) and (d) display the excess region.}
\label{fig:datavsrewmasssumdiffdur}
\end{figure}

\begin{figure}[h]
\begin{center}
\subfigure[]{\includegraphics[width=2.3in,bb=80 150 520 720]{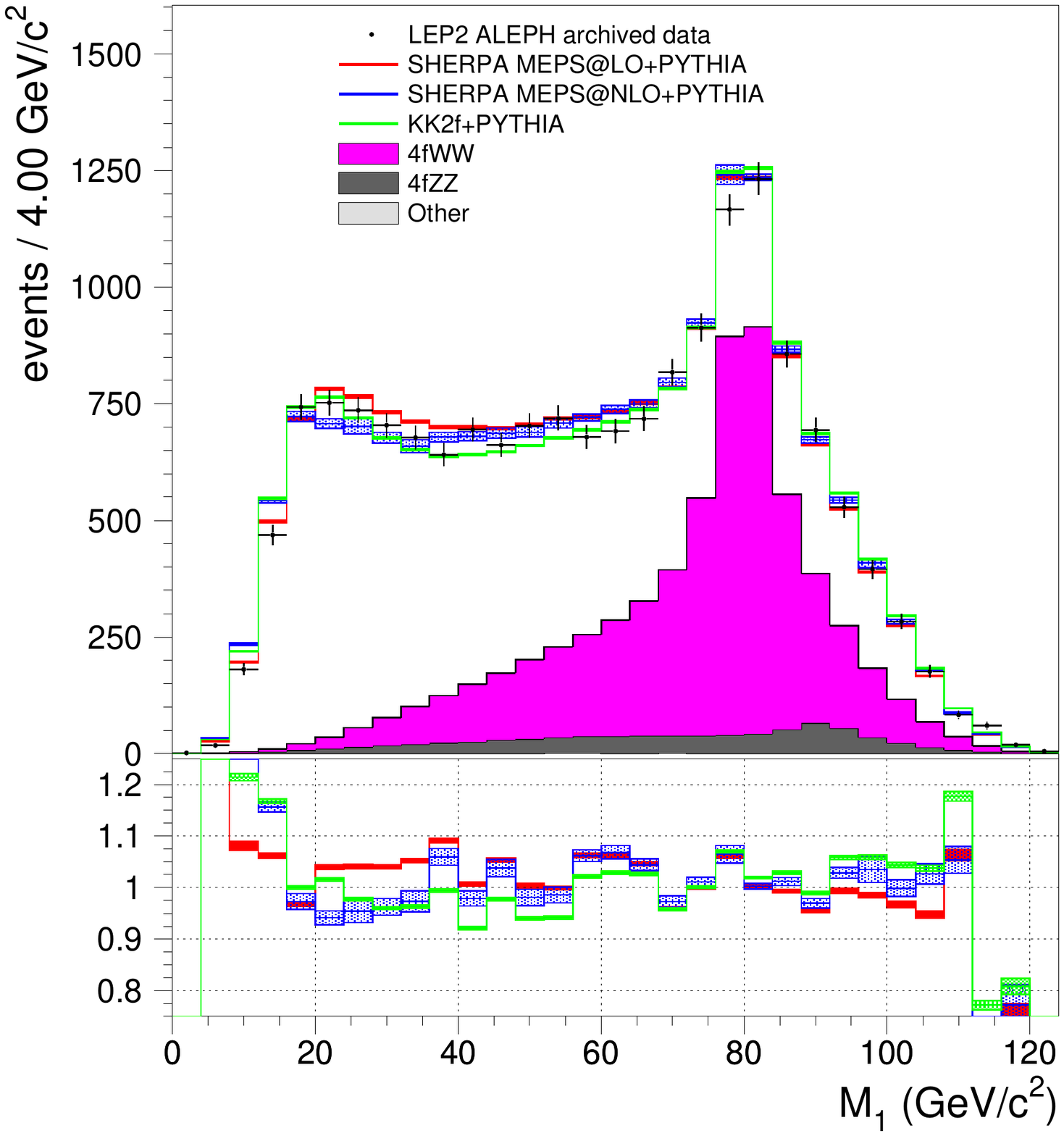}}\hspace{.35in}
\subfigure[]{\includegraphics[width=2.3in,bb=80 150 520 720]{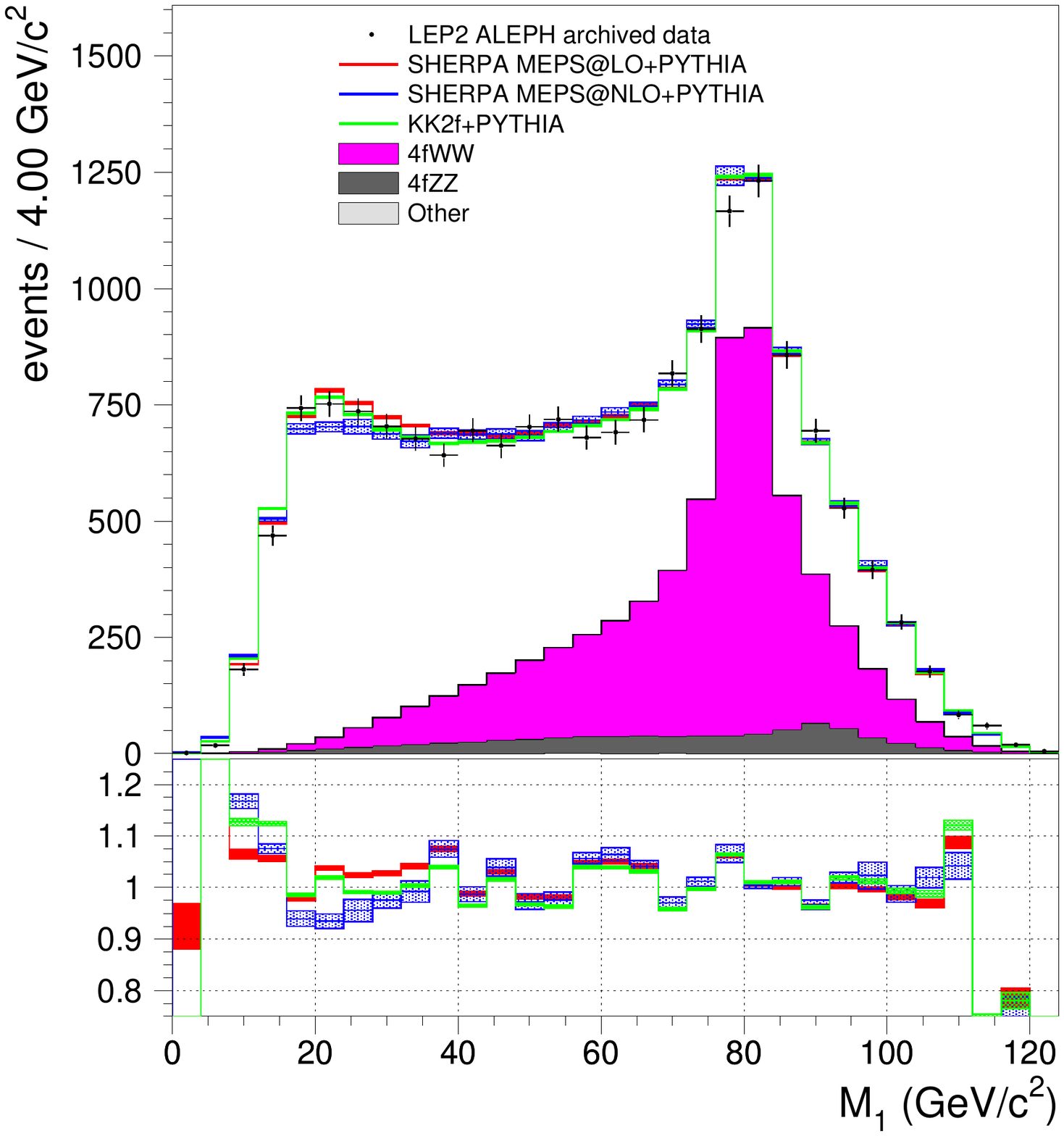}}\\
\subfigure[]{\includegraphics[width=2.3in,bb=80 150 520 720]{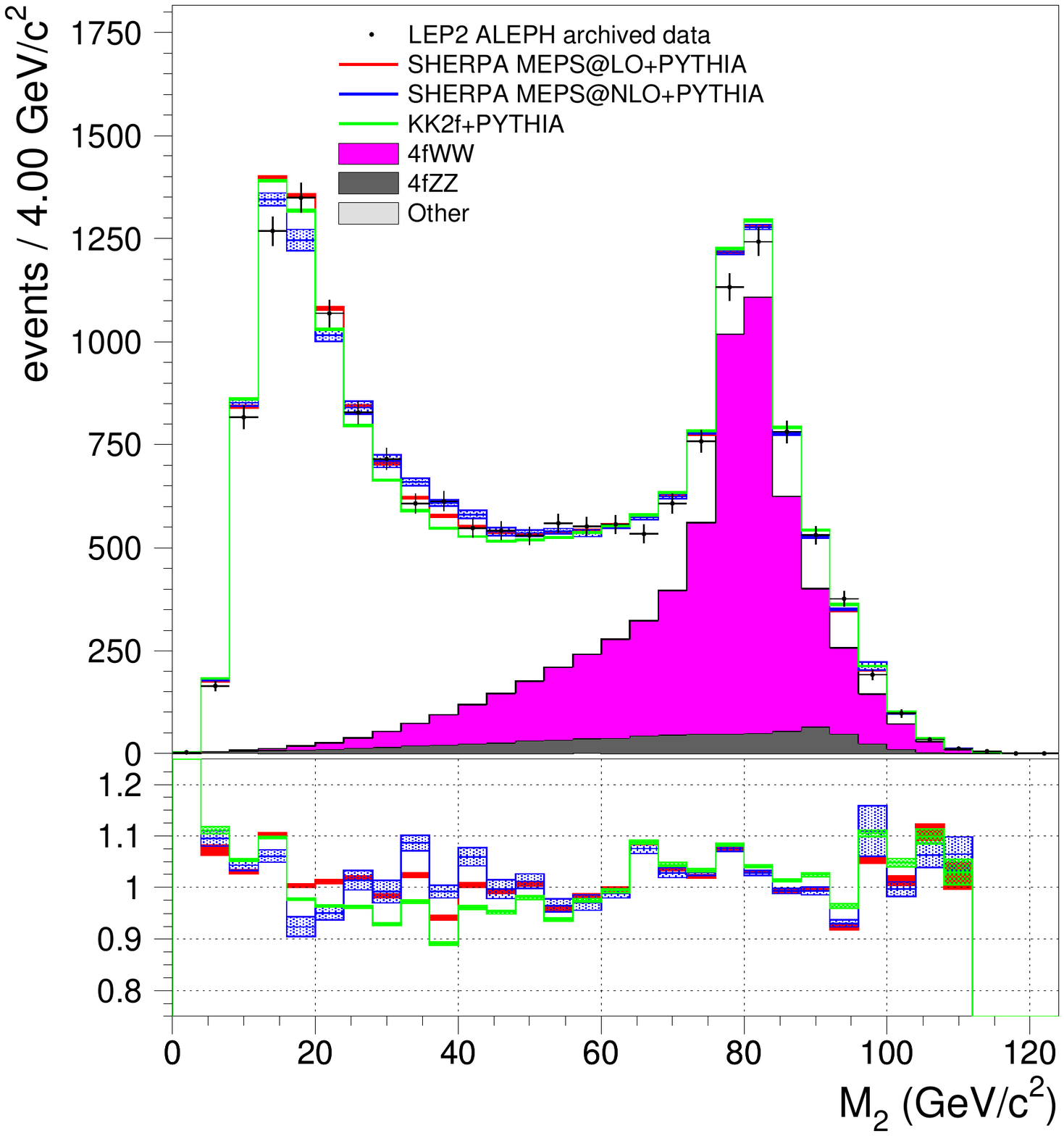}}\hspace{.35in}
\subfigure[]{\includegraphics[width=2.3in,bb=80 150 520 720]{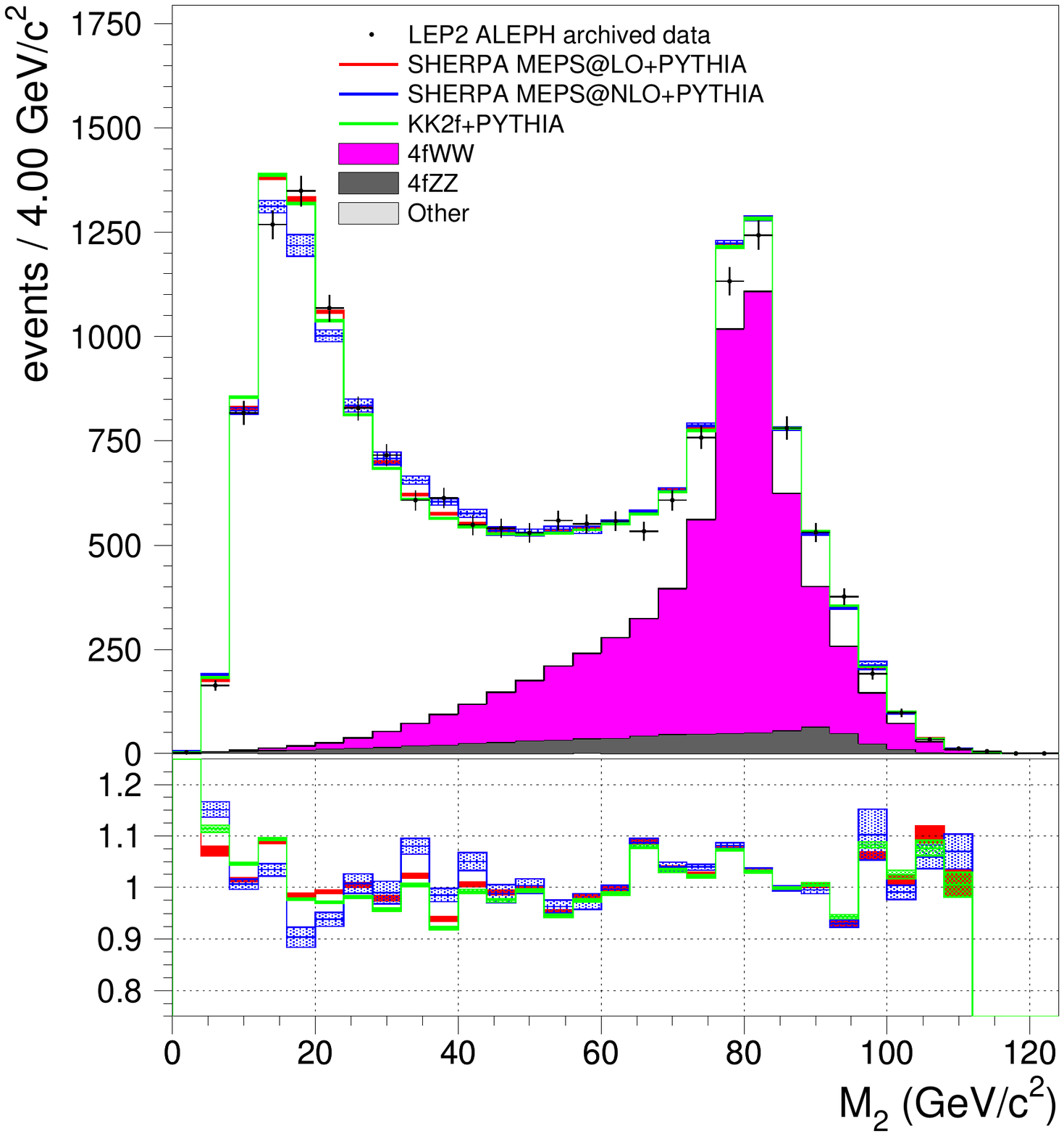}}
\end{center}
\caption{Comparison between LEP2 data and MC samples for (top) $M_1$ and (bottom) $M_2$ for DURHAM-clustered jets.  Plots include all LEP2 energies and show the LO SHERPA, NLO SHERPA, and KK2f predictions (left) before reweighting and (right) after reweighting.}
\label{fig:datavsrewm1m2dur}
\end{figure}

\begin{figure}[h]
\begin{center}
\subfigure[]{\includegraphics[width=2.5in,bb=80 150 520 720]{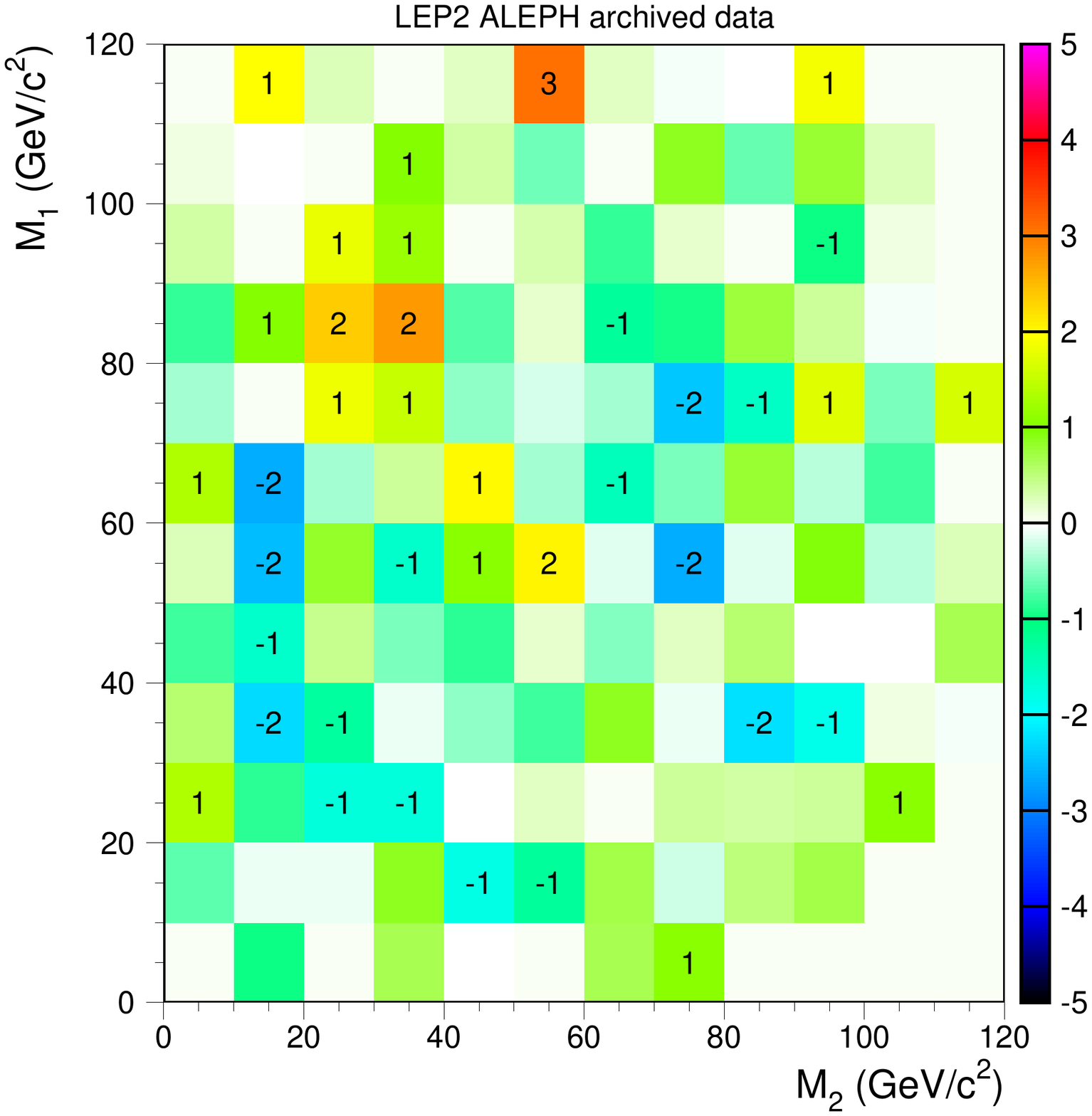}}\hspace{.6in}
\subfigure[]{\includegraphics[width=2.5in,bb=80 150 520 720]{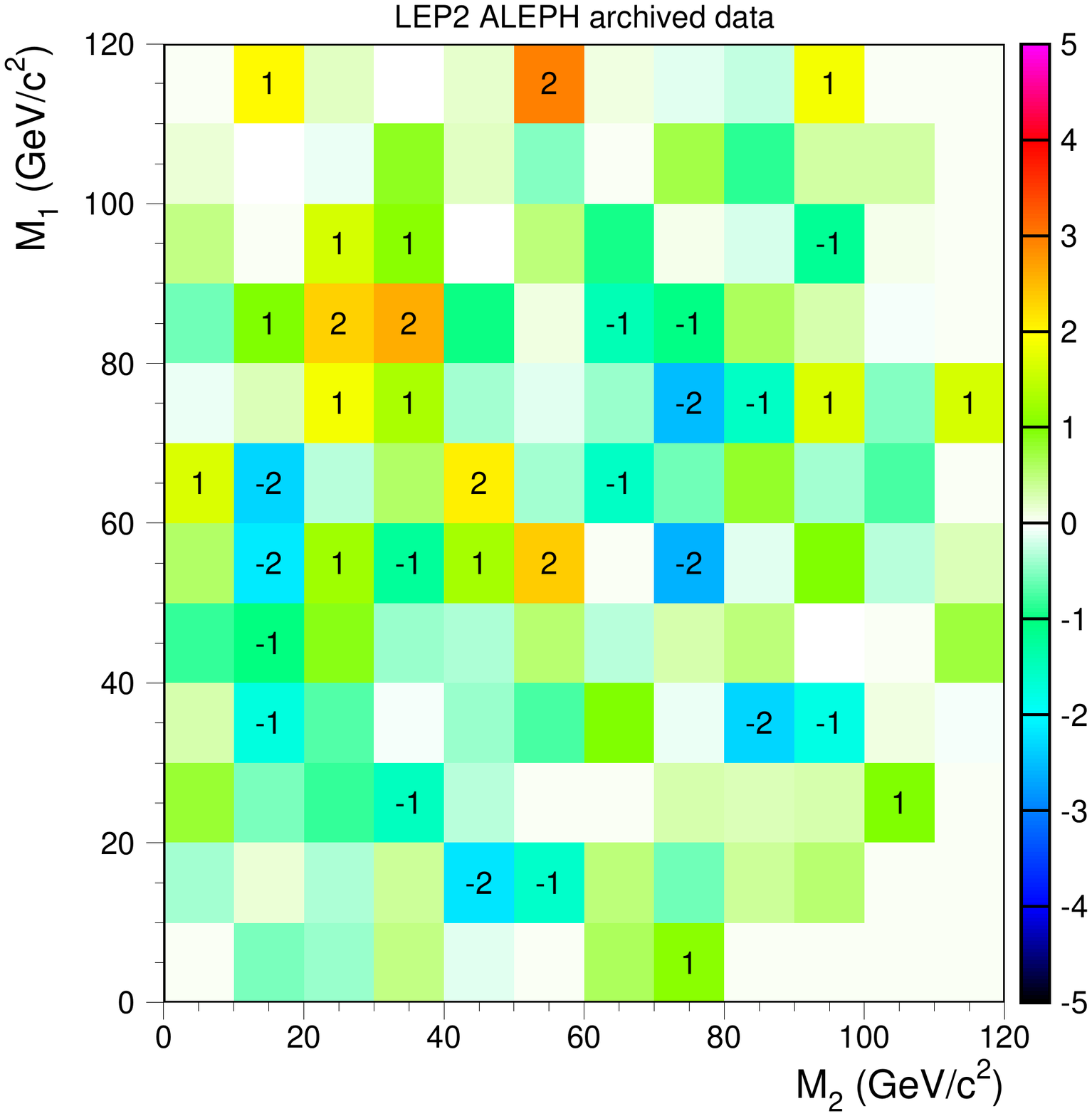}}
\end{center}
\caption{Comparison between LEP2 data and LO SHERPA in the $M_1$-$M_2$ plane (a) before reweighting and (b) after reweighting for DURHAM-clustered jets.  Significances are marked as in Fig. \ref{fig:datavsrewlo2d}.}
\label{fig:datavsrewlo2ddur}
\end{figure}

We now give the analogous plots for DURHAM-clustered jets in Figs. \ref{fig:datavsrewmasssumdiffdur}-\ref{fig:datavsrewlo2ddur}.  We again see excesses in the regions $45\mbox{ GeV}\lesssim\Sigma\lesssim 60\mbox{ GeV}$ and $\Delta\sim 55 \mbox{ GeV}$ for both the unreweighted and reweighted plots, as shown in Fig. \ref{fig:datavsrewmasssumdiffdur}.  We note also that agreement between the MC samples has again improved with reweighting; this can also be seen in the $M_1$ and $M_2$ plots shown in Fig. \ref{fig:datavsrewm1m2dur}.  Lastly, we also look at the significance of the difference between data and the LO MC in the $M_1$-$M_2$ plane; this is shown in Fig. \ref{fig:datavsrewlo2ddur}.  We see that the signficance of the excess near $M_1\sim 80$ GeV, $M_2\sim 25$ GeV has been reduced relative to that observed with LUCLUS-clustered jets.  This difference, as well as the behavior of other jet-clustering algorithms, will be studied in great detail in Ref. \cite{paper3}.

\section{Discussion}
\label{disc}

Several broad conclusions can be drawn from our results thus far.  We have seen that the LO and NLO SHERPA samples perform comparably to KK2f for event-shape variables, while showing substantial improvement for distributions which rely on clustering events into four jets.  Of particular interest for both our purposes and for numerous four-jet analyses are the dijet masses, the modelling of which improves significantly in moving from KK2f to either of the SHERPA samples.  Additionally, the LO and NLO samples show good agreement in the $M_1$-$M_2$ plane, as seen in Fig. \ref{fig:mcratios}.  Overall, we see that the LO SHERPA sample models four-jet variables the most accurately and, of the three samples, shows the most promise for further use.  Additionally, we find that reweighting the MC samples brings them into much better agreement.  Despite this agreement between the MC samples, we see an excess in the data in the region $M_1+M_2\sim 110\mbox{ GeV}$. 

We see that the discrepancies between our MC samples, particularly between KK2f and the two SHERPA samples, are reduced considerably at LEP2 by reweighting using the LEP1 data.  The interpretation of this improvement, however, is somewhat subtle and is closely related to a discussion of systematic uncertainties.  The three MC samples which we used in this work differed in the matrix elements used, the showering, the merging scale in the case of the SHERPA samples, and the treatment of the $b$ quark mass.  It is reasonable to conclude that the reduction in discrepancies between the MC samples upon reweighting implies that reweighting eliminates some, but not all, of the systematic error in the MC samples resulting from these differences.

However, there are important potential sources of uncertainty which are shared among the three MC samples.  These include systematics related to hadronization, as all three samples use PYTHIA for hadronization (albeit with different parameter values), and detector effects, which are the same for all samples.  While reweighting may also reduce systematic errors associated with these effects, comparison of the MC samples among themselves does not address this.  However, we point out that the question is not whether or not reweighting is appropriate, but instead how reweighting translates from LEP1 to LEP2; in the limit that the center-of-mass energies at LEP2 approach that of LEP1, the reweighting technique would improve the MC samples, regardless of the reason for the systematic differences between data and MC\footnote{This would be of use, for example, in a scenario where the LEP1 and LEP2 center-of mass energies were similar, but the LEP2 center-of-mass energy was above the threshold for some new physics process of interest.  In such a case, the LEP1 data may function better than actual MC as an SM MC set for the higher energy data. Here we ignore any (presumably small) experimental effects which change with year of running.}.  

The translation of the reweighting from LEP1 to LEP2 is embodied in our mapping functions for $M_1$ and $M_2$; these mapping functions are primarily justified by the empirical observation that they improve the agreement of the MC samples.  It is encouraging, however, that the mapping functions for all of the MC samples are very close to linear, except at small values of $M_1$ and $M_2$, where non-perturbative effects become more important, as shown in Fig. \ref{fig:mapfuncs}.  Were the evolution of both the data and the MC perfectly linear with $\sqrt{s}$, the reweighting factors applied at LEP1 would also be the exact reweighting factors to apply at LEP2.

For future studies, we take systematic errors from sources which vary from one MC sample to another (the different uses of the matrix elements, the $b$ quark mass, etc.) to be approximately covered by the differences in the MC samples after reweighting.  The systematic errors from the variation of the renormalization scale were studied briefly in Ref. \cite{paper1}, although this is partially redundant given that we can compare our LO and NLO SHERPA samples.  Systematic effects in the hadronization, shared among the MC samples, may also be complicated by the expectation that such effects would be most pronounced at low energy scales and may not be reduced by our reweighting procedure as effectively as other sources of systematic uncertainty.  These effects, as well as shared detector systematics, will need separate systematic studies.


Lastly, it is reasonable to ask why we did not obtain with the NLO SHERPA sample data-MC agreement as good or better as that which we obtained with the LO SHERPA.  Obtaining a definitive answer to this question is likely prohibitively difficult.  However, we can make a few comments.

It should be pointed out that the LO and NLO tunes did not differ only in the perturbative order of the matrix elements used.  There are several differences between the two tunes, and some of these differences arise because of the computational difficulty of the NLO calculation.  First of all, the version of SHERPA used for the NLO tune can only accomodate massless quarks in NLO calculations; thus, the $b$ quark mass is set to zero in both the matrix elements and parton shower.  The LO simulation, however, uses massive $b$ quarks in both the matrix element and the parton shower.  Additionally, the value of the $Q_{cut}$ had to be set higher for the NLO calculation relative to that of the LO generation, to decrease the phase space determined by the NLO matrix elements, in order to make the NLO generation calculationally feasible.  Third, the LO tune used quartic interpolation in Professor, but obtaining a stable NLO tune required resorting to quadratic interpolation.  We suspect that this was due to the use of partially-unweighted (with both positive and negative) events as well as the reduced sample size in the generation of NLO samples, choices which were both necessary in order to generate NLO events at a rate suitable for use.  (At the same time, studies of LO tunes with quadratic interpolation often did not result in successful tunes, so we suspect that quadratic interpolation is simply not sufficient to obtain a high-quality tune.)  In addition to these differences, the LO and NLO tunes also used different versions of Sherpa and different weight files; these last two issues are not directly related to the computational complexity of the NLO calculation.

The computational difficulty of the NLO calculation makes extensive studies of NLO tuning infeasible.  However, we performed several rudimentary LO tunes varying the above features to compare the size of their effects on $\sim 50$ distributions in Rivet.  We find that no single change in the tuning procedure dominates the difference between the LO and NLO results.  There were hints that the choice of SHERPA version and the weight file played smaller roles than the other considerations.  However, it must be stated that these are general statements and that there is much variation among the distributions studied.  

In summary, while we do not find a single cause for the degraded performance of the NLO tune relative to that of the LO tune, the computational complexity of the NLO generation forces a number of changes in the tuning procedure which can plausibly outweigh the improvement expected from going to the next order in the perturbative calculation.  Nonetheless, we reiterate that both the LO and NLO tunes also show substantial improvement over the KK2f MC for the simulation of observables directly related to four-jet states, and we encourage future $e^+e^-\rightarrow\mbox{hadrons}$ tuning studies at both LO and NLO.

\section{Hadronization Uncertainties}
\label{hadunc}

We reserve a complete discussion of systematic uncertainties for Ref. \cite{paper3} and here focus on hadronization uncertainties.  In Ref. \cite{paper3}, we perform some rudimentary comparisons comparing PYTHIA to HERWIG \cite{Corcella:2000bw}, both interfaced to KK2f.  However, PYTHIA and HERWIG differ not just in the hadronization scheme used, but also in their parton showers; this is especially important given that, when interfaced to KK2f, the parton shower determines the main structure of the event.  For this reason, we had reason to suspect that differences in QCD expectations observed when comparing PYTHIA and HERWIG overestimated the uncertainties from hadronization applicable to the LO SHERPA sample.  We thus quoted our final significances for the two excess regions in terms of a hadronization uncertainty, which we took to be between zero and $3\%$ for Region A and between zero and $2\%$ for Region B.  

\begin{figure}[h]
\begin{center}
\subfigure[]{\includegraphics[width=2.5in,bb=80 150 520 720]{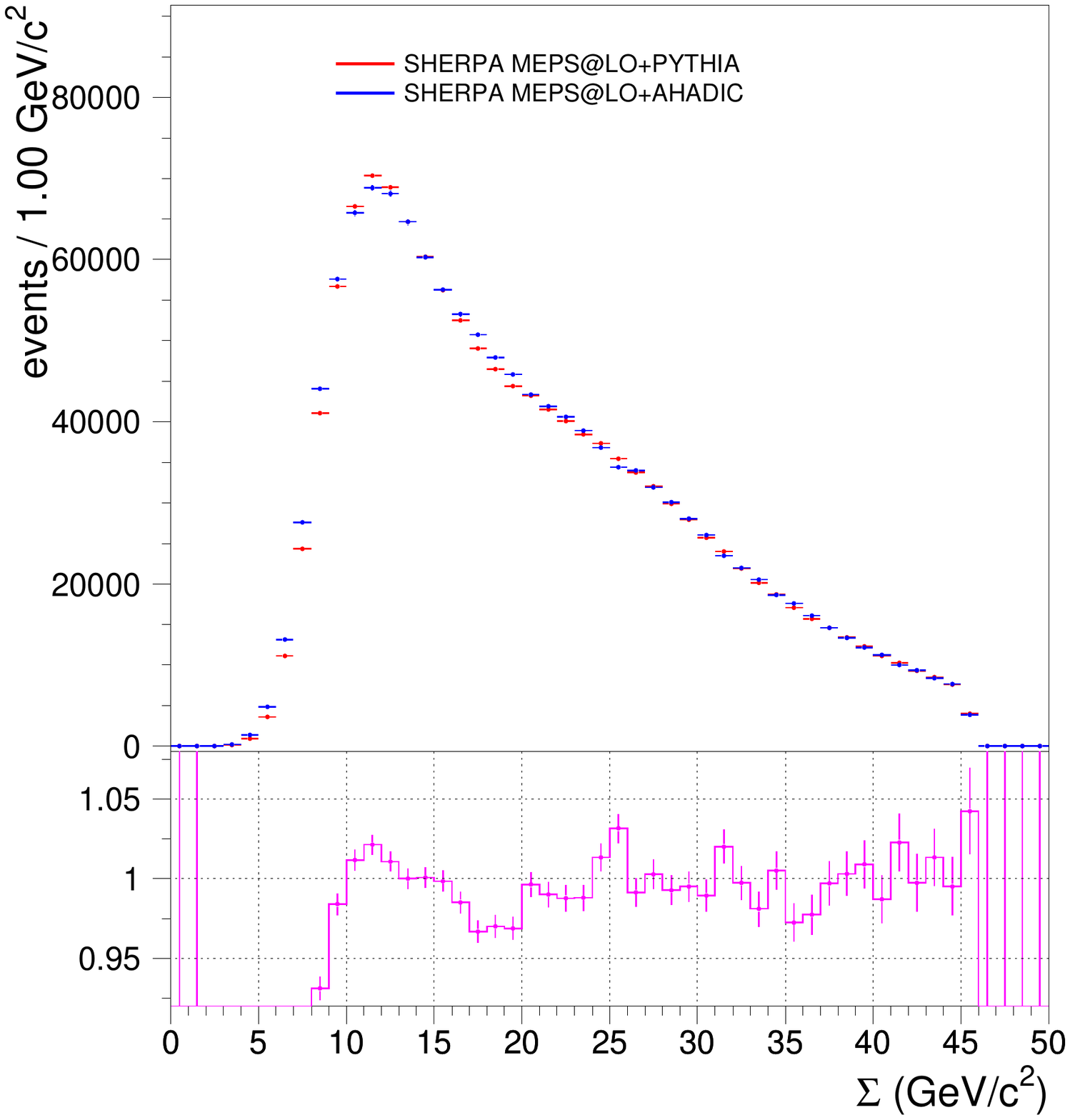}}\hspace{.6in}
\subfigure[]{\includegraphics[width=2.5in,bb=80 150 520 720]{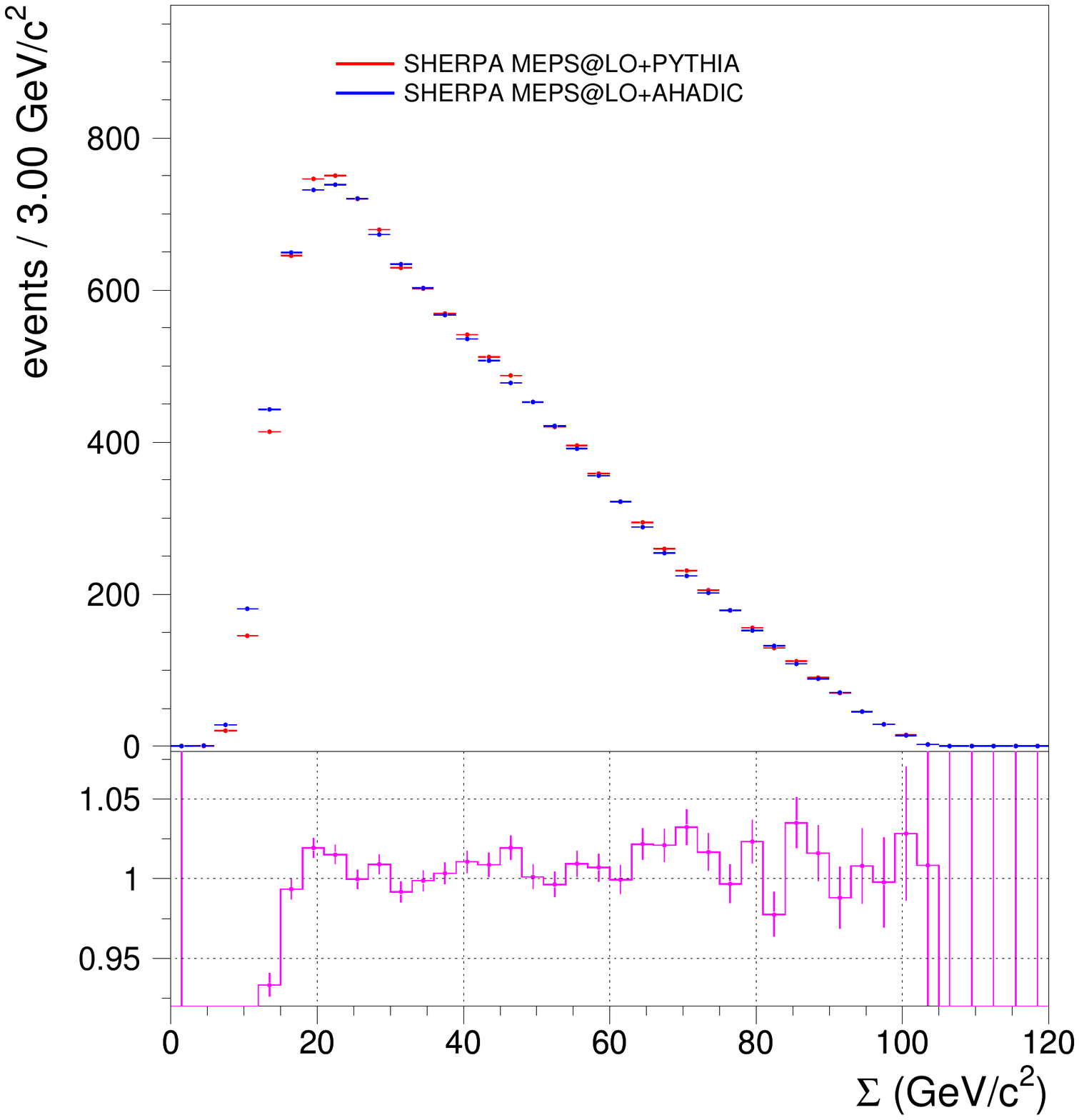}}
\end{center}
\caption{Comparison of SHERPA expectation of $\Sigma$ utilizing PYTHIA and AHADIC++ for hadronization (a) at LEP1 and (b) at LEP2.  The bottom panel of each plot shows the ratio of the PYTHIA expectation to that using AHADIC++.  The error bars include only the effect of MC statistics.}
\label{fig:pythvahad}
\end{figure}

Here, we give a comparison between our LO SHERPA sample and MC resulting from another LO tune of SHERPA, this time using AHADIC++ \cite{Winter:2003tt} for hadronization.  Details of our tune using AHADIC++ are provided in Appendix \ref{ahadicapp}.  Overall, in soft regions where effects of hadronization are expected to be important, we find the PYTHIA hadronization to have better performance than AHADIC++, while the two tunes are more similar in perturbative regions.  We compare the expectations for $\Sigma$ resulting from the two hadronization schemes in Fig. \ref{fig:pythvahad}.  In each plot, the top panel shows the PYTHIA and AHADIC++ expectations, and the bottom panel shows their ratio.  As may be expected, we see somewhat larger differences between the schemes at LEP1 than at LEP2, but the deviation is nonetheless $\mathcal{O}(\mbox{few}\%)$ in both cases.  In particular, in the region of interest for the LEP2 excess, $\Sigma\sim 40-60$ GeV, the difference is $\mathcal{O}(1\%)$.

We give the differences, in percent, between the sample using PYTHIA and that using AHADIC++ in Table \ref{tab:ahadicpythia} for Regions A and B.  Numbers are given for the ellipses which would contain $68\%$, $90\%$, and, in the case of Region A, $95\%$ of the nominal signal regions defined in Ref. \cite{paper3}, both before and after reweighting with LEP1 data and MC\footnote{The effect on hadronization uncertainties of reweighting LEP2 MC using data and MC from LEP1 contains some subtleties.  First, as hadronization is more important at LEP1, reweighting has the potential to over-correct for hadronization effects at LEP2.  Additionally, as hadronization is primarily important at low energy scales, its effects should not be expected to scale closely with $\sqrt{s}$ as implied by our mapping functions from LEP1 to LEP2.  Therefore, the improvement in agreement between the PYTHIA and AHADIC++ tunes upon reweighting in Regions A and B may indicate that the difference between these two tunes in these regions is not actually dominated by hadronization.  As the AHADIC++ sample is generated from an independent tune, the differences between the two samples will not be limited to only the difference in hadronization scheme.  However, as the differences shown in Table \ref{tab:ahadicpythia} are within the range of uncertainties considered in Ref. \cite{paper3} both before and after reweighting, these considerations are not important for our conclusions.}.  (We do not quote a number for the $95\%$ ellipse for Region B, as this would overlap significantly with Region A.)  We see that the ranges of uncertainties used in Ref. \cite{paper3}, up to $3\%$ in Region A and up to $2\%$ in Region B, are compatible with the differences observed here.

\begin{table}
\begin{tabular}{| c| c| c| c| c| c|}
\hline
& \multicolumn{3}{c|}{Region A} & \multicolumn{2}{c|}{Region B} \\
\hline
& $68\%$ & $90\%$ & $95\%$ & $68\%$ & $90\%$\\
\hline
Unreweighted & $2.9\pm1.0$ & $2.4\pm 0.7$ & $2.5\pm 0.6$ & $0.8\pm 0.7$ & $0.5\pm 0.5$\\
\hline
Reweighted & $-0.5\pm1.0$ & $0.1\pm 0.7$ & $0.8\pm 0.6$ & $1.1\pm 0.7$ & $0.5\pm 0.5$\\
\hline
\end{tabular}
\caption{Differences, in percent, between the QCD expectations using PYTHIA and AHADIC++ for hadronization, given for LEP2 excess Regions A and B.  Top line is for LEP2 MC without reweighting with LEP1 data and MC.  The bottom line is after reweighting.  Numbers are given for ellipses which would include $68\%$, $90\%$, and, in the case of Region A, $95\%$ of the signals defined in Ref. \cite{paper3}.}
\label{tab:ahadicpythia}
\end{table}

\section{Conclusions}
\label{conc}

In this paper, we demonstrate the application of a modern QCD MC generator, with proper merging and matching, to LEP analyses.  We apply the ALEPH detector simulation to the SHERPA LO and NLO $e^+e^-\rightarrow\mbox{hadrons}$ MC samples described in Ref. \cite{paper1} and compare the samples to LEP1 and LEP2 ALEPH data as well as to QCD samples generated with KK2f.  We present numerous data-MC comparisons at both LEP1 and LEP2 energies.  We pay particular attention to four-jet variables and find that the SHERPA MC samples show significant improvement in these observables in comparison to KK2f.  In addition, the SHERPA MC samples perform similarly to KK2f for event-shape observables.  We also estimate the size of hadronization uncertainties by comparing samples using PYTHIA and AHADIC++ for hadronization. 

Our results reflect those obtained at Rivet level \cite{paper1}.  For many observables, we find that the SHERPA LO MC describes the data more accurately than the SHERPA NLO MC.  Additionally, both still display some discrepancies with the LEP1 and LEP2 datasets.  We thus explore the possibility of reweighting the SHERPA LO and NLO samples as well as KK2f with correction factors obtained at LEP1 to improve the data-MC agreement at LEP2, for distributions of the dijet masses which we wish to study in Ref. \cite{paper3}.  We find that such reweighting does, in fact, bring the dijet mass distributions of the three MC samples into better agreement with each other.  We thus decide to use the SHERPA LO MC, with reweighting factors applied, as our estimation of the SM QCD in Ref. \cite{paper3}; the SHERPA NLO and KK2f will be retained for systematic studies.  We additionally find estimating hadronization uncertainties by comparing PYTHIA and AHADIC++ leads to estimates compatible with those explored in Ref. \cite{paper3}.


It is also worth pointing out that the tuning of the MC samples studied here was done largely without variables that depend on events being clustered into four jets\footnote{DURHAM resolution parameters $y_{ij}$ and jet rates were included, but these covered a range of jet multiplicities, not just four jets.}.  If one prioritizes the correct simulation of four-jet states, we speculate that it may be of value to do a tune of QCD MC which goes beyond the variables contained in the currently available Rivet analyses, perhaps including the LEP1 dijet masses explicitly.  We do not pursue this here.

While the SHERPA MC generation improves the simulation of four-jet states and MC reweighting improves the agreement between the MC samples, a statistically significant excess remains in the ALEPH LEP2 data for $45\mbox{ GeV}\lesssim\Sigma\lesssim 60\mbox{ GeV}$, with a particular concentration of events near $M_1\sim 80\mbox{ GeV}$, $M_2\sim 25\mbox{ GeV}$.  Additionally, this excess has no apparent analogue at LEP1.  Its significance and robustness under changes of QCD simulation and jet-clustering algorithm, as well as the characteristics of the events in the excess region, will be studied in great detail in Ref. \cite{paper3}.

Finally, we emphasize that, while our main purpose here is to obtain the optimal MC sample for study in Ref. \cite{paper3}, the correct simulation of four-jet states is of importance to many LEP analyses and will also be of importance at a future linear collider.  We have found marked improvement in the simulation of four-jet states in moving from KK2f to the LO SHERPA sample while maintaining the reasonable description of event-shape variables already present in KK2f.  The results here are very promising for future studies of hadronic states at LEP or future lepton colliders.

\begin{acknowledgments}

We would like to thank Thomas McElmurry for interesting and productive discussions throughout the duration of this project.
    
This work was performed using the ALEPH Archived Data. We would like to thank the members of the CERN accelerator divsion and the engineers and technical staff who made collection of the data possible.  We would also like to express our appreciation to the ALEPH leadership for their foresight in writing the ALEPH data policy, and to Marcello Maggi for helping secure access to the data.

The work of JK was supported in part by the Funda\c c\~ao para a Ci\^encia e a Tecnologia (FCT, Portugal), project UID/FIS/00777/2013.
\end{acknowledgments}


\appendix

\section{Run.dat details}
\label{sec:rundat}
For generating SHERPA samples for full simulation, certain particles must be defined as stable, as they are decayed by the ALEPH detector simulation software.  The following parameters in a SHERPA Run.dat file must be set as follows.\\
STABLE[310]=1; $\#$ K0\_S\\
STABLE[3122]=1; $\#$ Lambda\^{}0\\
STABLE[3222]=1; $\#$ Sigma+\\
STABLE[3112]=1; $\#$ Sigma-\\
STABLE[3322]=1; $\#$ Xi0\\
STABLE[3312]=1; $\#$ Xi-\\
STABLE[3334]=1; $\#$ Omega-\\

\section{NLO sample event weights}
\label{sec:nloevtwts}

\begin{figure}[h]
\begin{center}
\subfigure[]{\includegraphics[width=3in,bb=80 150 520 720]{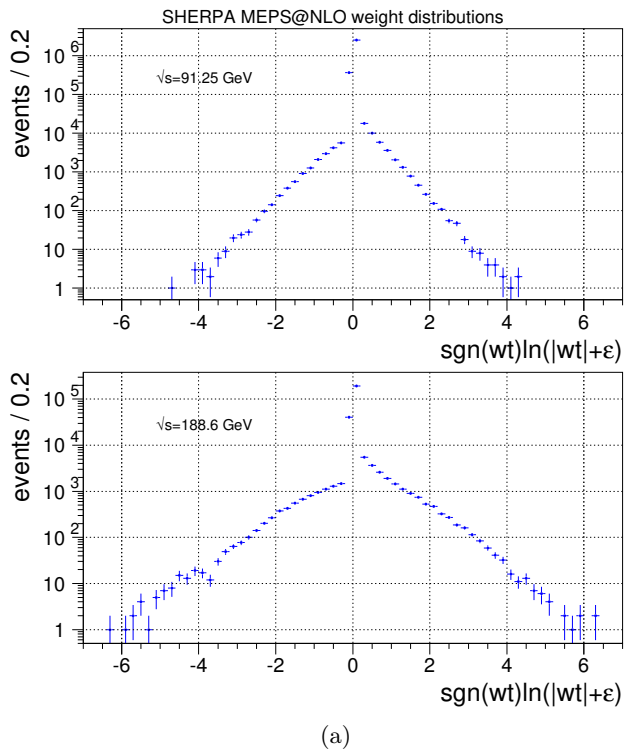}}
\end{center}
\caption{Event weights of the NLO SHERPA event sample.  A tiny offset $\epsilon$ has been applied to distinguish events with weights of $1$ and $-1$.}
\label{fig:nloweights}
\end{figure}

For the NLO SHERPA sample, partially unweighted events were generated.  In Fig. \ref{fig:nloweights}, we plot the resulting event weights.  The absolute values of the event weights in the plot have been given a tiny shift to allow the discrimination of events with weights of $1$ and $-1$.  There are no events with weights of magnitude $<1$. 

\section{Additional distributions}
\label{moredists}

\begin{figure}[h]
\begin{center}
\subfigure[]{\includegraphics[width=2.6in,bb=80 150 520 720]{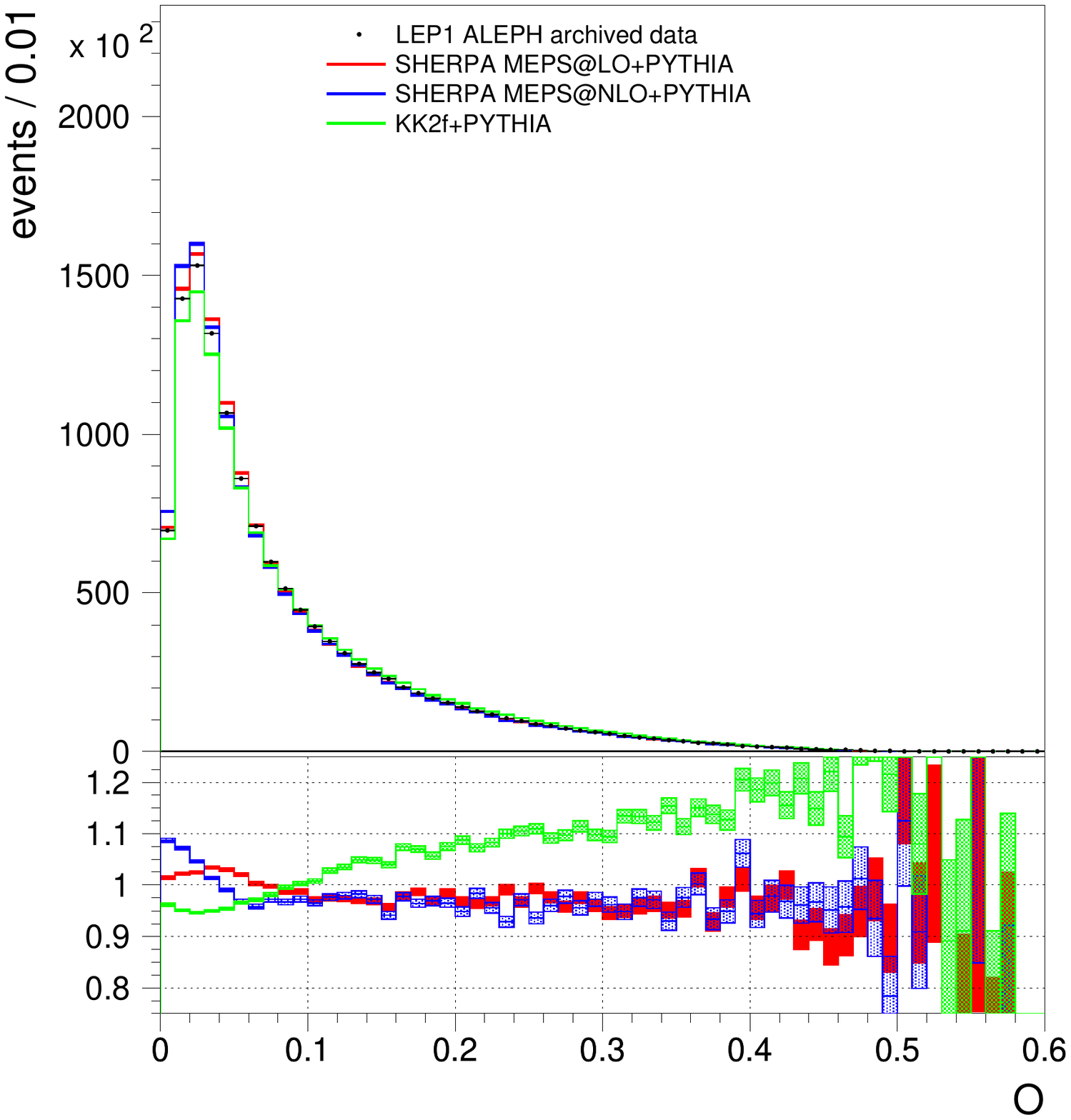}}\hspace{.4in}
\subfigure[]{\includegraphics[width=2.6in,bb=80 150 520 720]{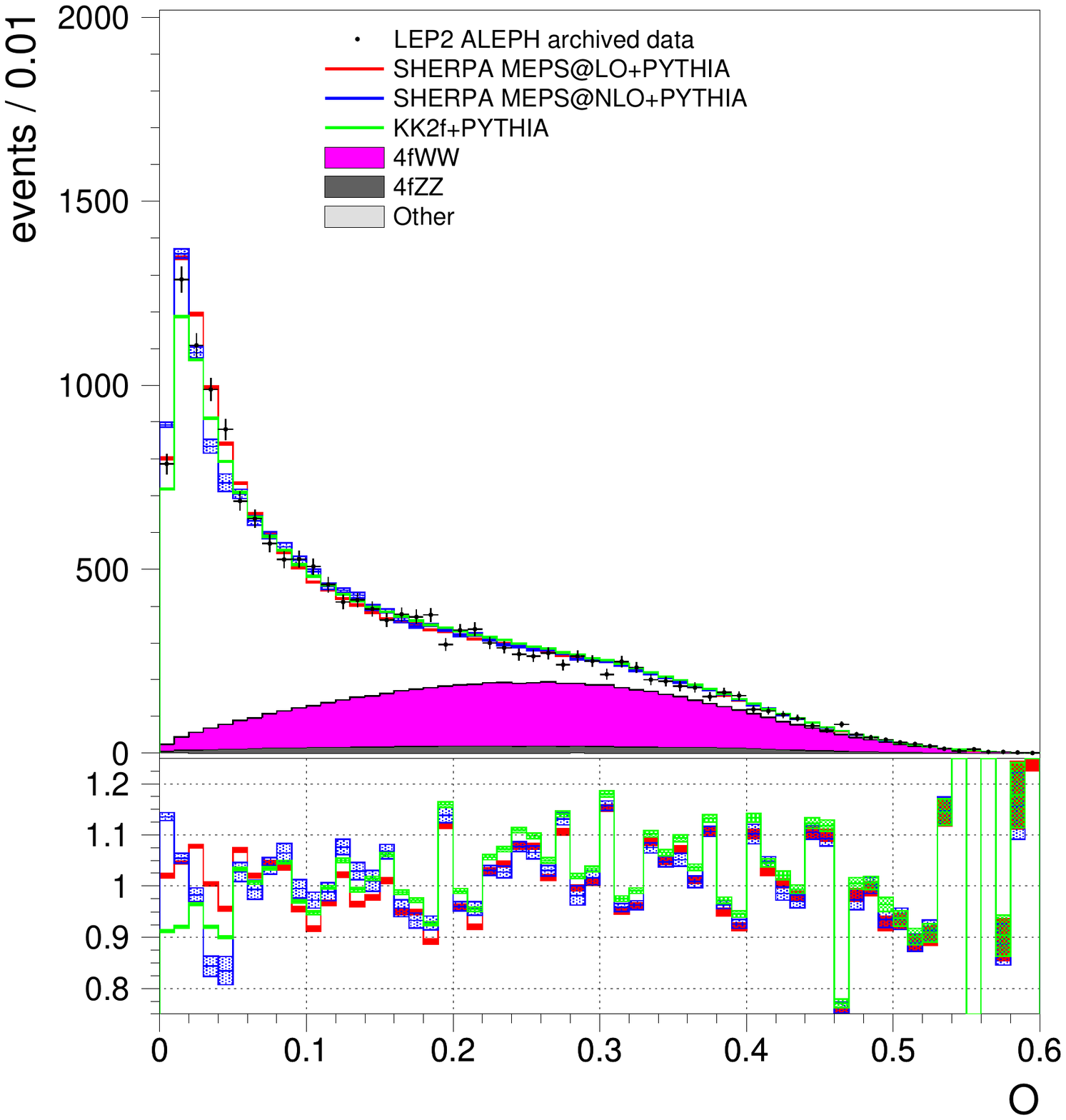}}\\
\subfigure[]{\includegraphics[width=2.6in,bb=80 150 520 720]{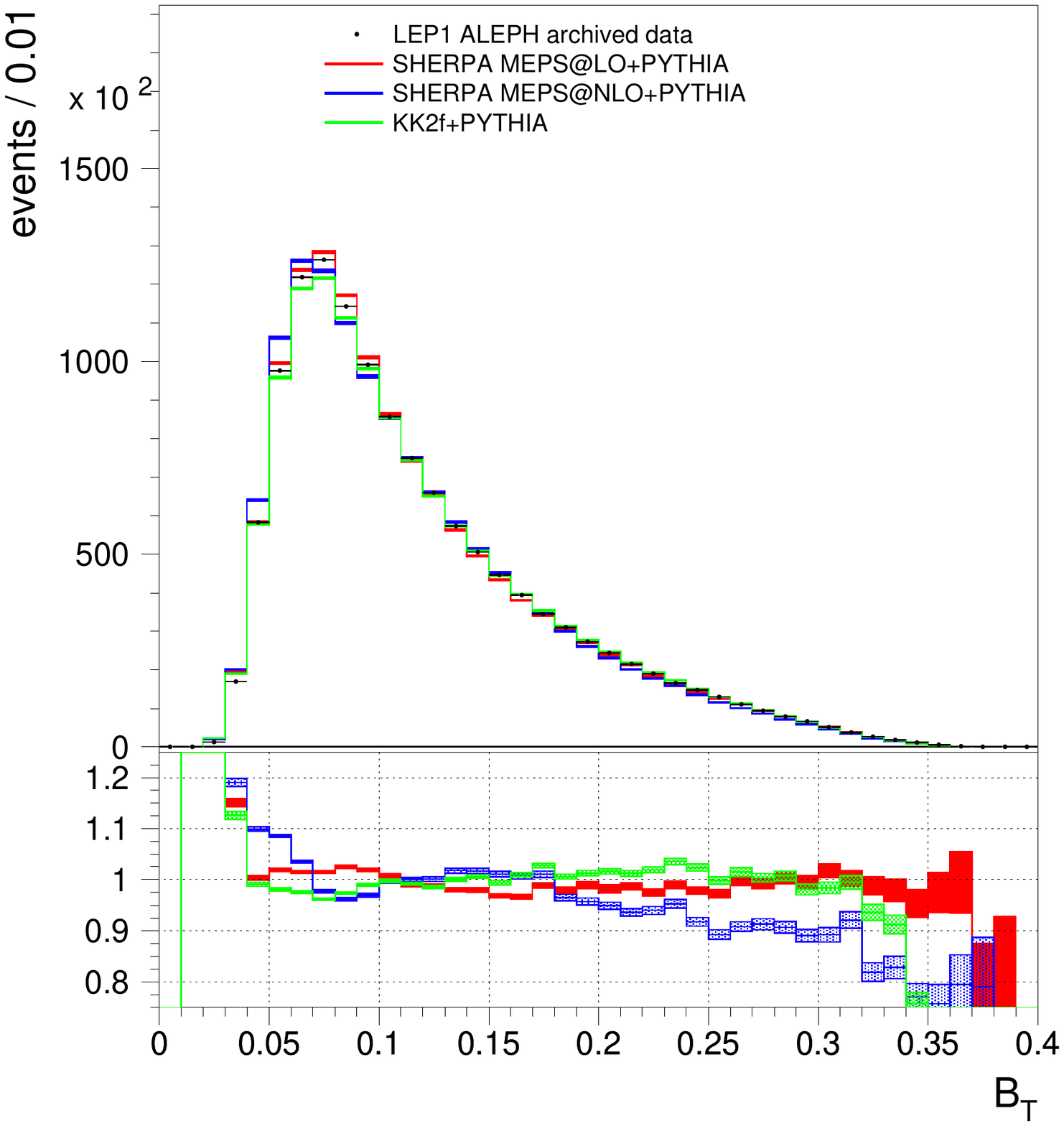}}\hspace{.4in}
\subfigure[]{\includegraphics[width=2.6in,bb=80 150 520 720]{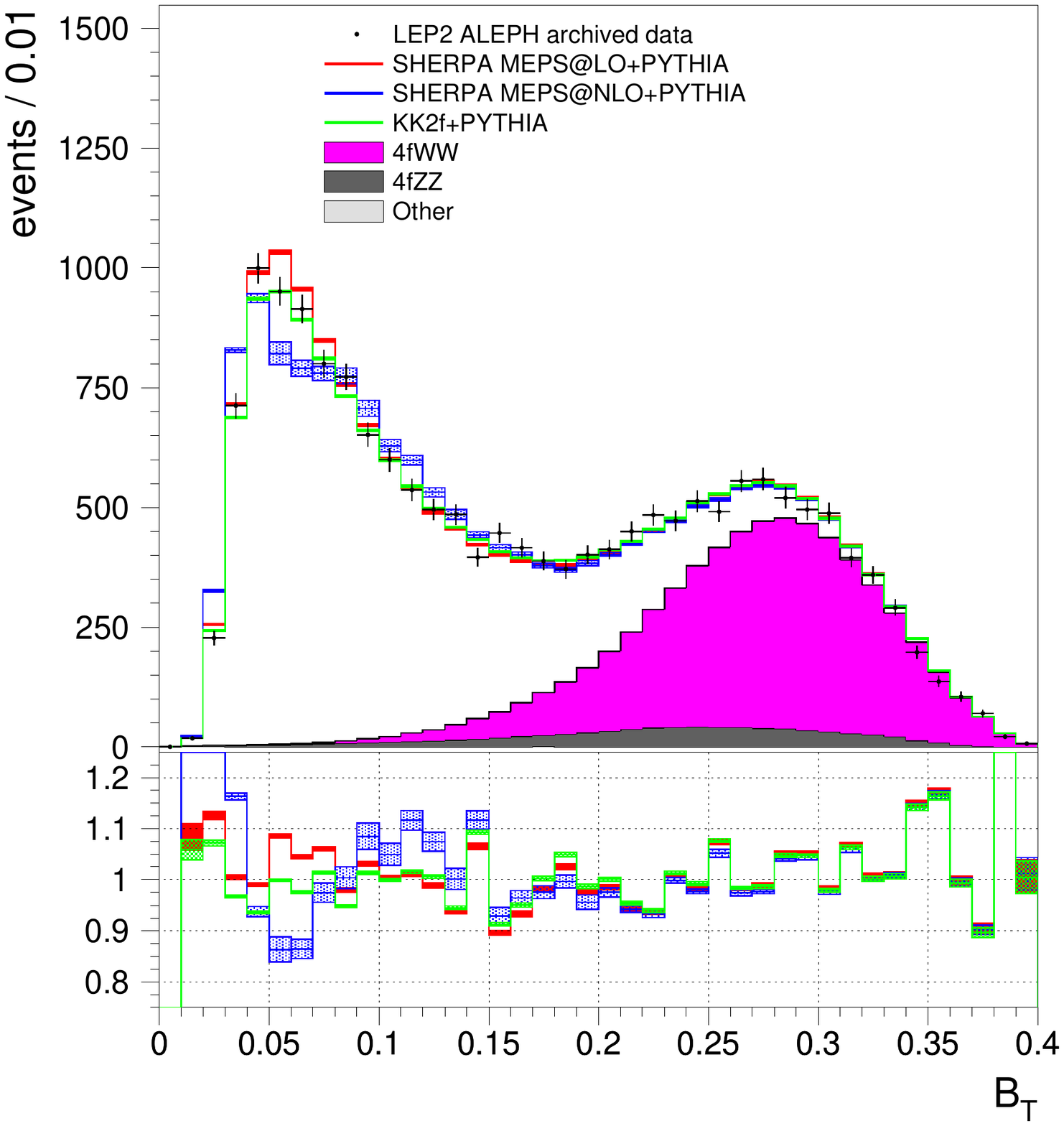}}
\end{center}
\caption{The (top) oblateness $O$ and (bottom) total jet broadening $B_T$ plotted (left) at LEP1 and (right) at LEP2.  Conventions are the same as those in Fig. \ref{fig:thr}.}
\label{fig:obl}
\end{figure}

Here, we present a few additional distributions not given in the main text.  First, we present two additional event-shape variables, oblateness $O$ and the total jet broadening $B_T$ for both LEP1 and LEP2, in Fig. \ref{fig:obl}.  We see at LEP1 that both SHERPA simulations of $O$ improve substantially over that of KK2f; this effect is diluted by the $W^+W^-$ background at LEP2.  The LO SHERPA simulation of $B_T$ and that from KK2f are similar at both LEP1 and LEP2, with both performing better than the NLO SHERPA sample.

\begin{figure}[h]
\begin{center}
\subfigure[]{\includegraphics[width=2.6in,bb=80 150 520 720]{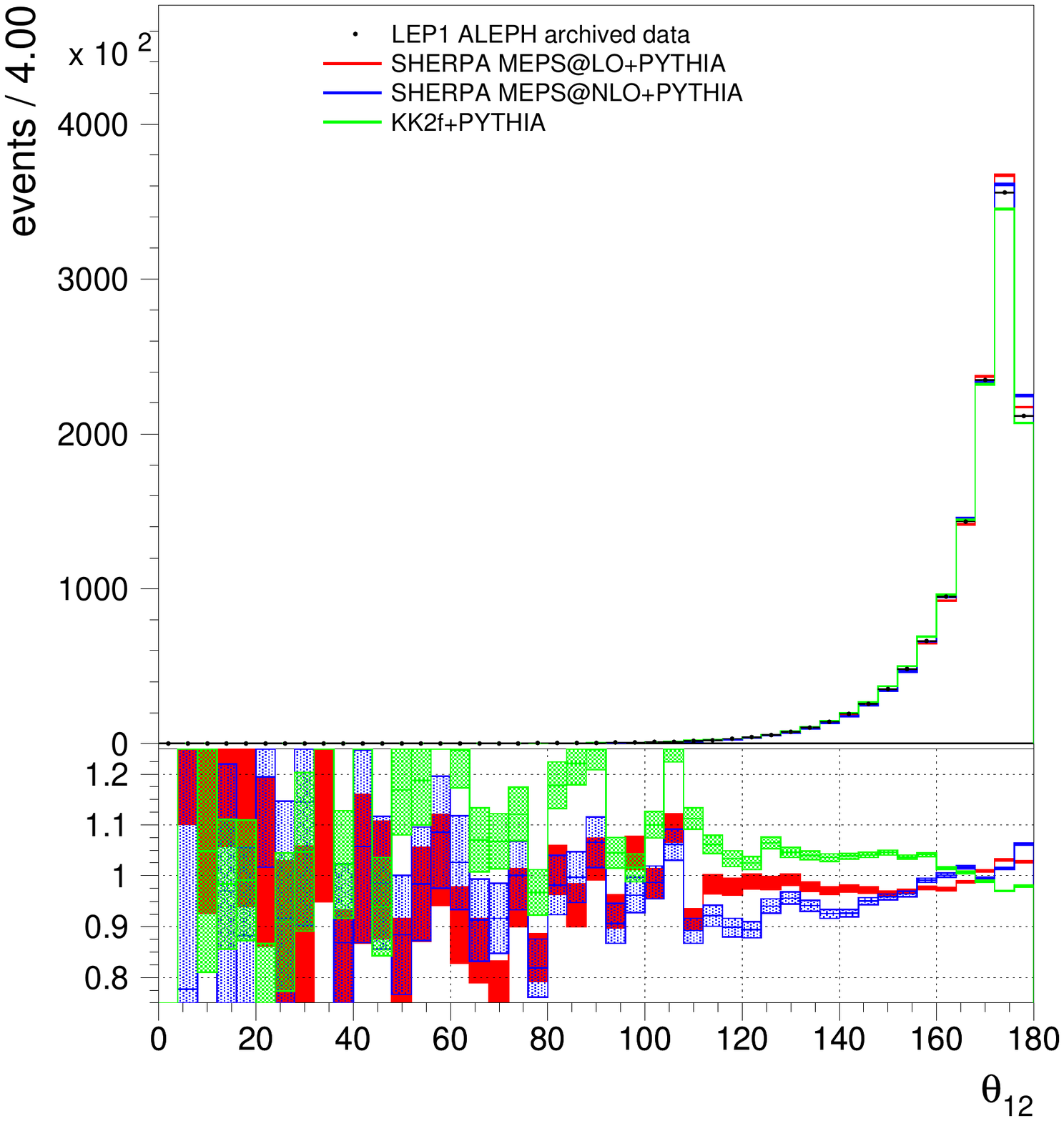}}\hspace{.35in}
\subfigure[]{\includegraphics[width=2.6in,bb=80 150 520 720]{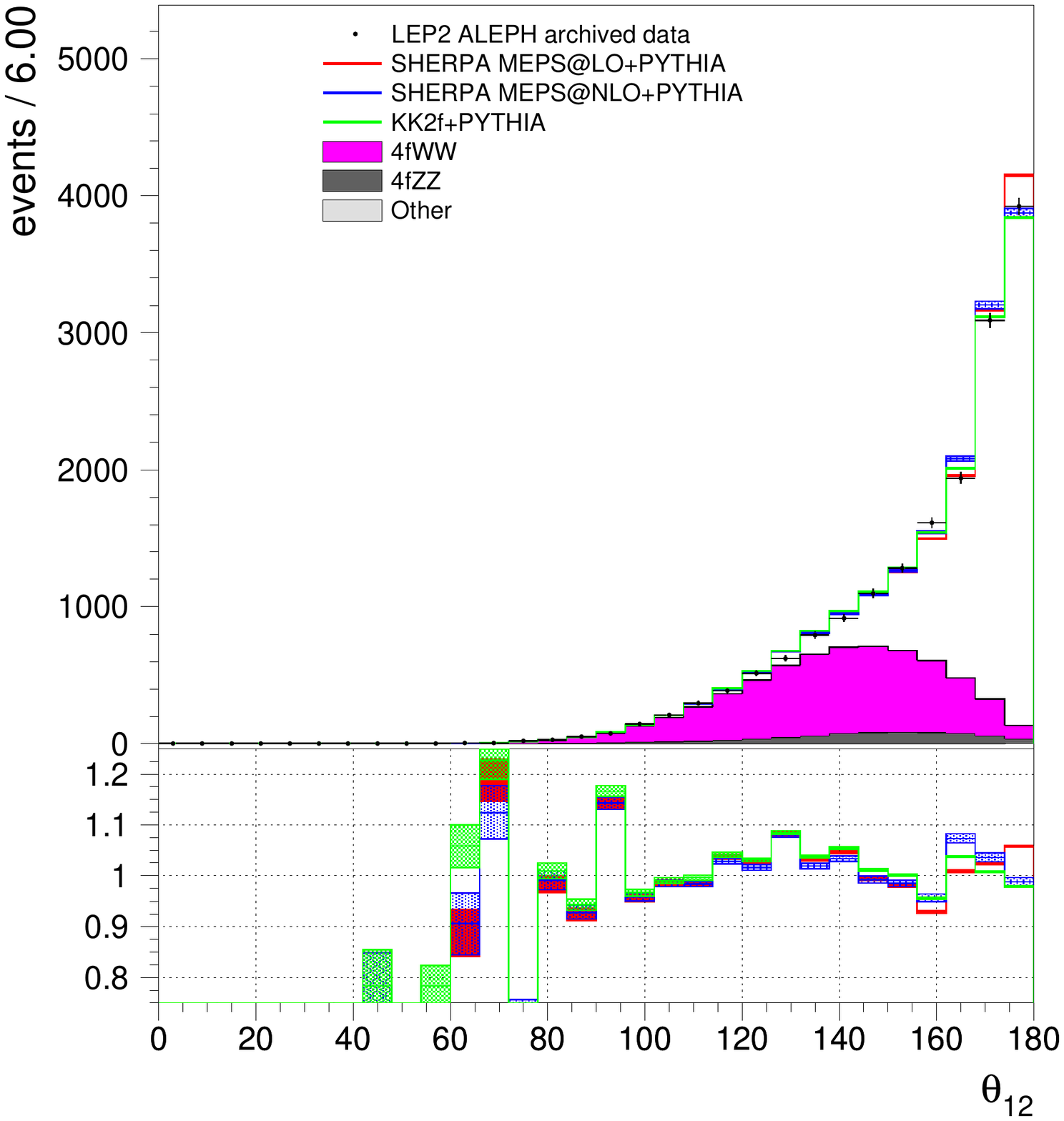}}\\
\subfigure[]{\includegraphics[width=2.6in,bb=80 150 520 720]{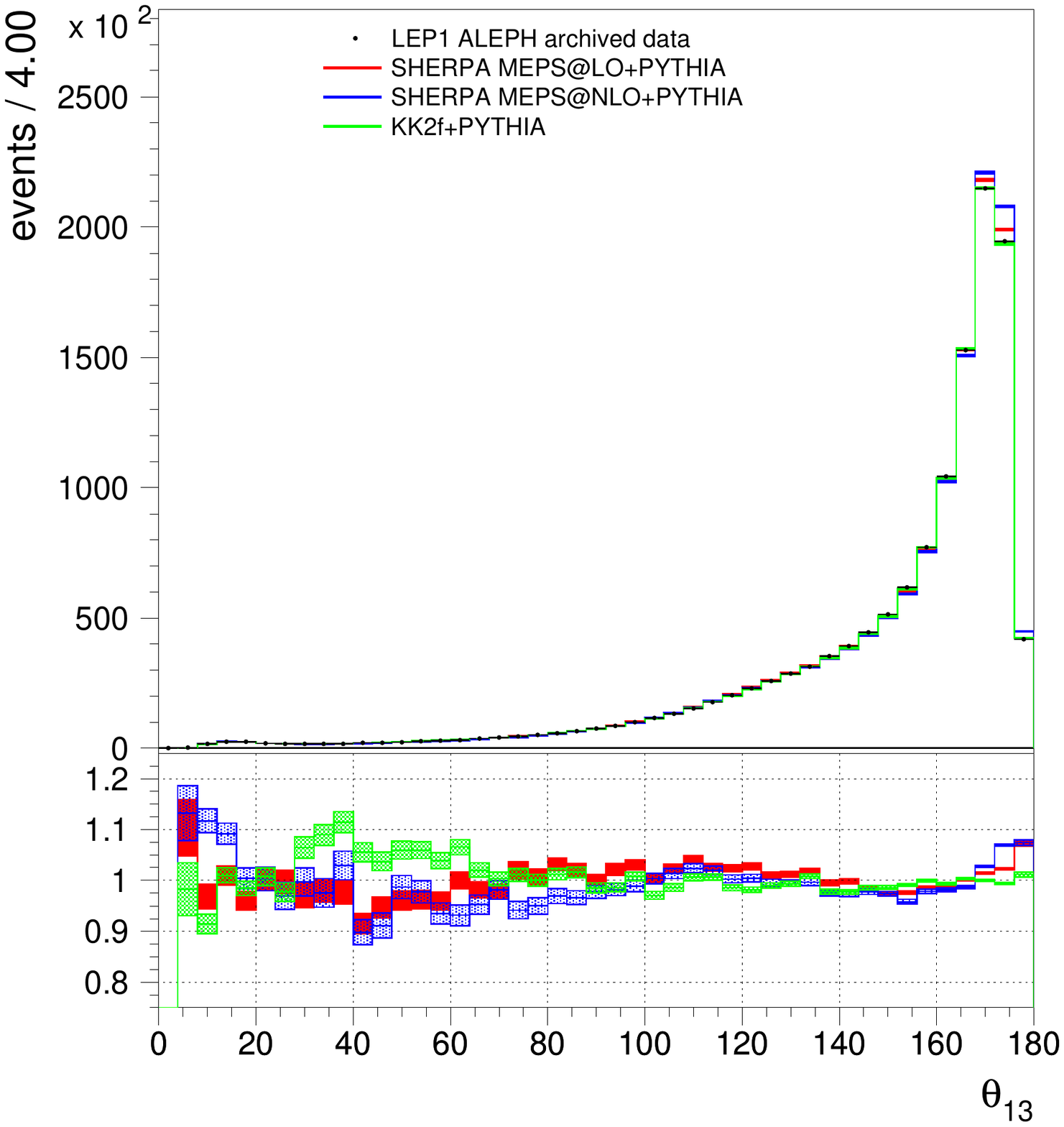}}\hspace{.35in}
\subfigure[]{\includegraphics[width=2.6in,bb=80 150 520 720]{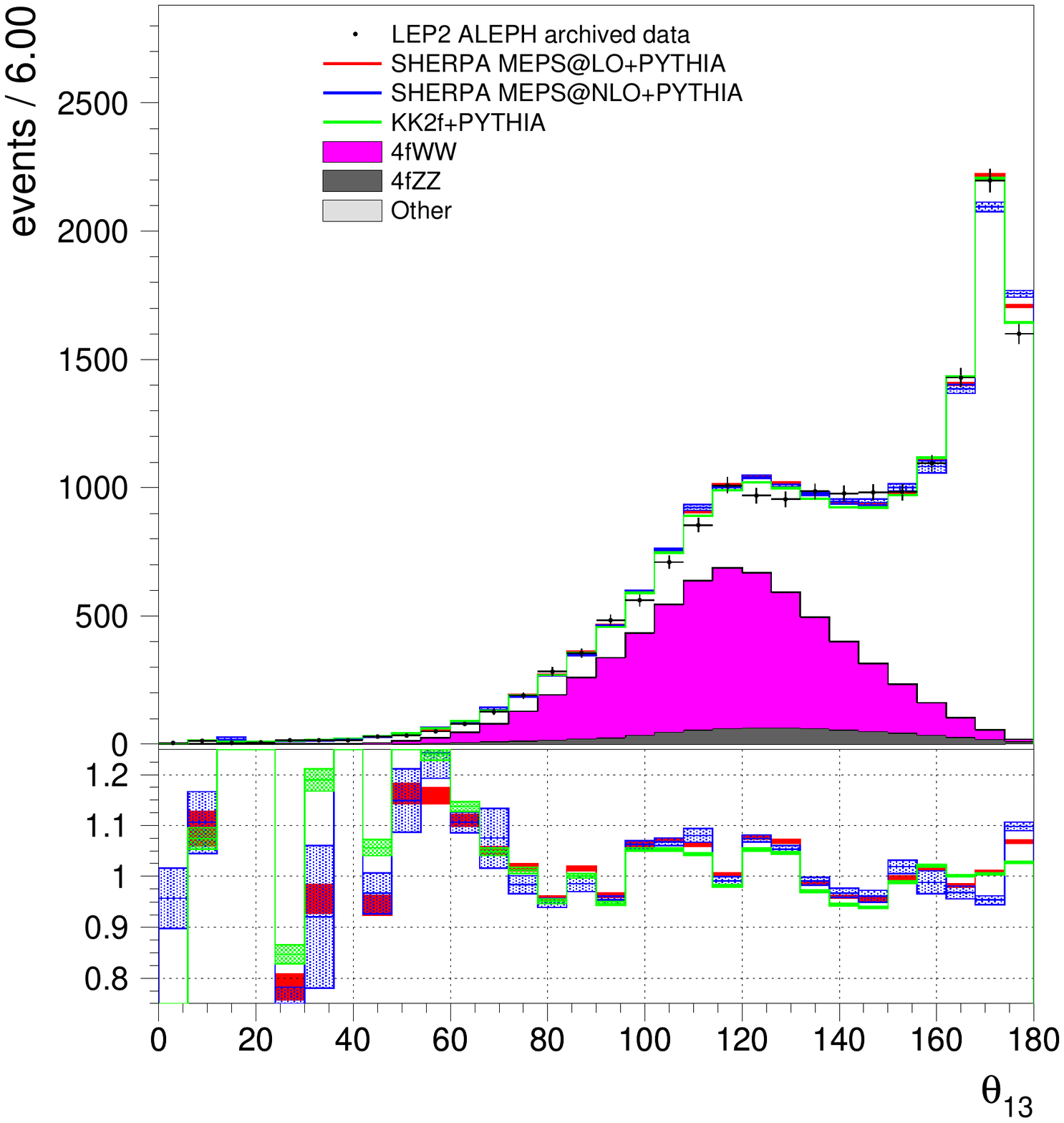}}
\end{center}
\caption{Inter-jet angles $\theta_{12}$ and $\theta_{13}$ at LEP1 and LEP2 using the DURHAM algorithm.  Data from ALEPH is compared to KK2f, SHERPA LO, and SHERPA NLO MC.}
\label{fig:t12t13dur}
\end{figure}

\begin{figure}[h]
\begin{center}
\subfigure[]{\includegraphics[width=2.6in,bb=80 150 520 720]{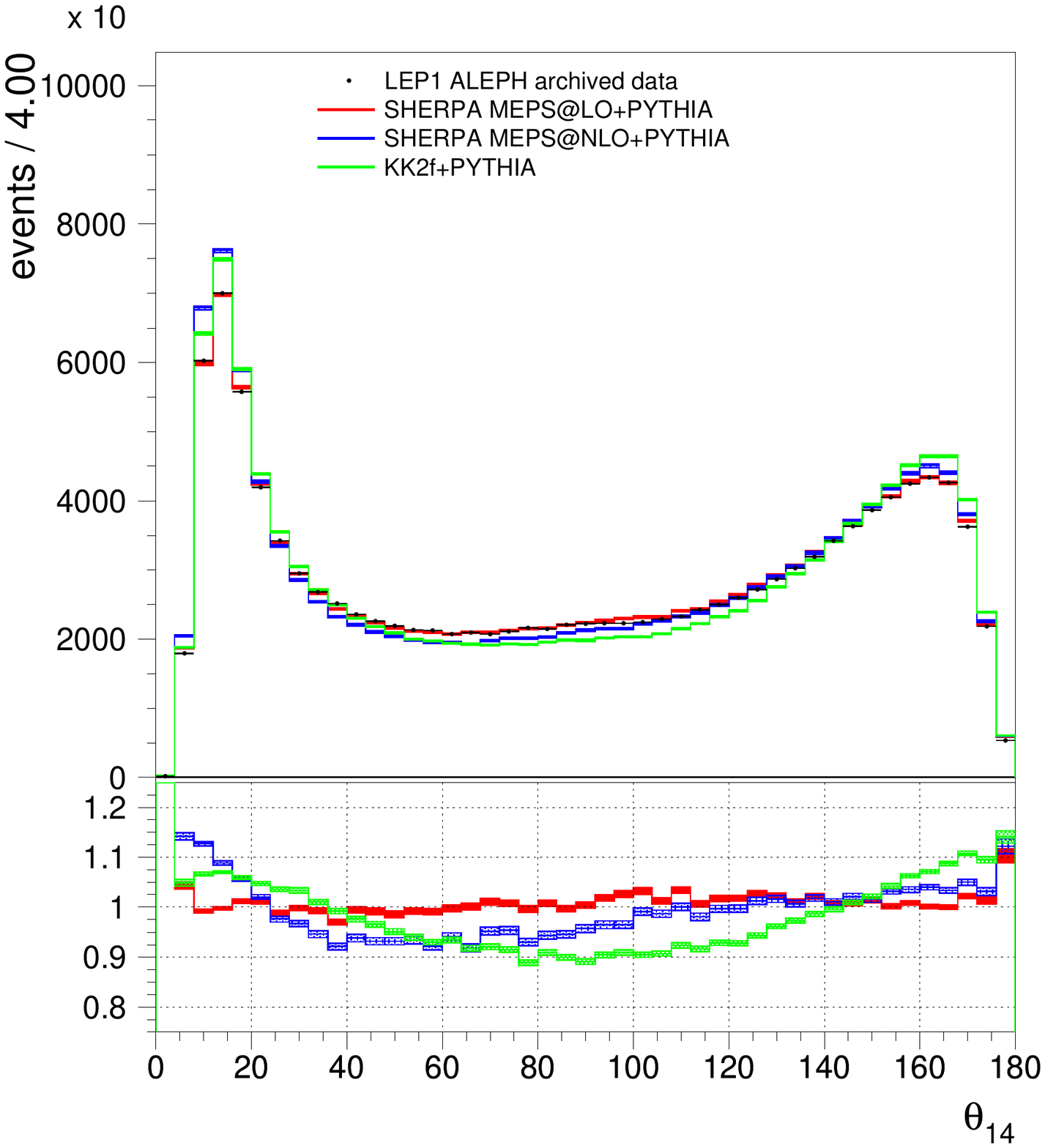}}\hspace{.35in}
\subfigure[]{\includegraphics[width=2.6in,bb=80 150 520 720]{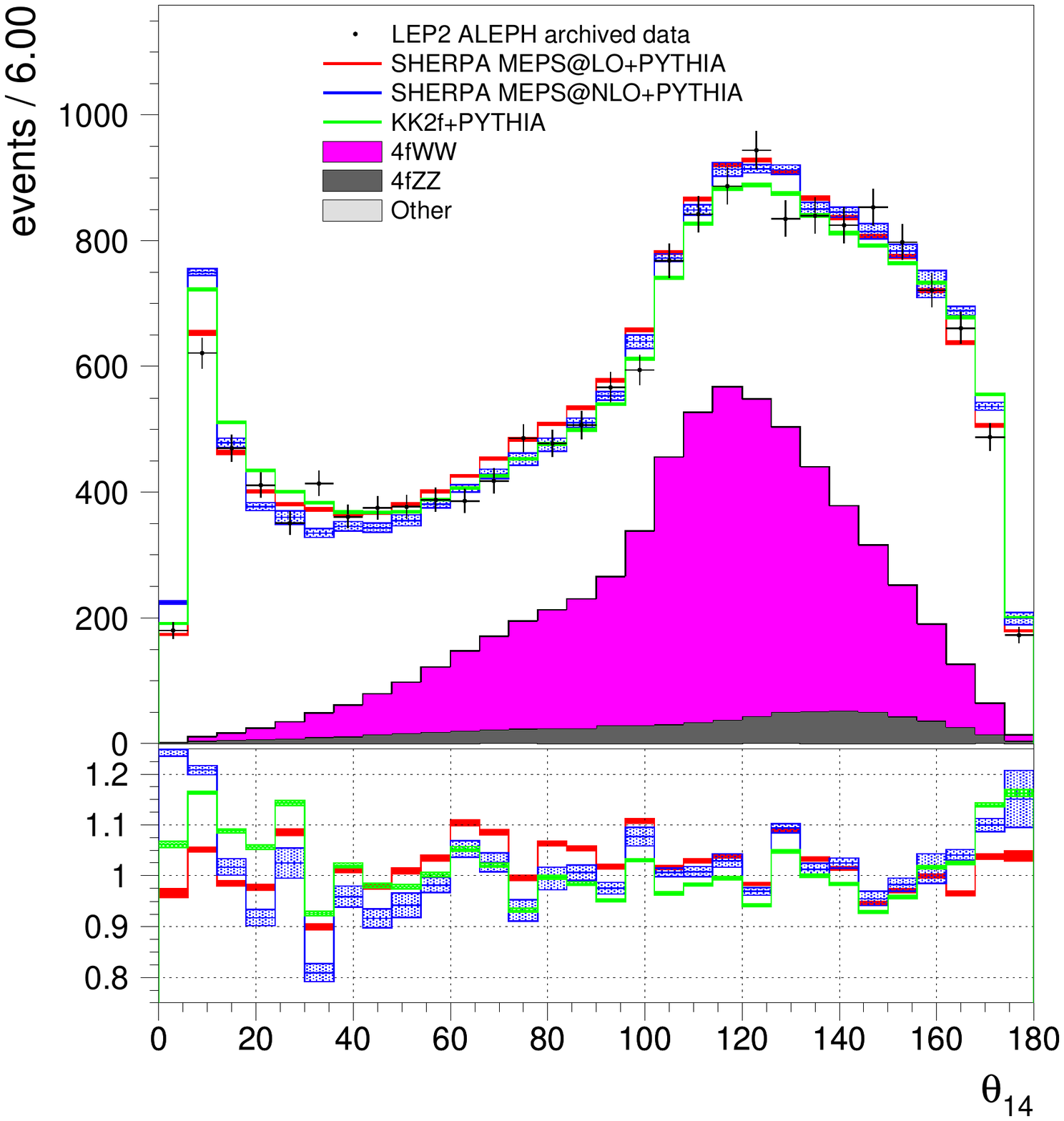}}\\
\subfigure[]{\includegraphics[width=2.6in,bb=80 150 520 720]{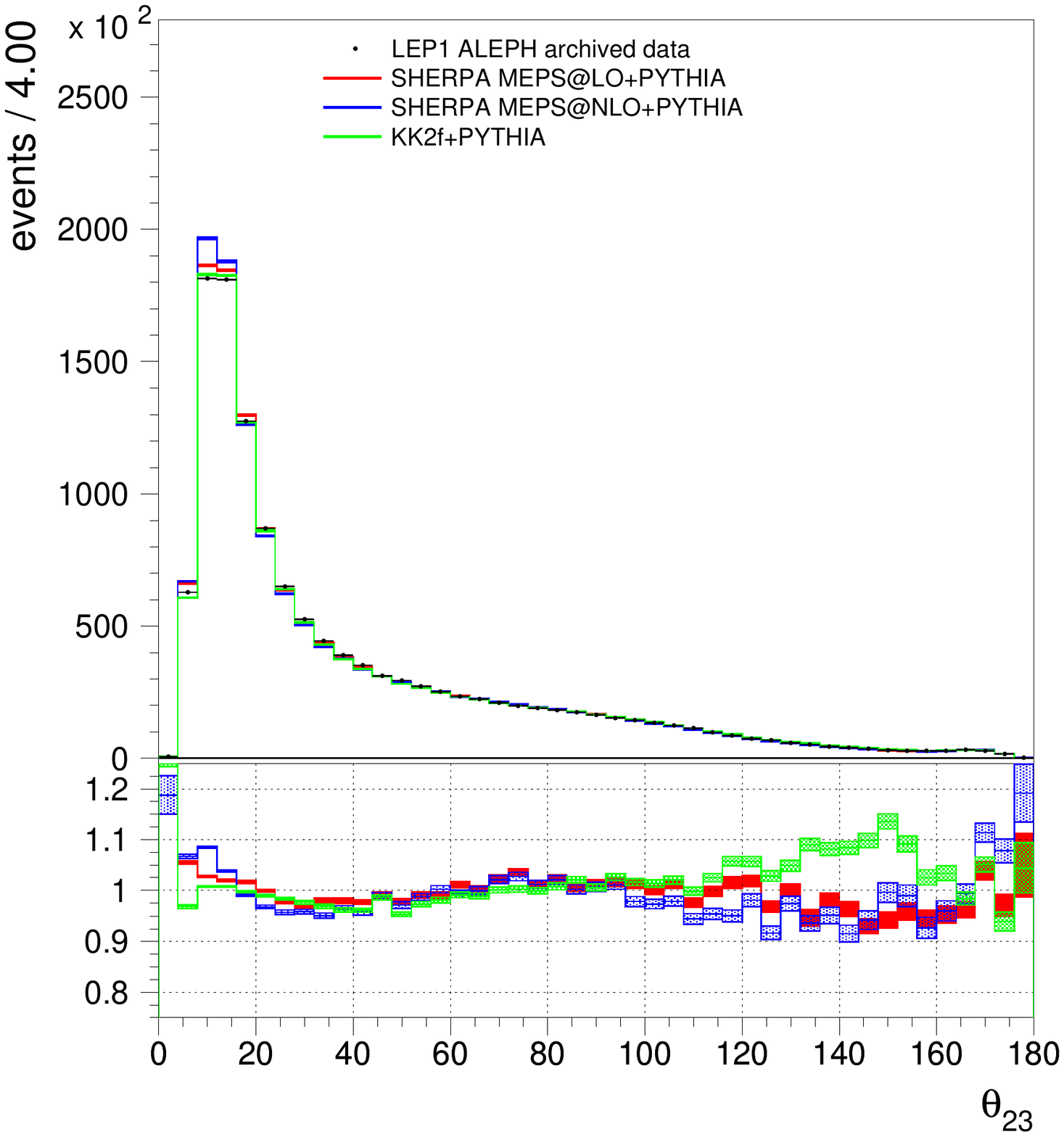}}\hspace{.35in}
\subfigure[]{\includegraphics[width=2.6in,bb=80 150 520 720]{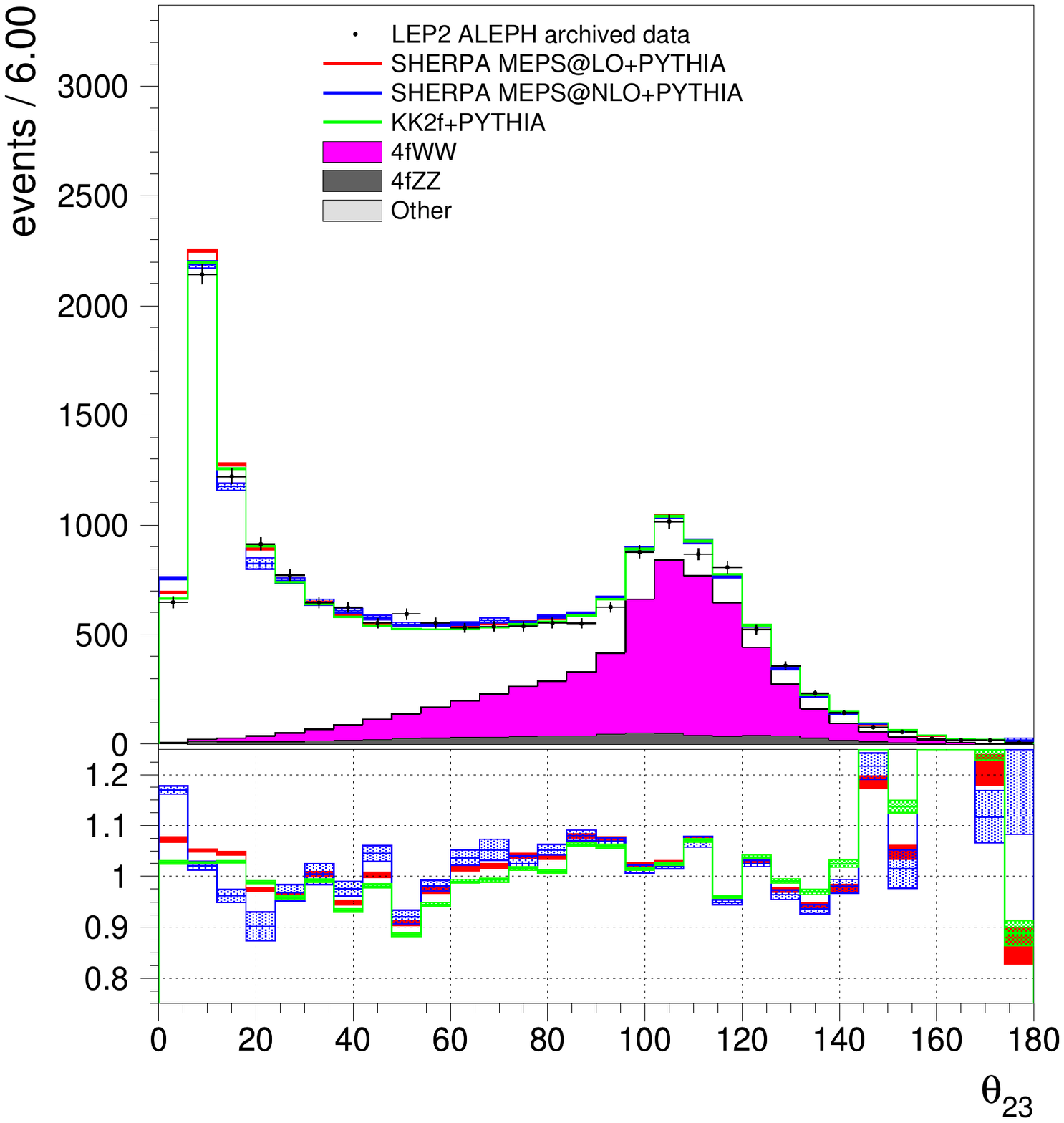}}
\end{center}
\caption{Inter-jet angles $\theta_{14}$ and $\theta_{23}$ at LEP1 and LEP2 using the DURHAM algorithm.  Data from ALEPH is compared to KK2f, SHERPA LO, and SHERPA NLO MC.}
\label{fig:t14t23dur}
\end{figure}

\begin{figure}[h]
\begin{center}
\subfigure[]{\includegraphics[width=2.6in,bb=80 150 520 720]{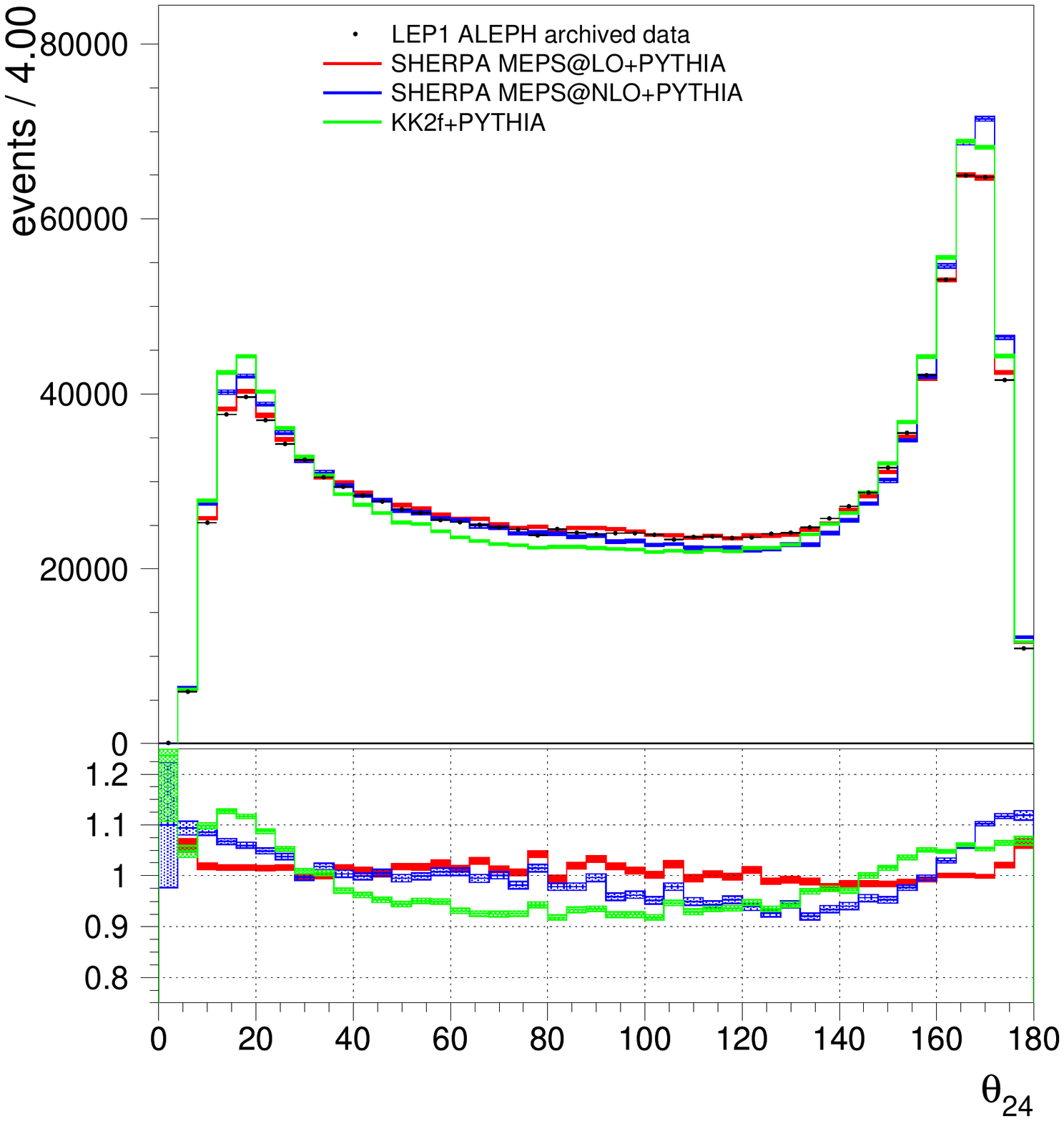}}\hspace{.35in}
\subfigure[]{\includegraphics[width=2.6in,bb=80 150 520 720]{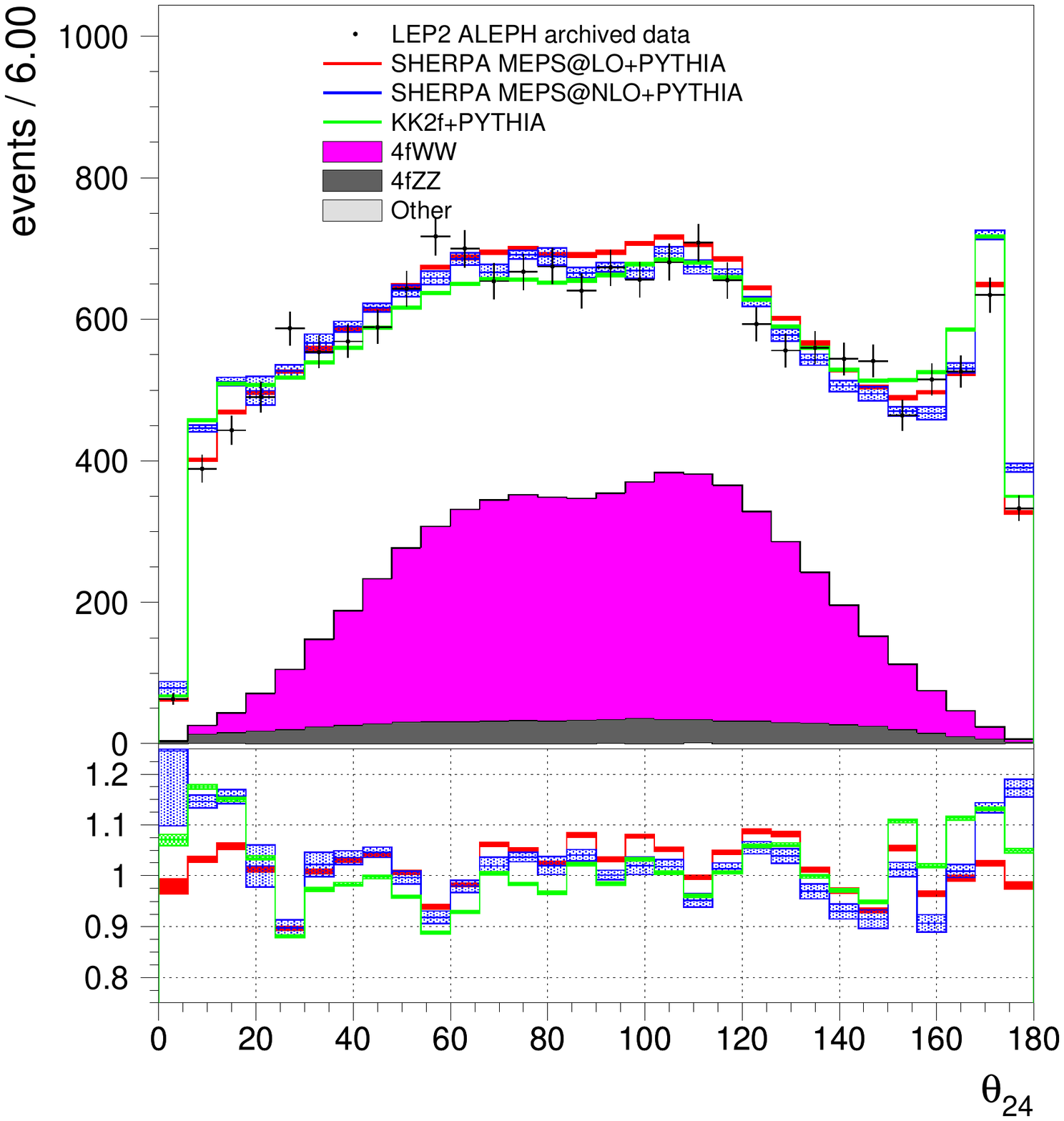}}\\
\subfigure[]{\includegraphics[width=2.6in,bb=80 150 520 720]{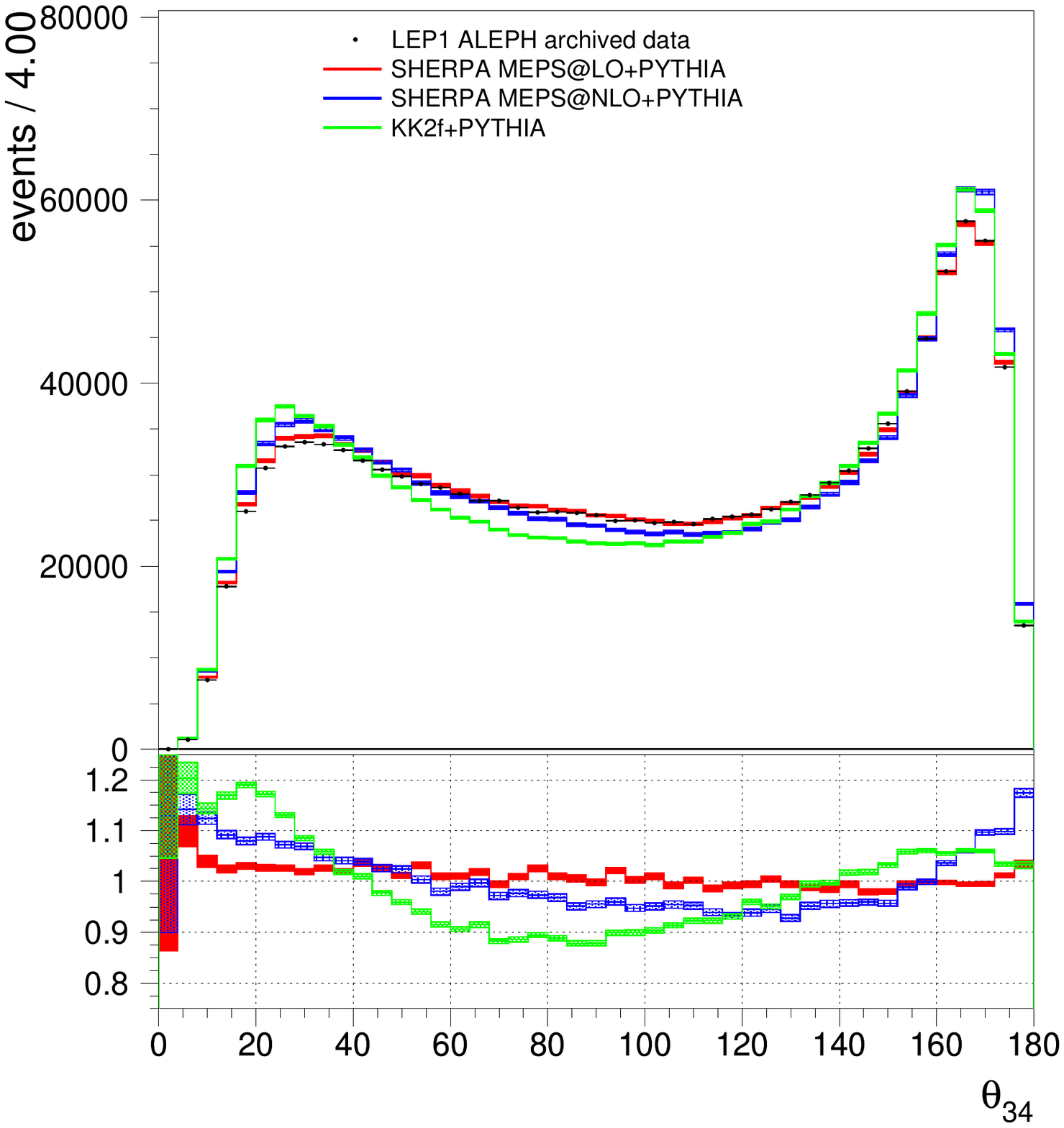}}\hspace{.35in}
\subfigure[]{\includegraphics[width=2.6in,bb=80 150 520 720]{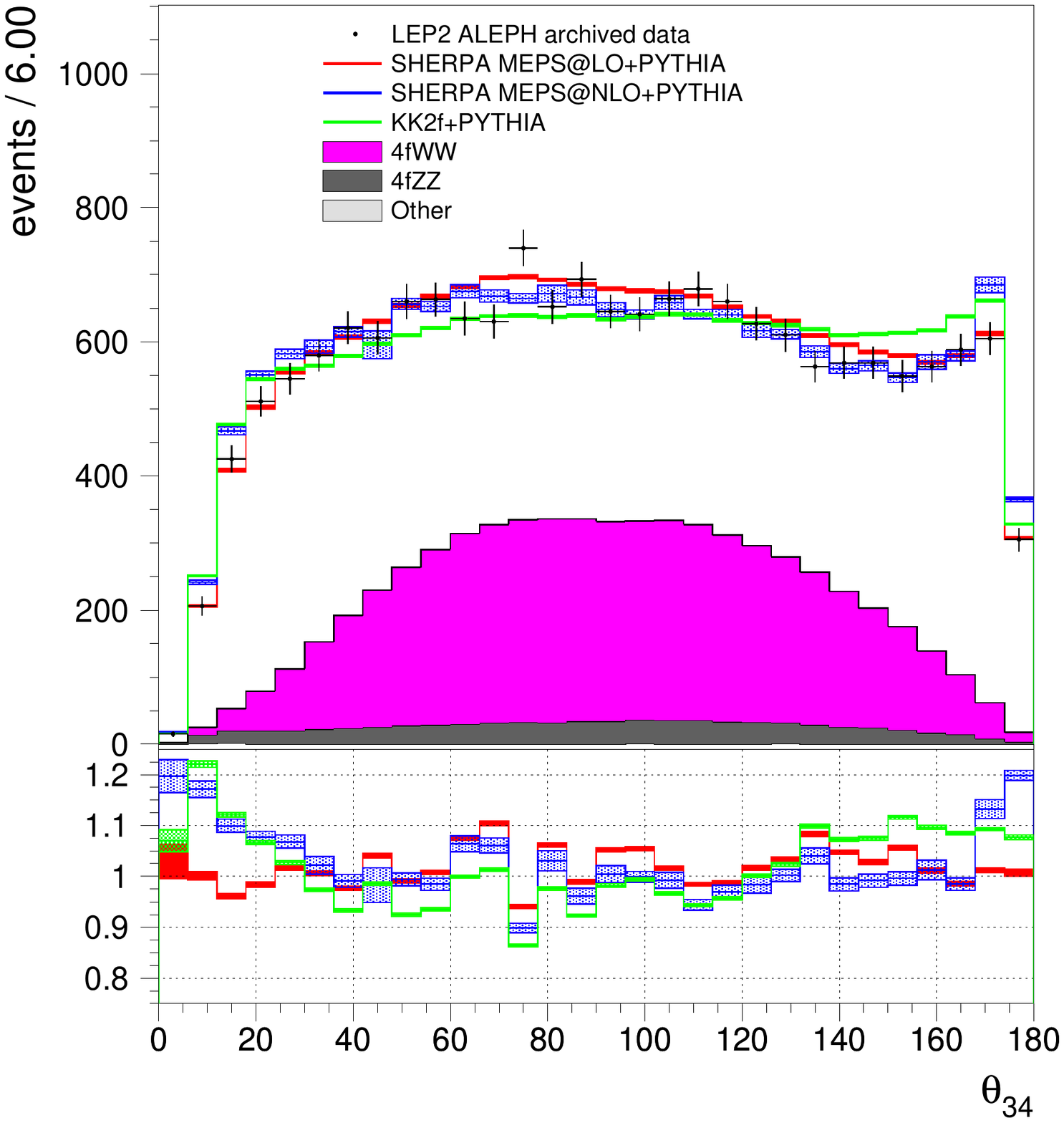}}
\end{center}
\caption{Inter-jet angles $\theta_{24}$ and $\theta_{34}$ at LEP1 and LEP2 using the DURHAM algorithm.  Data from ALEPH is compared to KK2f, SHERPA LO, and SHERPA NLO MC.}
\label{fig:t24t34dur}
\end{figure}

We present the inter-jet angles for jets clustered with the DURHAM algorithm in Figs. \ref{fig:t12t13dur}-\ref{fig:t24t34dur}.  These plots largely reflect the features of the analogous LUCLUS distributions.

\begin{figure}[h]
\begin{center}
\subfigure[]{\includegraphics[width=2.6in,bb=80 150 520 720]{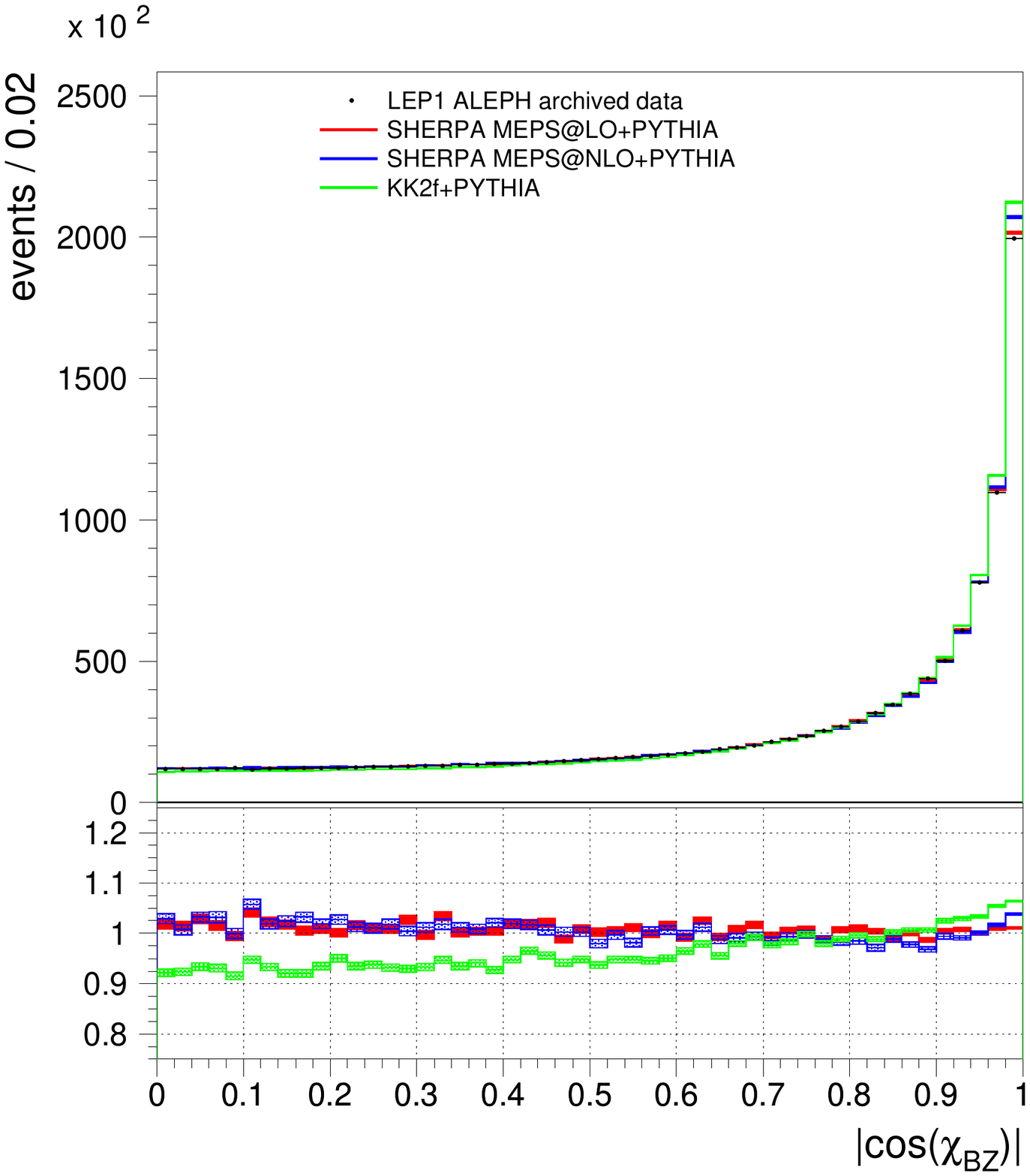}}\hspace{.35in}
\subfigure[]{\includegraphics[width=2.6in,bb=80 150 520 720]{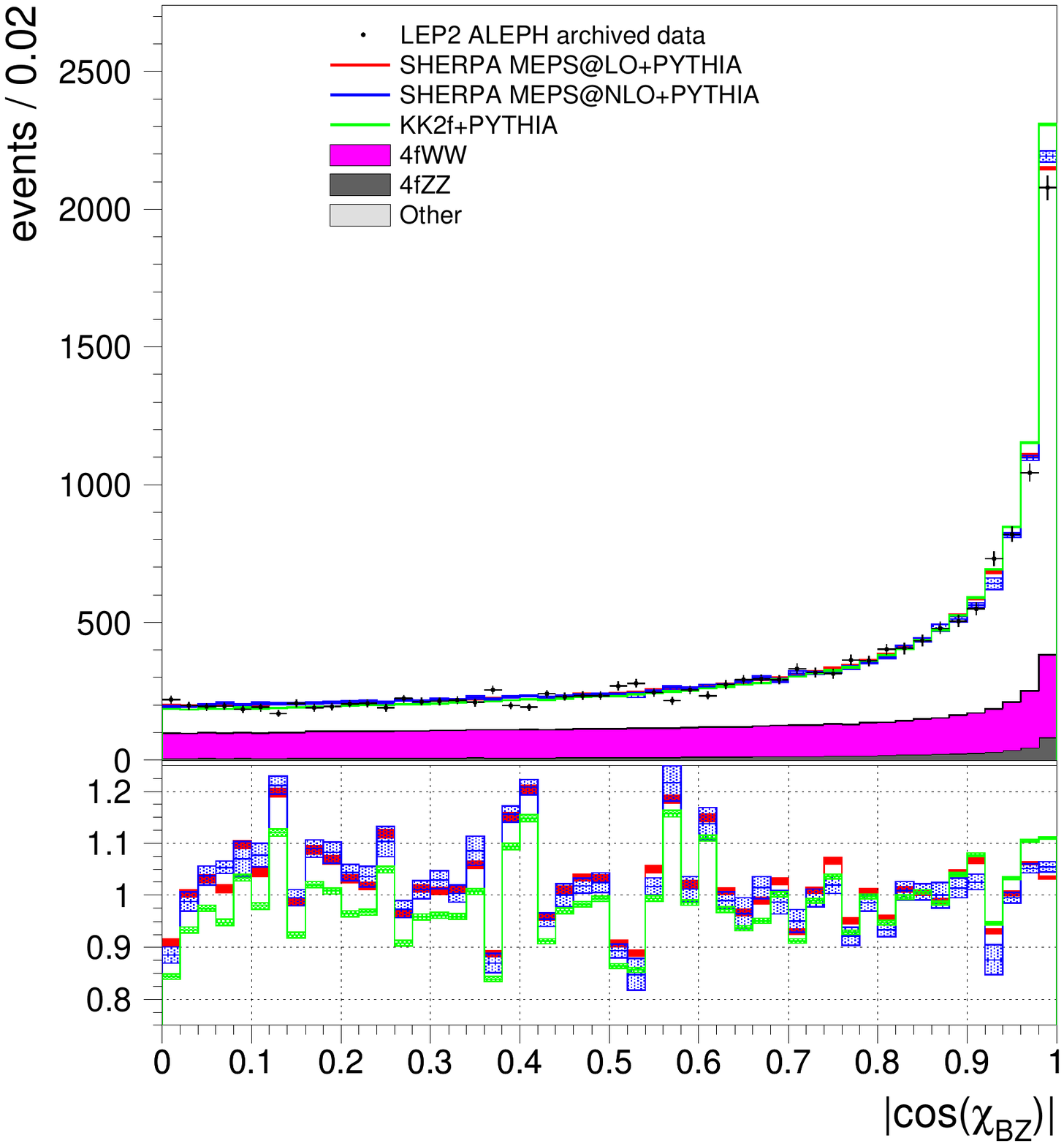}}\\
\subfigure[]{\includegraphics[width=2.6in,bb=80 150 520 720]{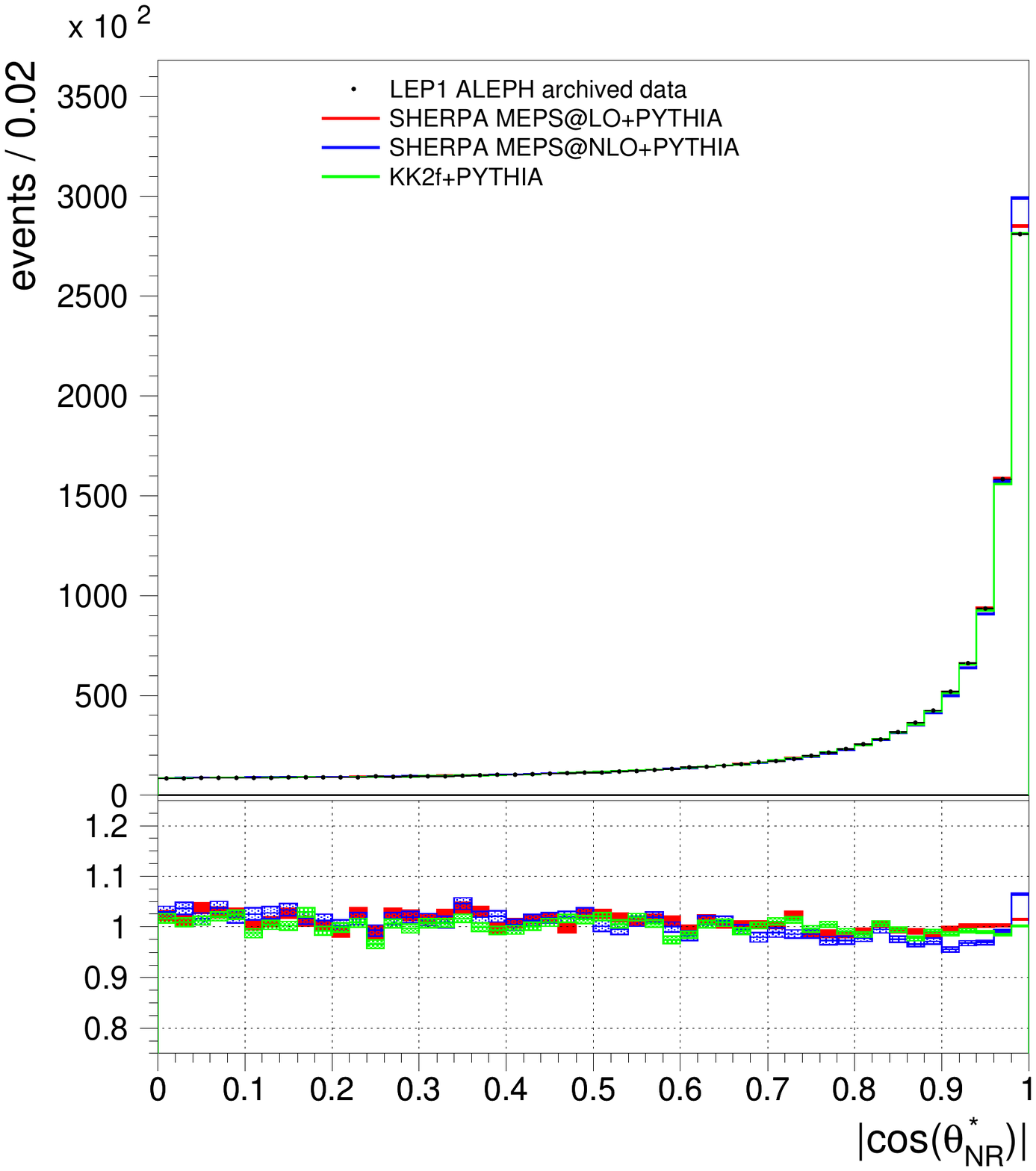}}\hspace{.35in}
\subfigure[]{\includegraphics[width=2.6in,bb=80 150 520 720]{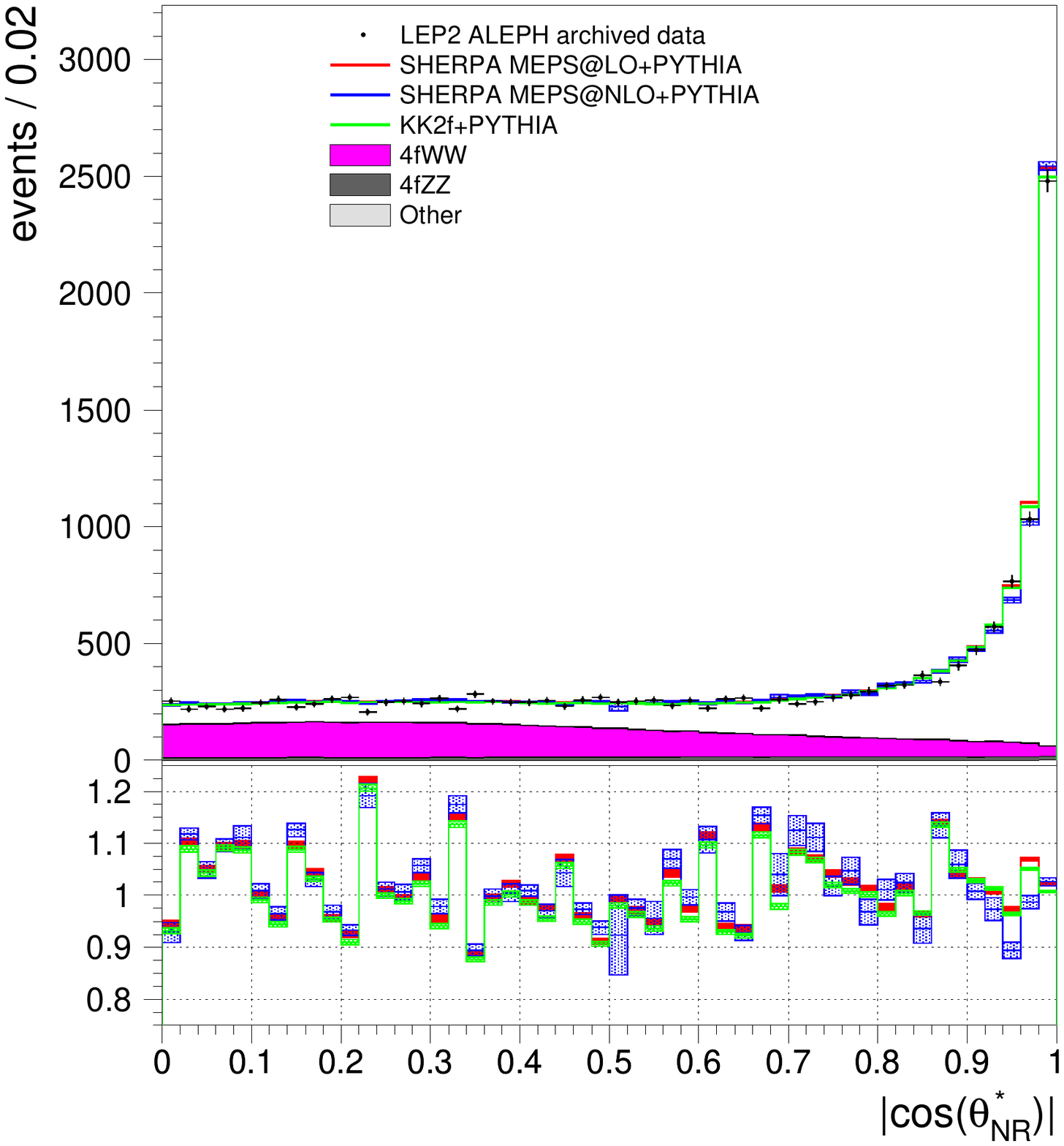}}
\end{center}
\caption{Plots of the Bengtsson-Zerwas (top) and the modified Nachtman-Reiter (bottom) angles obtained with DURHAM jet clustering.  Data from ALEPH is compared to KK2f, SHERPA LO, and SHERPA NLO MC.  Plots on the left are from LEP1, while those on the right are from LEP2.}
\label{fig:bznrangdur}
\end{figure}

\begin{figure}[h]
\begin{center}
\subfigure[]{\includegraphics[width=2.6in,bb=80 150 520 720]{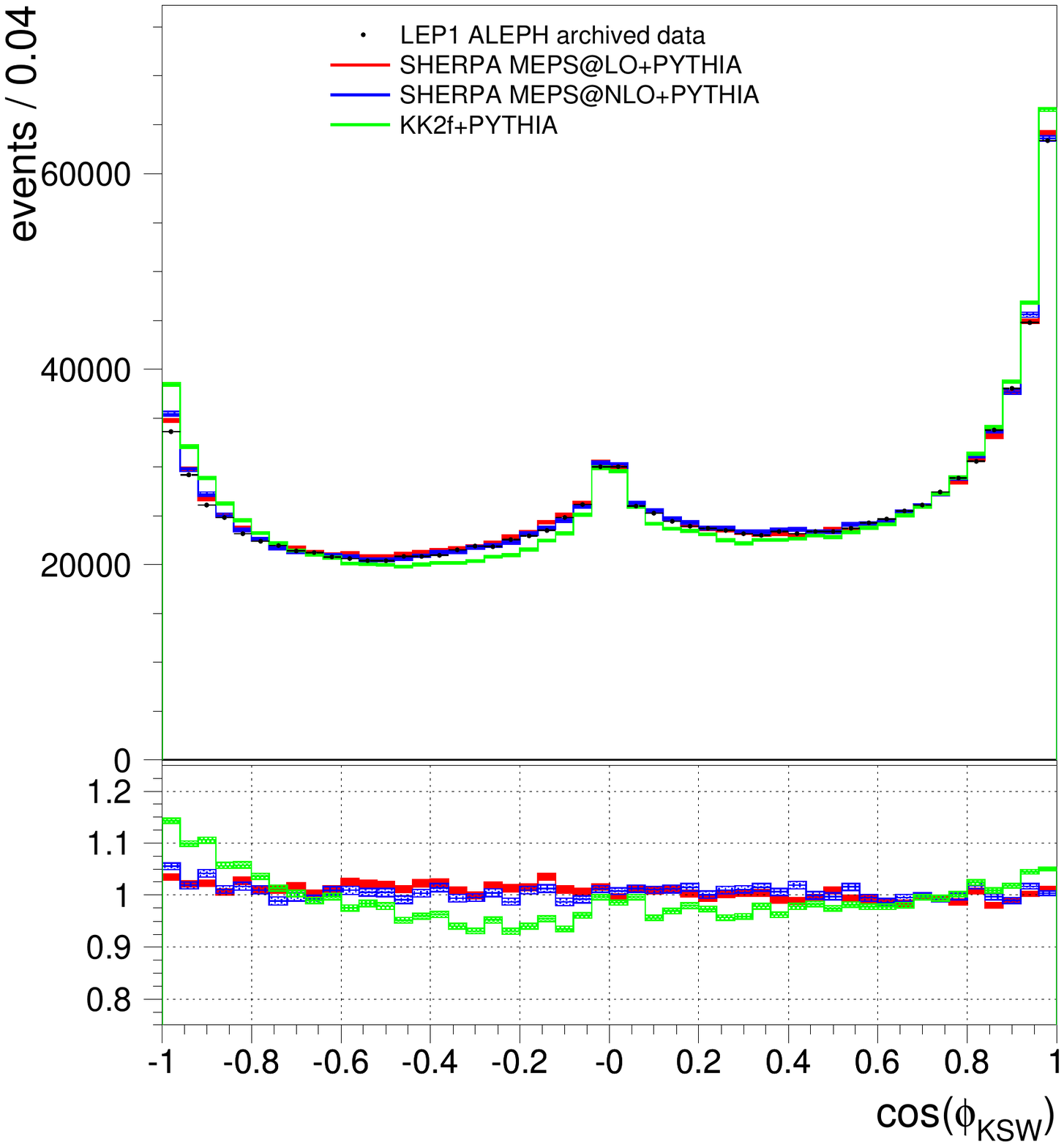}}\hspace{.35in}
\subfigure[]{\includegraphics[width=2.6in,bb=80 150 520 720]{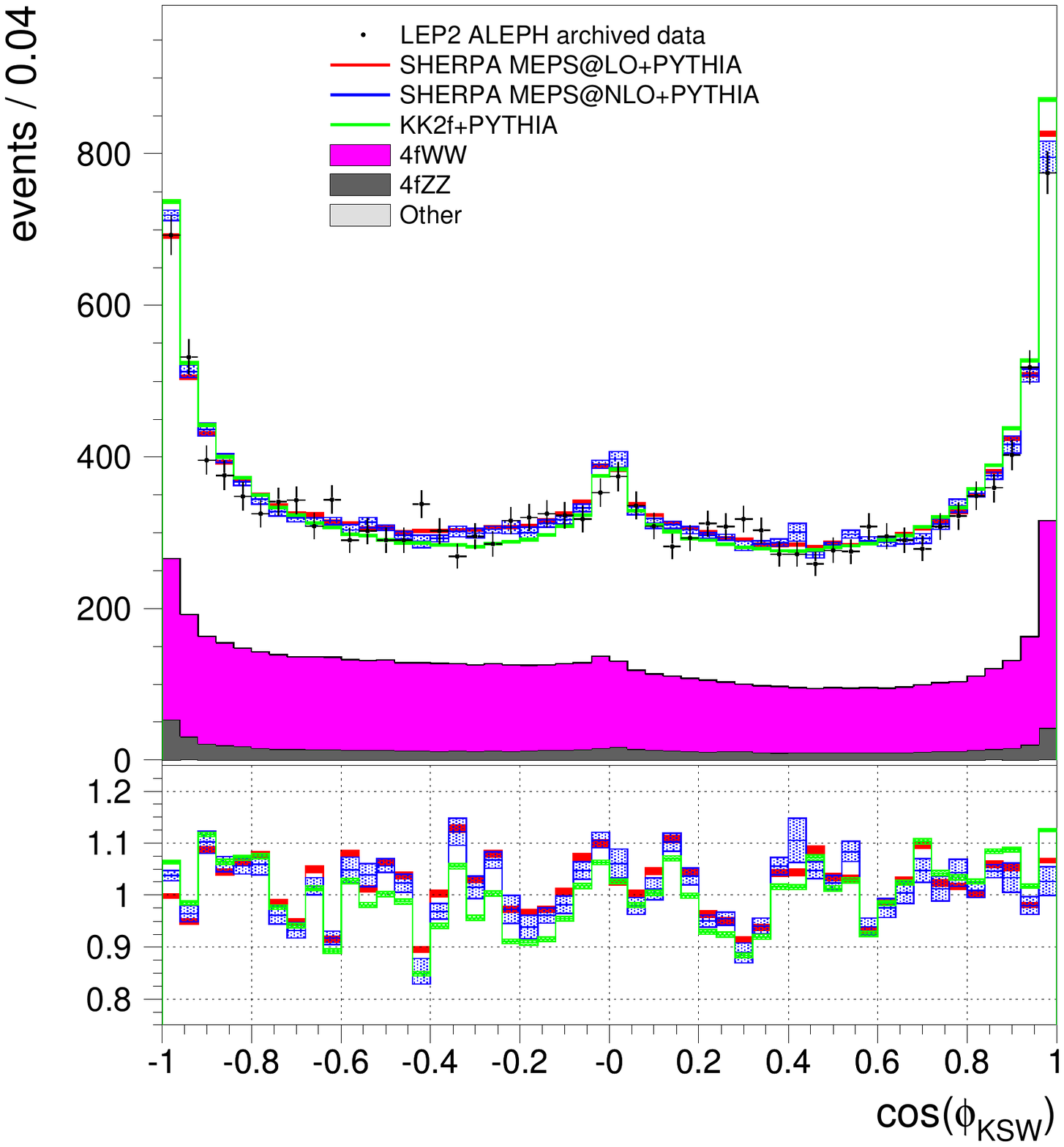}}
\end{center}
\caption{Plots of the K\"{o}rner-Schierholz-Willrodt angle constructed from DURHAM jets.  Data from ALEPH is compared to KK2f, SHERPA LO, and SHERPA NLO MC.  Plot (a) is from LEP1; plot (b) is from LEP2.}
\label{fig:kswangdur}
\end{figure}

We plot the Bengtsson-Zerwas and modified Nachtman-Reiter angles for both LEP1 and LEP2 in Fig. \ref{fig:bznrangdur} and the K\"{o}rner-Schierholz-Willrodt angle  in Fig. \ref{fig:kswangdur}, all for jets clustered with the DURHAM algorithm.  The plots again reflect the features of the analogous LUCLUS plots.

\section{AHADIC++ Tune Details}
\label{ahadicapp}

Here we give details of our LO tune using SHERPA v. 2.2.4 with AHADIC++ for hadronization.  We first do a preliminary tune over event-shape parameters, with flavor parameters held fixed.  We then hold the event-shape parameters fixed to the values obtained with this preliminary tune while we do a tune over flavor parameters, using cubic interpolation.  We then keep the values of the flavor parameters fixed and do a final event-shape tune, using quartic interpolation. 

\begin{table}
\begin{tabular}{|c |c| c| c|}
\hline
& Parameter & Tune Range & Tune Value \\
\hline
& DECAY\_EXPONENT & 0.5 - 5.0 & 3.9198\\
& DECAY\_OFFSET & 0.8 - 1.8 & 1.3683\\
& SINGLET\_SUPPRESSION & 0.35 - 0.85 & 0.7944\\
Flavor parameters & STRANGE\_FRACTION & 0.2 - 0.8 & 0.6440\\
& BARYON\_FRACTION & 0.5 - 3.0  & 2.5025\\
& P\_\{QQ\_1\}/P\_\{QQ\_0\} & 0.  - 1.3 & 1.2124\\
& P\_\{QS\}/P\_\{QQ\} & 0.  - 0.5 & 0.0600\\
& P\_\{SS\}/P\_\{QQ\} & 0.  - 0.5 & 0.0211\\
\hline
& CSS\_FS\_PT2MIN & 0.2 - 2.2 & 1.004347\\
& CSS\_FS\_AS\_FAC & 0.4 - 1.4 & 0.761703\\
& PT\_MAX & 0.  - 4.0 & 3.333018\\
Event-shape parameters & PT\_MAX\_FACTOR & 0.1 - 8.0 & 1.295605\\
& PT\^{}2\_0 & 0.  - 0.5 & 0.2642895\\
& DECAY\_EXPONENT & 1.0 - 4.5 & 2.234268\\
& DECAY\_OFFSET & 1.0 - 1.5 & 1.226938\\
\hline
\end{tabular}
\caption{Flavor and even-shape parameter results from a LO tune with AHADIC++ used for hadronization.}
\label{tab:ahadflavtun}
\end{table}

The parameters and their ranges used for our tune using AHADIC++ are given in Table \ref{tab:ahadflavtun}, along with their final values.  Throughout the tune, the value of $\alpha_s$ was set to $0.118$.  All other parameters were kept at their default values.

\subsection{AHADIC++ LO Flavor Tune Weight file}
\footnotesize
\begin{verbatim}
/ALEPH_1996_S3486095/d01-x01-y01  1  # Sphericity, $S$ (charged)
/ALEPH_1996_S3486095/d02-x01-y01  1  # Aplanarity, $A$ (charged)
/ALEPH_1996_S3486095/d03-x01-y01  1  # 1-Thrust, $1-T$ (charged)
/ALEPH_1996_S3486095/d04-x01-y01  1  # Thrust minor, $m$ (charged)
/ALEPH_1996_S3486095/d05-x01-y01  1  # Two-jet resolution variable, $Y_3$ (charged)
/ALEPH_1996_S3486095/d06-x01-y01  1  # Heavy jet mass (charged)
/ALEPH_1996_S3486095/d07-x01-y01  1  # $C$ parameter (charged)
/ALEPH_1996_S3486095/d08-x01-y01  1  # Oblateness, $M - m$ (charged)
/ALEPH_1996_S3486095/d09-x01-y01  1  # Scaled momentum, $x_p = |p|/|p_\text{beam}|$ (charged)
/ALEPH_1996_S3486095/d10-x01-y01  1  # Rapidity w.r.t. thrust axes, $y_T$ (charged)
/ALEPH_1996_S3486095/d11-x01-y01  1  # In-plane $p_T$ in GeV w.r.t. sphericity axes (charged)
/ALEPH_1996_S3486095/d12-x01-y01  1  # Out-of-plane $p_T$ in GeV w.r.t. sphericity axes (charged)
/ALEPH_1996_S3486095/d17-x01-y01  1  # Log of scaled momentum, $\log(1/x_p)$ (charged)
/ALEPH_1996_S3486095/d18-x01-y01  1  # Charged multiplicity distribution
/ALEPH_1996_S3486095/d19-x01-y01  100  # Mean charged multiplicity
/ALEPH_1996_S3486095/d20-x01-y01  20  # Mean charged multiplicity for rapidity $|Y| < 0.5$
/ALEPH_1996_S3486095/d21-x01-y01  20  # Mean charged multiplicity for rapidity $|Y| < 1.0$
/ALEPH_1996_S3486095/d22-x01-y01  20  # Mean charged multiplicity for rapidity $|Y| < 1.5$
/ALEPH_1996_S3486095/d23-x01-y01  20  # Mean charged multiplicity for rapidity $|Y| < 2.0$
/ALEPH_1996_S3486095/d25-x01-y01  1  # $\pi^\pm$ spectrum
/ALEPH_1996_S3486095/d26-x01-y01  1  # $K^\pm$ spectrum
/ALEPH_1996_S3486095/d28-x01-y01  1  # $\gamma$ spectrum
/ALEPH_1996_S3486095/d29-x01-y01  1  # $\pi^0$ spectrum
/ALEPH_1996_S3486095/d30-x01-y01  1  # $\eta$ spectrum
/ALEPH_1996_S3486095/d31-x01-y01  1  # $\eta'$ spectrum
/ALEPH_1996_S3486095/d32-x01-y01  1  # $K^0$ spectrum
/ALEPH_1996_S3486095/d34-x01-y01  1  # $\Xi^-$ spectrum
/ALEPH_1996_S3486095/d35-x01-y01  1  # $\Sigma^\pm(1385)$ spectrum
/ALEPH_1996_S3486095/d36-x01-y01  1  # $\Xi^0(1530)$ spectrum
/ALEPH_1996_S3486095/d37-x01-y01  1  # $\rho$ spectrum
/ALEPH_1996_S3486095/d38-x01-y01  1  # $\omega(782)$ spectrum 
/ALEPH_1996_S3486095/d39-x01-y01  1  # $K^{*0}(892)$ spectrum
/ALEPH_1996_S3486095/d40-x01-y01  1  # $\phi$ spectrum
/ALEPH_1996_S3486095/d43-x01-y01  1  # $K^{*\pm}(892)$ spectrum
/ALEPH_1996_S3486095/d44-x01-y02  10  # Mean $\pi^0$ multiplicity
/ALEPH_1996_S3486095/d44-x01-y03  10  # Mean $\eta$ multiplicity
/ALEPH_1996_S3486095/d44-x01-y04  10  # Mean $\eta'$ multiplicity
/ALEPH_1996_S3486095/d44-x01-y05  10  # Mean $K_S + K_L$ multiplicity
/ALEPH_1996_S3486095/d44-x01-y06  10  # Mean $\rho^0$ multiplicity
/ALEPH_1996_S3486095/d44-x01-y07  10  # Mean $\omega(782)$ multiplicity
/ALEPH_1996_S3486095/d44-x01-y08  10  # Mean $\phi$ multiplicity
/ALEPH_1996_S3486095/d44-x01-y09  10  # Mean $K^{*\pm}$ multiplicity
/ALEPH_1996_S3486095/d44-x01-y10  10  # Mean $K^{*0}$ multiplicity
/ALEPH_1996_S3486095/d44-x01-y11  10  # Mean $\Lambda$ multiplicity
/ALEPH_1996_S3486095/d44-x01-y12  10  # Mean $\Sigma$ multiplicity
/ALEPH_1996_S3486095/d44-x01-y13  10  # Mean $\Xi$ multiplicity
/ALEPH_1996_S3486095/d44-x01-y14  10  # Mean $\Sigma(1385)$ multiplicity
/ALEPH_1996_S3486095/d44-x01-y15  10  # Mean $\Xi(1530)$ multiplicity
/ALEPH_1996_S3486095/d44-x01-y16  10  # Mean $\Omega^\mp$ multiplicity

/ALEPH_1999_S4193598/d01-x01-y01  200  # Scaled energy of $D^{*\pm}$ in $e^+e^-\to Z\to\text{hadronic}$ at $\sqrt{s}=91.2$~GeV
/ALEPH_2001_S4656318/d06-x01-y01  1  # $b$ quark fragmentation function $f(x_B^\text{weak})$
/ALEPH_2001_S4656318/d07-x01-y01  1  # $b$ quark fragmentation function $f(x_B^\text{lead})$
/ALEPH_2001_S4656318/d08-x01-y01  10  # Mean of $b$ quark fragmentation function $f(x_B^\text{weak})$
/ALEPH_2001_S4656318/d08-x01-y02  10  # Mean of $b$ quark fragmentation function $f(x_B^\text{lead})$

/ALEPH_2004_S5765862/d157-x01-y01 10  # Durham jet resolution $3 \to 2$ ($E_\mathrm{CMS}=91.2$ GeV)
/ALEPH_2004_S5765862/d165-x01-y01 10  # Durham jet resolution $4 \to 3$ ($E_\mathrm{CMS}=91.2$ GeV)
/ALEPH_2004_S5765862/d173-x01-y01 10  # Durham jet resolution $5 \to 4$ ($E_\mathrm{CMS}=91.2$ GeV)
/ALEPH_2004_S5765862/d180-x01-y01 10  # Durham jet resolution $6 \to 5$ ($E_\mathrm{CMS}=91.2$ GeV)
/ALEPH_2004_S5765862/d54-x01-y01  10  # Thrust ($E_\mathrm{CMS}=91.2$ GeV)
\end{verbatim}
\subsection{AHADIC++ LO Event-Shape Tune Weight file}
\footnotesize
\begin{verbatim}
/ALEPH_1996_S3486095/d01-x01-y01  1  # Sphericity, $S$ (charged)
/ALEPH_1996_S3486095/d02-x01-y01  1  # Aplanarity, $A$ (charged)
/ALEPH_1996_S3486095/d03-x01-y01  10  # 1-Thrust, $1-T$ (charged)
/ALEPH_1996_S3486095/d04-x01-y01  1  # Thrust minor, $m$ (charged)
/ALEPH_1996_S3486095/d05-x01-y01  1  # Two-jet resolution variable, $Y_3$ (charged)
/ALEPH_1996_S3486095/d06-x01-y01  1  # Heavy jet mass (charged)
/ALEPH_1996_S3486095/d07-x01-y01  1  # $C$ parameter (charged)
/ALEPH_1996_S3486095/d08-x01-y01  1  # Oblateness, $M - m$ (charged)
/ALEPH_1996_S3486095/d09-x01-y01  1  # Scaled momentum, $x_p = |p|/|p_\text{beam}|$ (charged)
/ALEPH_1996_S3486095/d10-x01-y01  1  # Rapidity w.r.t. thrust axes, $y_T$ (charged)
/ALEPH_1996_S3486095/d11-x01-y01  1  # In-plane $p_T$ in GeV w.r.t. sphericity axes (charged)
/ALEPH_1996_S3486095/d12-x01-y01  1  # Out-of-plane $p_T$ in GeV w.r.t. sphericity axes (charged)
/ALEPH_1996_S3486095/d17-x01-y01  1  # Log of scaled momentum, $\log(1/x_p)$ (charged)
/ALEPH_1996_S3486095/d18-x01-y01  5  # Charged multiplicity distribution
/ALEPH_1996_S3486095/d19-x01-y01  200  # Mean charged multiplicity
/ALEPH_1996_S3486095/d20-x01-y01  40  # Mean charged multiplicity for rapidity $|Y| < 0.5$
/ALEPH_1996_S3486095/d21-x01-y01  40  # Mean charged multiplicity for rapidity $|Y| < 1.0$
/ALEPH_1996_S3486095/d22-x01-y01  40  # Mean charged multiplicity for rapidity $|Y| < 1.5$
/ALEPH_1996_S3486095/d23-x01-y01  40  # Mean charged multiplicity for rapidity $|Y| < 2.0$

/ALEPH_1999_S4193598/d01-x01-y01  50  # Scaled energy of $D^{*\pm}$ in $e^+e^-\to Z\to\text{hadronic}$ at $\sqrt{s}=91.2$~GeV

/ALEPH_2004_S5765862/d102-x01-y01   2  # Thrust minor ($E_\mathrm{CMS}=91.2$ GeV)
/ALEPH_2004_S5765862/d110-x01-y01   2  # Jet mass difference ($E_\mathrm{CMS}=91.2$ GeV)
/ALEPH_2004_S5765862/d118-x01-y01   2  # Aplanarity ($E_\mathrm{CMS}=91.2$ GeV)
/ALEPH_2004_S5765862/d141-x01-y01   2  # Sphericity ($E_\mathrm{CMS}=91.2$ GeV)
/ALEPH_2004_S5765862/d157-x01-y01 50  # Durham jet resolution $3 \to 2$ ($E_\mathrm{CMS}=91.2$ GeV)
/ALEPH_2004_S5765862/d165-x01-y01 20  # Durham jet resolution $4 \to 3$ ($E_\mathrm{CMS}=91.2$ GeV)
/ALEPH_2004_S5765862/d173-x01-y01 10 # Durham jet resolution $5 \to 4$ ($E_\mathrm{CMS}=91.2$ GeV)
/ALEPH_2004_S5765862/d180-x01-y01 10   # Durham jet resolution $6 \to 5$ ($E_\mathrm{CMS}=91.2$ GeV)
/ALEPH_2004_S5765862/d54-x01-y01   200  # Thrust ($E_\mathrm{CMS}=91.2$ GeV)
/ALEPH_2004_S5765862/d62-x01-y01   2 # Heavy jet mass ($E_\mathrm{CMS}=91.2$ GeV)
/ALEPH_2004_S5765862/d70-x01-y01   2  # Total jet broadening ($E_\mathrm{CMS}=91.2$ GeV)
/ALEPH_2004_S5765862/d78-x01-y01   2  # Wide jet broadening ($E_\mathrm{CMS}=91.2$ GeV)
/ALEPH_2004_S5765862/d86-x01-y01   2  # C-Parameter ($E_\mathrm{CMS}=91.2$ GeV)
/ALEPH_2004_S5765862/d94-x01-y01   20  # Thrust major ($E_\mathrm{CMS}=91.2$ GeV)
\end{verbatim}

\end{document}